\documentclass{pasj00}
\usepackage{graphicx}

\begin{document}
\SetRunningHead{Yagi et al.}{SDF/SXDS Re-Calibration}
\Received{2012/08/10}
\Accepted{2012/10/01}
\title{Re-calibration of SDF/SXDS Photometric Catalogs of 
Suprime-Cam with SDSS Data Release 8} 

\author{%
Masafumi \textsc{Yagi}\altaffilmark{1},
Nao \textsc{Suzuki}\altaffilmark{2},
Hitomi \textsc{Yamanoi}\altaffilmark{1},
Hisanori \textsc{Furusawa}\altaffilmark{3},
Fumiaki \textsc{Nakata}\altaffilmark{4},
and
Yutaka \textsc{Komiyama}\altaffilmark{1}
}

\altaffiltext{1}{Optical and Infrared Astronomy Division, 
National Astronomical Observatory of Japan, 
Mitaka, Tokyo, 181-8588, Japan}
\email{yagi.masafumi@nao.ac.jp}
\altaffiltext{2}{
E.O. Lawrence Berkeley National Lab, 
1 Cyclotron Rd., Berkeley, CA, 94720, USA}
\altaffiltext{3}{
Astronomy Data Center,
National Astronomical Observatory of Japan, 
Mitaka, Tokyo, 181-8588, Japan}
\altaffiltext{4}{
Subaru Telescope,
650 North A'ohoku Place, Hilo, Hawaii 96720, USA
}

\maketitle

\begin{abstract}
We present photometric recalibration of the Subaru Deep Field (SDF)
and Subaru/XMM-Newton Deep Survey (SXDS).
Recently, \citet{Yamanoi2012}
suggested the existence of a discrepancy between
the SDF and SXDS catalogs.
We have used 
the Sloan Digital Sky Survey (SDSS) Data Release 8 (DR8) catalog and
compared stars in common between SDF/SXDS and SDSS.
We confirmed that there exists a 0.12 mag offset in B-band
between the SDF and SXDS catalogs.
Moreover, we found that significant zero point offsets in i-band 
($\sim$ 0.10 mag) and z-band ($\sim$ 0.14 mag) need to be introduced to
the SDF/SXDS catalogs to make it consistent with the SDSS catalog. 
We report the measured zero point offsets of five filter bands 
of SDF/SXDS catalogs.
We studied the potential cause of these offsets, 
but the origins are yet to be understood.
\end{abstract}


\section{Introduction}

The Subaru Deep Field (SDF; \cite{SDF}) catalog 
and the Subaru/XMM-Newton Deep Survey (SXDS; \cite{SXDS2})
catalog are wide and deep photometric catalogs at 
high Galactic latitude using the Subaru Prime focus Camera 
(Suprime-Cam; \cite{Miyazaki2002}). 
They have been used for many studies
(e.g., \cite{Ouchi2005,Kashikawa2006,Hayashi2007,Furusawa2011,Toshikawa2012}).

Recently, \citet{Yamanoi2012} identified and reported
the discrepancy in B-band and R-band photometry
between SDF and SXDS catalogs
by comparing them with 
the Sloan Digital Sky Survey (SDSS) Data Release 8 (DR8; \cite{DR8})
photometric catalog%
\footnote{http://skyserver.sdss3.org/dr8/en/}.
The amount of the photometric zero point (ZP) difference 
is more than 0.1 mag.
\citet{SDF} wrote that
``In any case, the errors in the photometric zero
points of our final images would be less than 0.05mag.''
and 
\citet{SXDS2} wrote that
``... the uncertainties of calibrated photometric
zero points of the SXDS images are 0.03 -- 0.05 mag rms.''
The ZP difference of 0.1 mag is larger than 
the error that SDF/SXDS claimed.

In this paper, we 
measure the ZP offset 
of the SDF and SXDS catalog from SDSS, 
and investigate the potential causes.
It should be noted that 
the ZP offset also affects the catalog magnitude of extended sources,
though we only used photometry of point sources to measure it.
The investigation of the possible photometric errors of 
each object from flat-fielding,
sky subtraction, and/or coadding is beyond the scope of this paper,
and will be investigated elsewhere
\footnote{
In this work, we only estimated 
the total amount of errors, 
including our color conversion error,
in section \ref{sec:result}.
The total errors are
0.06, 0.02, 0.03, 0.05, and 0.06 mag for 
B, V, R, i, and z-band, respectively.}.

The official filter names of the Suprime-Cam, 
W-J-B, W-J-V, W-C-RC, W-S-I+ and W-S-Z+, 
are abbreviated as B, V, R, i, and z, respectively.
For the difference between the SDSS and Suprime-Cam band, 
we always subtract Suprime-Cam magnitude from the SDSS one.
If we write i-i in a figure, it means i(SDSS)-i(Suprime-Cam).
The AB magnitude system \citep{Oke1983} is used throughout the paper.

\section{Method}

\subsection{Problem of the Original Calibration of SDF/SXDS}

The original calibration of the SDF and SXDS catalogs 
was performed in two stages \citep{SDF,SXDS2}.
First, the photometric result is calibrated using standard stars.
The calibration of SXDS depends on the previous work
by \citet{Ouchi2001b} and \citet{Ouchi2004}.
Then, a slight ($\simeq 0.05$ mag) shift is applied 
to the photometric ZP 
so that the color-color diagram of stars 
matches the model color-color diagram.
The model color is constructed 
with the atlas of \citet{GS83}
spectral library(GS83),
assuming that the distribution of the color 
of the stars in GS83 is the same as those in the observed field.
SDF used 100 bright i$<$23 mag stars in the observed field,
and SXDS used 700-1100 of 20.5$<$R$<$23.5 mag stars. 

To verify the assumption that the color of GS83 is the same as 
the stars in the field, we used another calibrated catalog (SDSS DR8).
We retrieved stellar objects (type=6 in PHOTOOBJ table in 
SDSS Catalog Archive Server)
in 
$33.828958<\alpha$(deg)$<35.158429$, $-5.648833<\delta$(deg)$<-4.354703$ for SXDS
and 
$200.880078<\alpha$(deg)$<201.444238$, $27.181905<\delta$(deg)$<27.799080$ for SDF.
The areas are 1.714(SXDS) and 0.309(SDF) square degree,
and the difference in the area size is the main reason for 
the difference in the number of the stars.
We use psfMag of SDSS hereafter, but our result is the same
if we use modelMag instead.
SDSS uses asinh magnitude $l$ \citep{Lupton1999}.
The Pogson magnitude\citep{Pogson1856} $m$ is calculated as 
\begin{equation}
m=-2.5 log_{10}\left[10^{-0.4 l} - b^2 10^{0.4 l}\right],
\label{eqn:laptitude}
\end{equation}
where $b$ is a softening parameter.
The parameter for each band is given on the SDSS webpage%
\footnote{http://www.sdss3.org/dr8/algorithms/magnitudes.php\#asinh}.
In $r<21$ range, however, the difference between asinh magnitude 
and Pogson magnitude is negligible ($<0.01$ mag)
as calculated from equation (\ref{eqn:laptitude}),
and we neglect the difference hereafter.
The difference between SDSS magnitude and AB magnitude is 
still under debate. 
We adopted the offset used in Kcorrect\citep{Blanton2007} v4%
\footnote{http://howdy.physics.nyu.edu/index.php/Kcorrect};
$\Delta$ m = m$_{\rm AB}$ - m$_{\rm SDSS}$ = -0.036, 0.012, 0.010, 
0.028, 0.040 in u, g, r, i, and z bands.
Though the precision of the values is not explicitly given, 
we expect that it would be smaller than 0.01 mag,
since the values have a 0.001 digit.
In SDSS-DR8, some stars have multiple entries in the database,
as they were observed more than once.
We calculate the magnitude difference
in g,r,i and z-band between each pair of the entries of 
stars observed multiple times.
We adopted a threshold of difference as 3$\sigma$ 
($3\times\sqrt{{\rm psfMag\_Err}_1^2+{\rm psfMag\_Err}_2^2}$),
and if the difference is larger than the threshold,
the star is not used in our analysis,
since it might be a variable star.
If the difference is smaller than the threshold,
the psfMag of the multiple entries are averaged and used in 
the following analysis.

\begin{figure}
\FigureFile(80mm,60mm){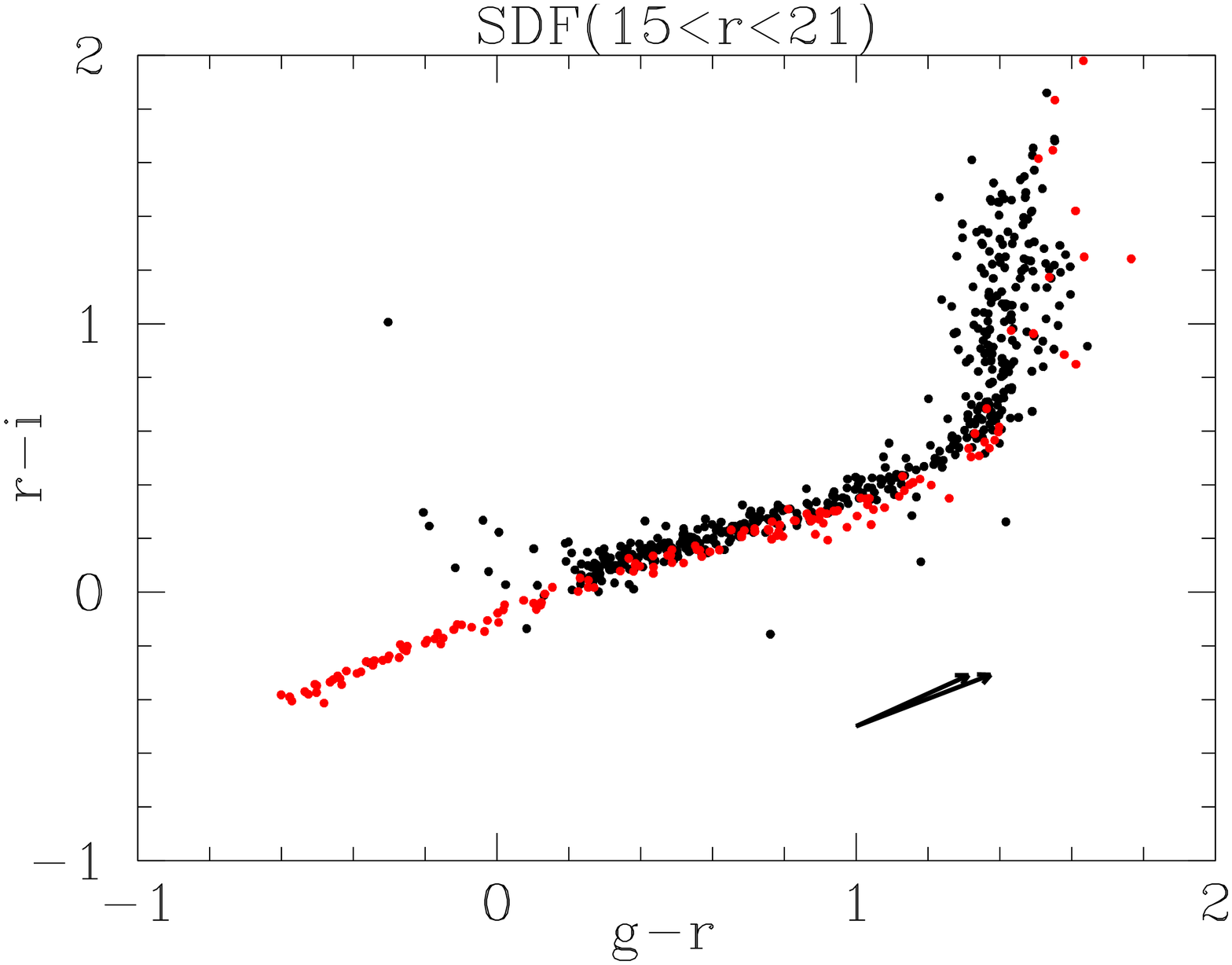}
\FigureFile(80mm,60mm){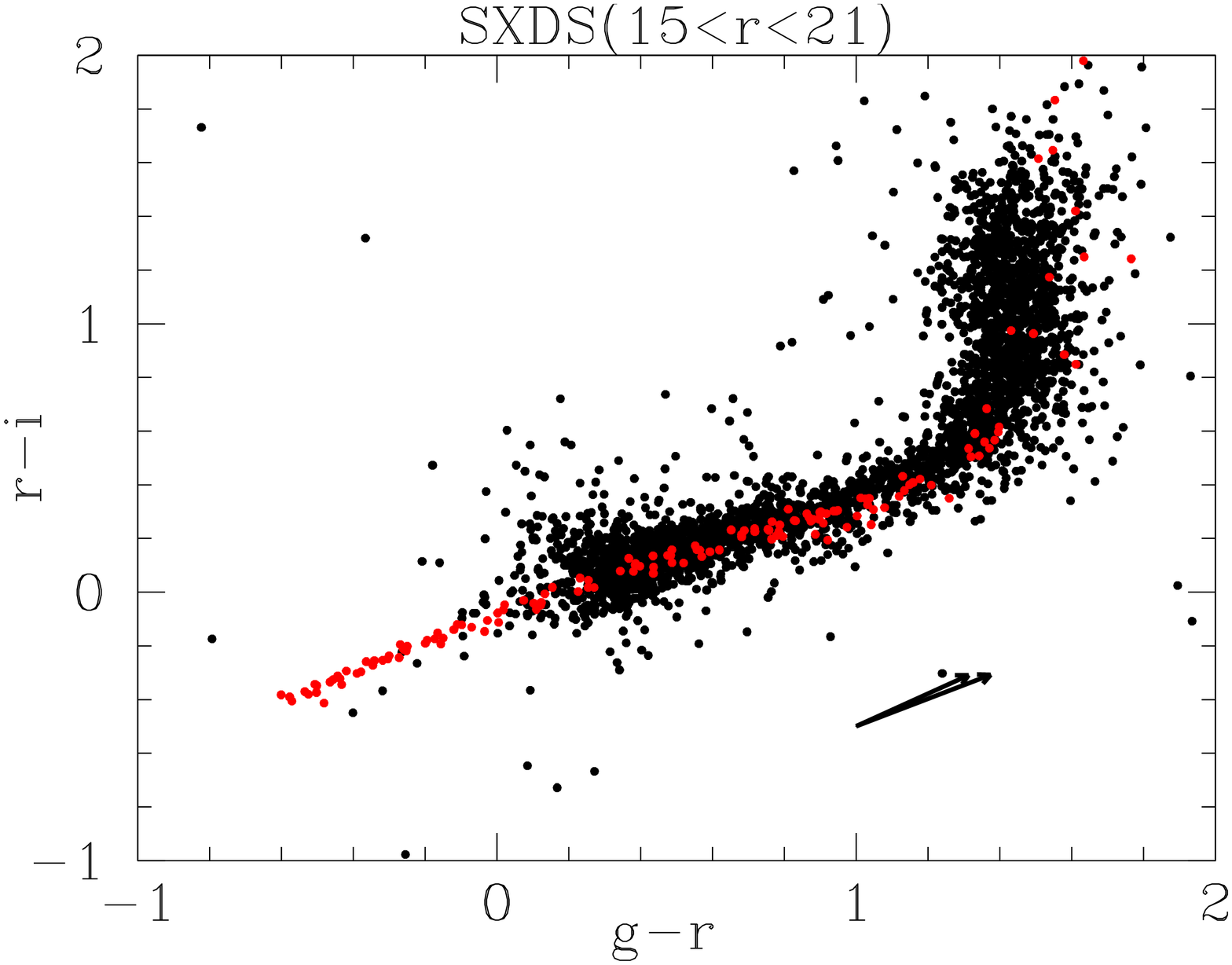}  
\caption{Color-color diagram of SDSS DR8 stars of $15<r<21$ in 
SDF(left) and SXDS(right) field. 
The color of SDSS stars is calculated from psfMag.
The offset of m$_{\rm AB}$-m$_{\rm SDSS}$
given by Kcorrect \citep{Blanton2007} v4 is applied.
The filled red circles represent synthetic colors of GS83 stars
calculated with the transmission by \citet{Doi2010}.
The arrows indicate the direction of 
Av=1 reddening of Galactic extinction for O and M stars
in GS83.
}
\label{fig:colcol_GS83}
\end{figure}

The synthetic color of GS83 stars are calculated 
following the equation (9) of \citet{Fukugita1995} using
the transmission by \citet{Doi2010}, 
and overlaid on the catalog colors of SDSS stars 
in figure \ref{fig:colcol_GS83}.
\citet{Doi2010} measured the responses of correctors,
filters and CCDs in SDSS using a monochromatic illumination system.
Then it is multiplied with
the model reflectivity of the primary and the secondary mirrors
and given in their Table 4.
We multiplied it with the atmospheric transmission at 1.3 airmass
given in the table, and the product was used as 
the SDSS transmission in this study.
\citet{Doi2010} reported that the variation of the response function may 
amount to 0.01 mag in g,r,i and z band,
but it is cancelled by a calibration procedure, 
and does not appear in the final SDSS catalogs.
They concluded that the residual effects are smaller than 0.01 mag 
for all passbands.
We can therefore expect that the error of synthetic magnitudes
due to a possible error of the transmission curve would be $<$0.01 mag.

Since the GS83 spectral energy distribution (SED) adopts
the air wavelength while SDSS transmission uses the vacuum wavelength,
we converted the SED to the vacuum wavelength 
using the index of \citet{Ciddor1996}, which has very small 
($<0.1\%$) difference from the IAU standard by \citet{Morton1991}.
We omitted 22 stars from the plot since they
lack data at some wavelength within g, r, and/or i passband.

In figure \ref{fig:colcol_GS83},
it is clearly seen that GS83 stars have an offset from SDSS stars
to the right-bottom direction.
The possible effect of the Galactic extinction is checked with
the extinction curve by \citet{Cardelli1989} and \citet{ODonnell1994} 
with Rv=3.1.
The reddening vector is also written in figure \ref{fig:colcol_GS83}.
As the reddening is a bit different in different SED, 
we plotted the median reddening vectors for O, 
and M-type stars in GS83.
The direction of the GS83 offset from SDSS stars 
is perpendicular to the reddening,
and therefore the offset is not caused by the Galactic extinction.

\begin{figure}
\FigureFile(120mm,90mm){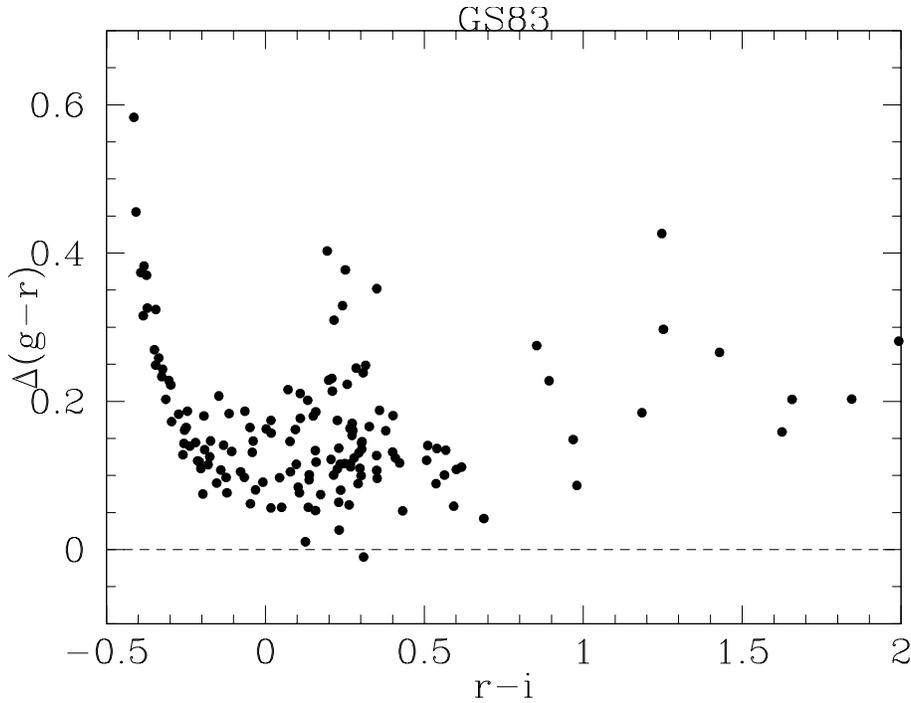}  
\caption{The difference of (g-r) color of GS83 stars 
from the analytic fit of SDSS stars by \citet{Juric2008}
as a function of (r-i).
$\Delta$(g-r)=0 is shown as a broken line for comparison.
}
\label{fig:compJuric}
\end{figure}

We calculate
the amount of the offset by comparing the distribution of GS83 stars
with an empirical fit of the distribution of SDSS stars.
\citet{Juric2008} gives an analytical expression of 
(g-r) color of SDSS stars as a function of their (r-i) color.
The amount of (g-r) offset of GS83 stars from the fit
is shown in figure \ref{fig:compJuric}. 
The median offset of (g-r) is 0.13 mag for the $-0.2<$(r-i)$<0.7$ clump.

This color offset between SDSS and GS83 was already recognized 
by previous studies (e.g.,\cite{Lenz1998,Fukugita2011}).
\citet{Fukugita2011} discussed that the offset is explained by 
the metallicity variance. They investigated the color of stars 
in the SDSS system using a photometric catalog of SDSS-DR6\citep{DR6}. 
The blue GS83 stars are metal-rich disk stars, while the SDSS stars, 
especially at high Galactic latitude, are metal-poor popII stars.
We discuss the metallicity effect in section \ref{sec:metal}.
Here we stress that 
the color distribution of GS83 stars 
would not be a good representation of 
that of the faint field stars that were used for the SDF/SXDS calibration.

\subsection{Color Conversion from SDSS to Suprime-Cam System}

In this study, we return to a classical color calibration method;
a color conversion between filter systems.
If we can convert the SDSS catalog magnitude 
to the AB magnitude in the Suprime-Cam system,
all the stars in SDSS would be used as photometric standards.
Though the photometric error of SDSS ($\sim 0.04$mag) is relatively larger 
than that of the well-calibrated photometric standard stars 
(e.g.,$<0.01$mag;\cite{Landolt2009}),
or that of the spectrophotometric standard stars 
(e.g., $<$0.5\%$\sim$0.005 mag ;\cite{Bohlin2010})
the large number of SDSS stars, 
$>$50 stars in a Suprime-Cam field
will make the systematic error smaller.
This method was not possible when the SDF or SXDS catalog was constructed,
since the SDSS catalog of the region did not exist.
Now this method is promising, thanks to the wide coverage of the SDSS 
DR8 catalog.
We fit the difference of Suprime-Cam magnitude and SDSS magnitude 
as a function of the SDSS color;
\begin{equation}
SDSS-Suprime = c_0 + c_1 (color) + c_2 (color)^2 + ...
\label{eqn:colorconv}
\end{equation}

\subsubsection{Model SEDs}
In our previous work \citep{Yagi2010}, 
we fit Bruzual-Persson-Gunn-Stryker atlas (BPGS)%
\footnote{http://www.stsci.edu/hst/observatory/cdbs/bpgs.html}%
\footnote{ftp://ftp.stsci.edu/cdbs/grid/bpgs/}
colors with quadratic functions.
BPGS is an extrapolation of the original GS83 data,
and is also used in this study instead of GS83, hereafter.
In this work, we first use the model spectral 
energy distributions (SEDs) to cover a wider variation of stars,
such as low-metal stars.
We retrieved several flux calculation of ATLAS9 models
\citep{Castelli2004} on the web (ATLAS9 grids)%
\footnote{http://wwwuser.oat.ts.astro.it/castelli/grids.html}.
The input parameters are the metallicity ([Fe/H]), 
the alpha enhancement ([$\alpha$/Fe]),
the temperature (T), the surface gravity (log g), 
mixing length parameter (l/H),
the Helium enhancement (DY), and the perturbation velocity(v$_{\rm turb}$).
We adopted l/H=1.25, 
v$_{\rm turb}$=2km s$^{-2}$, and DY=0 models,
because these parameters are used as default values in the ATLAS9 grids.
The temperature range is $3750 K\le T\le 50000 K$.
Since the relation between the temperature and the surface gravity 
changes according to [Fe/H], the possible combinations are taken from 
Yonsei-Yale(Y$^2$) isochrone version 2 \citep{Yi2001,Demarque2004},
which covers 0.4$<$M$_{initial}/$M$_{\odot}<$5 stars.
In the isochrone, stars of all ages are used.
We used [$\alpha$/Fe]=0.3 models in Y$^2$ for [$\alpha$/Fe]=0.4
models to set constraints on the combinations of [Fe/H], T and log g.
Note that low mass (M$_{initial}/$M$_{\odot}<$0.4) stars,
subdwarfs and white dwarfs are not included in the model.

The set of [Fe/H] and [$\alpha$/Fe] we used are
(+0.5,0.0), (+0.5,+0.4), (0.0,0.0), (0.0,+0.4),
(-0.5,0.0), (-1.5,0.0), (-1.5,+0.4), and (-2.5,+0.4).
The set of the temperature and surface gravity are 
taken from Y$^2$ isochrone \citep{Yi2001,Demarque2004}.
We do not apply Galactic extinction at this stage.
The effect of the Galactic extinction is investigated later.

\subsubsection{System Responses}

For SDSS synthetic magnitude, we adopted the transmission by 
\citet{Doi2010}. It includes the atmospheric effect at airmass=1.3 
at the SDSS site.
The Suprime-Cam response is calculated as the product of 
the quantum efficiency(QE) of MIT/Lincoln Laboratory (MIT/LL) CCDs%
\footnote{http://www.naoj.org/Observing/Instruments/SCam/ccd\_mit.html},
the filter responses%
\footnote{http://www.naoj.org/Observing/Instruments/SCam/sensitivity.html},
the transmittance of the primary focus corrector%
\footnote{http://www.naoj.org/Observing/Telescope/Parameters/PFU/},
and the reflectivity of Primary mirror%
\footnote{http://www.naoj.org/Observing/Telescope/Parameters/Reflectivity/}.
The extinction of a model atmosphere at airmass=1 is then multiplied. 
The total throughput curves are slightly different from the one used in 
SDF and SXDS photometric calibrations, 
because of the update of the responses of 
the telescopes, and difference of the airmass.
However, the difference is negligible ($<$0.01mag) in this study.
It should also be noted that the SDSS system is at 
airmass=1.3, and Suprime-Cam system is at airmass=1. 
The possible effect of the change in airmass on the Suprime-Cam 
system is investigated later.

The synthetic AB magnitude is calculated by multiplying 
the model SEDs and the system response.
The wavelength of model SEDs by ATLAS9 are in a vacuum,
and the Suprime-Cam response is in the air.
We corrected the SED to the air wavelength and measured the color.
Note that the difference of the wavelength between in the vacuum and 
in the air is 1 -- 2\AA, and makes 0.007 mag difference at most.

\subsubsection{Conversion Function}

We then fit the equation (\ref{eqn:colorconv}) to 
the model color distribution.
The internal calibration error of SDSS DR8 is claimed to be 0.01 mag 
in g,r,i, and z-band \citep{DR8}. 
DR8 adopted ubercal method \citep{Padmanabhan2008},
and the absolute zero point is calibrated against DR7\citep{DR7}.
The absolute zero point error of DR6 is evaluated 
by \citet{Fukugita2011}.
They compared photometry against spectroscopic data,
and concluded that the error is smaller than 0.04 mag.
Since the data handling of DR7 is the same as DR6,
the error of the absolute calibration of DR7 and DR8 
would be smaller than 0.04 mag.
We therefore aim to obtain a fit whose 
systematic error $\lesssim$ 0.04 mag.
The order of the polynomial is set so that Akaike's
Information Criterion is minimal.
The fit results are shown as figure \ref{fig:fit0},
and the coefficients are presented in table \ref{tab:coeff0}.
In the color range shown in the table, 
the deviation of the model color from the fitting polynomial
is smaller than 0.04 mag.

\subsection{Application to the Empirical SEDs}

We collected four empirical SEDs;
BPGS, HILIB\citep{Pickles1998},
STELIB\citep{LeBorgne2003},
and The Indo-U.S. Library of Coud\'e Feed Stellar Spectra
(CFLIB; \cite{Valdes2004})
to see whether the color conversions also work well for them.
We did not interpolate the SEDs, and the data which lacks 
some part in the filter coverage were omitted.
The result is shown in bottom panels of figure \ref{fig:fit0}.
Most of the color deviations are within the $-0.04<\Delta<0.04$ range,
except for a few ($\sim 1\%$ in B, at most) outliers.

\subsection{The Effect of Galactic Extinction}

The color of the stars is changed by the Galactic extinction,
and we can only know the upper limit of the extinction.
We adopted the model first developped by \citet{Cardelli1989} 
and updated by \citet{ODonnell1994},
and applied an Av=1 extinction with Rv=3.1 to the model SEDs
to see how the deviation changes.
The reddened SDSS color versus residual is plotted in figure \ref{fig:Av1}.
Because of the nonlinear fit of the equation (\ref{eqn:colorconv}),
the nearly linear shift of the distribution
by Galactic extinction makes the distribution winding.
Even with the extreme Av=1 reddening, however,
the fit is better than $\pm 0.04$ mag in the selected color range.
For the SDF and SXDS field, the total Galactic extinction is 
Av=0.049 and 0.058, respectively,
from \citet{Schlafly2011} via NASA/IPAC Extragalactic Database(NED)%
\footnote{http://ned.ipac.caltech.edu/forms/calculator.html}.
We can therefore expect that the effect of Galactic extinction on
SDF and SXDS catalogs does not affect the color conversion much.

\subsection{The Effect of Atmospheric Extinction}
\label{sec:airmass}

The atmospheric extinction is another factor to change the
color conversion.
The SDF and SXDS were observed in various airmasses.
We estimated the exposure time weighted mean of the airmass using 
the Subaru STARS archive \citep{Takata2000}.
The result is shown in table \ref{tab:airmass}.
The mean airmass is between 1. and 1.6.
We can also see a trend that longer wavelength data 
are observed in larger airmass.
Observers know that
seeing size in shorter wavelength is likely to be  
affected by the airmass, while
lights of longer wavelength are less dimmed by atmospheric extinction,
observers prefer to observe shorter wavelength at smaller airmass.
This may introduce some systematic biases in the calibration.

\begin{figure}
\FigureFile(120mm,90mm){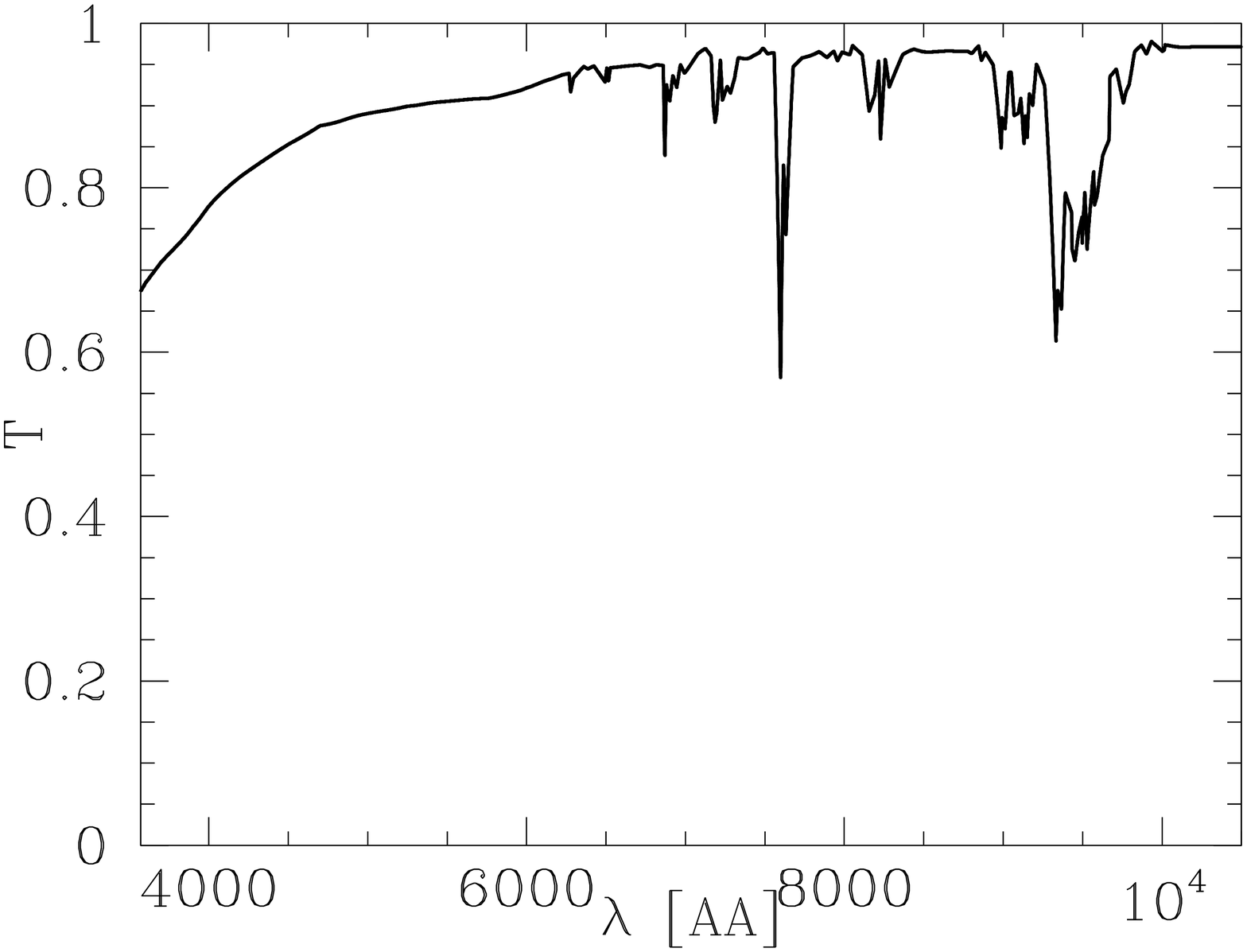}
\caption{The airmass=1 sky extinction model at Subaru used in this study.}
\label{fig:skyext}
\end{figure}

\begin{figure}
\FigureFile(80mm,60mm){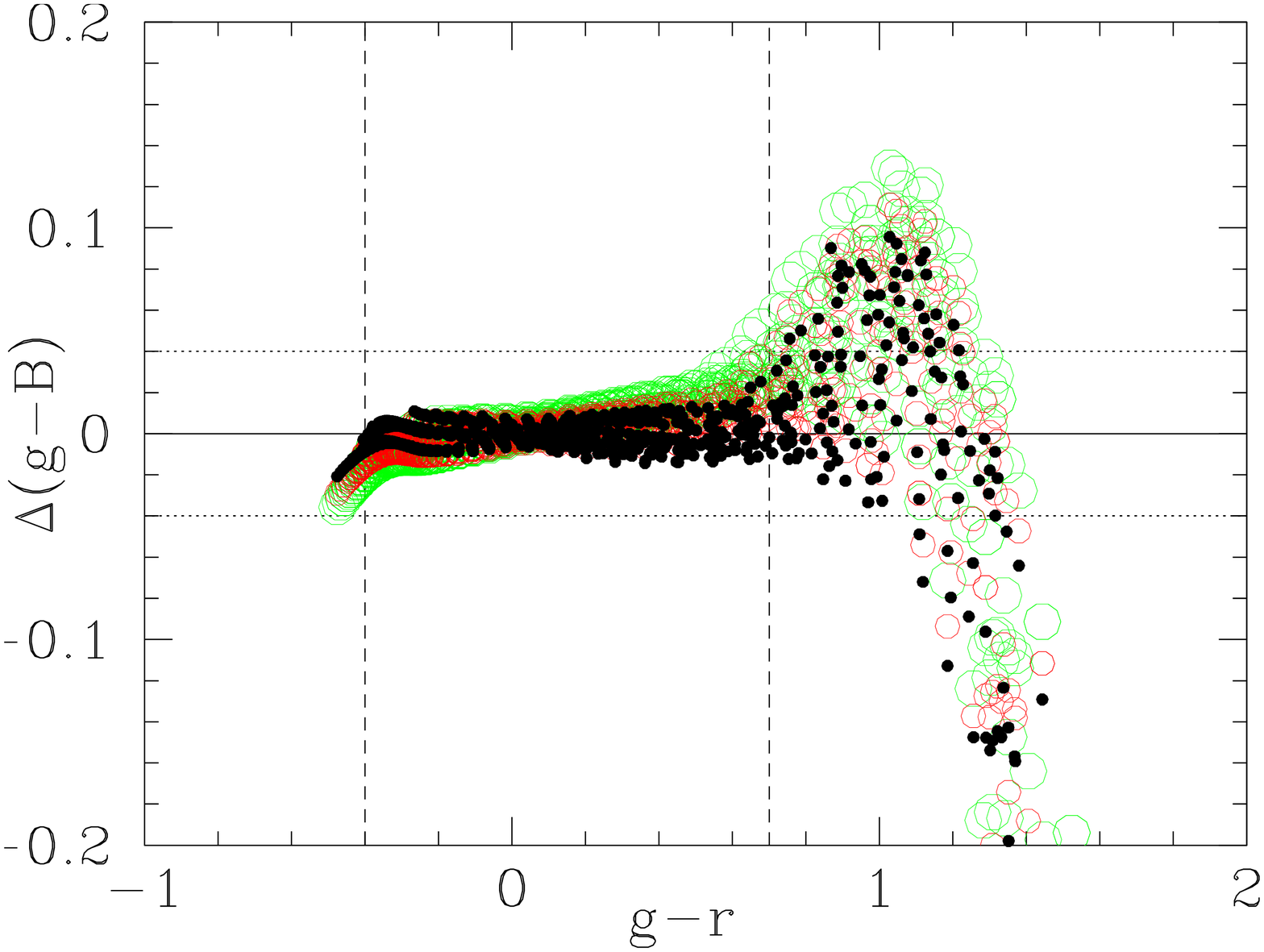}
\FigureFile(80mm,60mm){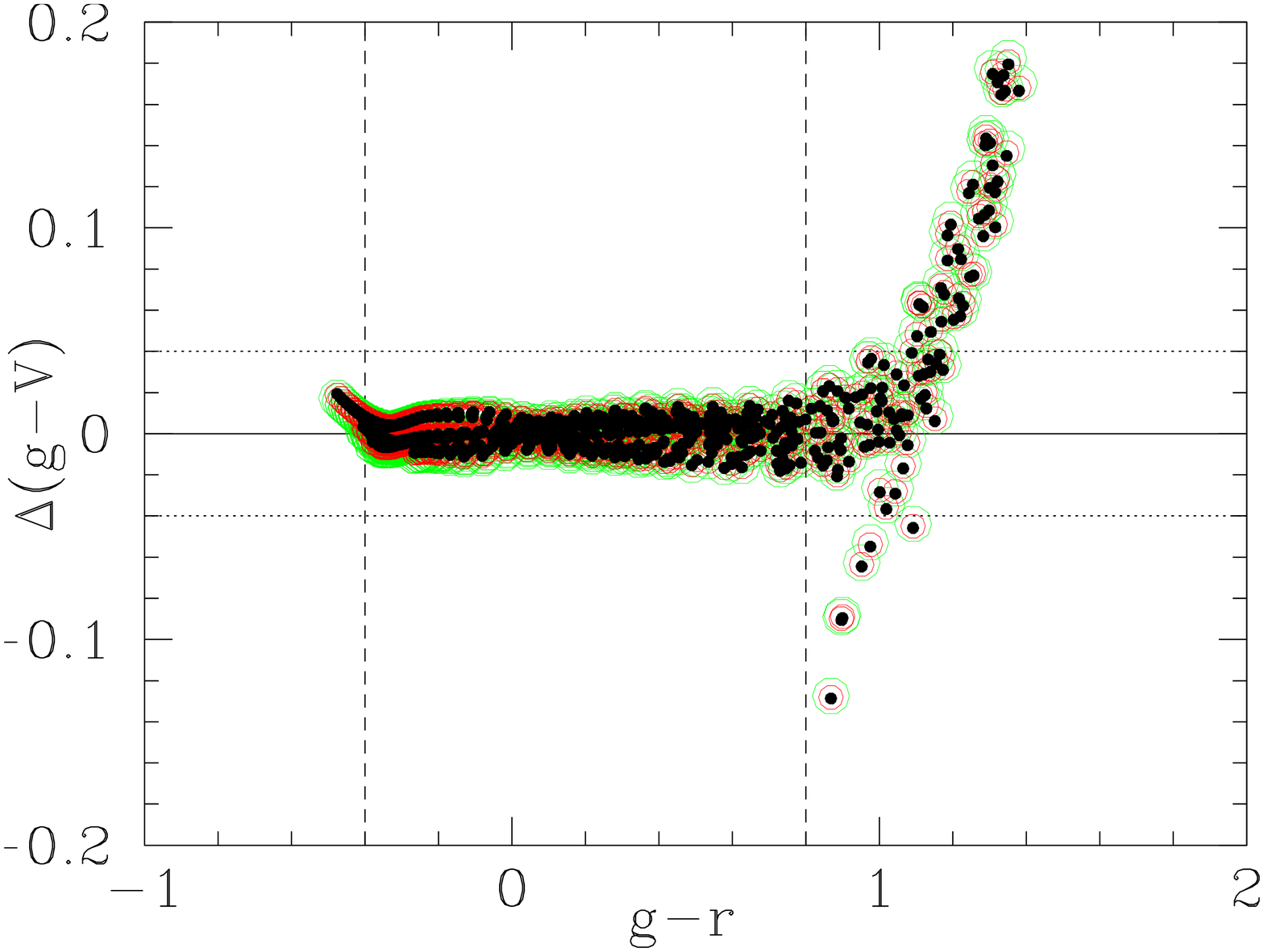}\\
\FigureFile(80mm,60mm){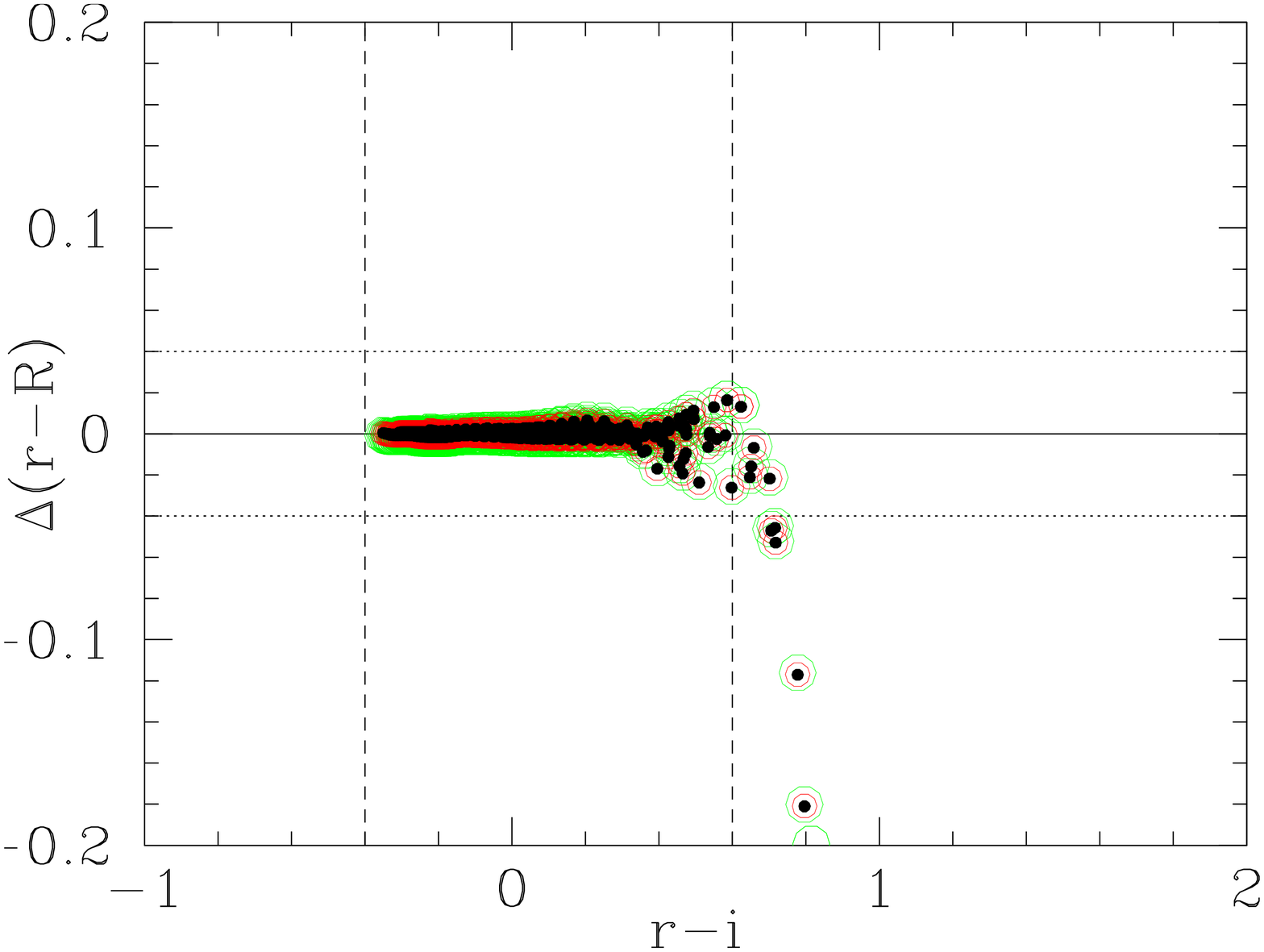}
\FigureFile(80mm,60mm){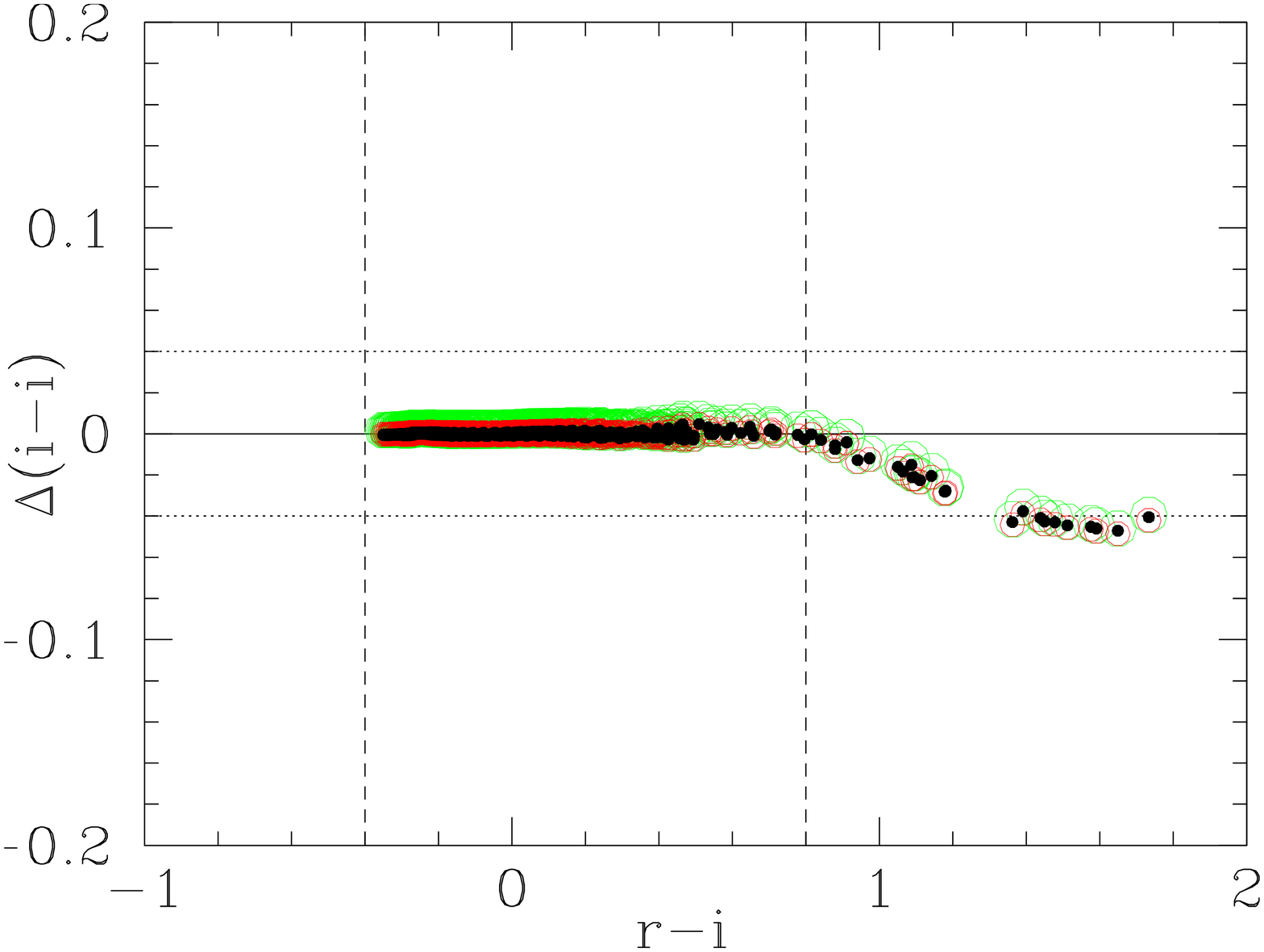}\\
\FigureFile(80mm,60mm){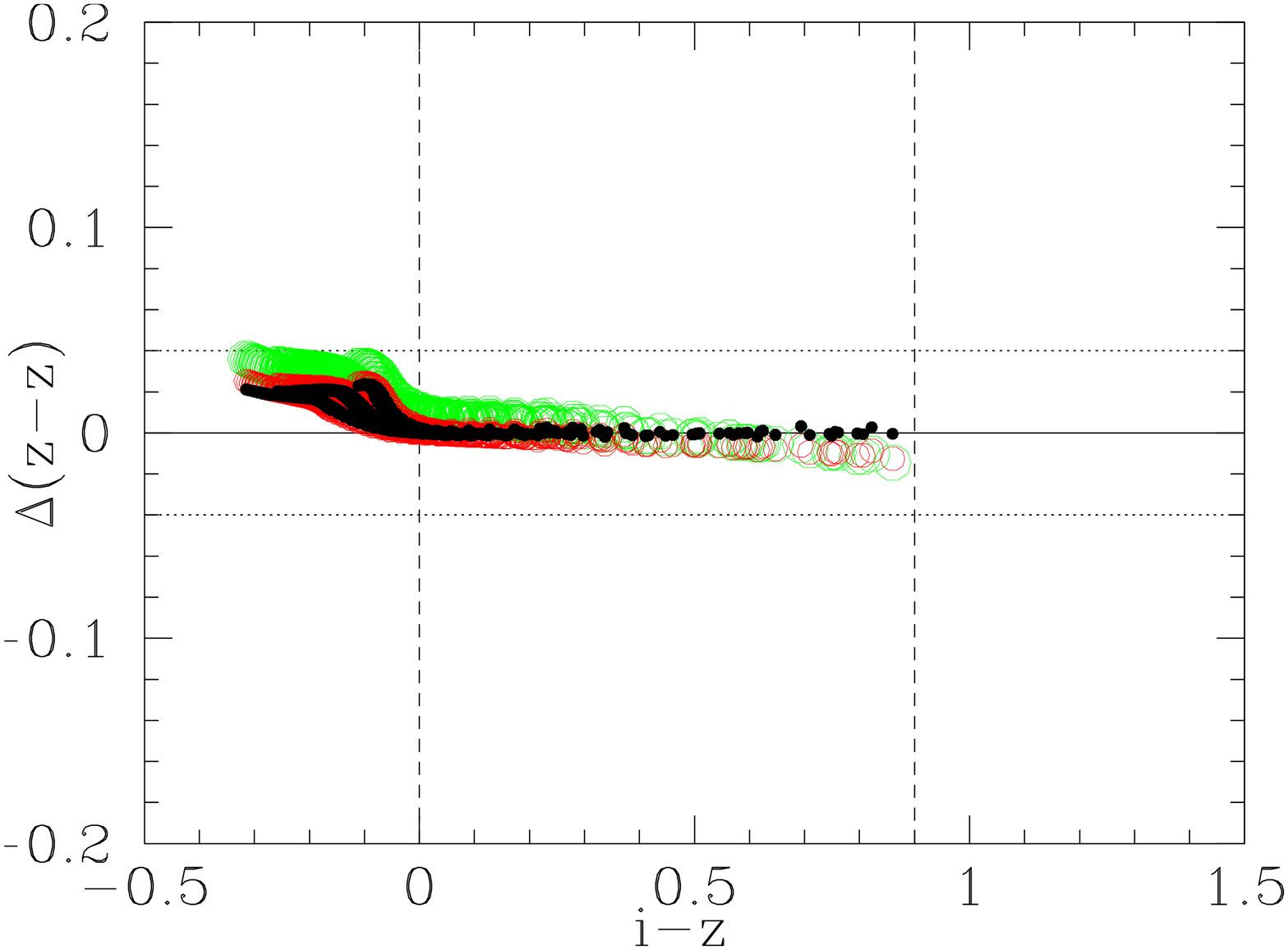}
\caption{
Residual of model color from the best-fit function. 
Vertical lines represent the fitting color range,
and horizontal lines are 0 and $\pm$0.04 mag for reference.
The different airmass models are plotted;
airmass=1(black filled circles), airmass=2(red open circles) and 
airmass=3(green open circles).
A linear term of airmass $k_1$(airmass-1) is corrected 
for the airmass=2 and 3 models.
}
\label{fig:airmass}
\end{figure}

We used an atmospheric extinction curve shown in figure \ref{fig:skyext}.
The model is constructed using the high resolution line extinction data 
by \citet{Stevenson1994} and the Mauna Kea extinction model.
The change of the SDSS color vs $\Delta$ color relation
is shown in figure \ref{fig:airmass}.
We fit a function
\begin{equation}
\Delta m = k_1 \times (airmass) + k_2 \times (airmass) \times (color)
\label{eqn:airmass}
\end{equation}
to the difference of the synthetic magnitude between 
airmass=1 and 2 models.
The coefficients are shown in table \ref{tab:extcoeff}.
In figure \ref{fig:airmass}, only $k_1(airmass-1)$ is corrected.
After the correction of the linear airmass term,
we cannot see any difference among the different airmass models 
in V, R, and i-band relations, as suggested by small $k_2$.
In B-band, a (g-r)=1.5 object would brighter by 
$\sim 0.02$ mag at airmass=2.
In z-band, an (i-z)=1.5 object would brighter by 
$\sim 0.01$ mag at airmass=2.
It should be noted that the extinction in magnitude is not a linear
function of the airmass when a cross-term of color and airmass exists,
and it makes the airmass=3 model in z-band slightly shifted.
%
%

In airmass$<$3, the relations are in the $\pm 0.04$ mag range.
If we use this SDSS color versus (SDSS)-(Suprime-Cam) plot, 
the airmass effect is small, 
and the difference of
the mean airmass among the different bands would have no effect.

\subsection{The Effect of Recession Velocity}

\begin{figure}
\FigureFile(80mm,60mm){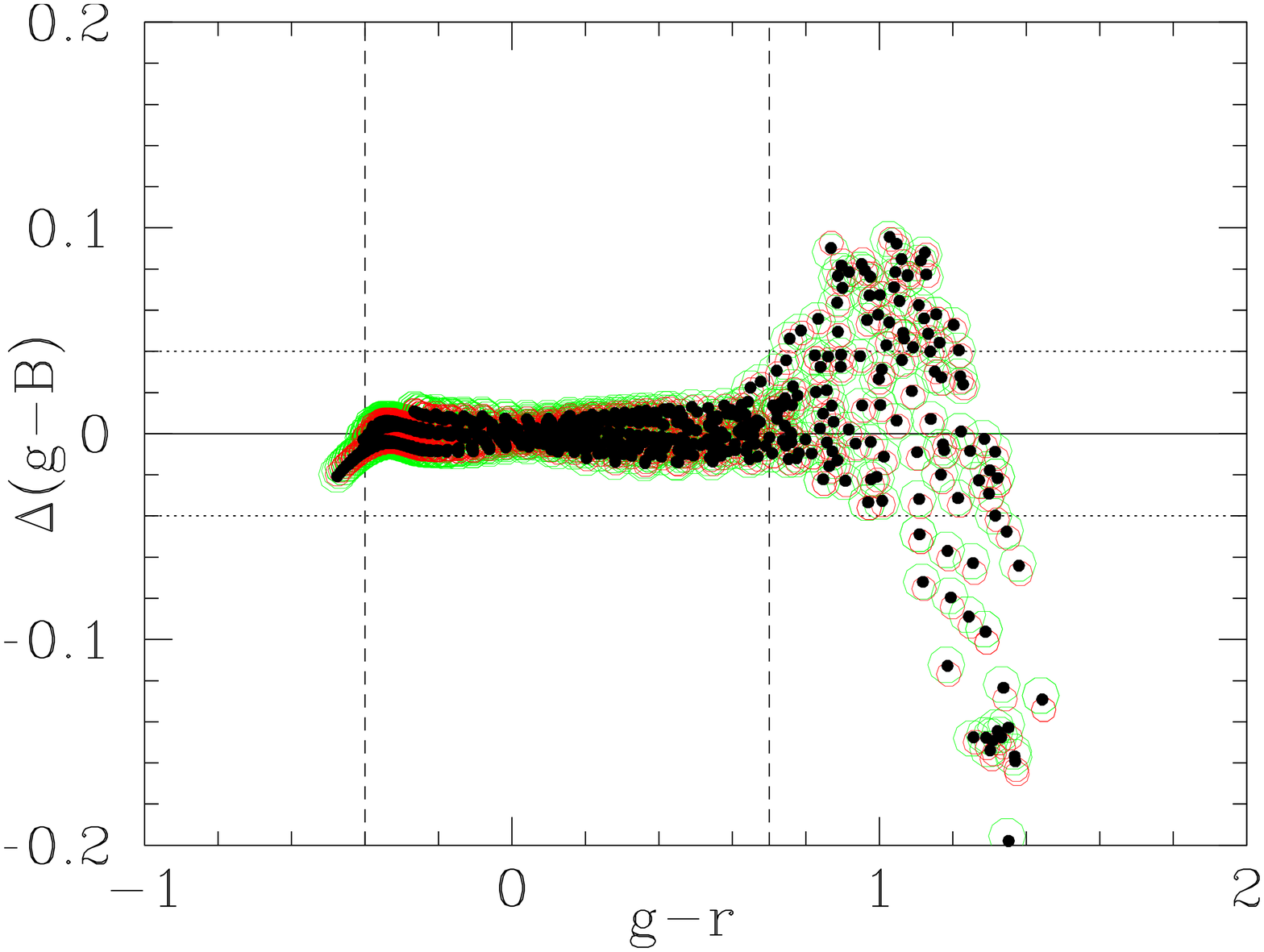}
\FigureFile(80mm,60mm){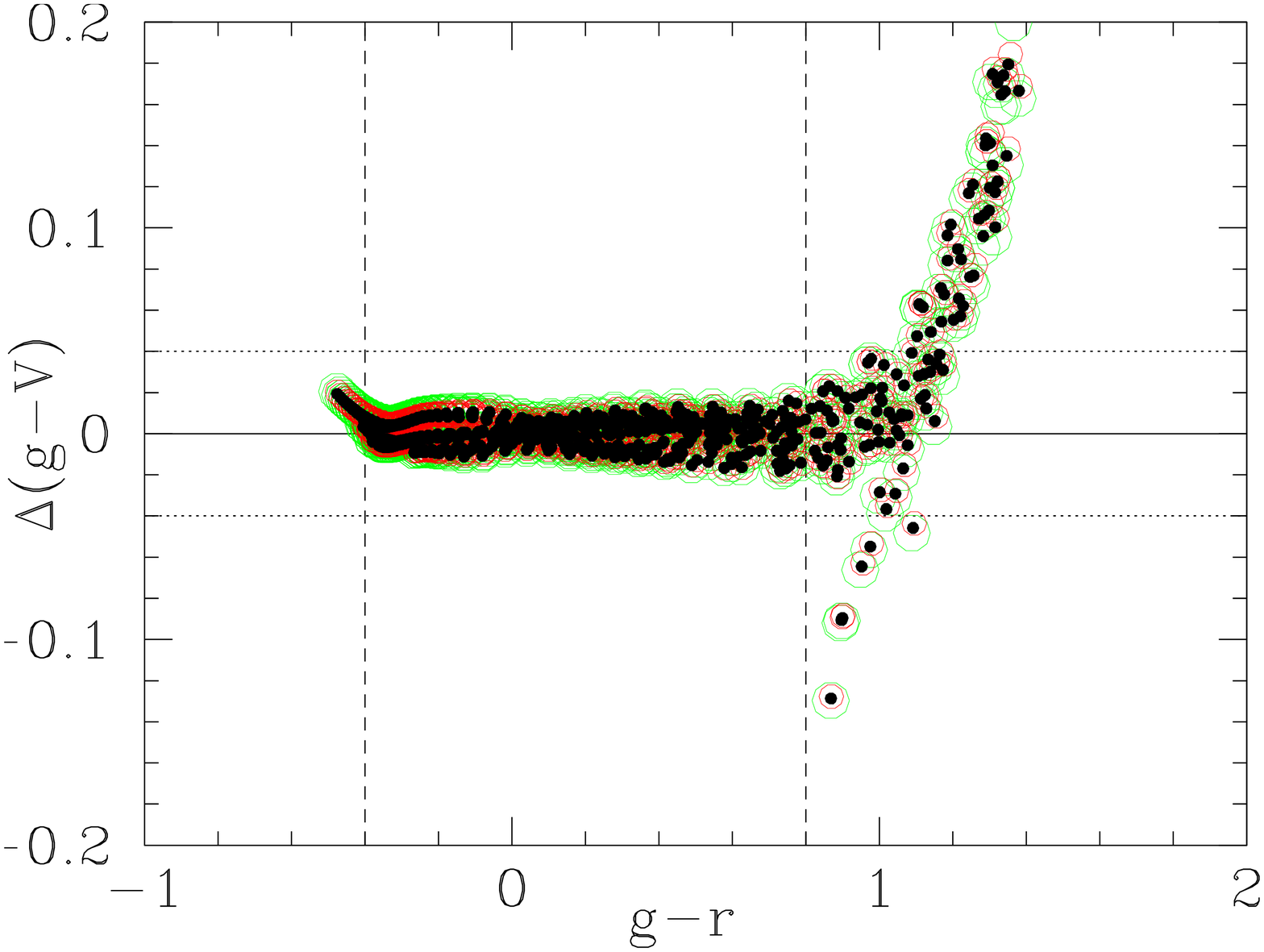}\\
\FigureFile(80mm,60mm){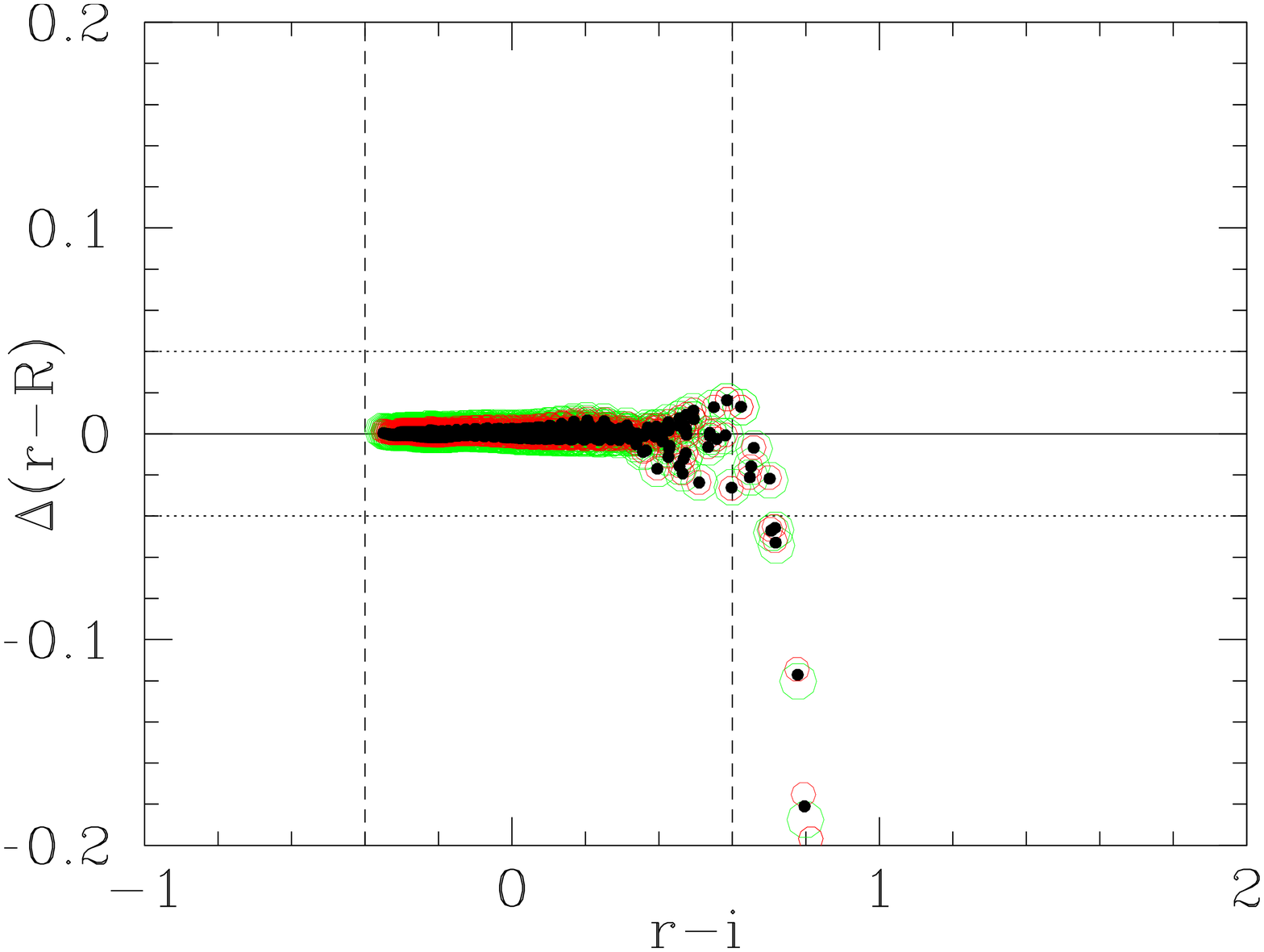}
\FigureFile(80mm,60mm){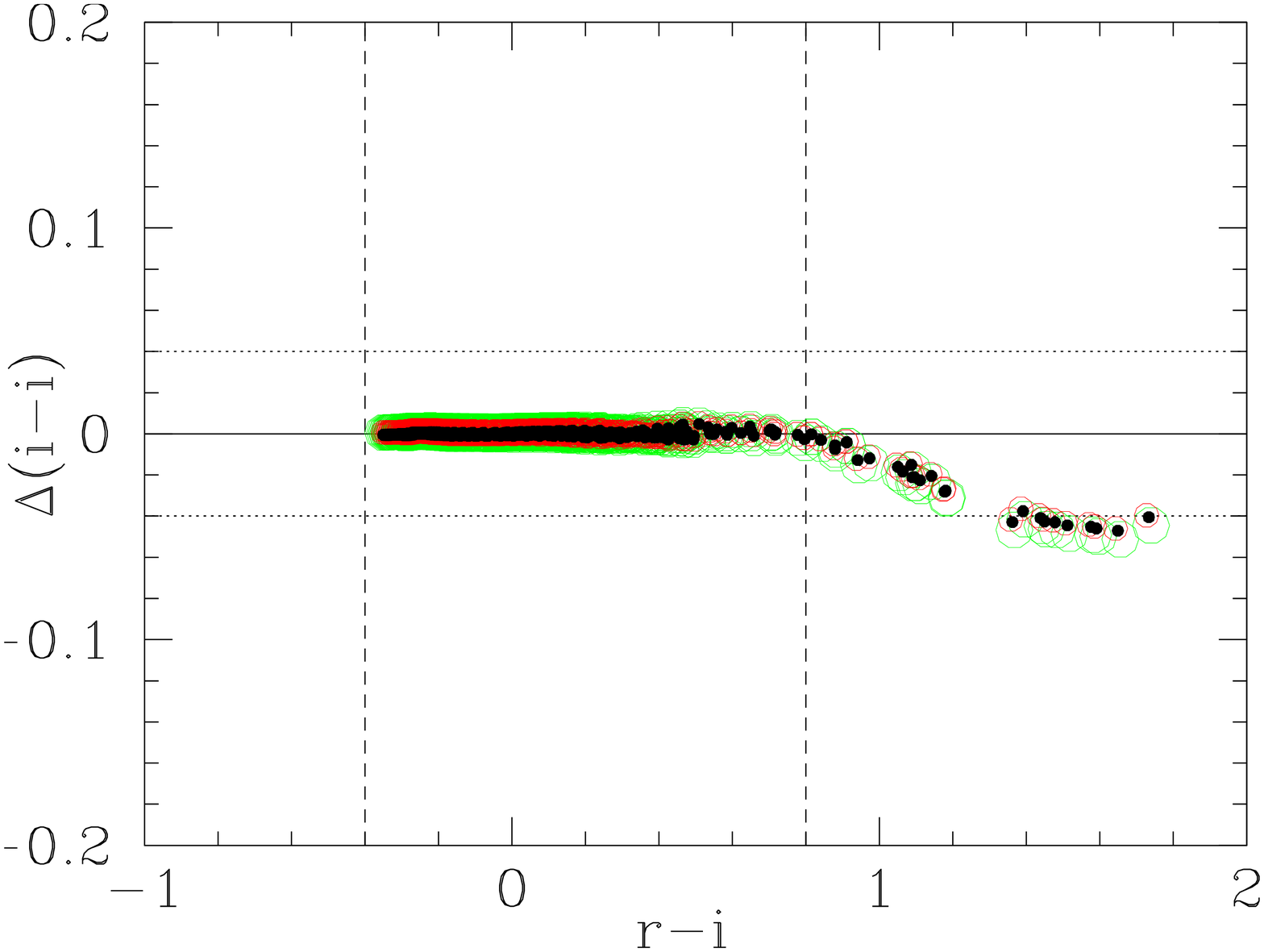}\\
\FigureFile(80mm,60mm){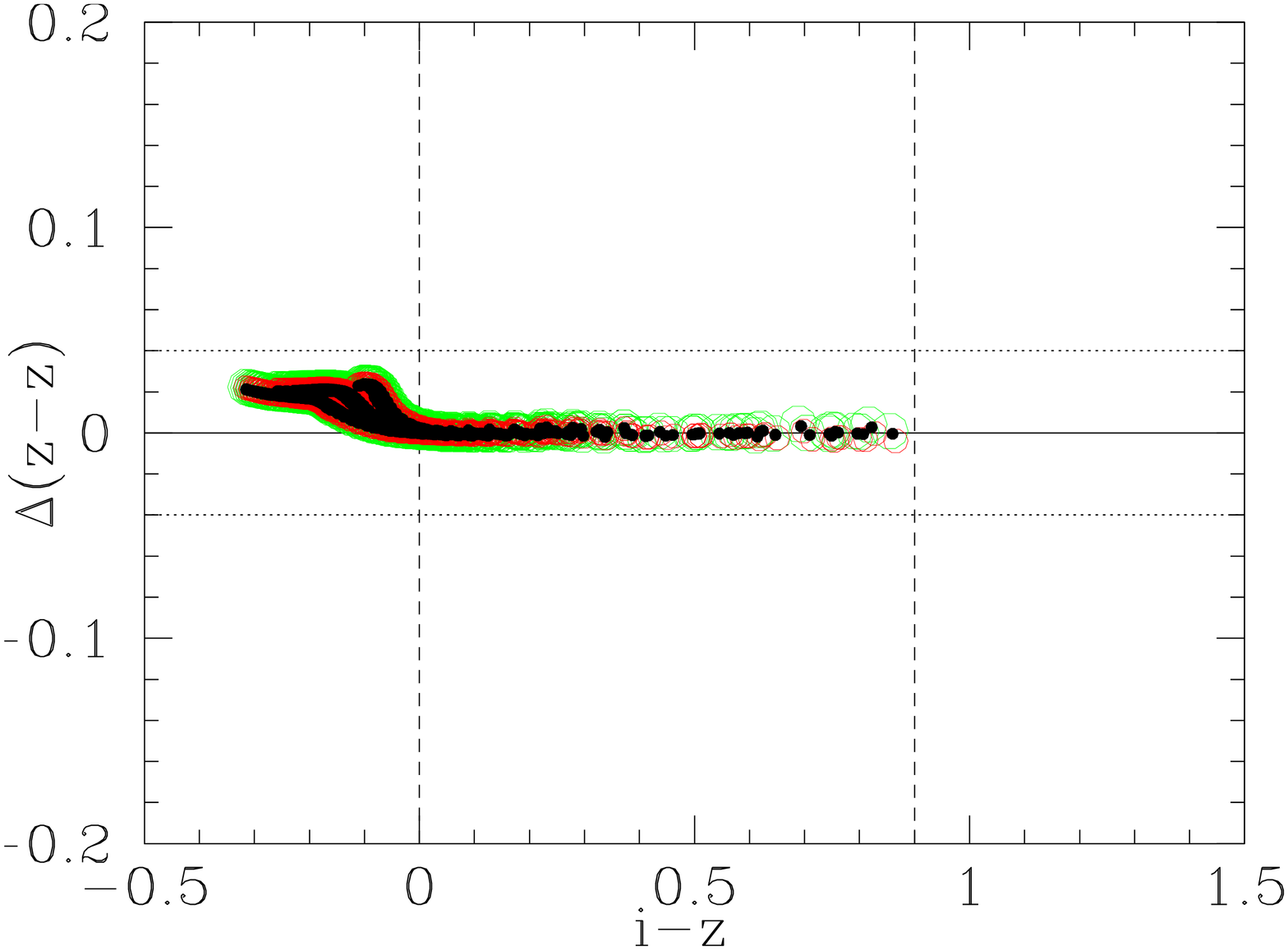}
\caption{
Same as figure \ref{fig:airmass}
but airmass=1 with different recession velocities;
$v_r$=0 km s$^{-1}$ model in black filled circles,
$v_r$=+300 km s$^{-1}$ model in red open circles,
and $v_r$=-300 km s$^{-1}$ model in green open circles.
}
\label{fig:pecvel}
\end{figure}

The model spectra are calculated in the rest frame.
The recession velocity of a star makes the spectrum redshifted/blueshifted,
and may change the color.
In faint magnitude, the majority of stars will be halo stars, and they
would have a large recession velocity in some field,
because of the rotation of the sun around the Galaxy.
We constructed $\pm$300 km s$^{-1}$ model spectra and made 
SDSS color vs $\Delta$color plots.
The result is shown in figure \ref{fig:pecvel}.
In z-band calibration, we can see a slight offset,
but the difference is 0.006 mag at most.
We can conclude that the effect of the recession velocity on 
the color conversion is negligible for Galactic stars.

\section{Result}
\label{sec:result}

We then re-calibrate the ZP of SDF and SXDS catalogs using the 
color conversion in the previous section.
We used the public catalog of SDF Data Products version 1%
\footnote{http://soaps.nao.ac.jp/SDF/v1/index.html}
and SXDS Data Release 1 (DR1)%
\footnote{http://soaps.nao.ac.jp/SXDS/Public/DR1/index\_dr1.html}.
SXDS consists of 5 fields, 
SXDS-C, SXDS-N, SXDS-S, SXDS-E, and SXDS-W, 
which are defined in \citet{SXDS2}.
Both catalogs used Suprime-Cam \citep{Miyazaki2002} with MIT/LL CCDs.
The MAG\_AUTO parameter is used for the magnitude.
The Galactic extinction is not corrected in the catalog.
We set constraints so that FWHM is smaller than 8 pixels (1.6arcsec),
and brighter than 24 mag.
For both catalogs, all combinations of the detection band and 
the measurement band are available, but 
we used catalogs whose detection band is the same as the measurement
band.

We cross-matched the SDF and SXDS catalogs with stars in SDSS DR8.
The object within 2 arcsec from the SDSS star is regarded as the 
corresponding object.
We first select the nearest SDSS star within 2 arcsec 
from each SDF/SXDS star.
If an SDSS star is assigned to more than one SDF/SXDS star,
the nearest pair is kept for identification.

\begin{figure}
\FigureFile(80mm,60mm){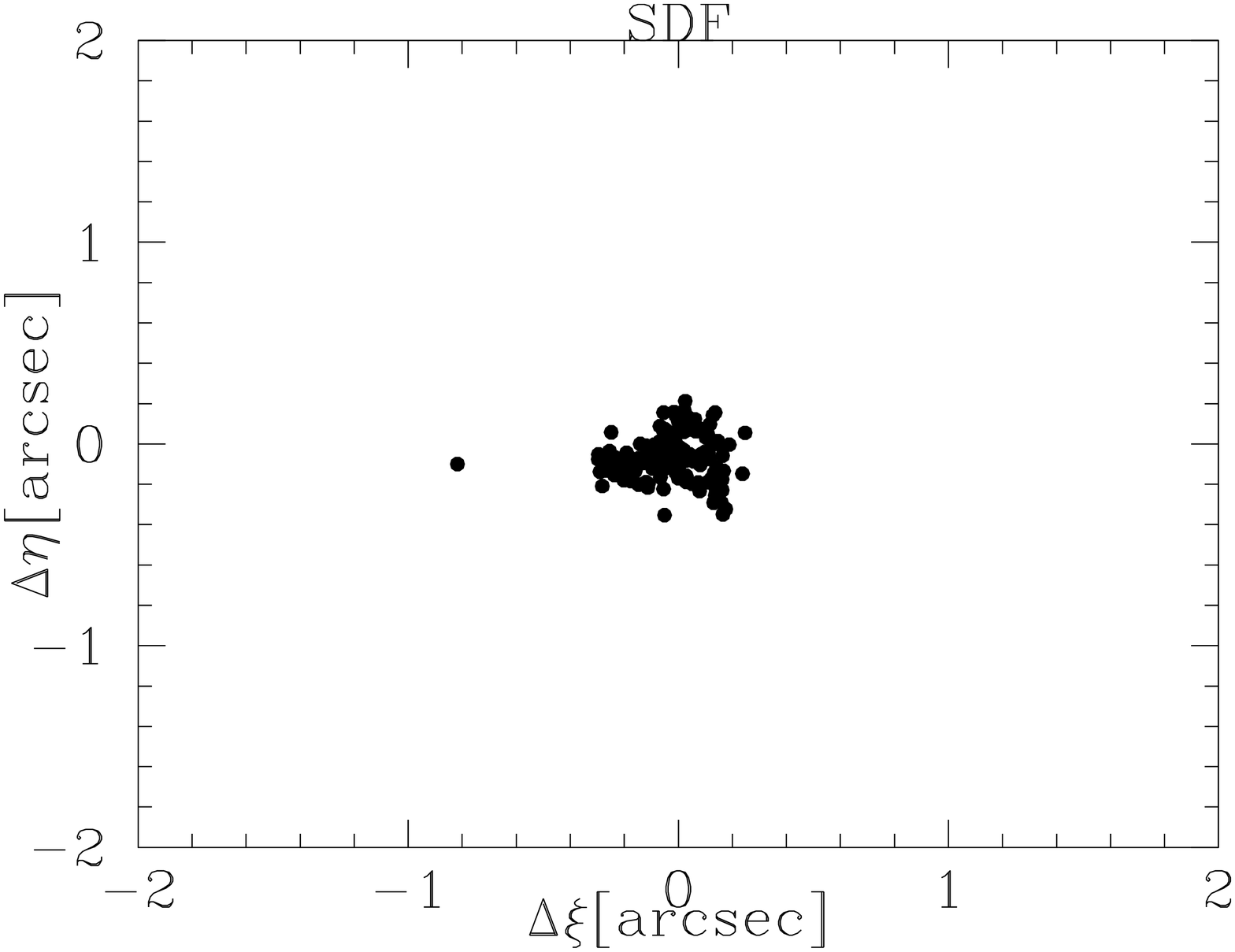}
\FigureFile(80mm,60mm){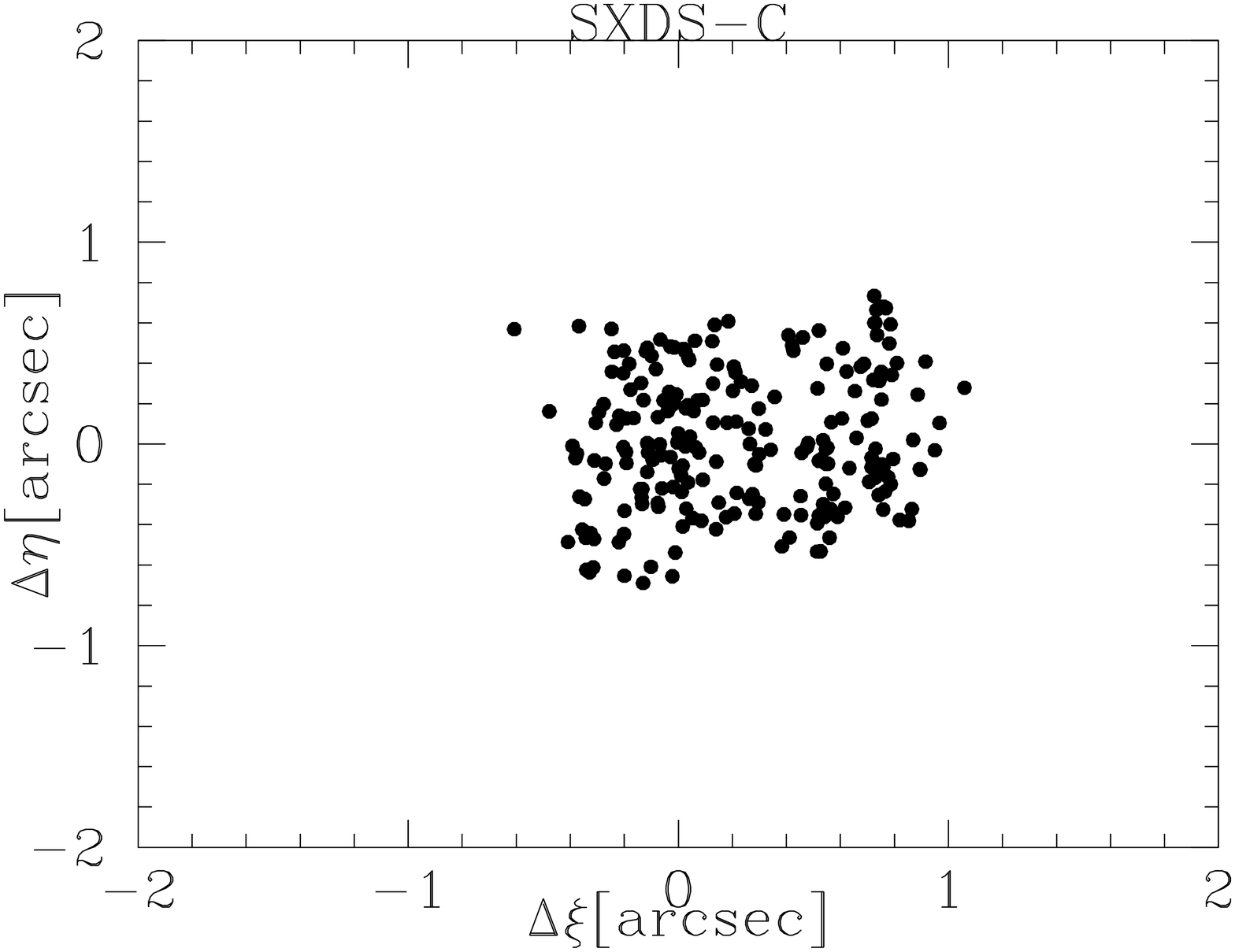}\\
\FigureFile(80mm,60mm){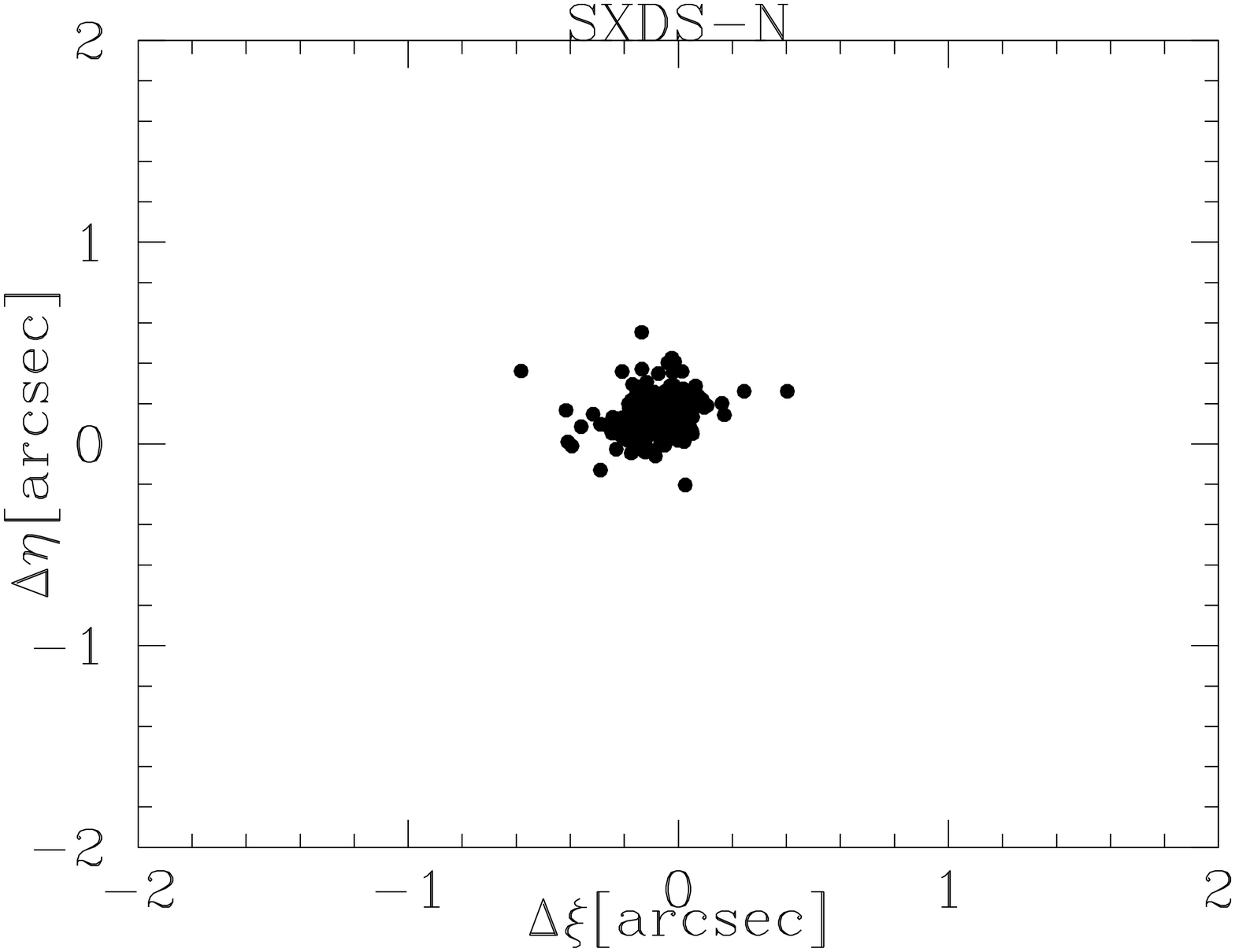}
\FigureFile(80mm,60mm){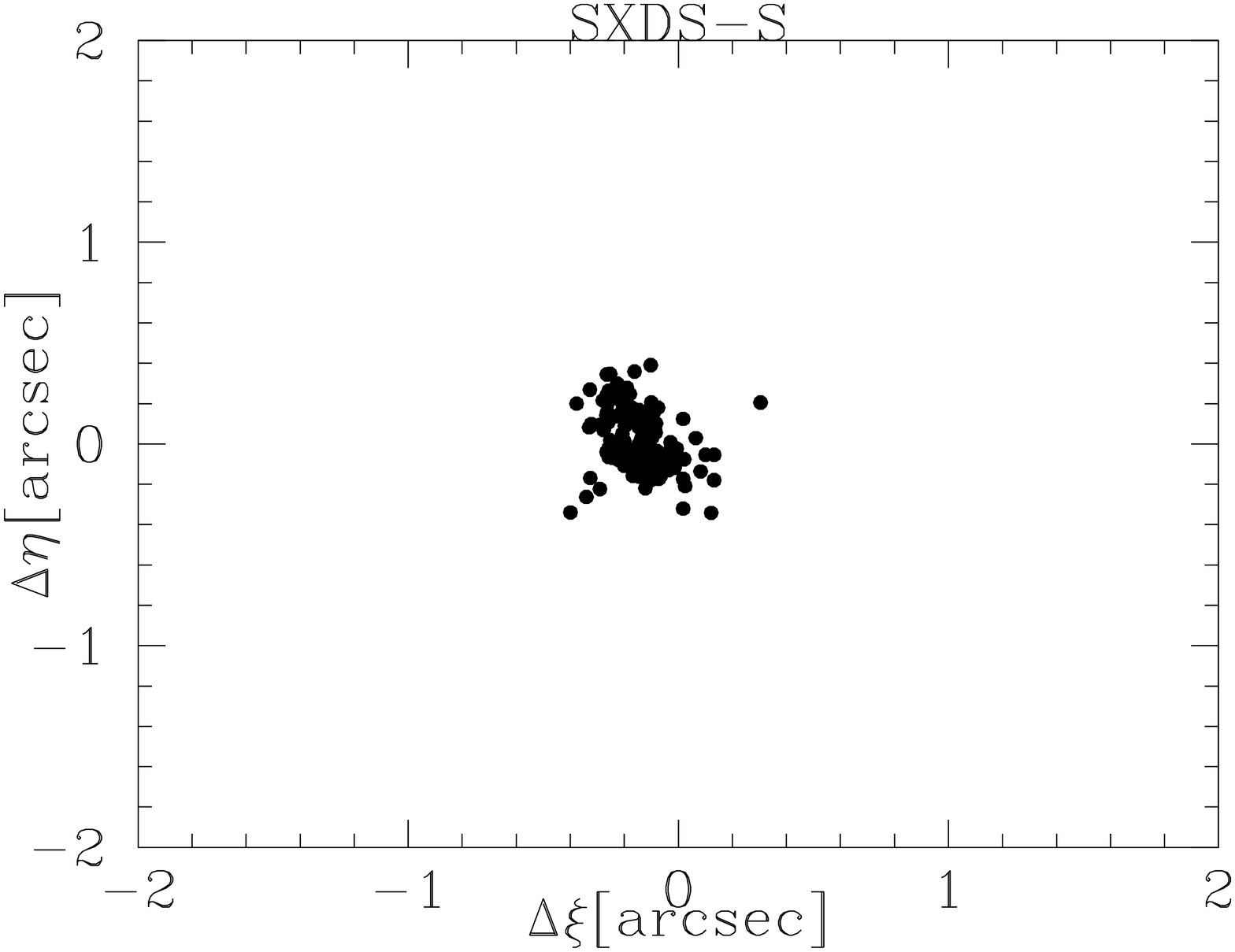}\\
\FigureFile(80mm,60mm){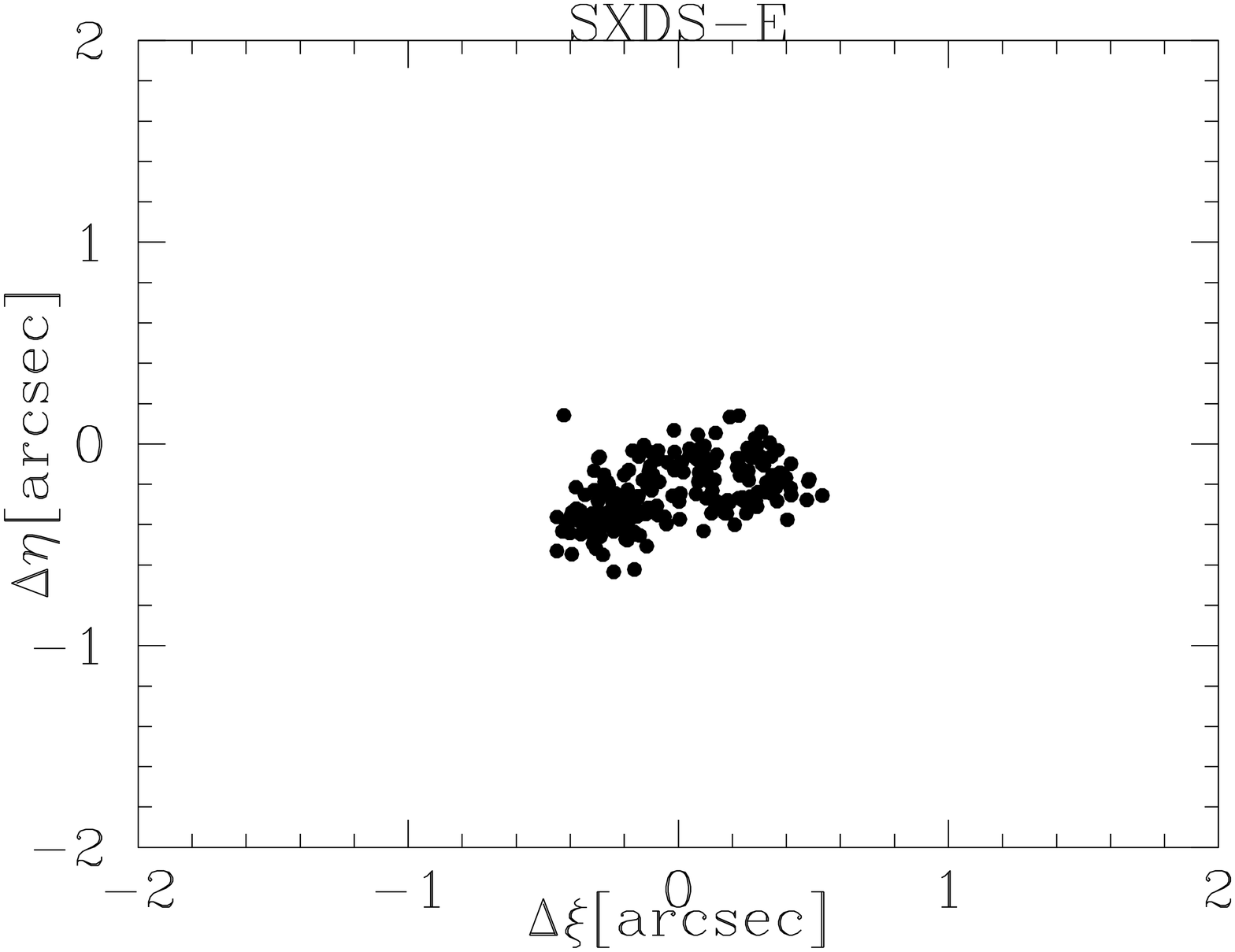}
\FigureFile(80mm,60mm){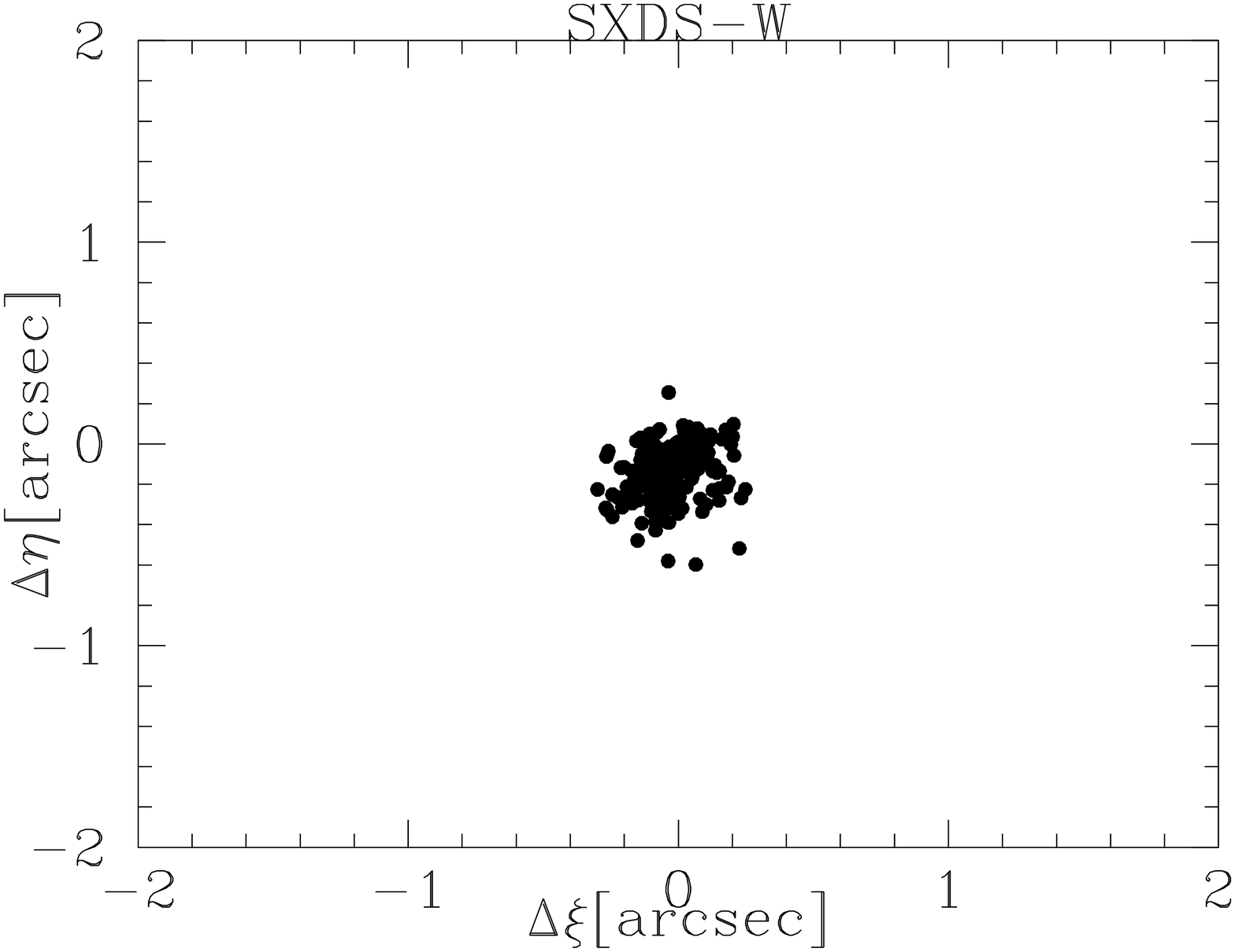}\\
\caption{
The difference of the position of matched stars in $20<r<21$ 
magnitude range. $\Delta\xi=\alpha\cos(\delta)({\rm SDSS})-
\alpha\cos(\delta)({\rm Suprime-Cam})$,
and $\Delta\eta=\delta({\rm SDSS})-\delta({\rm Suprime-Cam})$.}
\label{fig:pos0}
\end{figure}

The residual of coordinates of stars 
in r-band magnitude of $20<r<21$ are shown in figure \ref{fig:pos0},
where we simply converted $\xi=\alpha \cos(\delta)$ and $\eta=\delta$,
Though the astrometry of SXDS-C shows a larger scatter,
the radius of 2 arcsec of cross-match covers the difference.

\begin{figure}
\FigureFile(80mm,60mm){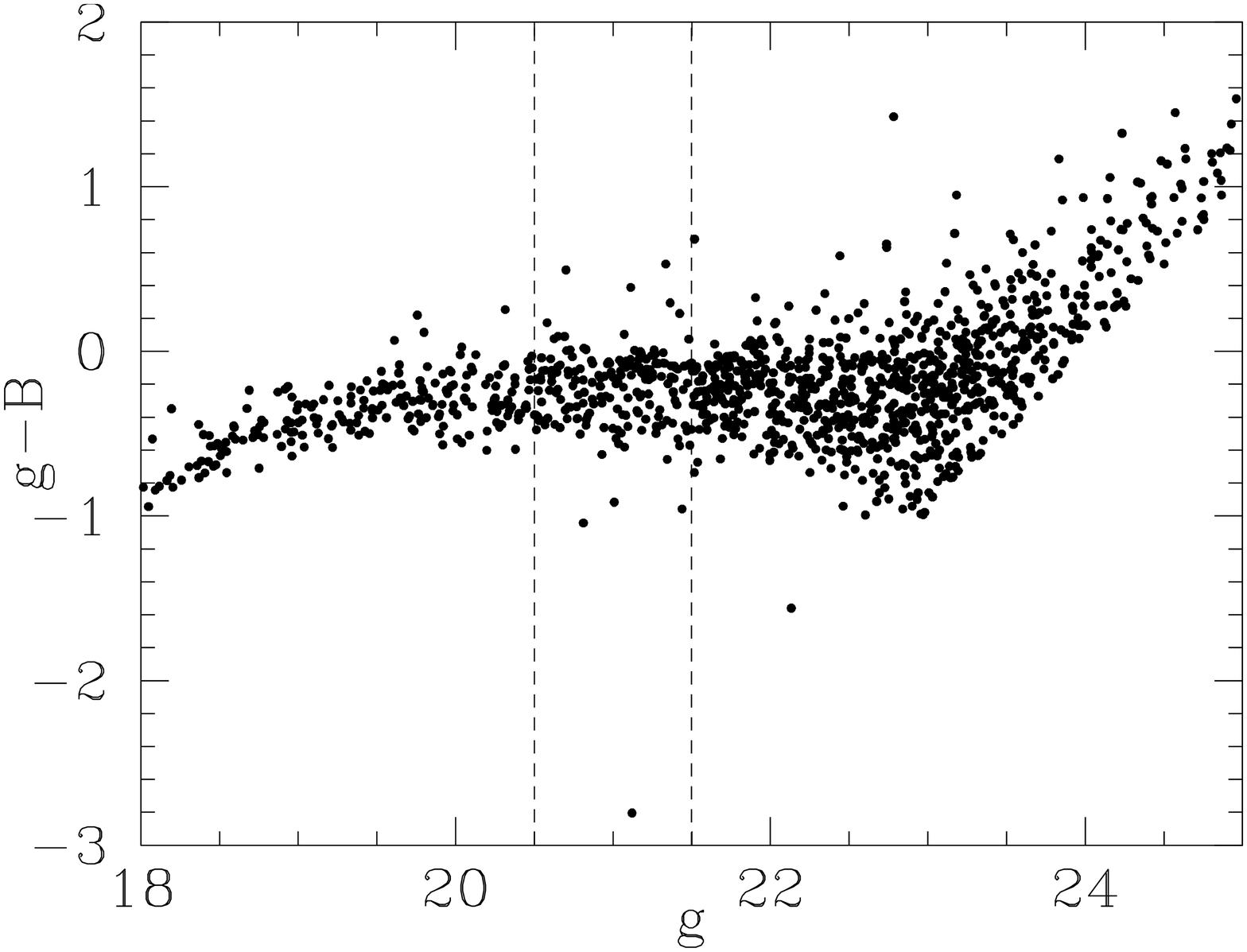}
\FigureFile(80mm,60mm){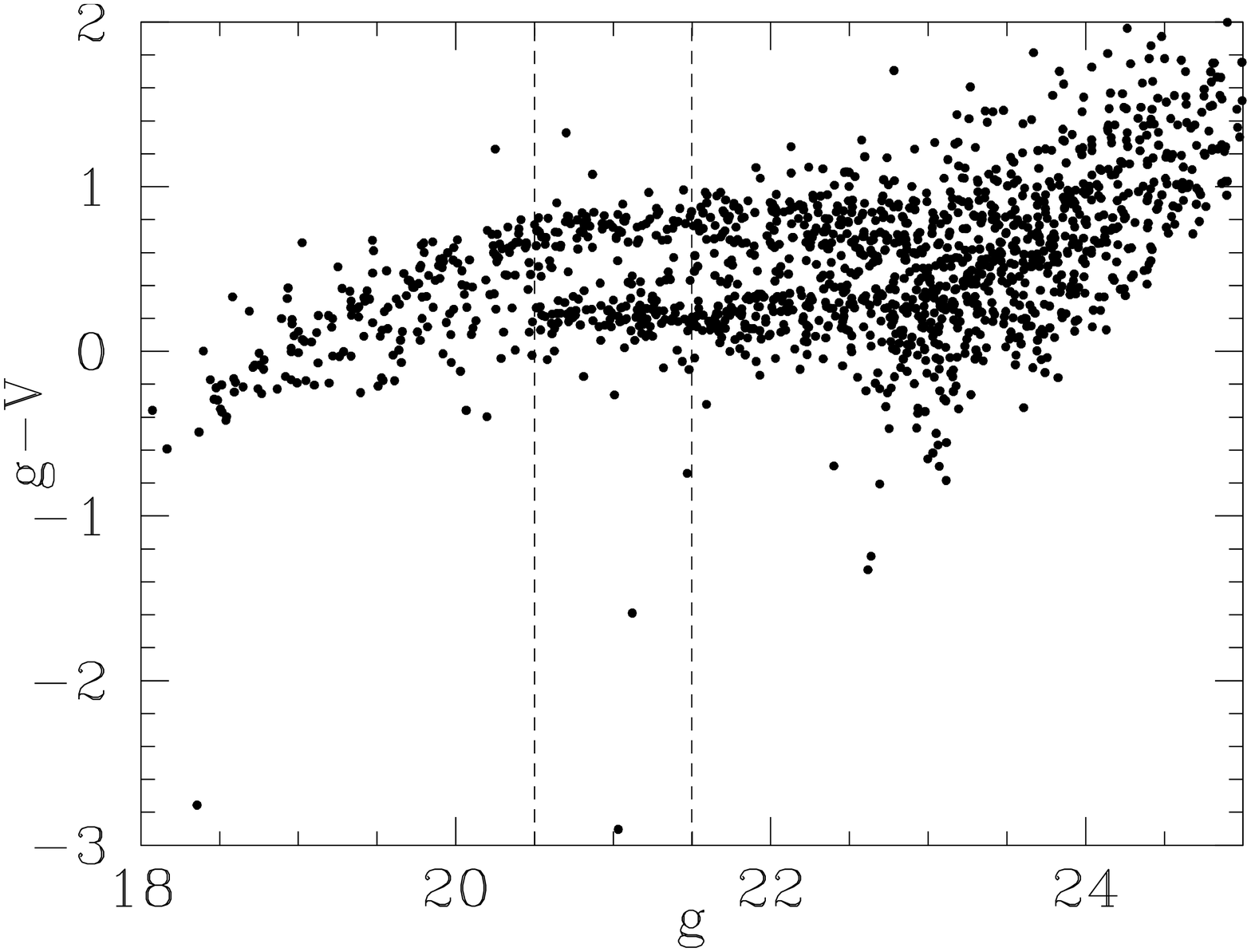}\\
\FigureFile(80mm,60mm){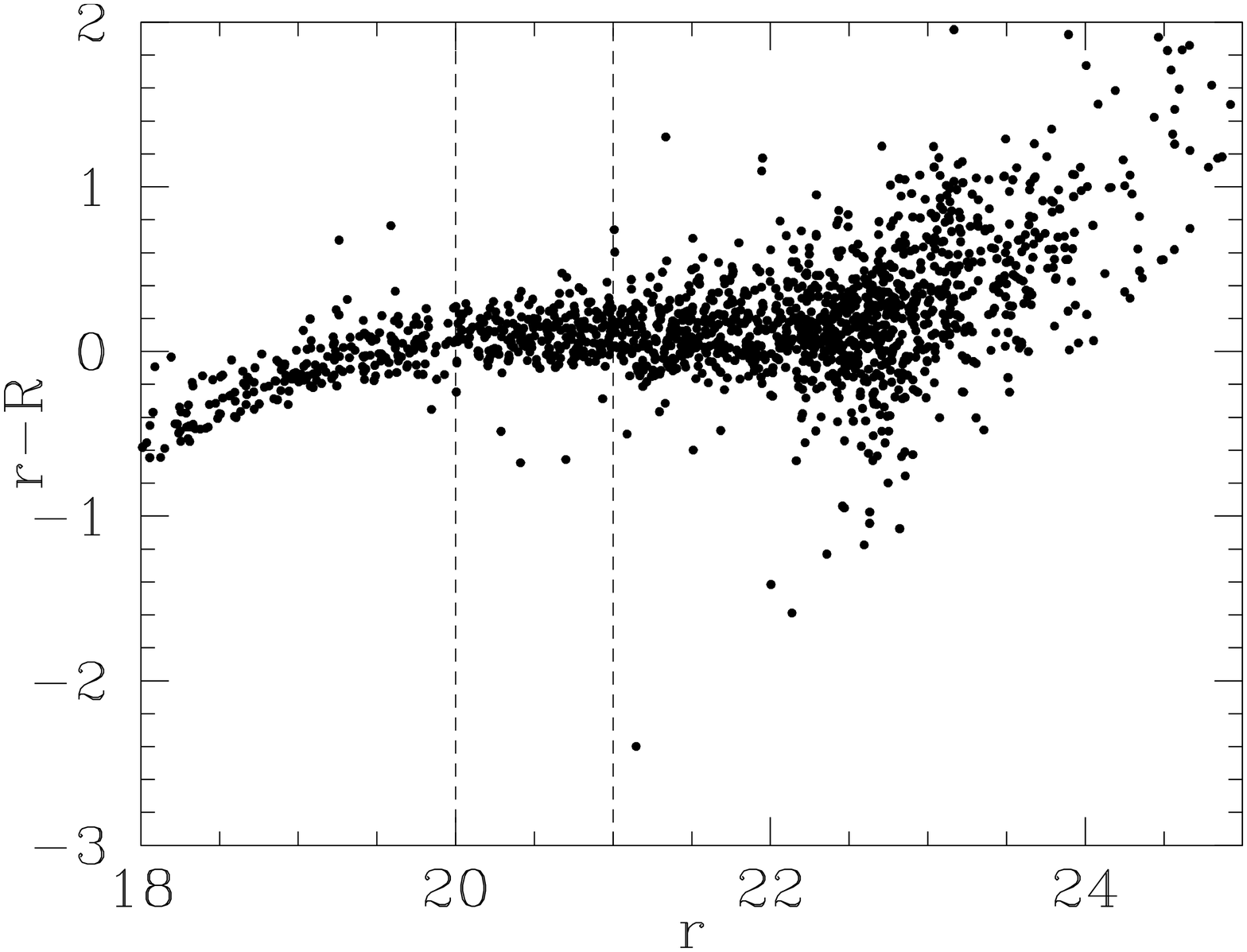}
\FigureFile(80mm,60mm){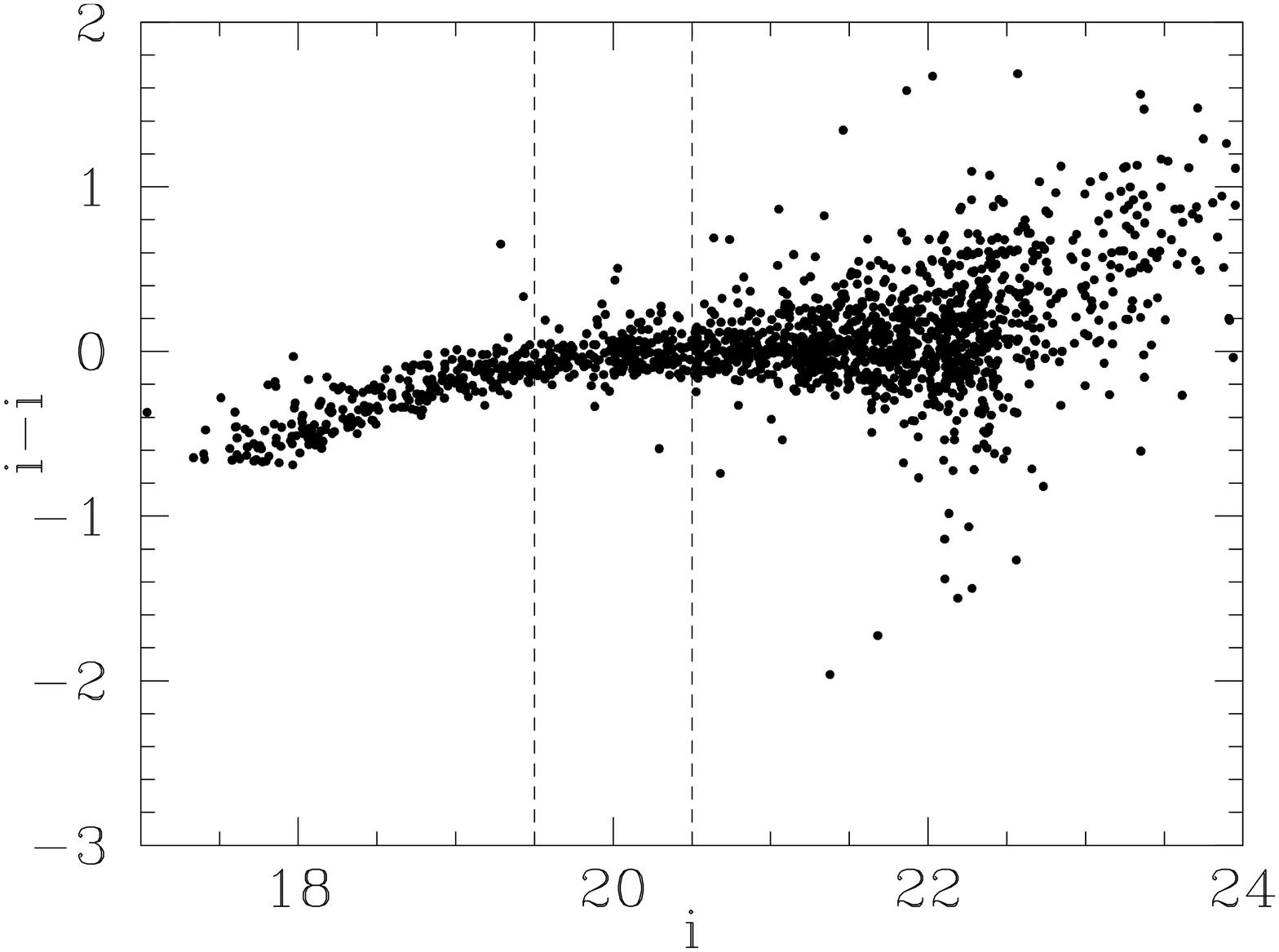}\\
\FigureFile(80mm,60mm){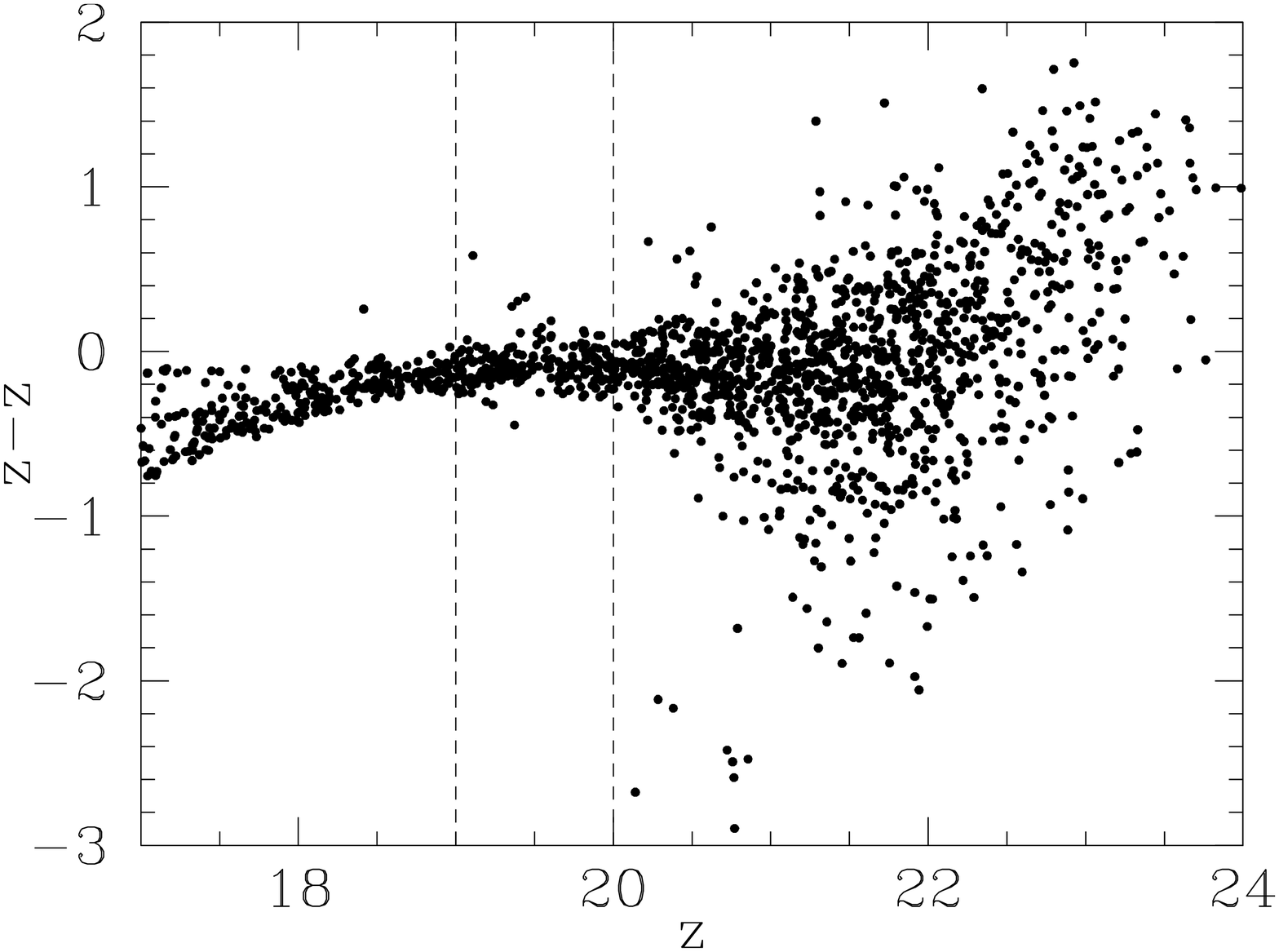}
\caption{
Color magnitude diagram of (SDSS)-(Suprime-Cam) matched stars in
SDF field. The cutoff of the left bottom side is due to the
Suprime-Cam magnitude cut of mag$<$24.
The vertical lines show the adopted magnitude range for the
calibration.}
\label{fig:CMD0}
\end{figure}

The SDSS magnitude versus the difference color of the SDSS and Suprime-Cam 
is shown as figure \ref{fig:CMD0}.
In the figure, the object whose Suprime-Cam magnitude is fainter has
a negative difference in y. The tendency that bright objects has
a negative difference is due to a saturation in Suprime-Cam data.
As it is not simple to exclude saturated objects from the catalog 
correctly, we did not apply any selection of saturation 
in this study.
Note that the difference in y is not corrected for 
the color dependence, and therefore the relation is broad.
For example, 
the two loci in V-band correspond to two different populations,
distant blue stars and closer red stars.

From figure \ref{fig:CMD0}, we adopted the magnitude range for the 
calibration; $20.5<g<21.5$ for B and V, 
$20<r<21$ for R,
$19.5<i<20.5$ for i, and $19<z<20$ for z.
In the magnitude range, 
the saturation of Suprime-Cam stars has not effect.
We also set a constraint that the error of the SDSS photometry,
psfMag\_Err$<$0.1, to 
avoid poor photometry data.
Then, the SDSS color versus the difference color of the SDSS and Suprime-Cam 
in the magnitude range are plotted in figure \ref{fig:cal0}.
They correspond to the top panels of figure \ref{fig:fit0}.
In some panels, we can see a significant offset from the 
model color distribution.
The green filled circles with errorbars show 
the median of the (SDSS)-(Suprime-Cam) color
in 0.2 mag bin of SDSS color. 
The errorbar represents 
the root mean square (rms) of the bin
estimated from the median of the absolute deviation (MAD) as
\begin{equation}
{\rm rms}={\rm MAD} \times 1.4826.
\end{equation}
The trend is almost parallel to the distribution 
of the model distribution.
The apparent difference in the slope between the model and the catalog 
in z-band is a fake.
It was made by the relatively larger 
dispersion of z(SDSS) ($\sim 0.07$ mag)
compared with that of i(SDSS) ($\sim 0.04$ mag) 
and z(Suprime-Cam) ($<0.01$ mag).
The distribution is elongated from top-left to
bottom-right by the error of z(SDSS),
and taking a median of z(SDSS)-z(Suprime-Cam) after binning
in i(SDSS)-z(SDSS) changed the apparent slope.
If we plot i(SDSS)-z(Suprime-Cam) vs z(SDSS)-z(Suprime-Cam),
the z(SDSS) error is perpendicular to the binning axis,
and the model and the catalog is almost parallel.
The result suggests that the difference between the SDSS
and Suprime-Cam catalog would be the offset of the ZP of 
Suprime-Cam catalogs.

\begin{figure}
\FigureFile(90mm,60mm){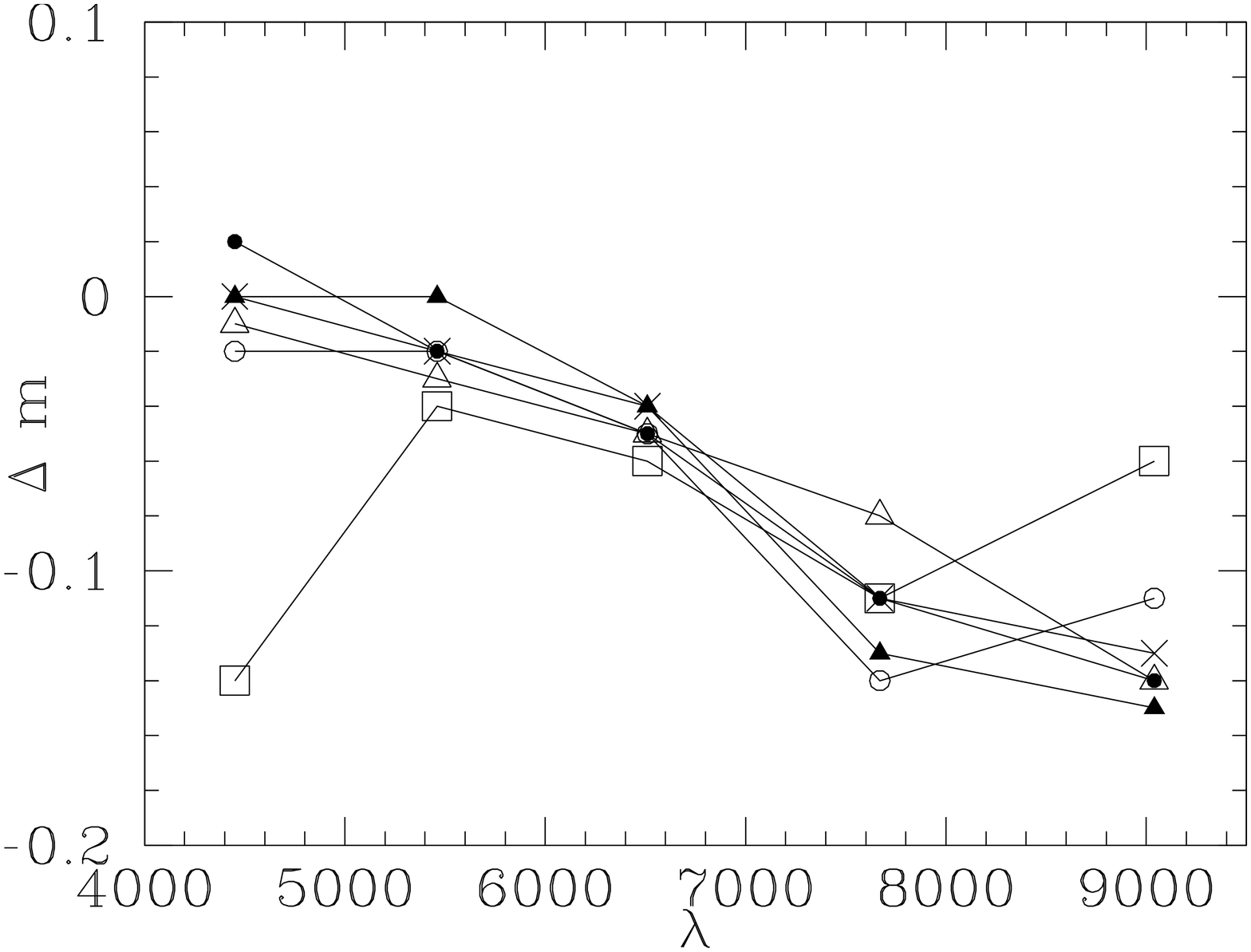}
\caption{
The ZP difference between the catalog and the
estimated value from SDSS as a function of the wavelength of
filters. 
The symbols represent field as
open squares (SDF), the crosses (SXDS-C), 
the filled circles (SXDS-N), the open circles (SXDS-S),
the filled triangles (SXDS-E), and the open triangles (SXDS-W).
The data of the same field are connected by a line.
}
\label{fig:wavdepend}
\end{figure}

We then estimate the offset of the ZP of the SDF and SXDS catalogs 
from the SDSS based on an estimation using 
the parameters in table \ref{tab:coeff0}.
The number of stars, the median of the offset, 
and the rms of the distribution ($\sigma$) calculated from the MAD
are shown in table \ref{tab:ZPdiff}.
The offset is also plotted as a function of the wavelength as 
figure \ref{fig:wavdepend}, 
where open squares represent SDF,
and other symbols represent SXDS fields.
For SXDS, a clear correlation between the wavelength and the offset 
is recognized.

In table \ref{tab:ZPdiff},
the $\sigma$ (5th column) is the rms of the distribution and not the error.
The determination error of the offset from statistics 
is roughly approximated by $\sigma \sqrt{\pi/2N} \sim 0.01$ mag,
where $\sqrt{\pi/2}$ is the factor for the rms of the median.
The $\sigma$ mainly comes from the photometric error of SDSS.
We calculated the contribution of the SDSS error as
\begin{equation}
\sigma_{\rm SDSS}={\rm median}({\rm psfMag\_Err}),
\end{equation}
and is shown in the 6th column in table \ref{tab:ZPdiff}.

The $\sigma$ is partly explained by the SDSS error.
If we subtract the effect, the mean residuals of the 6 fields
are 0.06, 0.02, 0.03, 0.05, and 0.06 mag for
B, V, R, i, and z-band, respectively.
The residuals include the intrinsic dispersion around our best fit,
which is the error of our color conversion, 
and the error of the data reduction and the photometry.

Some Suprime-Cam data show an offset even larger than the 
sigma of the distribution, and the offset is significant.
These values are shown in bold in table \ref{tab:ZPdiff}.
In B-band, the offset is different between SDF and SXDS by
$0.13\pm0.01$.
On the other hand,
no difference is seen among the fields in V-band and R-band.
The systematic difference between SDF and SXDS seems to be
marginal in z-band, $0.07\pm0.01$.
In i-band, the variation of the estimated offset is large
(peak-to-peak 0.06 mag).

In summary, we found a significant offset of ZP 
in some of the SDF/SXDS catalogs at bright ($r\sim20$) magnitude stars 
from AB magnitude of SDSS. 
The amount of the offset is significantly larger than 
the estimated uncertainty from the catalogs.
If we correct the ZP offset, the distribution of 
the (SDSS)-(Suprime-Cam) color follows the model sequence well.
Examples are shown as figure \ref{fig:cal1}.

\section{Discussion}

\subsection{The Color Histogram of Faint Objects in SDF and SXDS}

The different offset of B-band and z-band (figure \ref{fig:wavdepend})
makes the color distribution of SDF and SXDS different.
We made a color histogram of faint objects in SDF and SXDS-C
and compared them.
Most of the faint objects are galaxies.
We used the R-band selected catalog, 
the R-band based object extraction and MAG\_AUTO.

\begin{figure}
\FigureFile(80mm,60mm){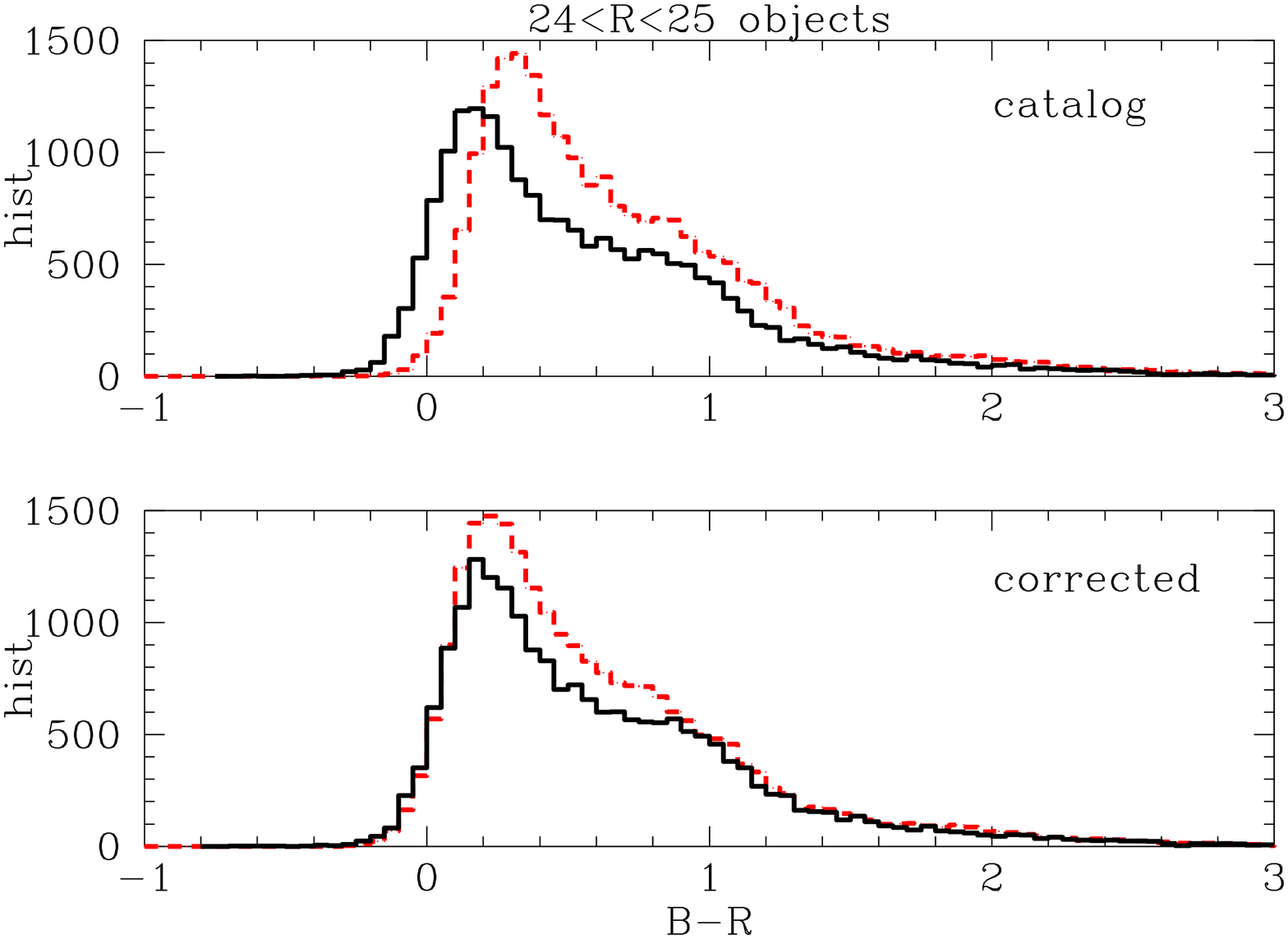}
\FigureFile(80mm,60mm){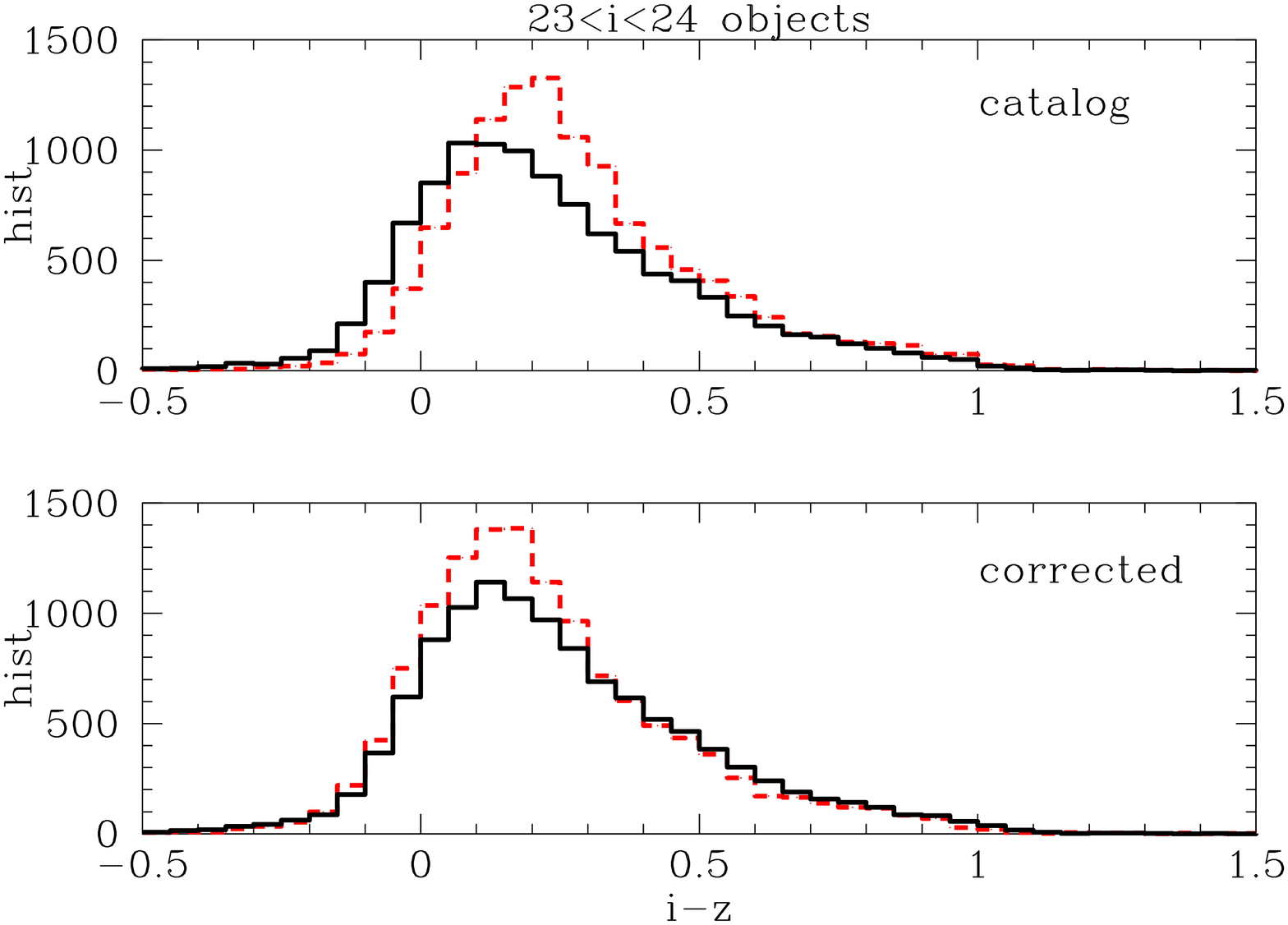}
\caption{
(left)
(B-R) color histogram of 24$<$R$<$25 objects in SDF (dashed red)
and SXDS (solid black). The top panel uses the catalog value, 
and the bottom panel uses the value with the correction
in table \ref{tab:ZPdiff}.
Galactic extinction is corrected.
(right) Same as the left panel but of (i-z) color of 23$<$i$<$24 objects.
}
\label{fig:colhist0}
\end{figure}

For the B-band check, we used (B-R) color,
because \citet{Yamanoi2012} calibrated and checked B and R-bands.
The Galactic extinction is assumed to be
$(A_B, A_R)$=(0.07,0.04) for SDF%
\footnote{http://soaps.nao.ac.jp/SDF/v1/common/galactic\_extinction},
and (0.091,0.056)  for SXDS-C\citep{SXDS2}.
The result is shown as the left panel of figure \ref{fig:colhist0}.
The slight difference in the size of the observed area is not corrected.
In the original catalogs, the blue end of the 
color histogram is different(figure \ref{fig:colhist0} top-left).
The difference gets smaller 
if we use our recalibrated catalogs
(figure \ref{fig:colhist0} bottom-left).

The (i-z) color histogram is shown as the right panel of 
figure \ref{fig:colhist0}. The Galactic extinction is 
$(A_i, A_z)$=(0.03,0.02), and
$(A_i, A_z)$=(0.044,0.031) for SDF and SXDS-C, respectively.
Our correction again decreases the apparent difference of the
color distribution.

This result suggests that the different ZP offset 
between SDF and SXDS in B-band and z-band 
would be due to a calibration error,
and not coming from possible inhomogeneity of SDSS DR8.
The result also shows that the offset of magnitude 
found around R$\sim$20 mag stars would be similar in 
R$\sim$24 mag objects. 
It supports our assumption that the difference between
the SDSS-based magnitude and the original SDF/SXDS magnitude 
would be an offset of the ZP in SDF/SXDS catalogs.

\subsection{Possible Origins of the ZP Offset}

\subsubsection{Metallicity Effect on the Color-color Diagram}
\label{sec:metal}

\begin{figure}
\FigureFile(80mm,60mm){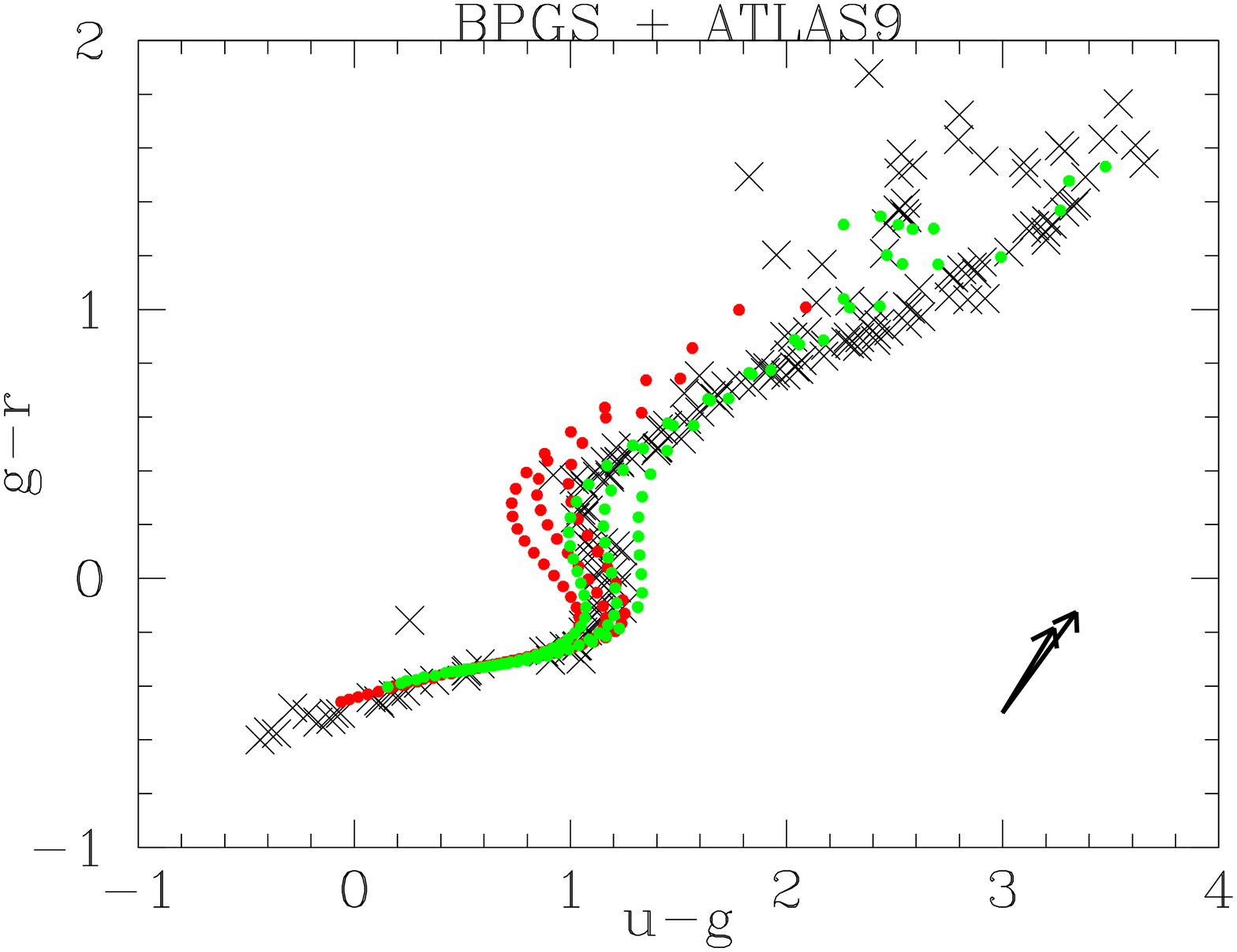}
\FigureFile(80mm,60mm){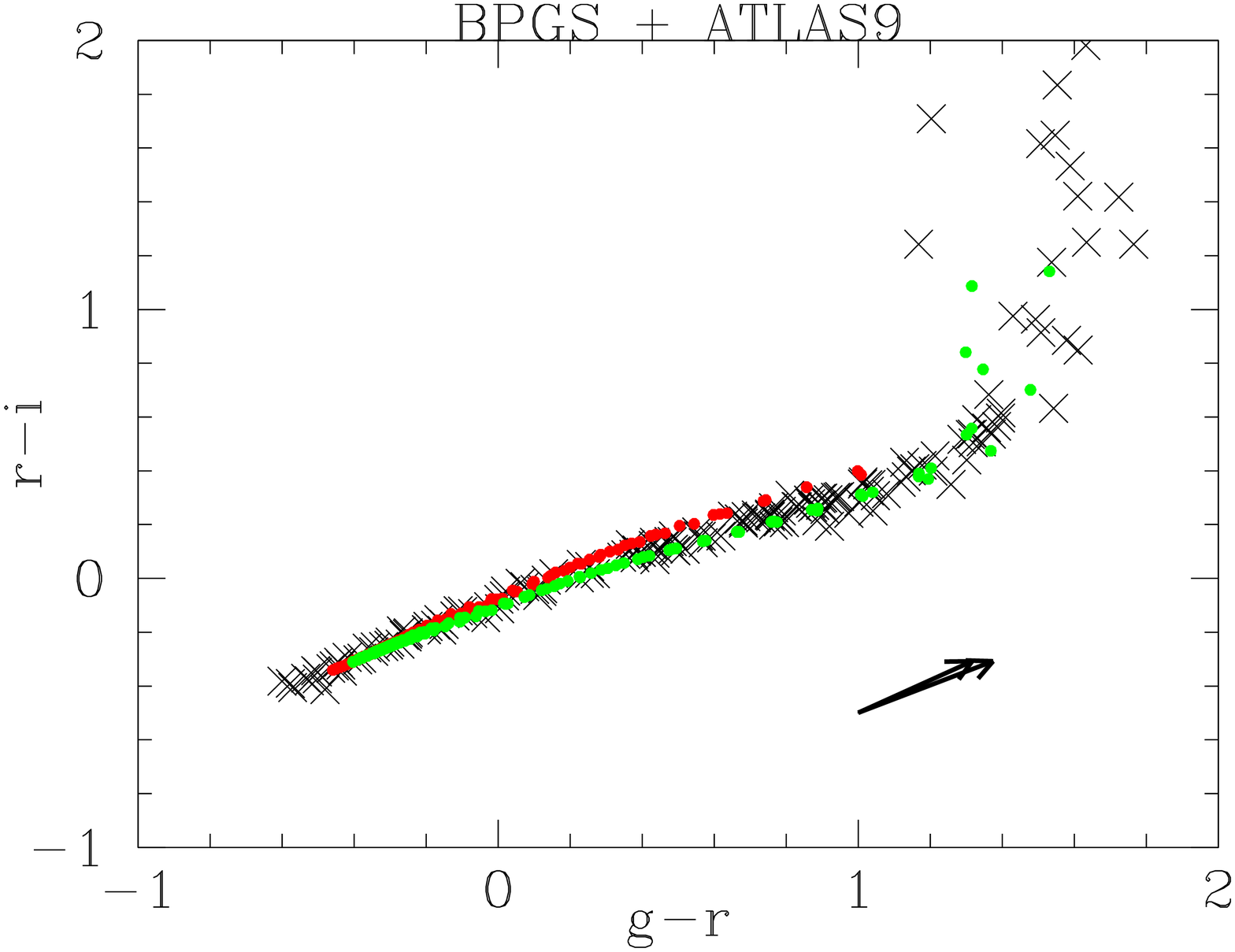}\\
\FigureFile(80mm,60mm){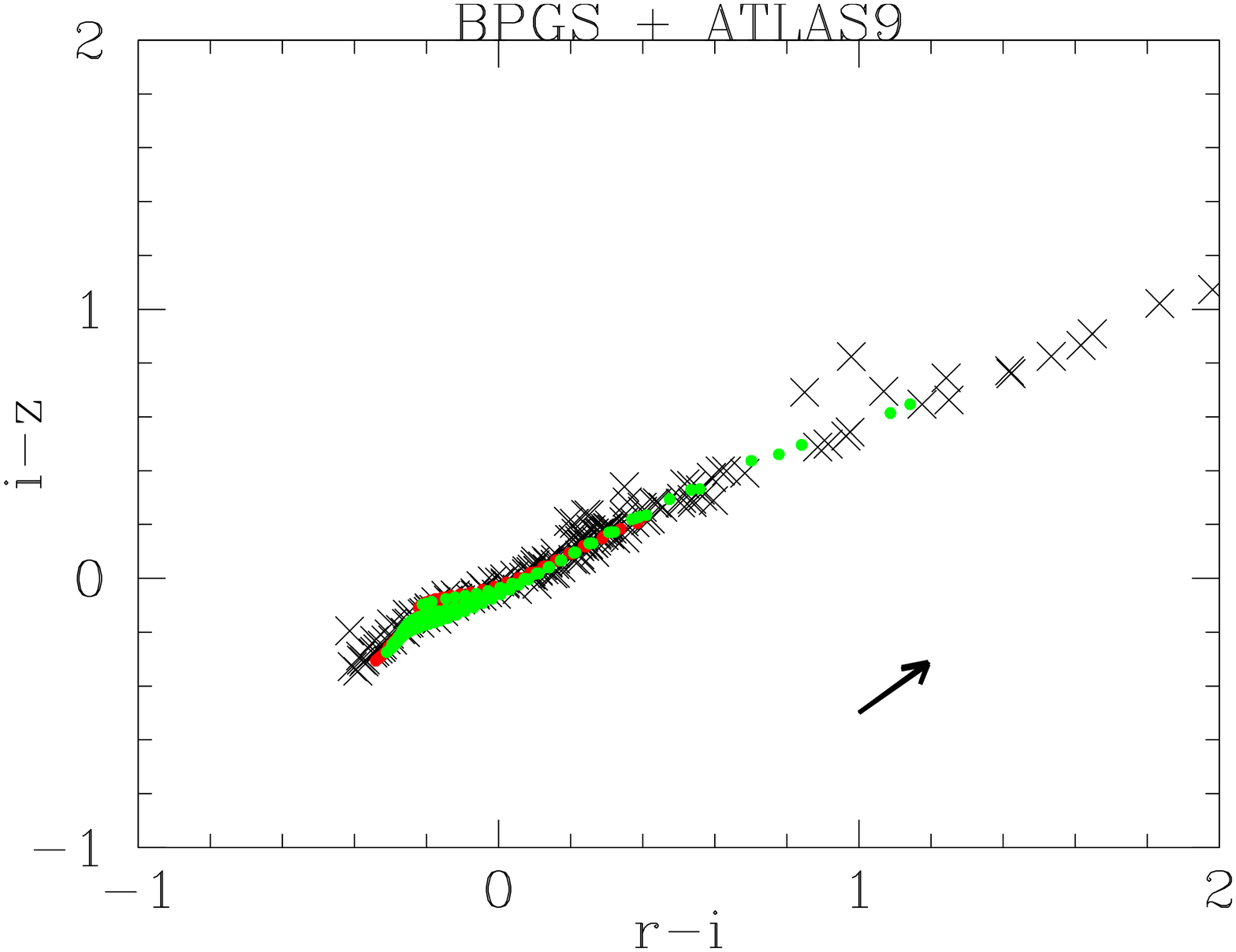}
\caption{
Color-color diagrams in SDSS AB magnitude. 
The crosses are synthetic colors calculated from BPGS SEDs.
The filled red circles are 
[Fe/H]=-2.5 with [$\alpha$/Fe]=+0.4 models,
and the filled green circles are 
[Fe/H]=0 with [$\alpha$/Fe]=0 models.
Models are calculated from the flux data of ATLAS9 grid,
and the sets of the temperature and the surface gravity 
are taken from Yonsei-Yale isochrone.
The arrows indicate the direction of 
Av=1 reddening of Galactic extinction for O and M stars in BPGS.
}
\label{fig:colcolmodel_BPGS}
\end{figure}

\begin{figure}
\FigureFile(80mm,60mm){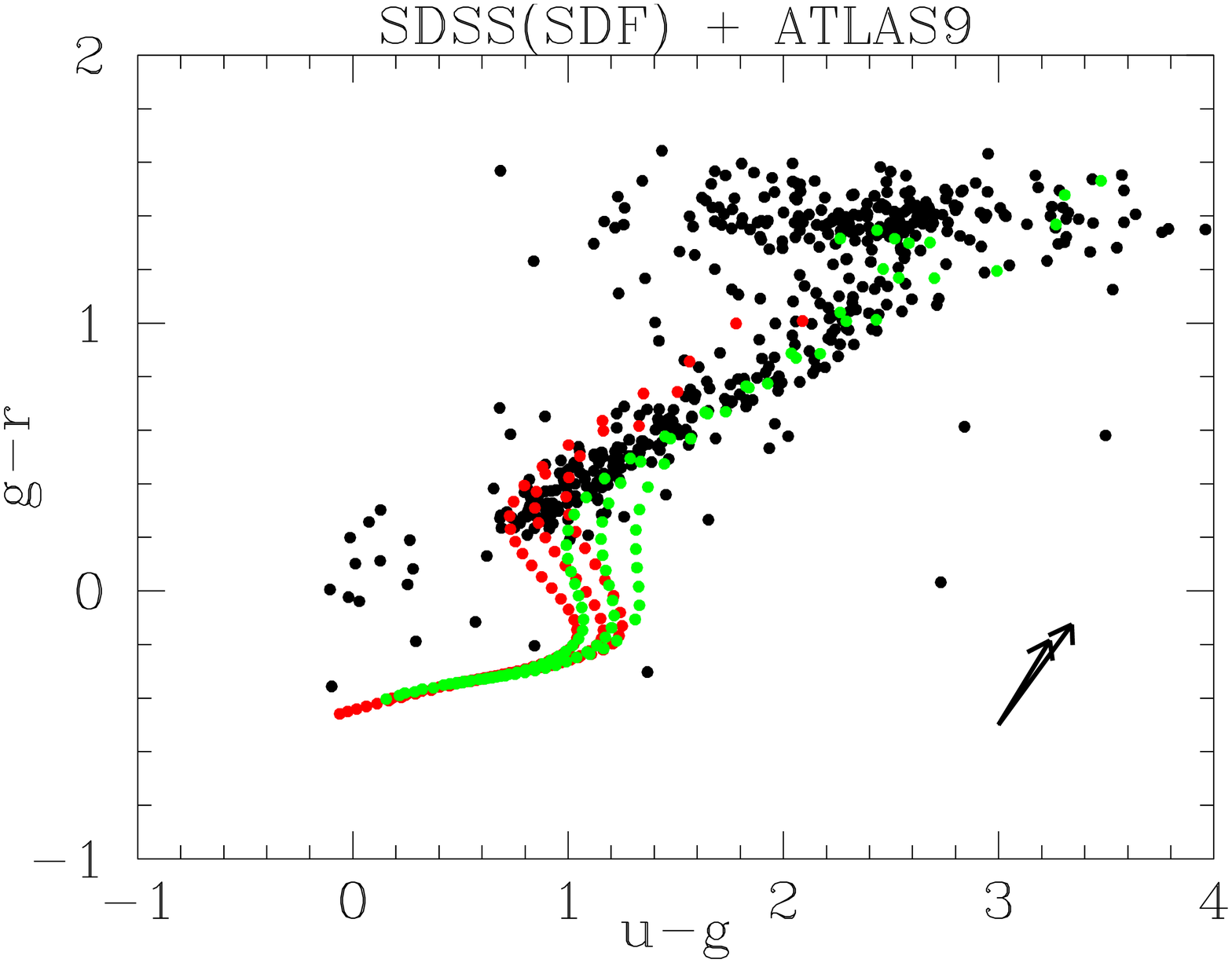}
\FigureFile(80mm,60mm){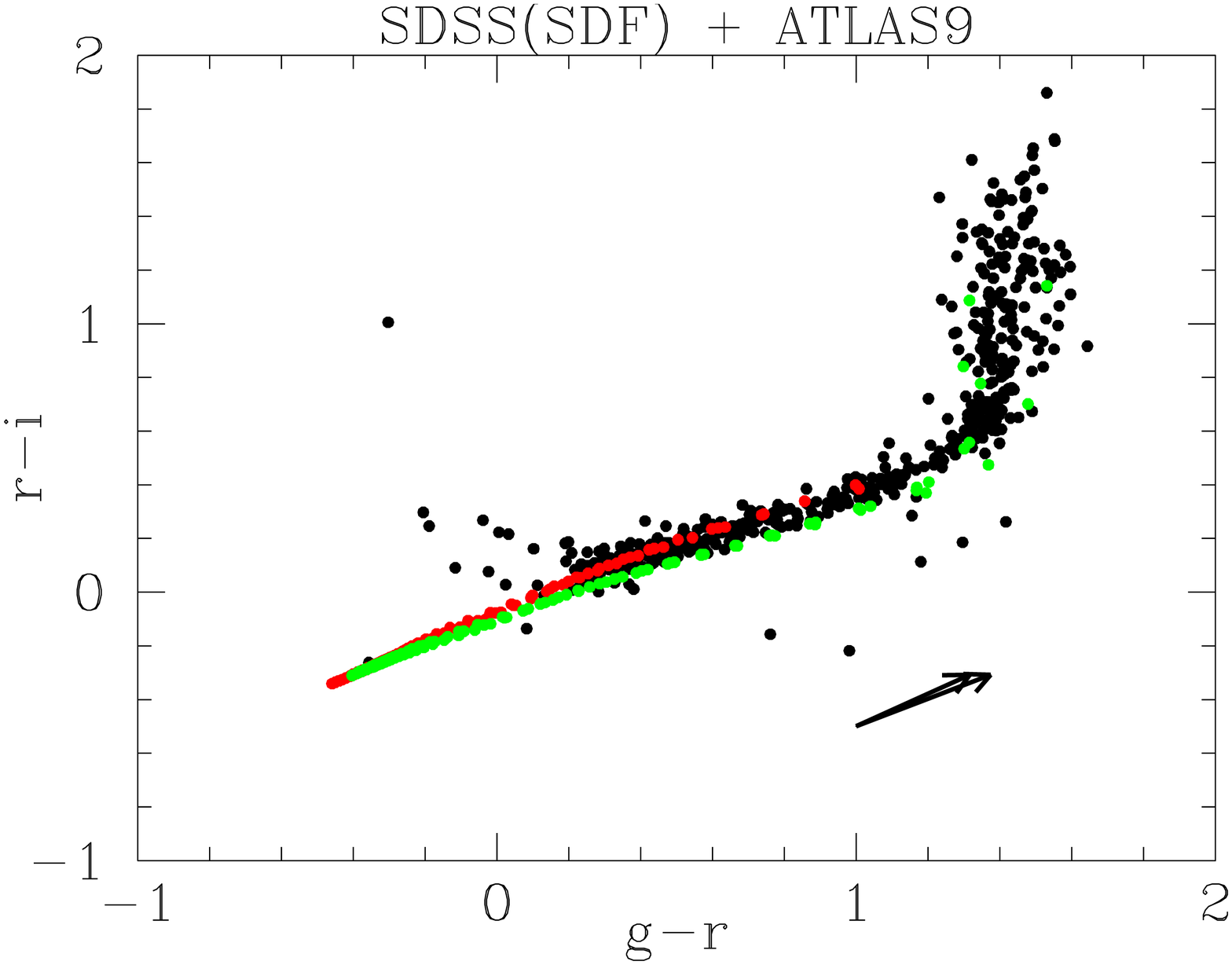}\\
\FigureFile(80mm,60mm){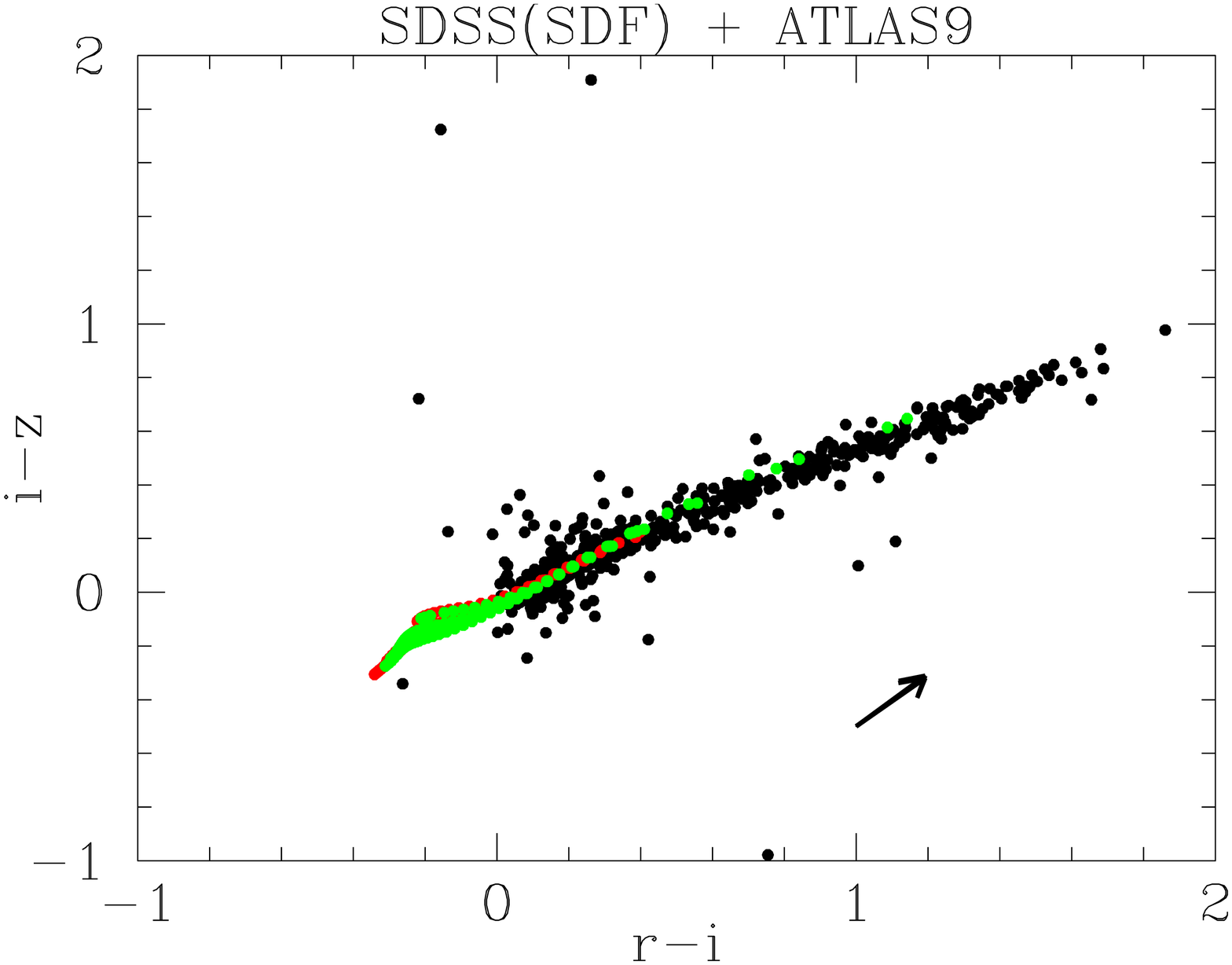}
\caption{
Same as figure \ref{fig:colcolmodel_BPGS}, but on SDSS stars.
The filled black circles are SDF stars of $15<r<21$ taken from SDSS DR8, 
and are corrected the offset adopted in Kcorrect\citep{Blanton2007} v4.
}
\label{fig:colcolmodel_SDSS}
\end{figure}

In figure \ref{fig:colcolmodel_BPGS}, we plotted BPGS stars
and ATLAS9 models.
We plotted model points with [Fe/H]=0 and [$\alpha$/Fe]=0 
(hereafter [Fe/H]=0 model) as filled green circles 
and those with [Fe/H]=-2.5 and [$\alpha$/Fe]=0.4 (hereafter [Fe/H]=-2.5a model)
as filled red circles.
The [Fe/H]=0 model seems suitable in all the color combinations in ATLAS9.
For a comparison, we plotted 
SDSS stars of ($15<r<21$) in SDF and overlaid 
models in figure \ref{fig:colcolmodel_SDSS}.
The [Fe/H]=-2.5a models well covers the SDSS color distribution 
on the bluer side, and [Fe/H]=0 model is somewhat better on the redder side.

The reason why the color-color diagram (figure \ref{fig:colcolmodel_SDSS})
shows the offset between different metallicities but 
not in figure \ref{fig:fit0} is simply the difference in the 
central wavelength of the filters used in the y-axis.
The variation of slope and curvature of the spectrum 
has less of an effect on the y-axis for a pair of passbands
which have close central wavelengths.

\begin{figure}
\FigureFile(85mm,60mm){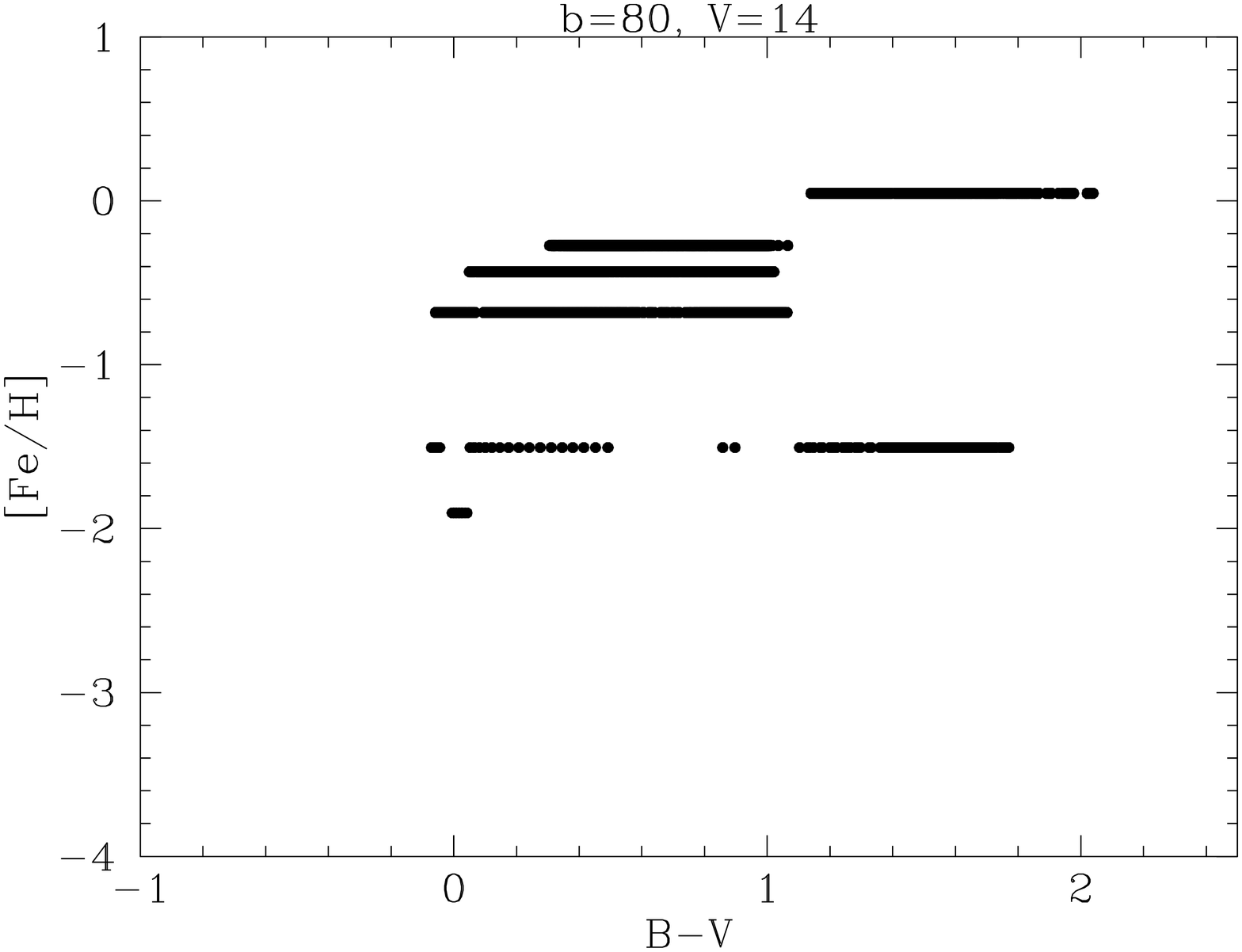}
\FigureFile(85mm,60mm){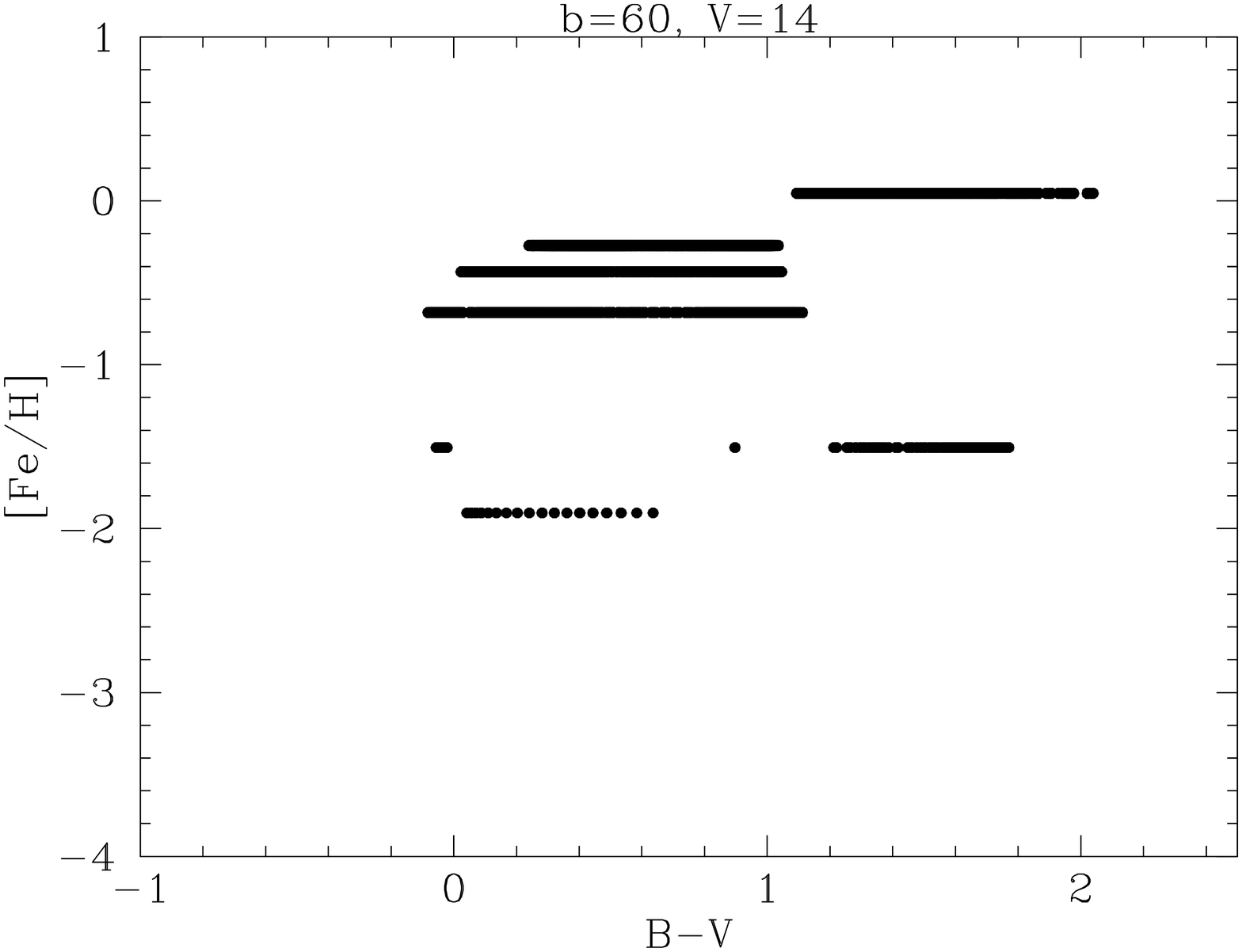}\\
\FigureFile(85mm,60mm){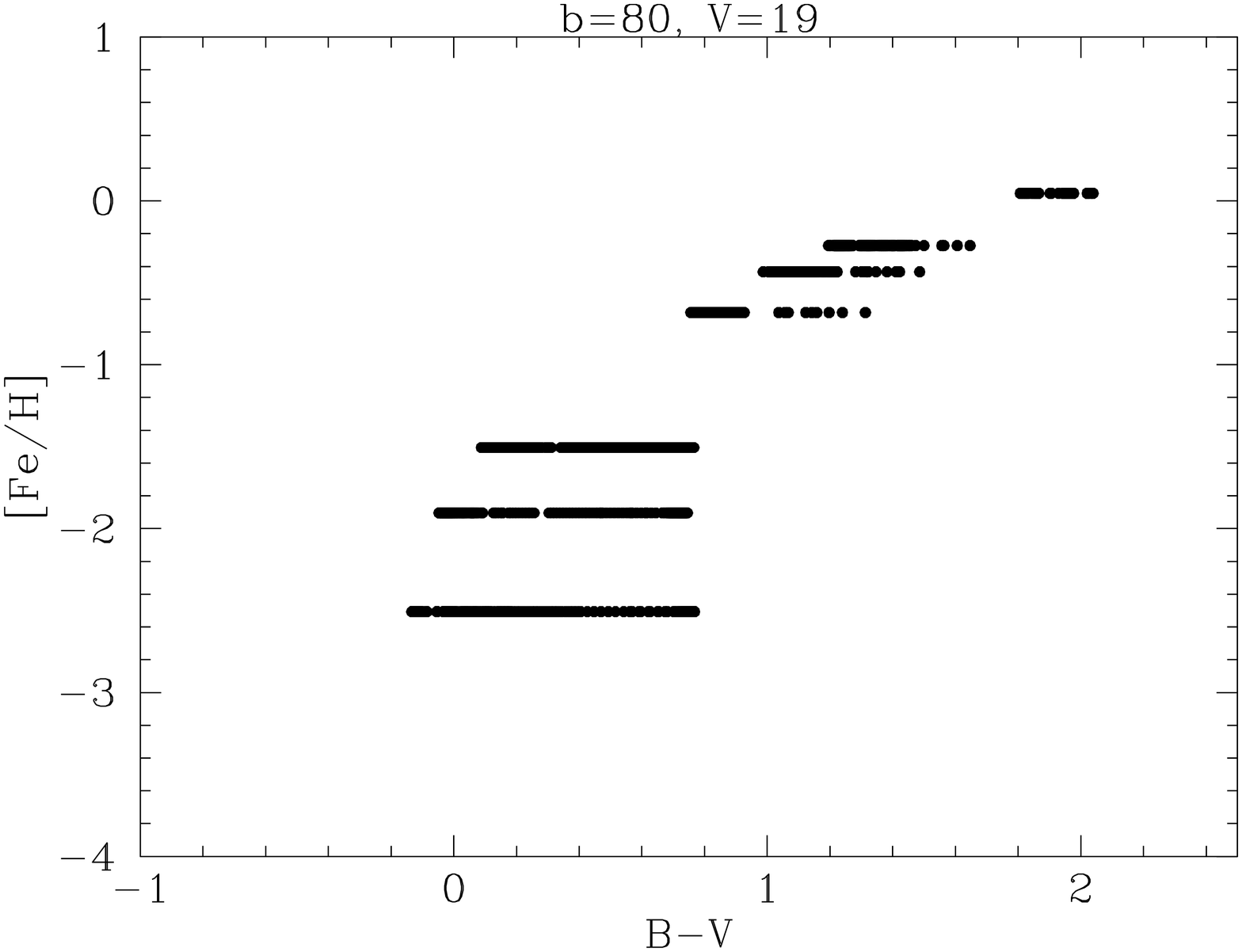}
\FigureFile(85mm,60mm){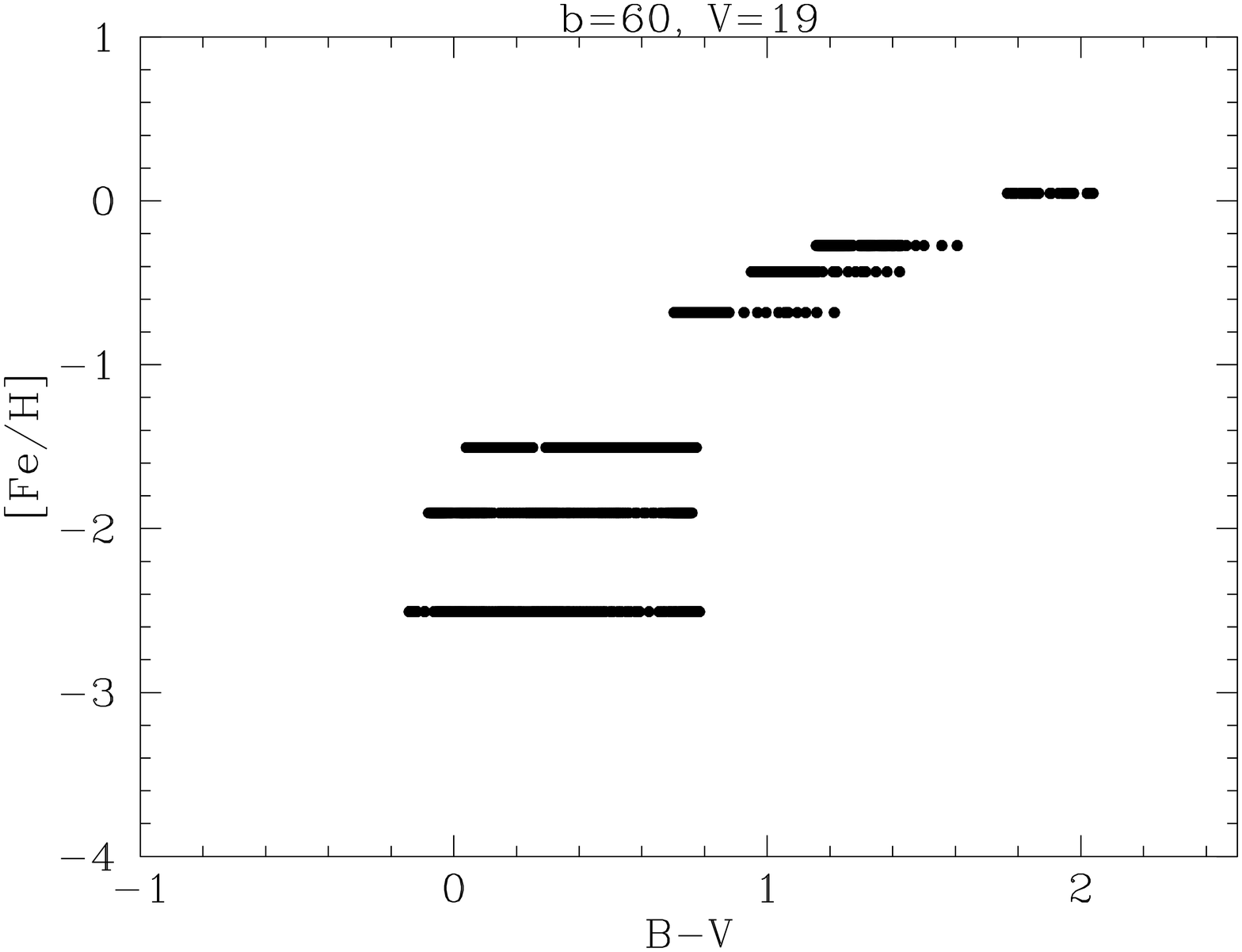}
\caption{
An estimation of the relation of color and metallicity.
(Top left) V=14 mag stars at Galactic latitude b=80 model.
(Top right) V=14 mag stars at b=60.
(Bottom left) V=19 mag stars at b=80.
(Bottom right) V=19 mag stars at b=60.
The metallicity gradient model is taken from \citet{Peng2012},
and the color-magnitude combination is taken from Y$^2$ isochrones.
}
\label{fig:colZ}
\end{figure}

The trend that blue stars are metal poor 
is understood by the color-magnitude relation of 
the stars, and the metallicity gradient along 
the distance from the Galactic disk.
We checked this effect using a simple model.
We adopted a metallicity distribution by \citet{Peng2012}, 
and connected them with intercepts as
\begin{eqnarray}
{\rm [Fe/H]} &=& -0.21 |z| ~~(|z|<2)\nonumber\\
&& -0.16 |z|-0.1 ~~(2\leq|z|<5)\nonumber\\
&& -0.05 |z|-0.65 ~~(5\leq|z|<48)\nonumber\\
&& -3 ~~(48\leq|z|),
\label{eqn:Zzmodel}
\end{eqnarray}
where $z$ is the distance from the Galactic plane in kilo parsec unit.
Then the correlation of the color and the metallicity is
calculated for a certain apparent magnitude.
We used the color and the magnitude from Y$^2$ isochrone
with \citet{Green1987} color models \citep{Yi2001,Demarque2004},
as the model SEDs do not give the radius of the star,
and therefore the calculation of the absolute magnitude is 
difficult to perform.
We adopted [$\alpha$/Fe]=0.3 for [Fe/H]$\leq$-1, and 
[$\alpha$/Fe]=0 for [Fe/H]$>$1.
Note that the Y$^2$ model prepares [$\alpha$/Fe]=0.3 and 0.6,
while the ATLAS9 grids prepare [$\alpha$/Fe]=0.4.
Then, the metallicity, the color and the absolute magnitude 
are obtained from the isochrone.
When we set an apparent magnitude,
the absolute magnitude is converted to the distance,
and then it is converted to $z$ using the Galactic latitude.
If the metallicity is within $\pm 0.1$ of 
equation (\ref{eqn:Zzmodel}), the point is plotted.
In figure \ref{fig:colZ},
we plotted the (B-V) color versus metallicity relation of 
V=14 and V=19 magnitude stars in Galactic latitude of 60 and 80 degree 
fields, which correspond to SXDS and SDF fields.
Note that the color of \citet{Green1987} is in the Johnson UBV system,
and (B-V)$_{\rm AB}$=(B-V)$_{\rm Vega}$+0.10 \citep{Fukugita1995}.
We corrected the color for the magnitude system difference
in figure  \ref{fig:colZ}.
The trend is thus reproduced by a simple model,
and the effect is stronger in fainter magnitude. 

We can therefore think that
``If one tries to match the distribution of faint stars,
which should have low metallicity,
with that of higher metallicity stars (such as BPGS and GS83)
by tuning the ZP,
incorrect shifts might be introduced.''
The second step of the SDF and SXDS calibration 
may have suffered this effect.

In figure \ref{fig:Supcolcol1}
we plot color-color diagrams in Suprime-Cam color
for SDF and SXDS-C.
The stars matched with the SDSS catalog are used, 
and magnitude cut based on SDSS magnitude is applied;
$20.5<g<21.5$ for (X-V) vs (V-Y),
$20<r<21$ for (X-R) vs (R-Y), and 
$19.5<i<20.5$ for (X-i) vs (i-Y), where X=(B,V,R) and Y=(R,i,z).
The magnitude range is the same as that for
the calibration in the previous section.
The filled black circles are the values in the catalog.
The estimated ZP offset in table \ref{tab:ZPdiff} is 
shown as a blue arrow.
The filled circles represent the model colors;
[Fe/H]=0 model in green and [Fe/H]=-2.5a model in red,
as in figure \ref{fig:colcolmodel_BPGS}.

SDF B-band offset is understood by this metallicity effect.
The original calibration follows the [Fe/H]=0 models
and the arrow shows that it should rather follow 
[Fe/H]=-2.5a models.
Since \citet{SXDS2} did not shift the B-band of SXDS,
the distribution is free from the metallicity effect,
and it would be the reason the B-band ZP offset of SXDS 
from SDSS is small ($\leq 0.02$ mag).

\subsubsection{Previous Catalog of SXDS Region}

The calibration of SXDS depends on the calibration by \citet{Ouchi2004}.
We refer to the catalog by \citet{Ouchi2004} as the 
GT-SXDF-catalog hereafter.
The B,V,R and i-band data of GT-SXDF-catalog were obtained 
in November and December 2000. 
Additional R-band data and z-band data
were obtained in September 2001.
We investigated the GT-SXDF-catalog which was used 
for the calibration of SXDS DR1 catalog, whether 
they also show an offset from SDSS.

In November and December 2000, 
the CCD configuration of Suprime-Cam was heterogeneous;
four SITe CCDs, three MIT/LL engineering CCDs with a brickwall pattern
(we call them MIT0),
and two MIT/LL scientific-grade CCDs without a brickwall pattern (MIT1).
The three different types of CCDs have different QE functions.
In April 2001, the MIT0 and SITe chips were replaced with
MIT1 CCDs. The SDF and SXDS DR1 data re-calibrated in this study
were obtained with the MIT1 CCDs.

In figure \ref{fig:GT0},
the SDSS color versus (SDSS)-(Suprime-Cam) magnitude are plotted
using the GT-SXDF-catalog.
We simply classified the stars in the GT-SXDF-catalog
to the nearest CCD group. Therefore, misclassifications and/or
hybrids may contaminate.
In \citet{Ouchi2004}, they adopted the 2 arcsec aperture magnitude 
with 0.2 mag aperture correction, but we adopt MAG\_AUTO 
as in the plot as \citet{SXDS2}.
For figure \ref{fig:GT0}, 
we adopted a slightly different magnitude selection 
from figure \ref{fig:CMD0} as 
$20.5<g<21.5$ for B and V, 
$20<r<21$ for R,
$20.5<i<19$ for i, and $19.5<z<18.5$ for z.
As R-band is a coadd of data taken with old CCDs and MIT1 CCDs,
and MIT1 model is overplotted, 
though the difference is indistinguishable.

The GT-SXDF-catalog was calibrated as the AB magnitude 
with the SITe response as done in \citet{Ouchi2001a},
which is slightly different from the MIT1 response adopted in SXDS.
It might have contributed partly to the offset of ZP of SXDS
in B-band.
In the B-band MIT0 plot, the color slope is different between the model and 
the catalog, suggesting that the CCD response used in the calculation
was wrong at shorter wavelength.
It is reported that the linearity of MIT/LL CCDs was not so good
from October 2000 to December 2000%
\footnote{http://smoka.nao.ac.jp/about/subaru.jsp}%
\footnote{http://anela.mtk.nao.ac.jp/suprime/report/linearity.pdf}.
The effect of the correction on the ZP is unclear, though.

In i and z-band, 
the offset in the ZP from the estimation is already seen
in the GT-SXDF-catalog.
It should be noted that the CCDs used in z-band are the same as 
those for the SXDS catalog, and the offset is not from 
the difference of CCD responses.

Unfortunately, the details of the first step of the 
standard star calibration for the GT-SXDF-catalog were lost 
because of a hardware failure (Ouchi, M., 2012, private communication).
We cannot unveil the history further.

\subsubsection{Standard Star for SXDS i-band}

\begin{figure}
\FigureFile(120mm,90mm){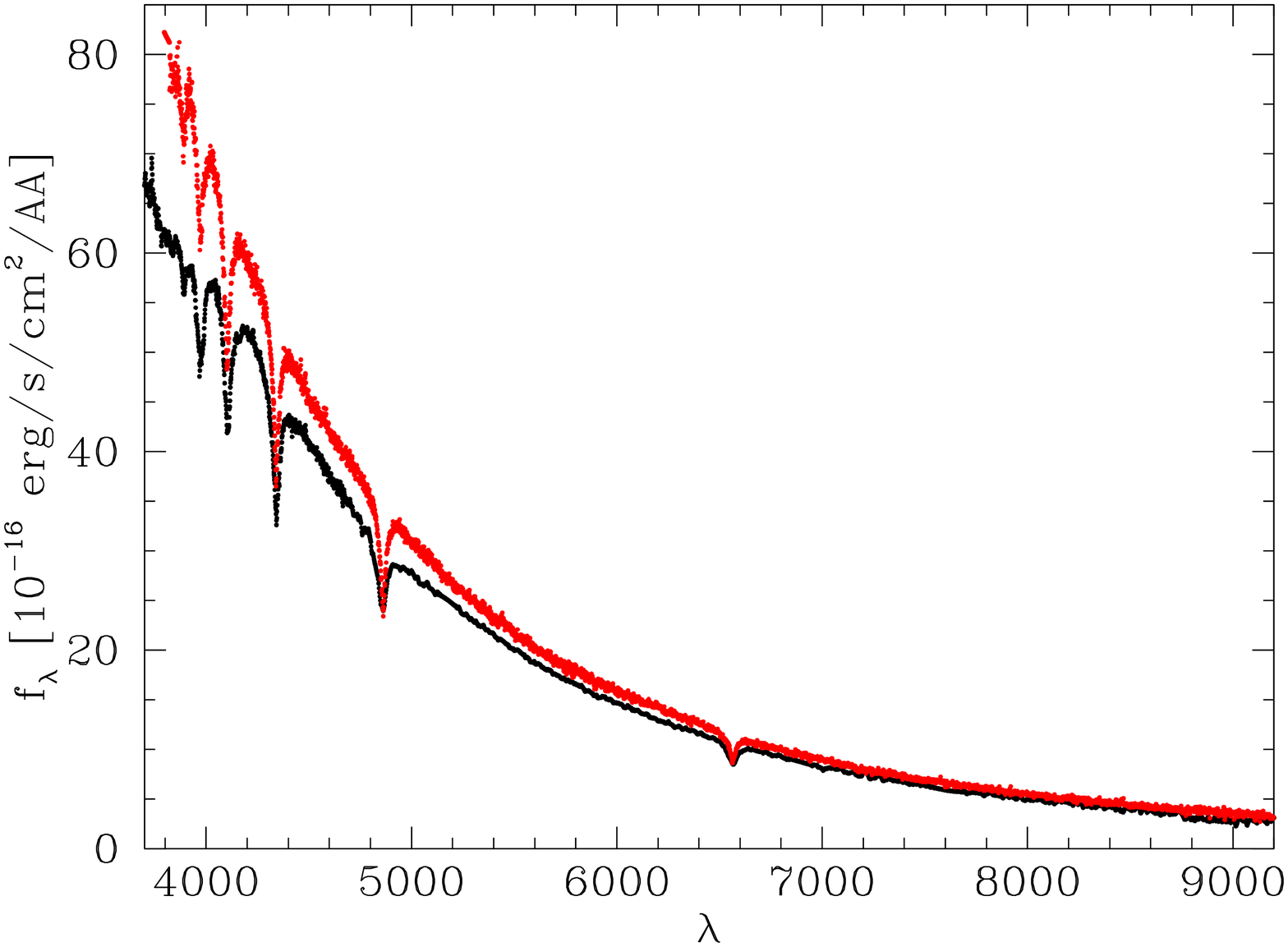}
\caption{
Spectrophotometric spectra of SA95-42,
Black dots are taken from \citet{Oke1990},
and red dots are taken from SDSS DR8. 
}
\label{fig:specSA95-42}
\end{figure}

For i-band in SXDS%
\footnote{
For z-band calibration of SXDS,
\citet{Ouchi2004} and \citet{SXDS2} wrote that
SA95-42 is used. We however found 
that SA95-42 was not observed in z-band when the field was observed.
We guess that GD71 should actually be used in \citet{Ouchi2004}.}, 
synthetic magnitude of SA95-42 
calculated from spectrophotometric data by \citet{Oke1990} 
was used\citep{Ouchi2001a,Ouchi2004,SXDS2},
and the synthetic magnitude of 16.23 was adopted.
We noticed that Oke's spectrum is different from
that of SDSS as shown in figure \ref{fig:specSA95-42}.
The SDSS i-band magnitude calculated from SDSS spectrum (16.11) 
is 0.12 mag brighter.
In SDSS DR8 photometric data, meanwhile,
the magnitude of SA95-42 is 16.16 after the AB-SDSS correction
of 0.028 by Kcorrect v4.
The SDSS i-band AB magnitude of SA95-42 
has thus not converged.
It might be an origin 
of the $\sim$ 0.1 mag dimming of SXDS catalog and GT-SXDF-catalog 
from SDSS.
And since both i-band and z-band had offsets in the ZP in the SXDS catalog, 
the offsets in both bands
would not have 
been fully corrected in the comparison with the GS83 distribution.

\subsubsection{The i-band and z-band Magnitude of Standard Stars for SDF}

The spectrophotometric standard stars used 
in SDF for i and z-band are
HZ21, HZ44, GD153, P177D, and P330E \citep{SDF}.
They are originally from \citet{Oke1990,Bohlin2001,Bohlin2003},
and are available at STSci sites%
\footnote{http://www.stsci.edu/hst/observatory/cdbs/calspec.html}%
\footnote{ftp://ftp.stsci.edu/cdbs/oldcalspec/}.
The stars do not have SDSS spectra, but have psfMag in SDSS DR8.
The difference of the i and z-band PSF magnitudes
of SDSS DR8 and the synthetic magnitudes using \citet{Doi2010}
transmission and the SED
are shown in table \ref{SDFstandards}.
The offset given by Kcorrect v4 is applied to SDSS magnitude 
in the table.
The error shows the SDSS catalog error (psfMag\_err).
The version of the SED used for the analysis of SDF
and adopted magnitude were lost 
and we cannot investigate the historical detail.
However, the difference is mostly too small ($<$0.04 mag, 
except for z of HZ44 and i of GD153)
to explain the 0.1 mag offset we found.
Moreover, the sign of the GD153 offset in i-band (+0.14 mag) is 
opposite to the ZP offset we found.
In our analysis, SDF stars are fainter than SDSS, 
while the synthetic magnitude of GD153 is brighter than SDSS.
Therefore, the i-band and z-band magnitude difference of SDF 
may not be explained only 
by some error of spectrophotometric standards.

\section{Summary}

Using SDSS,
we found an offset of the ZP of SDF and SXDS.
We present a robust color conversion from SDSS to
the Suprime-Cam system. The color conversion is robust against 
the difference in metallicity, Galactic extinction,
atmospheric extinction, and recession velocity of the stars.
And the offset of ZP is then calculated against SDSS.

If we applied the correction of the offset we obtained,
the difference in the color distribution of faint ($R\sim 24$) objects in 
SDF and SXDS disappears. The result supports that 
the relative color difference between SDF and SXDS is 
corrected well by our result.
And since the relative color offset of SDF and SXDS 
of bright stars ($R\sim 20$) and at the faint object 
are consistent, it should be explained by a simple offset 
of the ZP.

The B-band offset of SDF would be explained by the difference of 
the color of stars in the data and the adopted references,
due to the different metallicity.
The i and z-band of SXDS inherited the offset in the previous catalog.
The magnitude of i-band standard star SXDS adopted (SA95-42) 
has a offset ($\gtrsim$0.07) from SDSS,
which may be the origin of the SXDS i-band offset.
The origin of the offset of SDF i-band and
SDF/SXDS z-band are yet to be understood.

\bigskip

We thank the anonymous referee for 
valuable comments and suggestions.
We appreciate Masayuki Tanaka, Masami Ouchi, 
Nobunari Kashikawa, and Kazuhiro Shimasaku
for their suggestive comments, indications, and information.
The SDF catalog and SXDS catalog are obtained from 
Astronomy Data Center (ADC) at the 
National Astronomical Observatory of Japan.
This work has made use of the SDSS database
and the computer systems at ADC.


\onecolumn

\clearpage 

\begin{figure}
\FigureFile(80mm,60mm){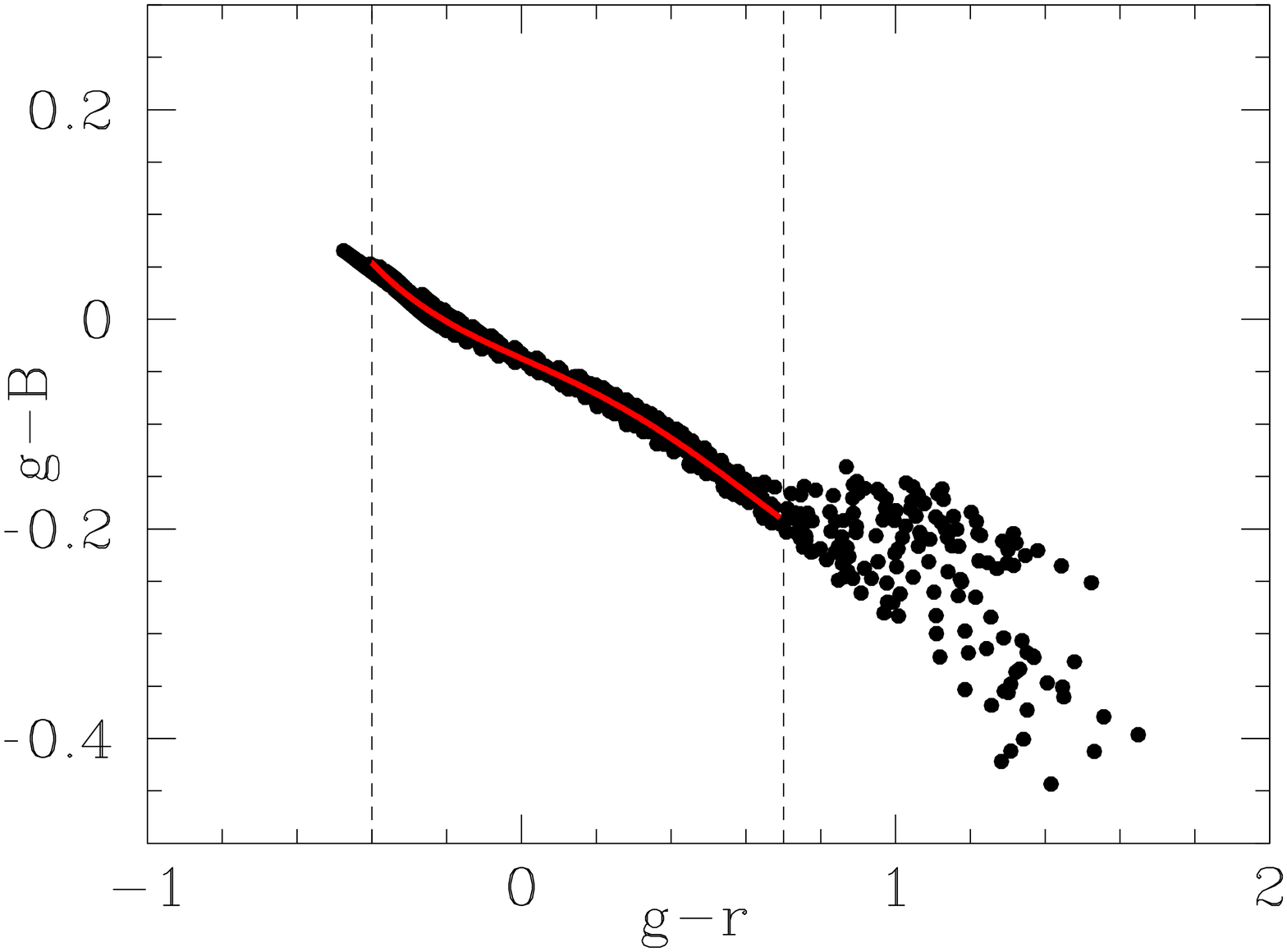}
\FigureFile(80mm,60mm){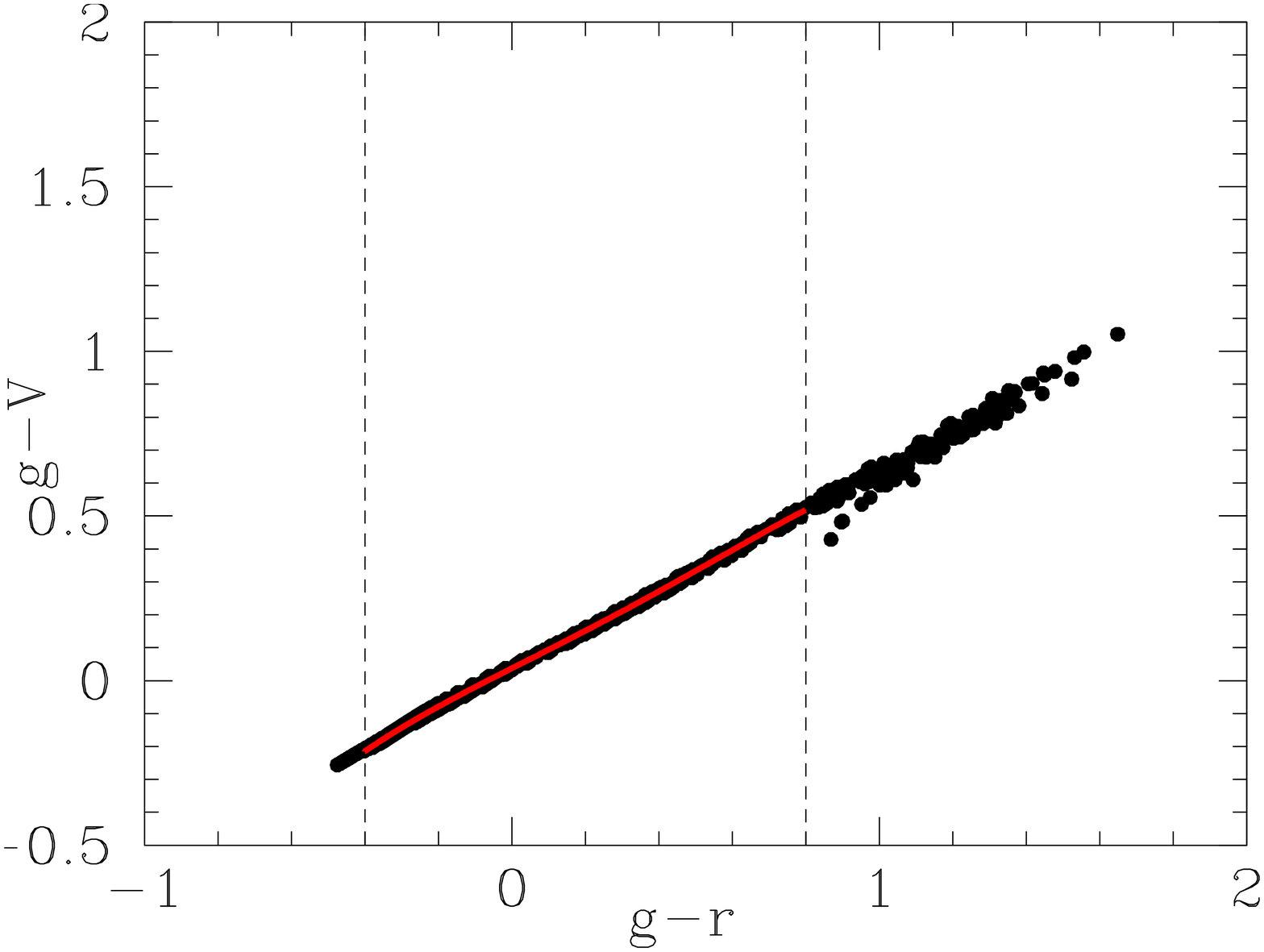}\\
\FigureFile(80mm,60mm){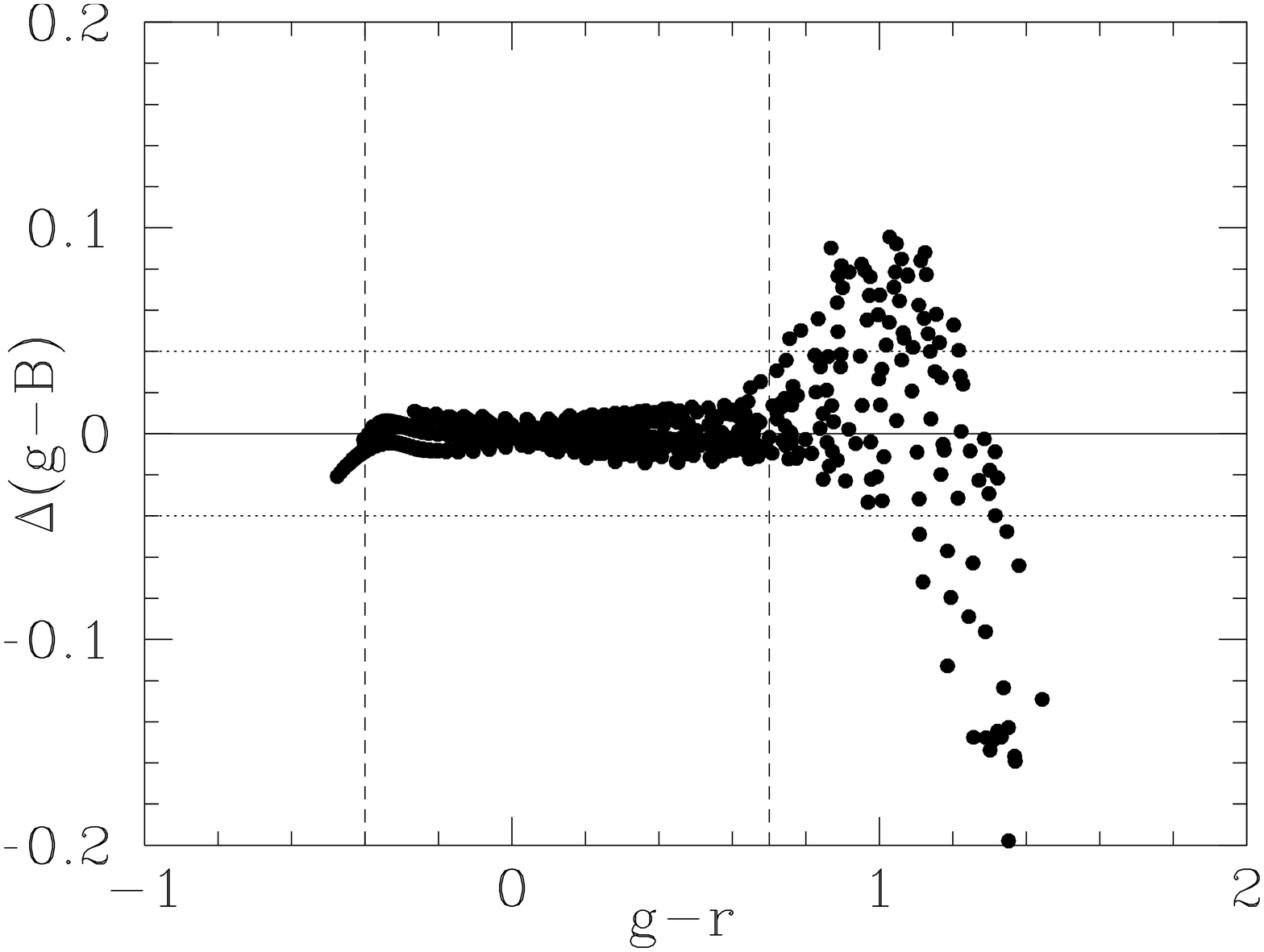}
\FigureFile(80mm,60mm){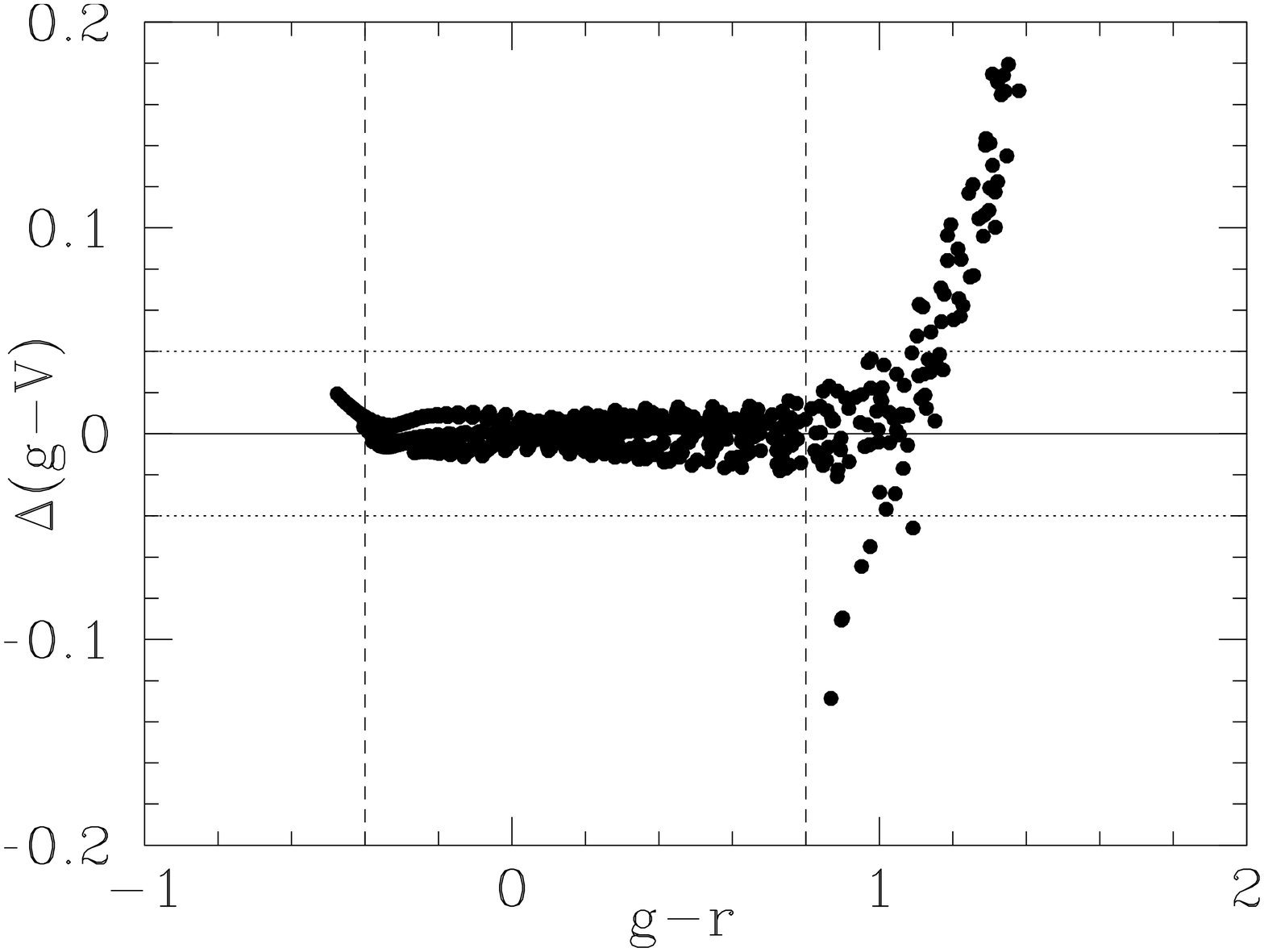}\\
\FigureFile(80mm,60mm){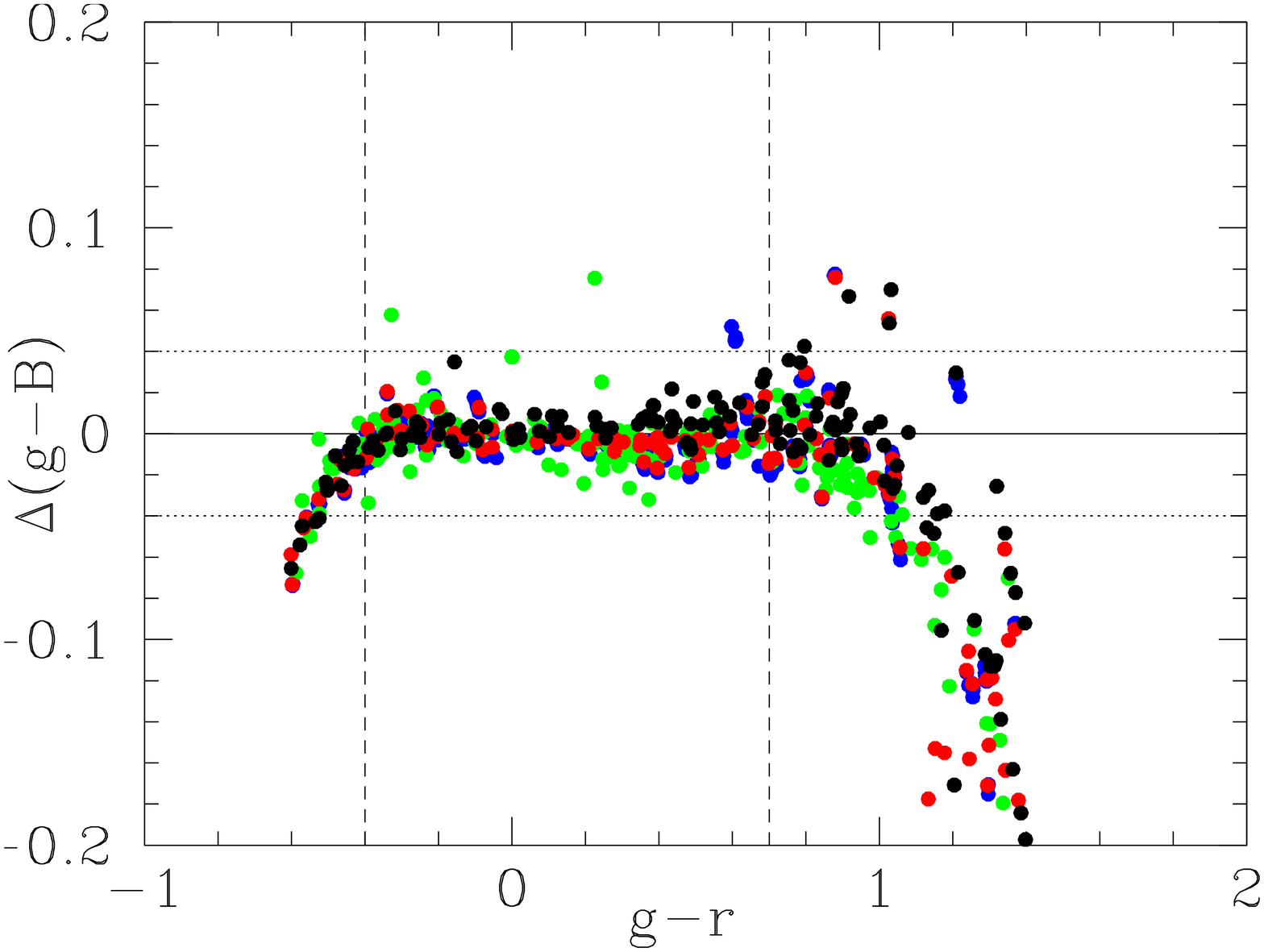}
\FigureFile(80mm,60mm){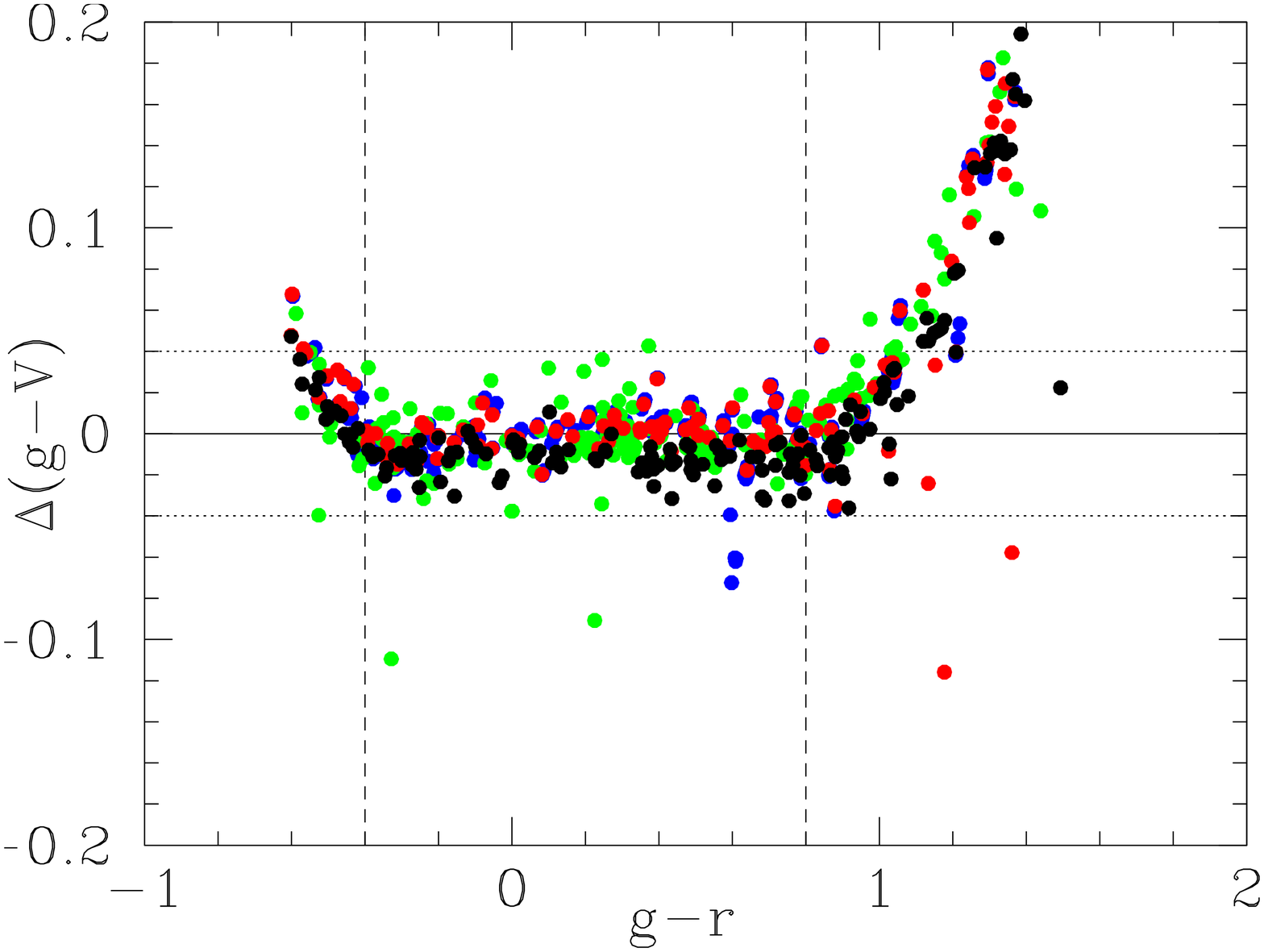}
\caption{
(top) Fitting of polynomial to the synthetic color of 
SDSS minus Suprime-Cam color as a function of SDSS color. 
The model adopts atmospheric extinction at airmass=1, 
and no Galactic extinction.
The red line represents the fit function 
and the dots are the model colors.
(middle) Residual of model color from the best-fit function.
Vertical lines represent the fitting color range.
Horizontal lines are 0 and $\pm$0.04 mag for reference.
The filled black circles are ATLAS9 models.
(bottom) Residual of SDSS minus Suprime-Cam color of various SEDs;
CFLIB, STELIB, HILIB, and BPGS in blue, green, red and  black, 
respectively. The vertical and the horizontal lines are the 
same as the middle panel.
}
\label{fig:fit0}
\end{figure}

\clearpage 

\begin{figure}
\FigureFile(80mm,60mm){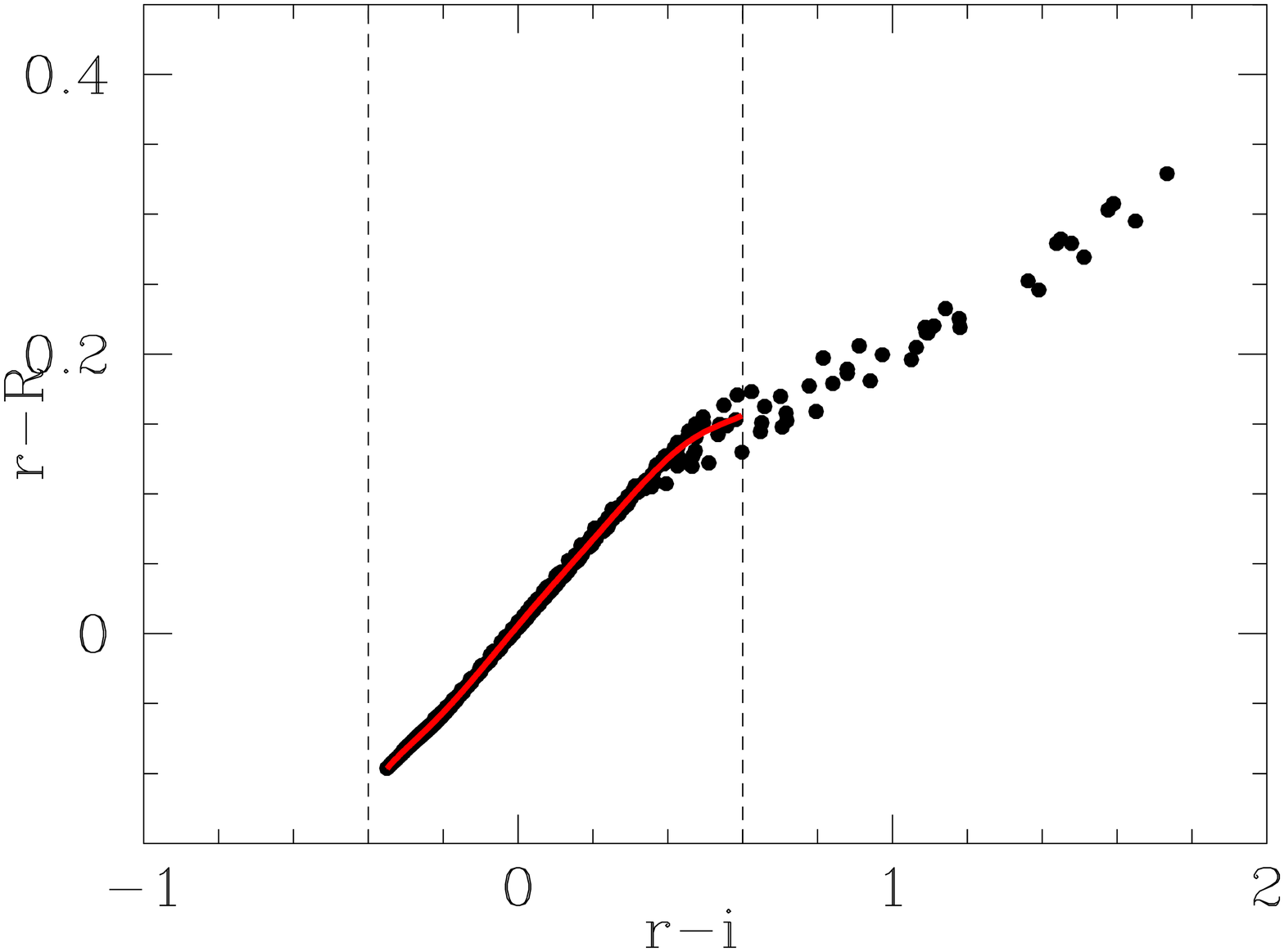} \FigureFile(80mm,60mm){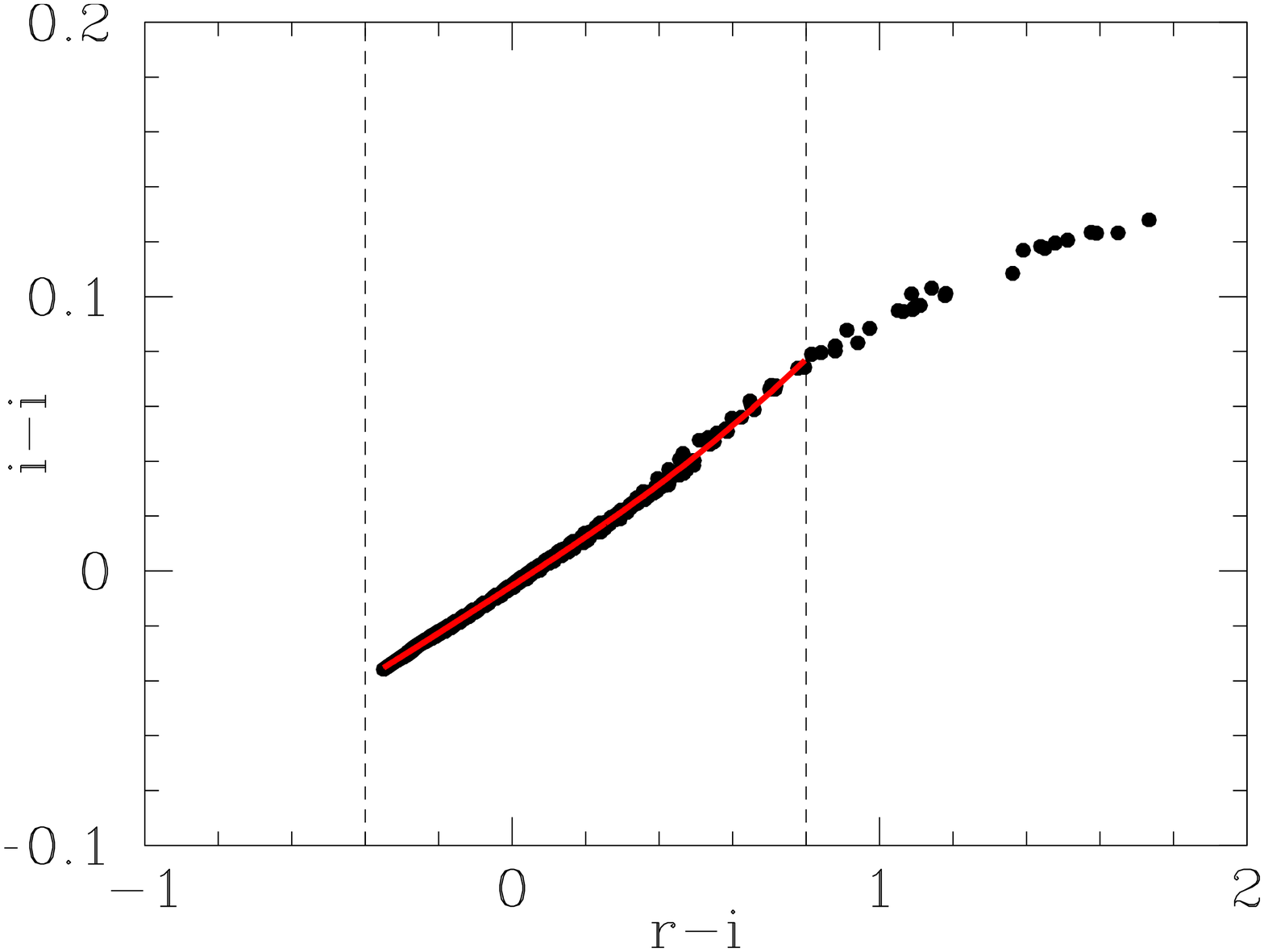}\\
\FigureFile(80mm,60mm){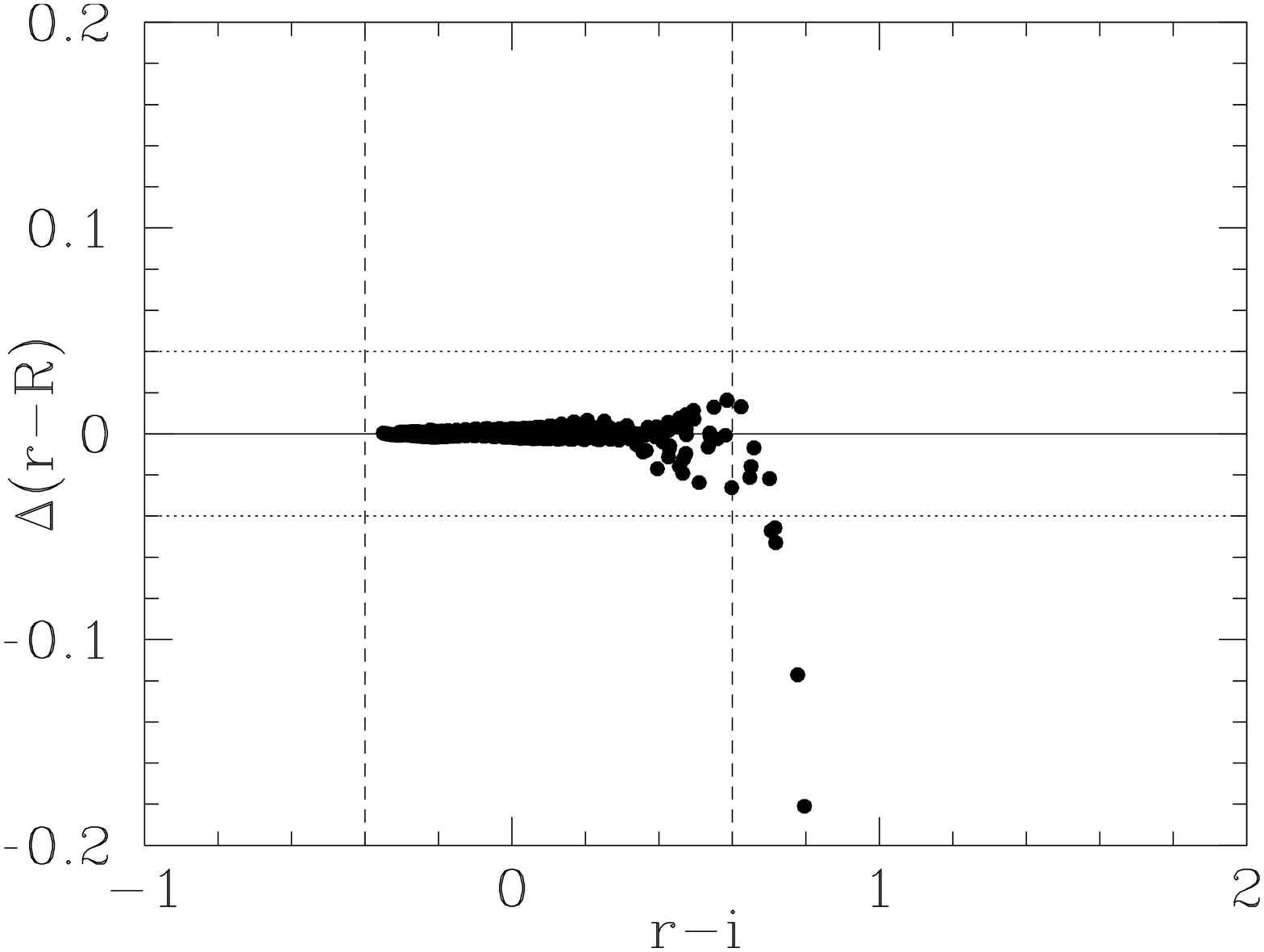} \FigureFile(80mm,60mm){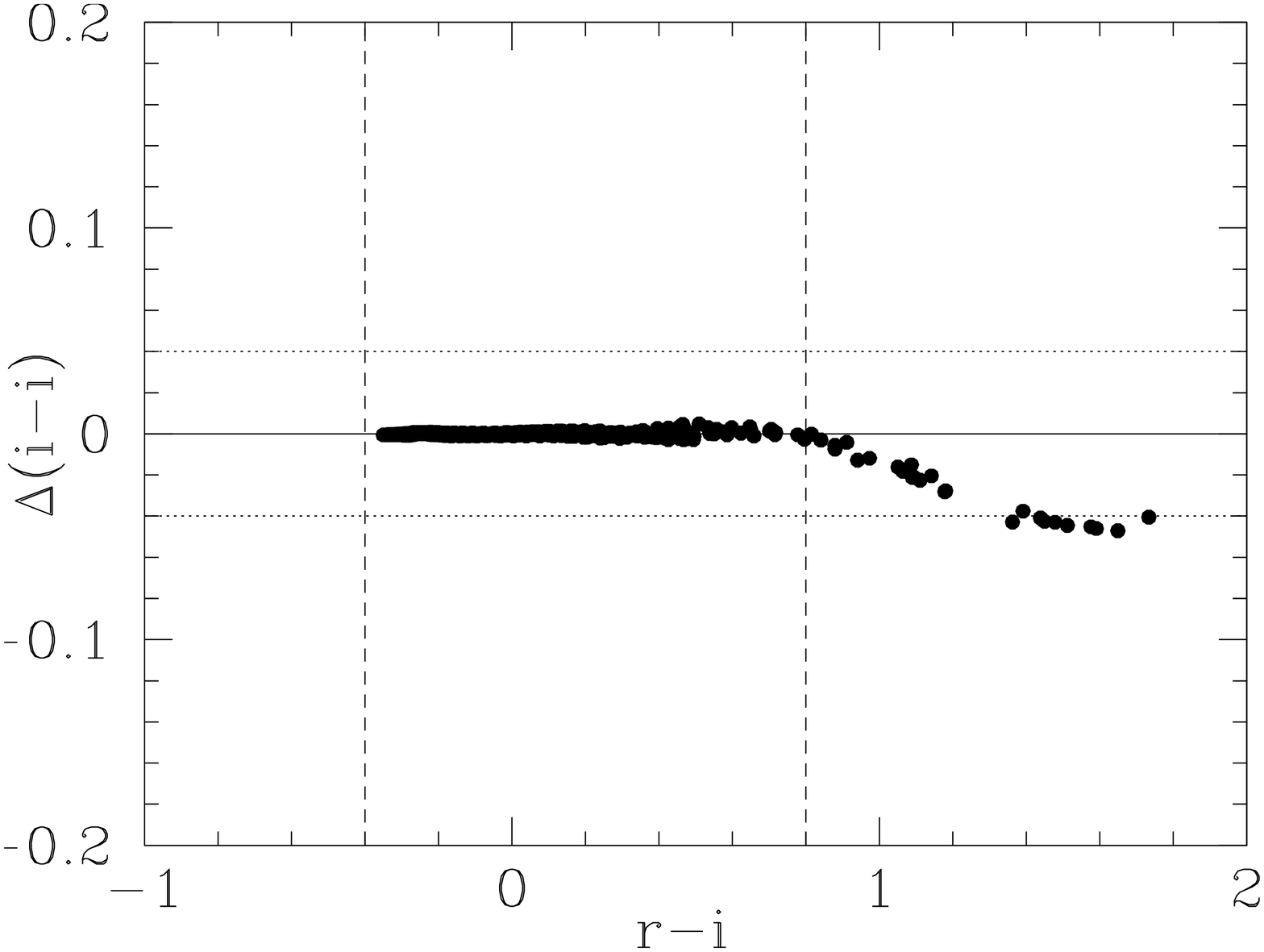}\\
\FigureFile(80mm,60mm){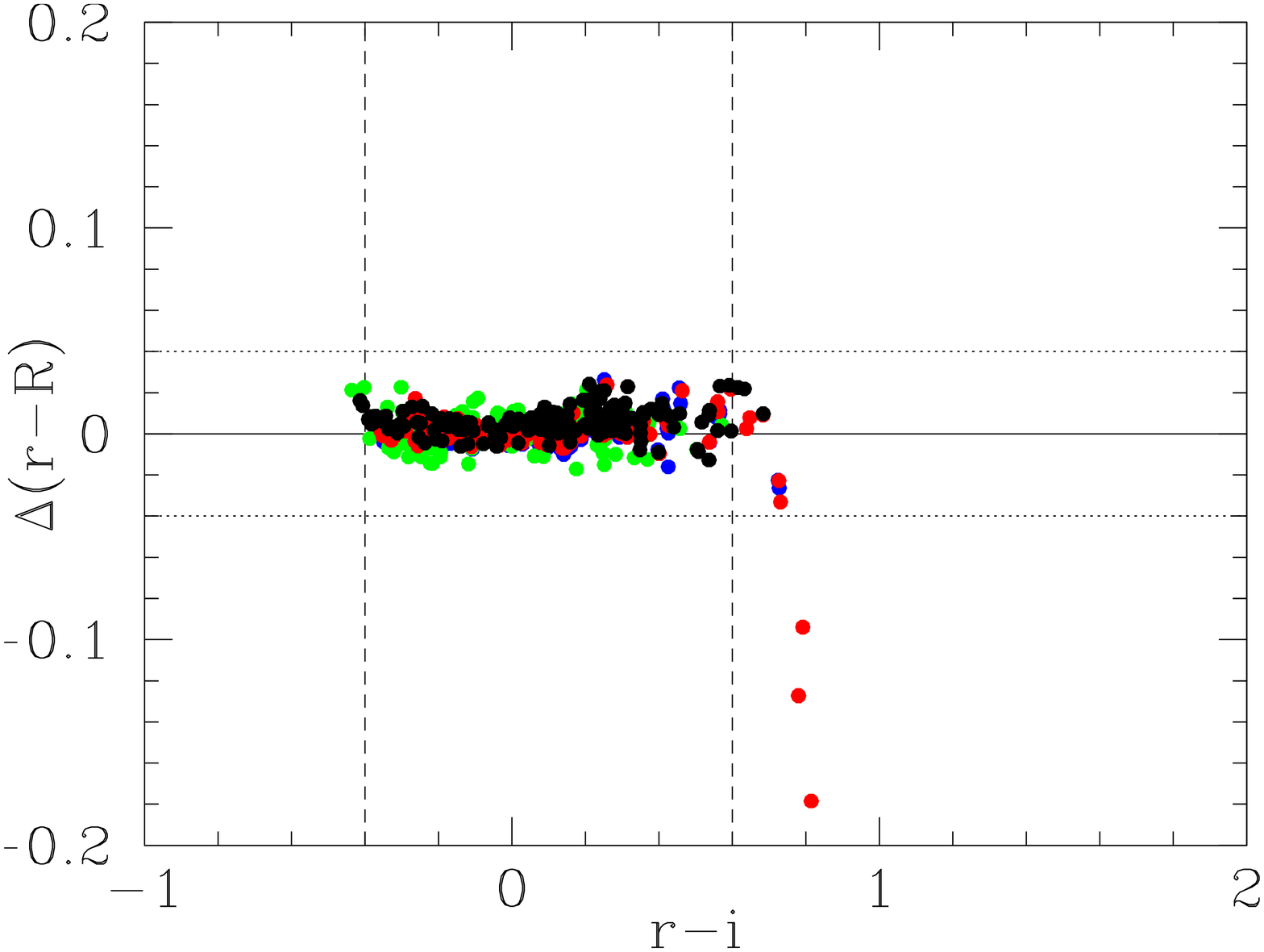} \FigureFile(80mm,60mm){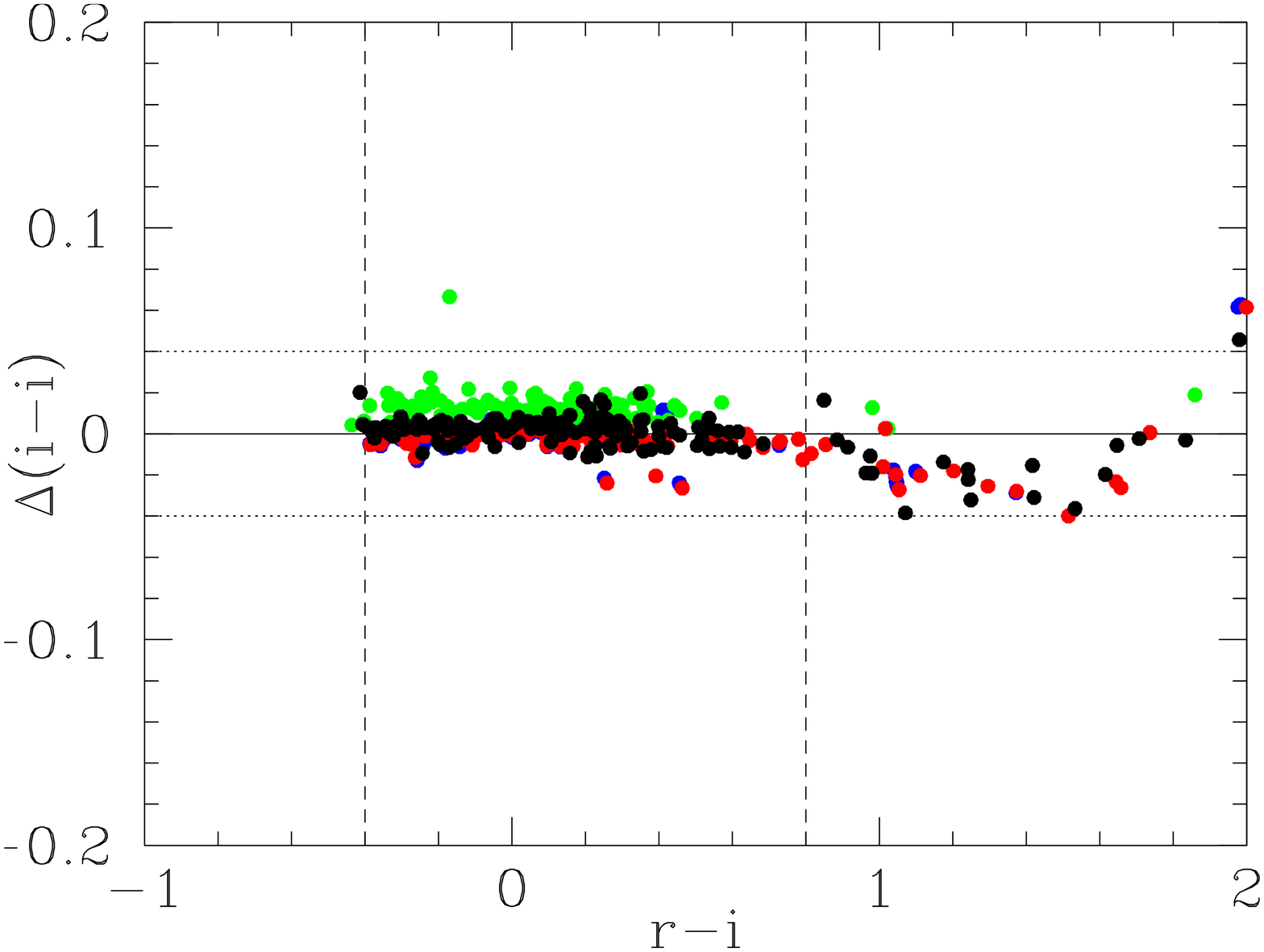}\\
\addtocounter{figure}{-1}
\caption{
Continued...
}
\end{figure}

\clearpage 

\begin{figure}
\FigureFile(80mm,60mm){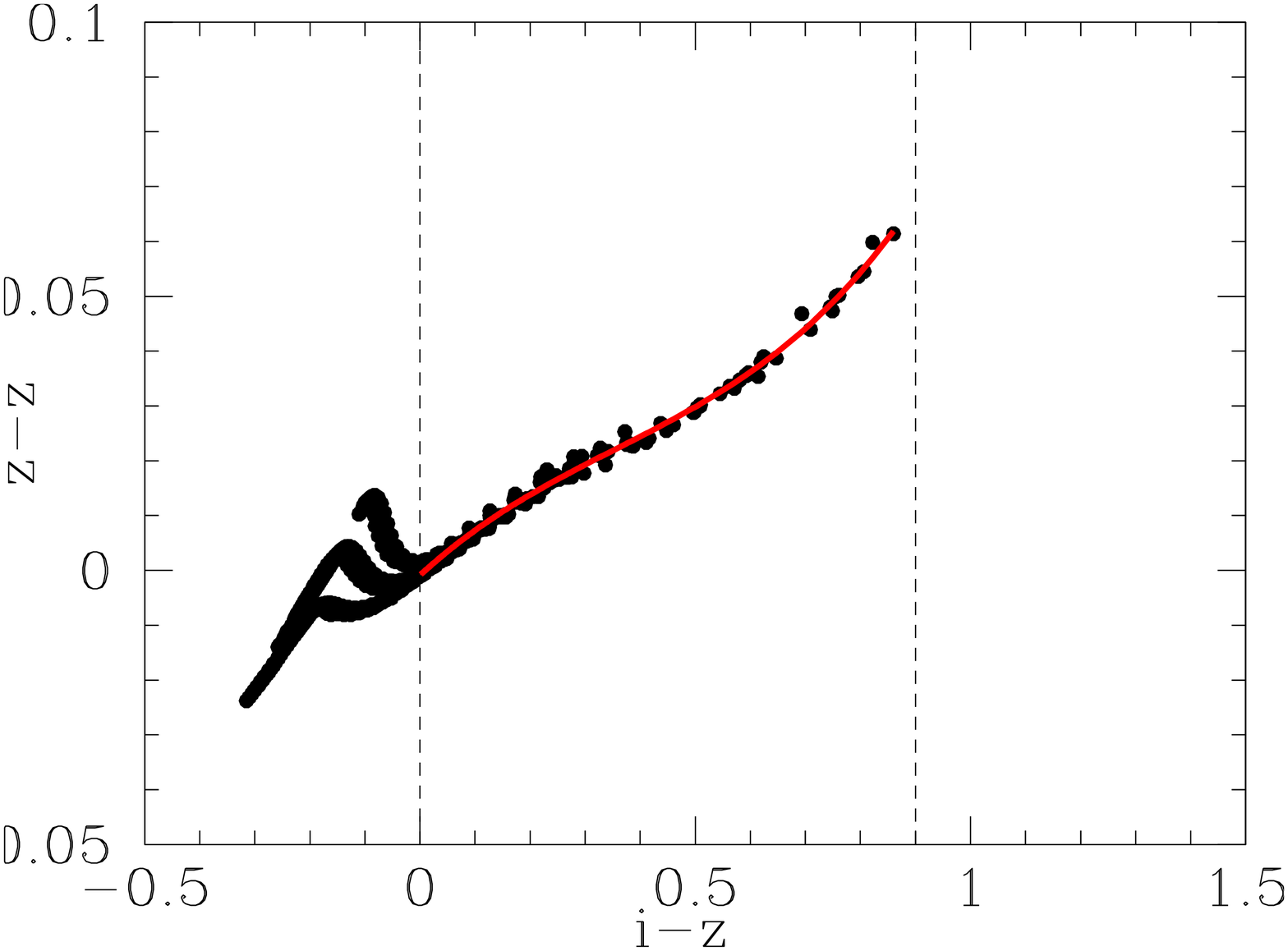}\\
\FigureFile(80mm,60mm){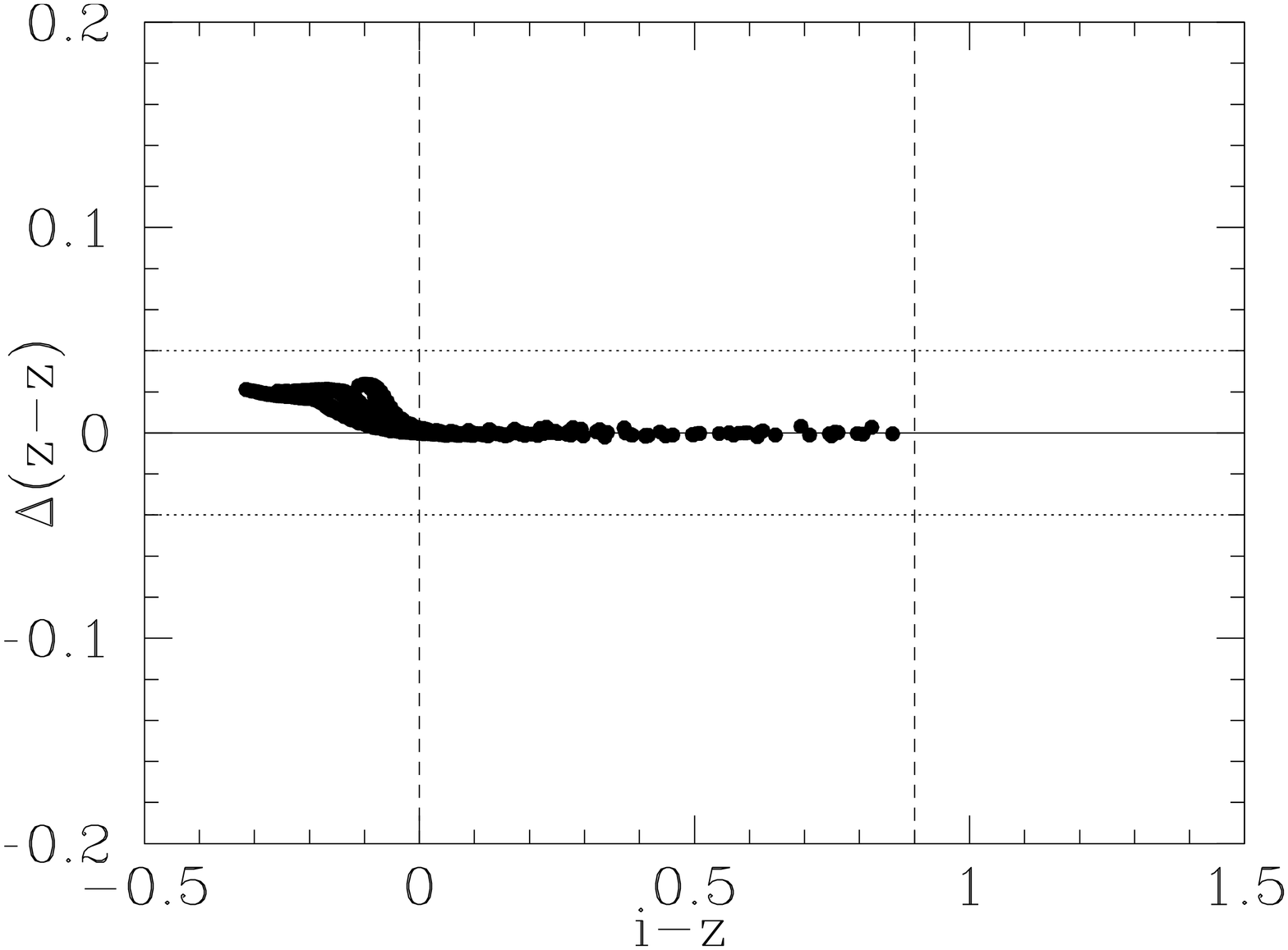}\\
\FigureFile(80mm,60mm){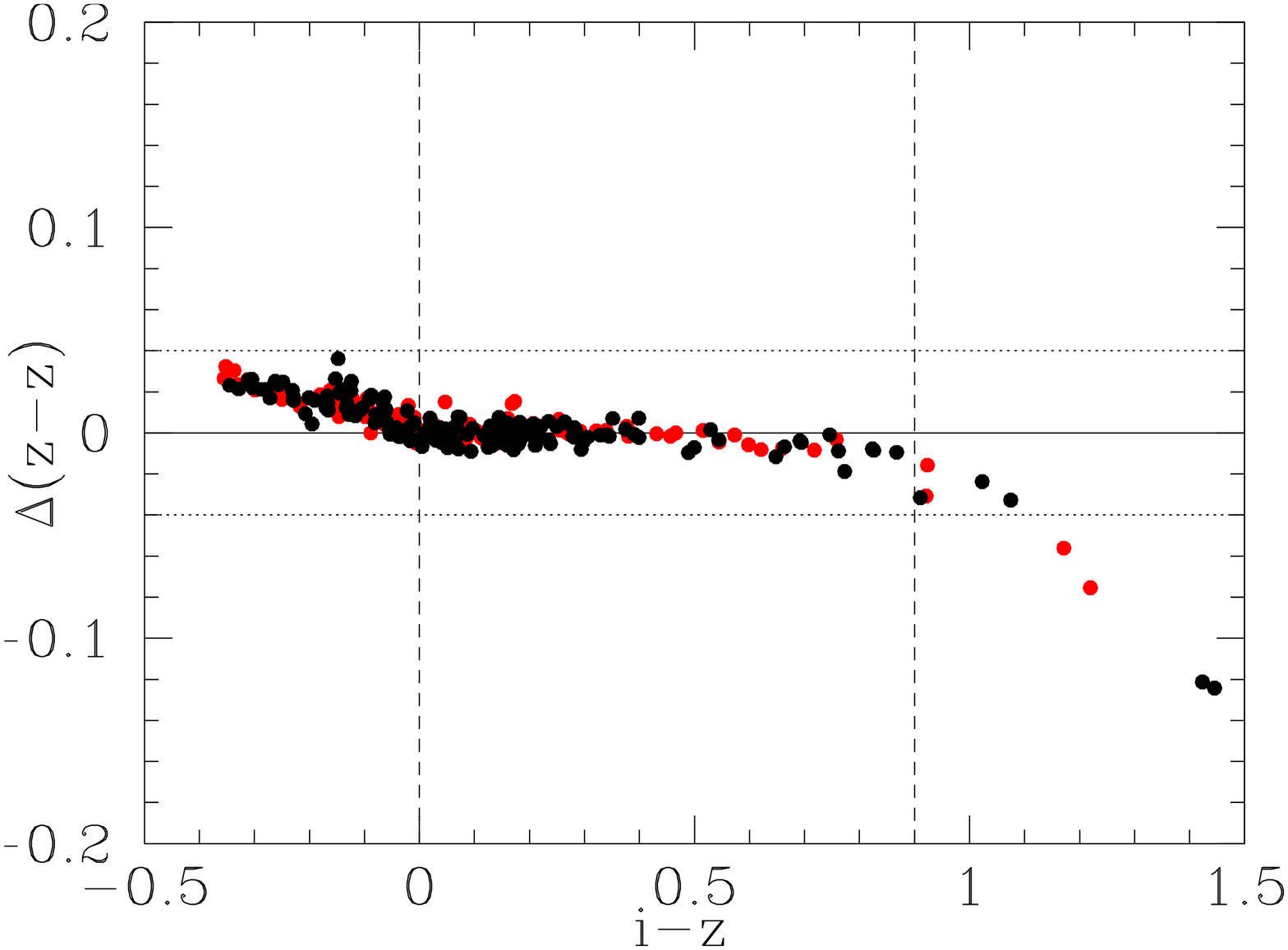}\\
\addtocounter{figure}{-1}
\caption{
Continued...
}
\end{figure}

\clearpage 

\begin{figure}
\FigureFile(80mm,60mm){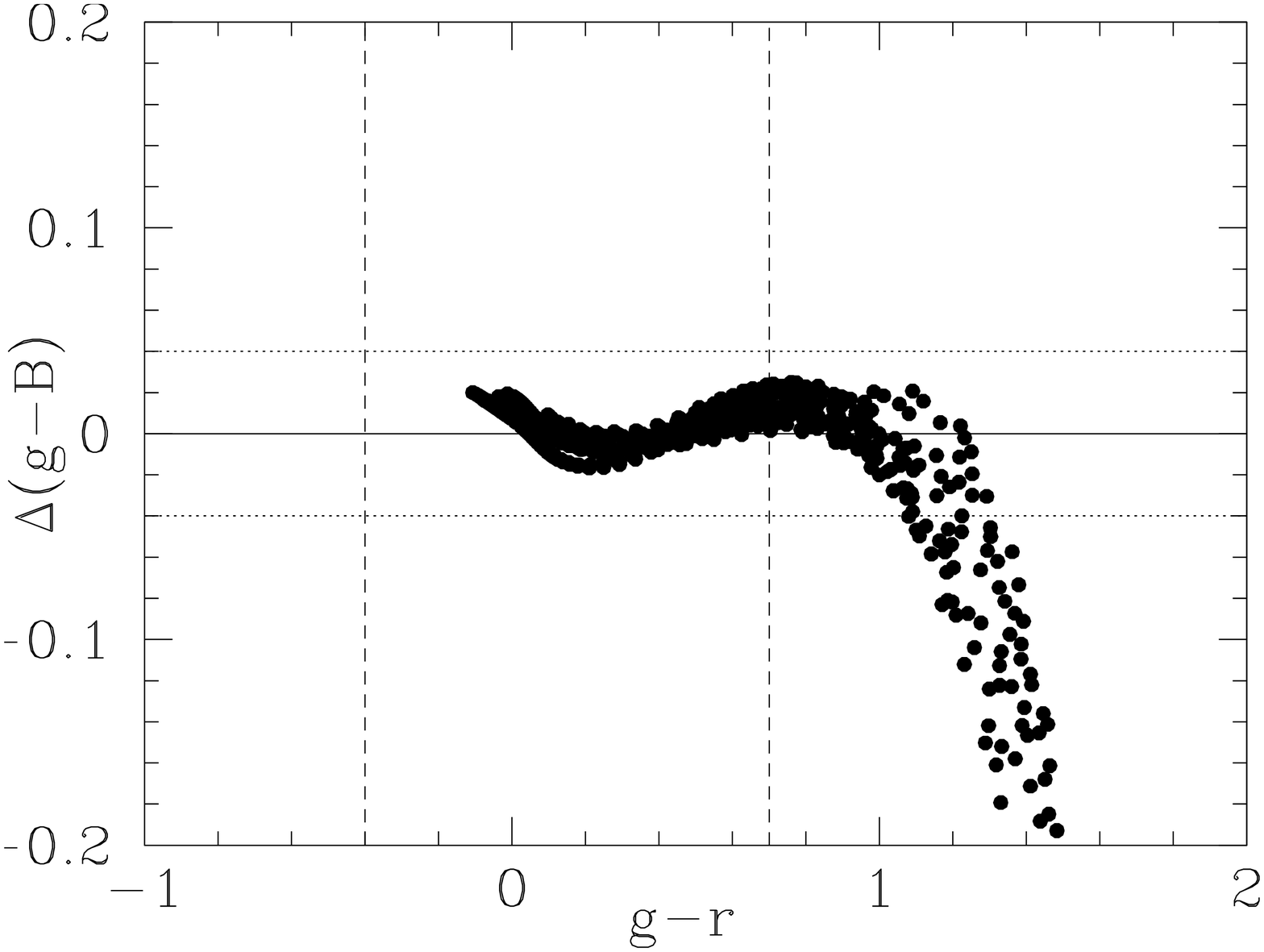}
\FigureFile(80mm,60mm){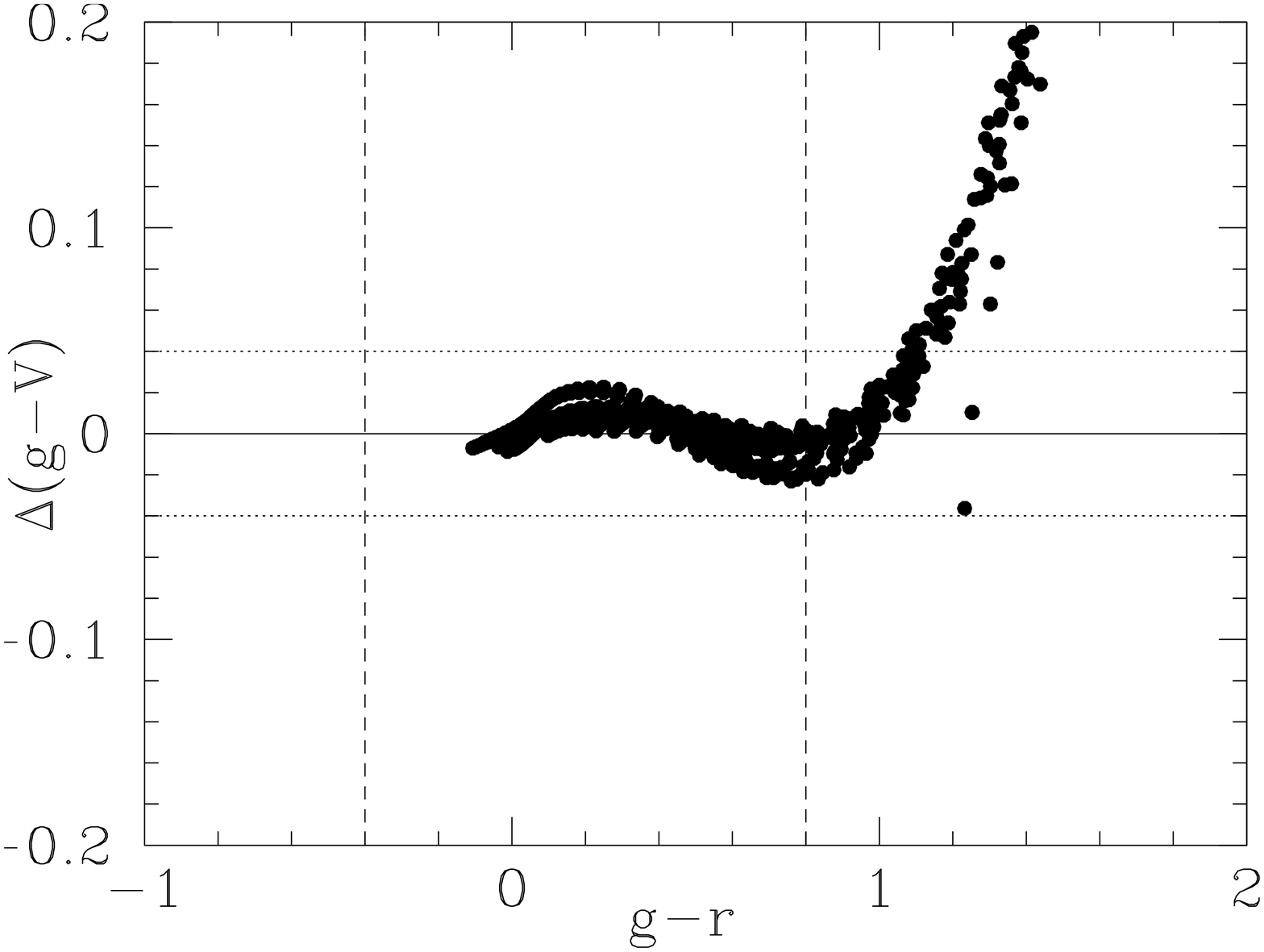}\\
\FigureFile(80mm,60mm){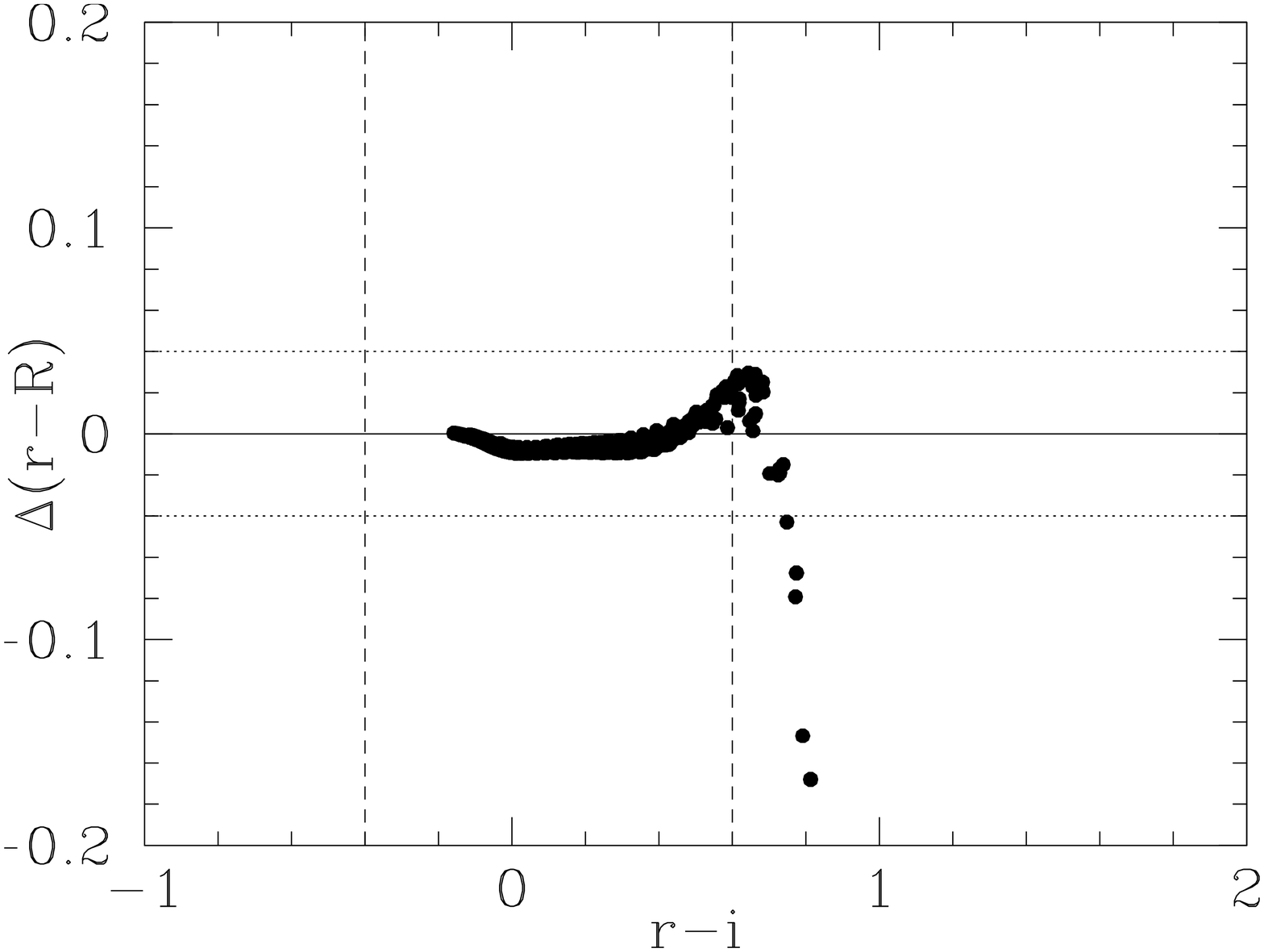}
\FigureFile(80mm,60mm){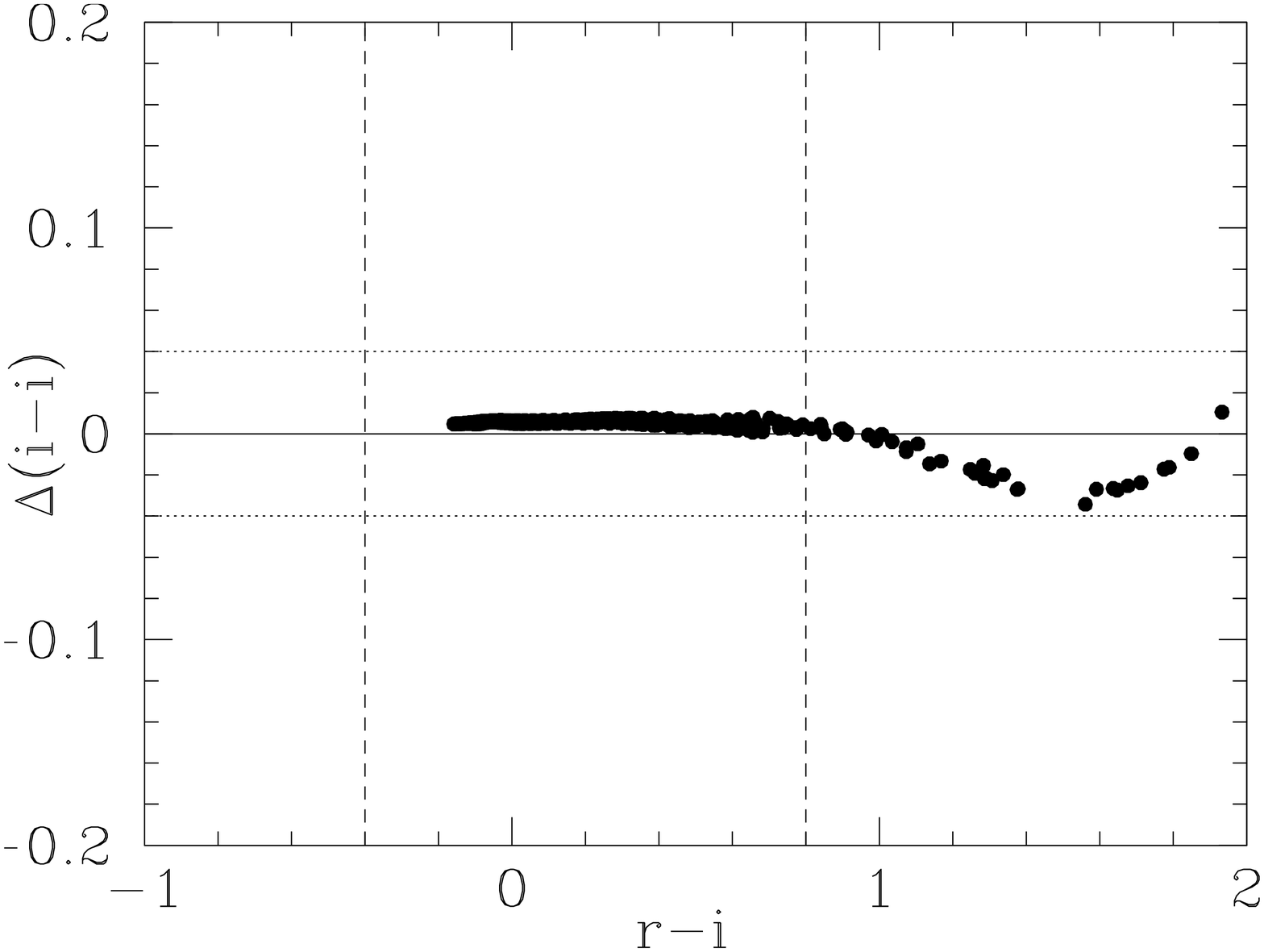}\\
\FigureFile(80mm,60mm){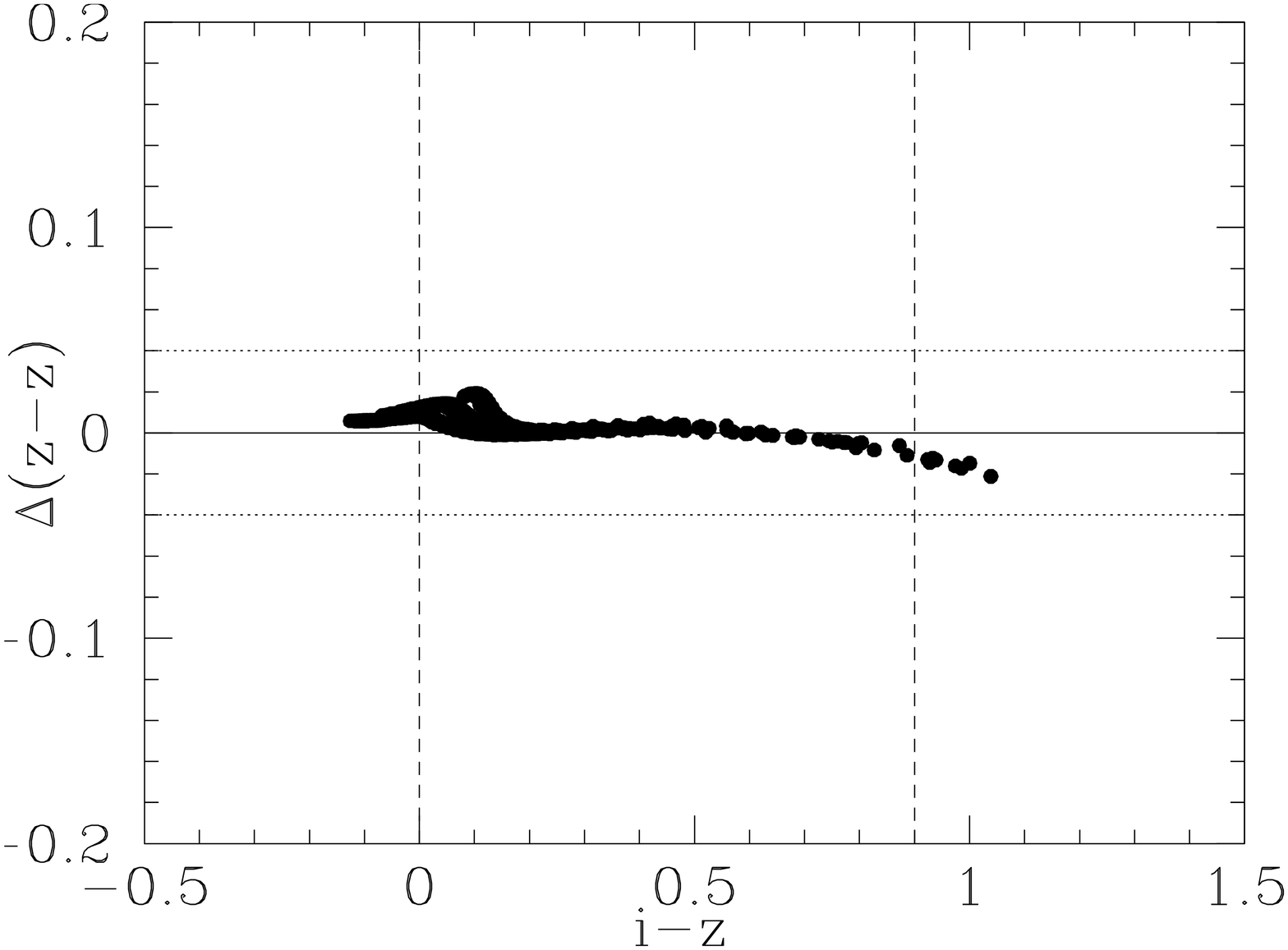}
\caption{
Same as the middle panel of figure \ref{fig:fit0}.
but an extreme Av=1 mag extinction is applied to the model.
}
\label{fig:Av1}
\end{figure}

\clearpage 

\clearpage 

\clearpage 

\clearpage 

\clearpage 

\clearpage

\begin{figure}
\FigureFile(80mm,60mm){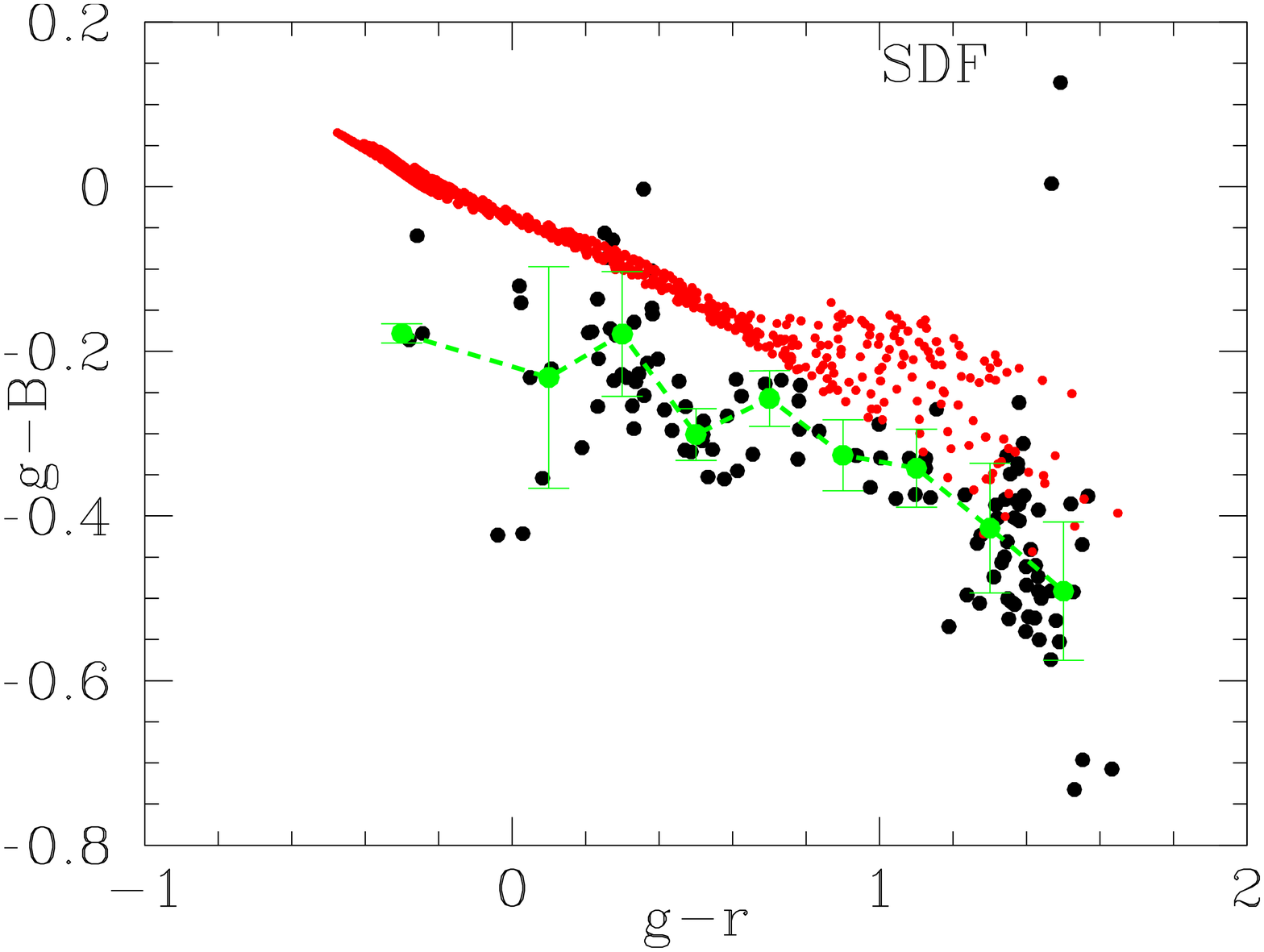}
\FigureFile(80mm,60mm){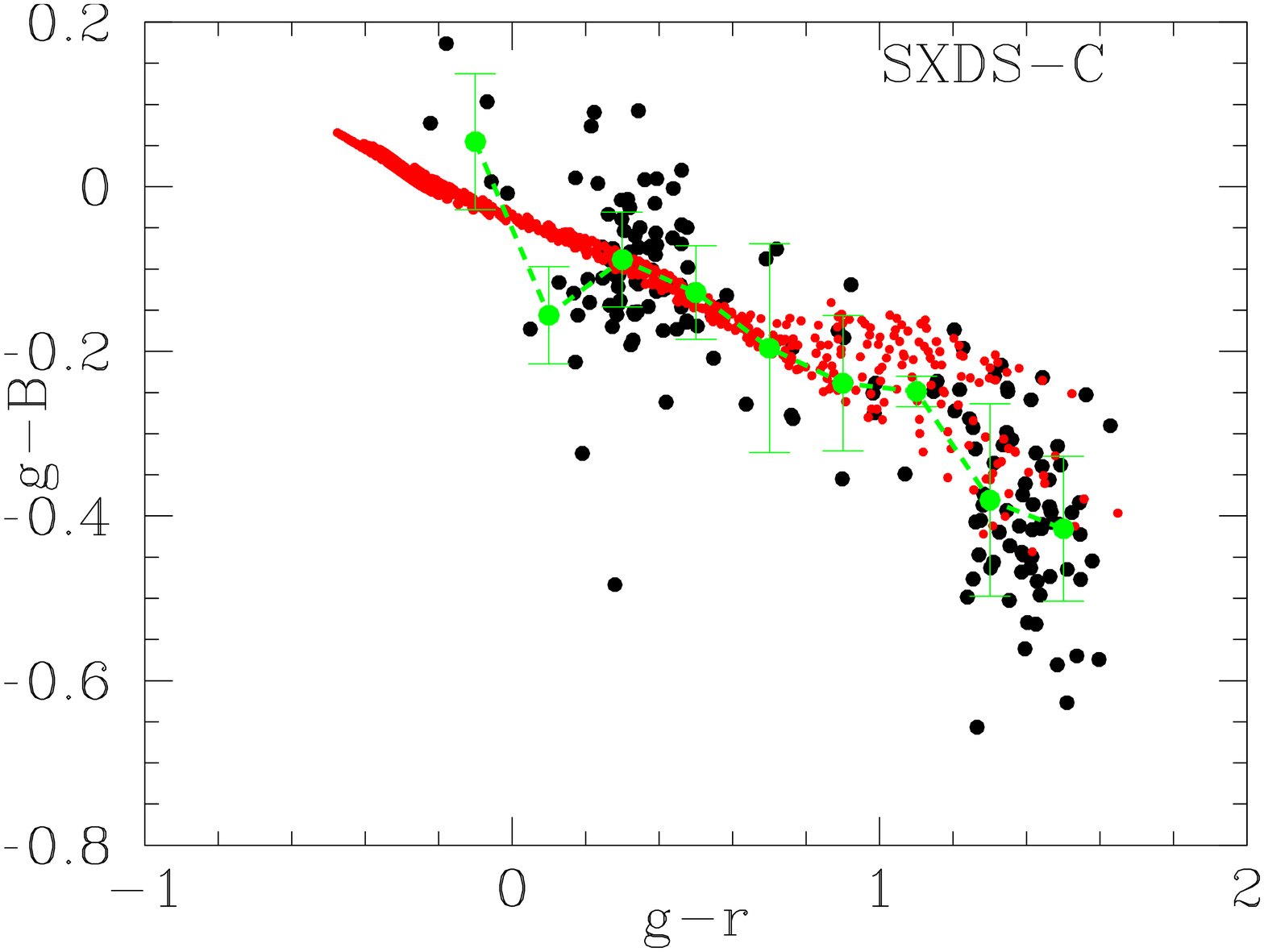}\\
\FigureFile(80mm,60mm){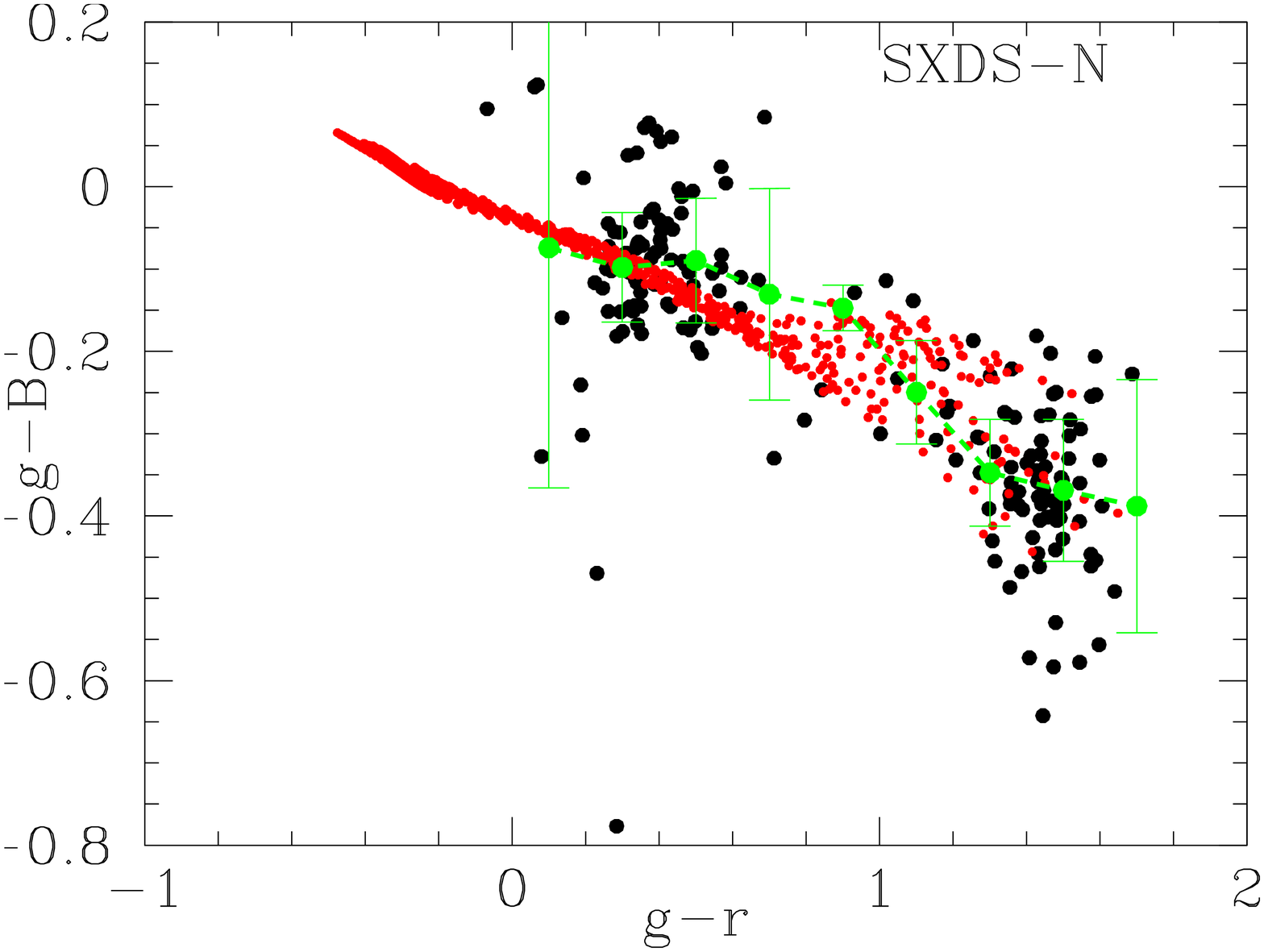}
\FigureFile(80mm,60mm){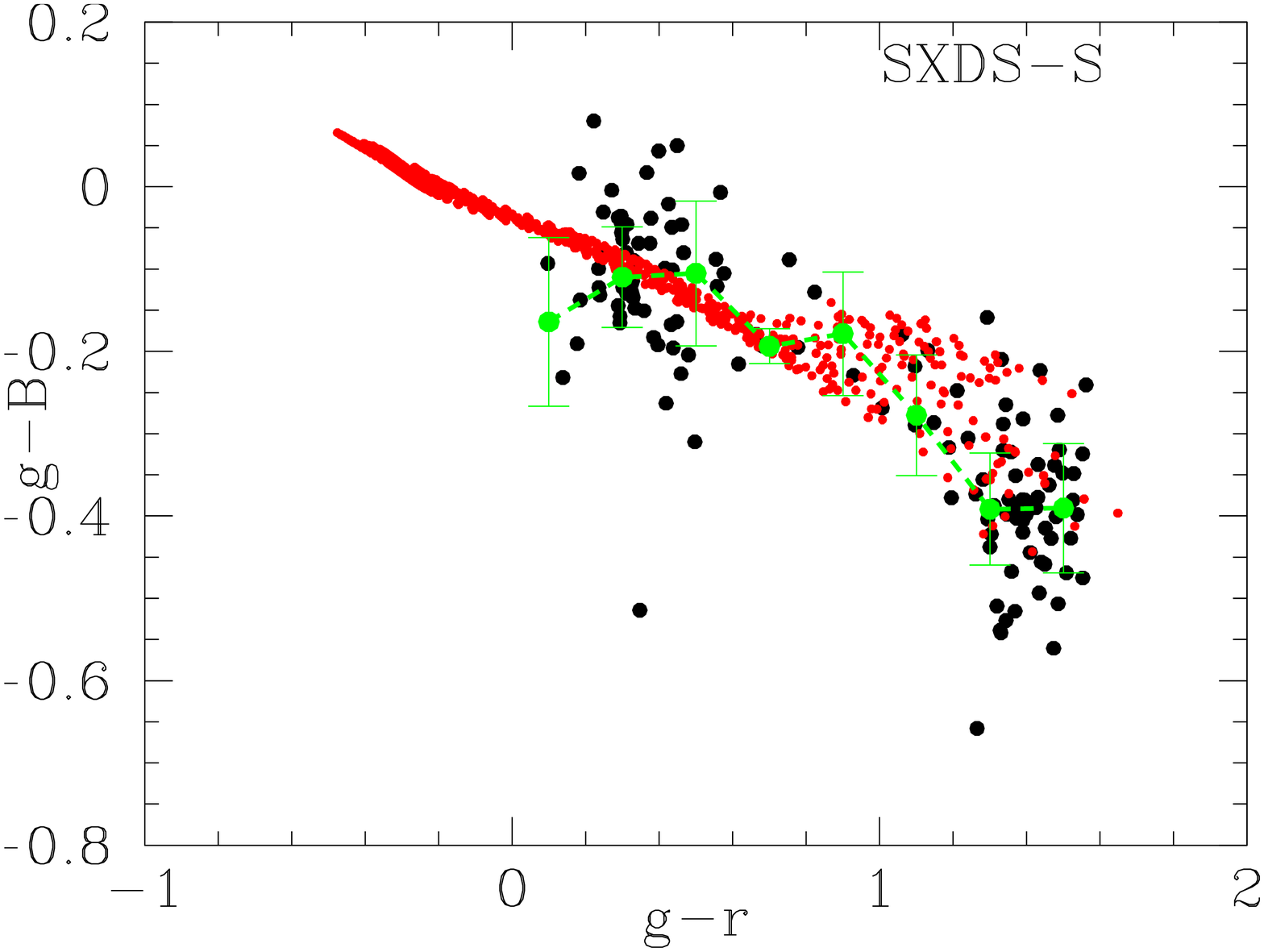}\\
\FigureFile(80mm,60mm){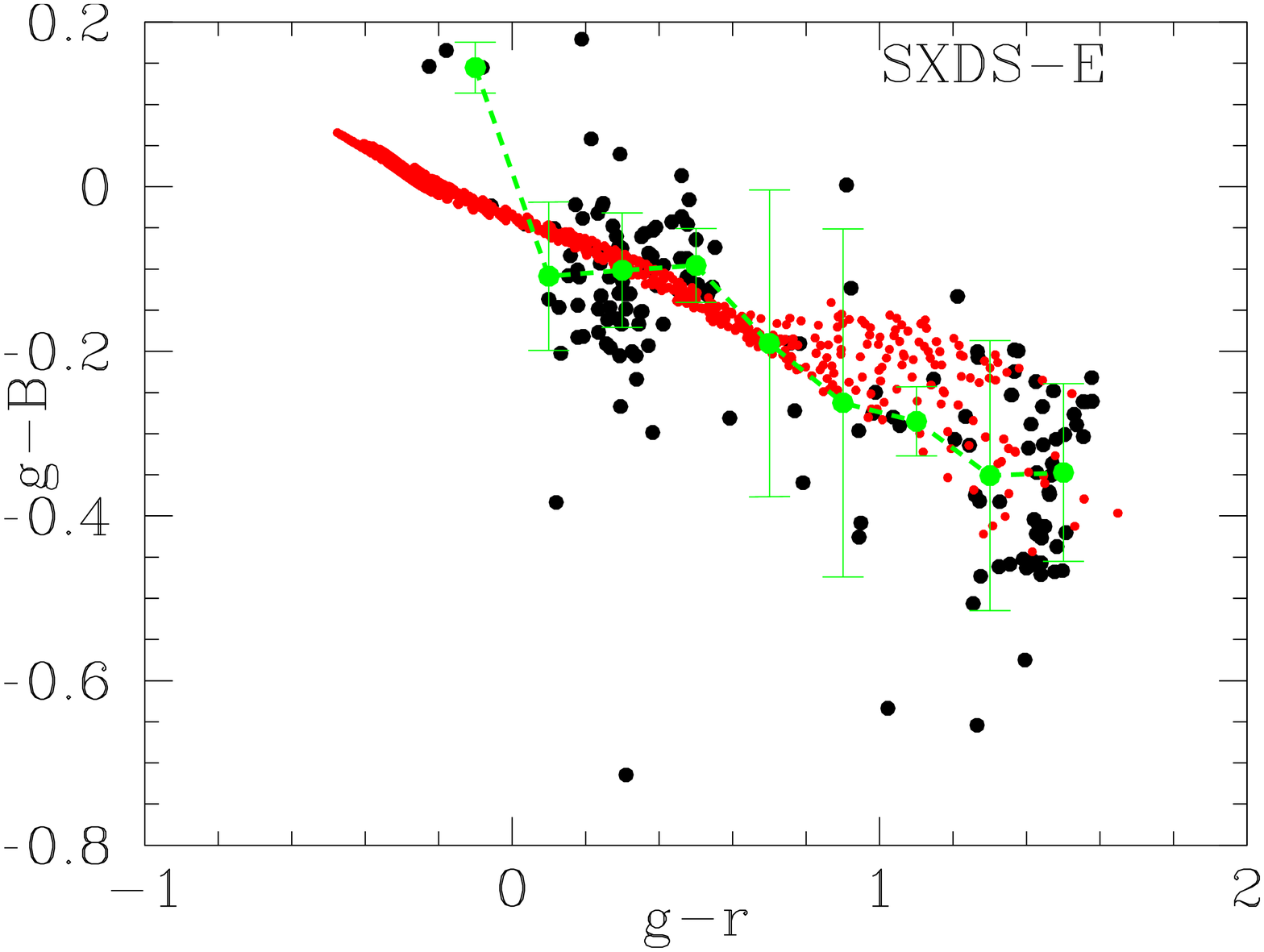}
\FigureFile(80mm,60mm){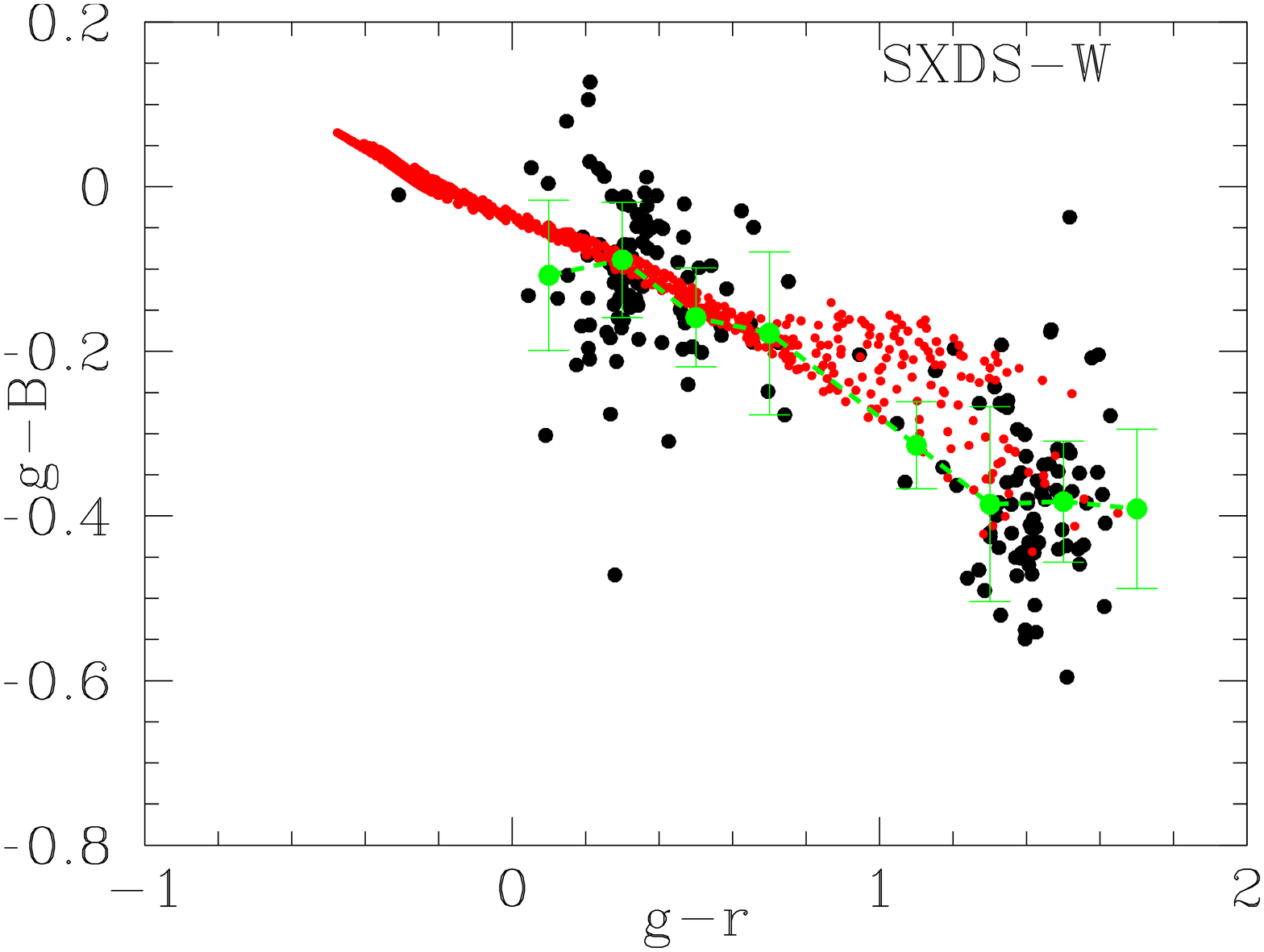}
\caption{
SDSS color versus the (SDSS)-(Suprime-Cam) color.
Top-left panel is SDF and other panels are SXDS.
The filled circles represent matched stars.
The magnitude ranges are 
$20.5<g<21.5$ for B and V, 
$20<r<21$ for R,
$19.5<i<20.5$ for i,
and $19<z<20$ for z.
The filled red circles are ATLAS9 model colors, which are shown as
filled black circles in left panels of figure \ref{fig:fit0}.
The filled green circles with errorbar show 
the median of the (SDSS)-(Suprime-Cam) color
in 0.2 mag bin of SDSS color. 
The errorbar represents rms estimated from MAD of the bin. 
}
\label{fig:cal0}
\end{figure}

\clearpage

\begin{figure}
\FigureFile(80mm,60mm){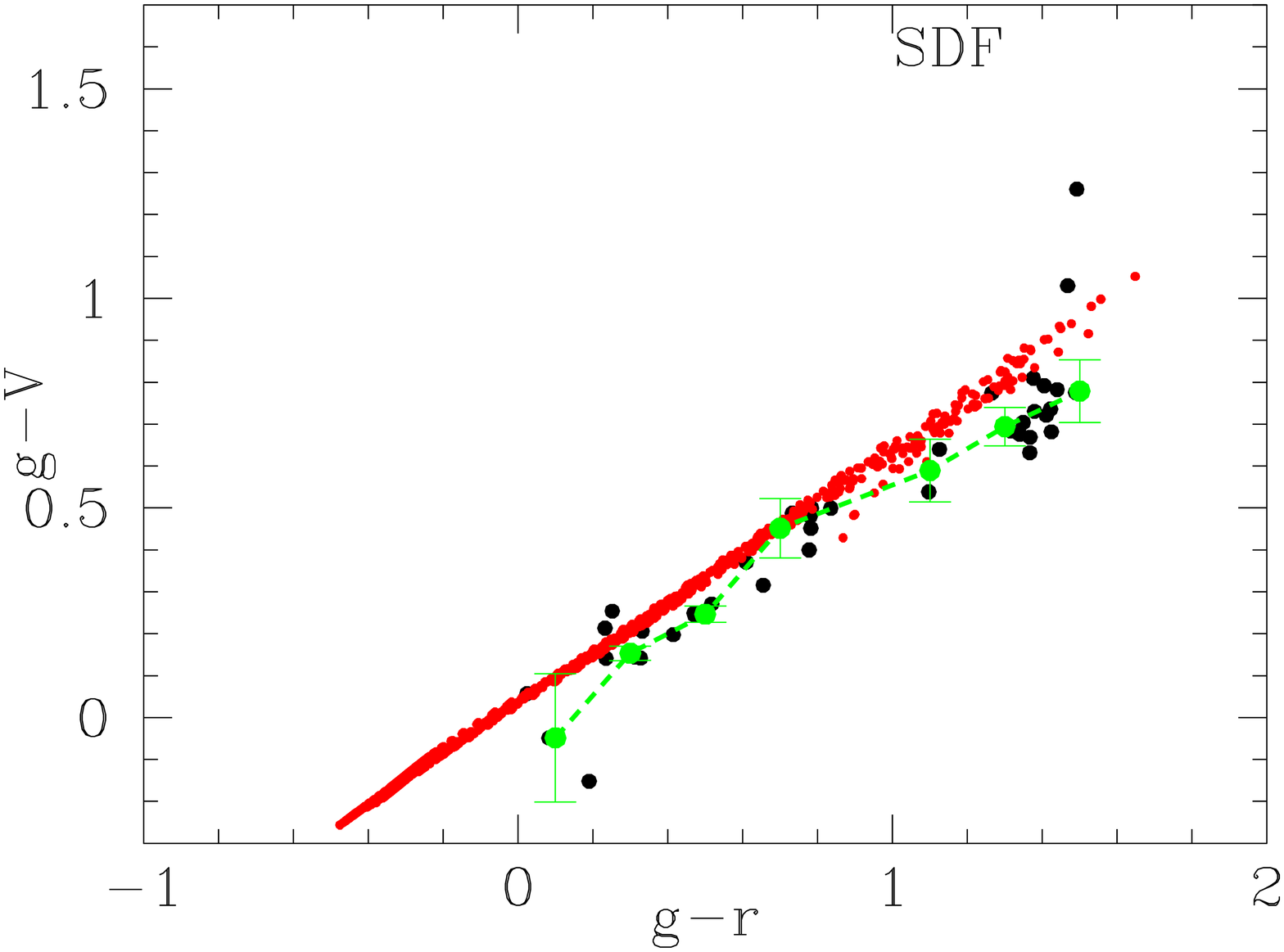}
\FigureFile(80mm,60mm){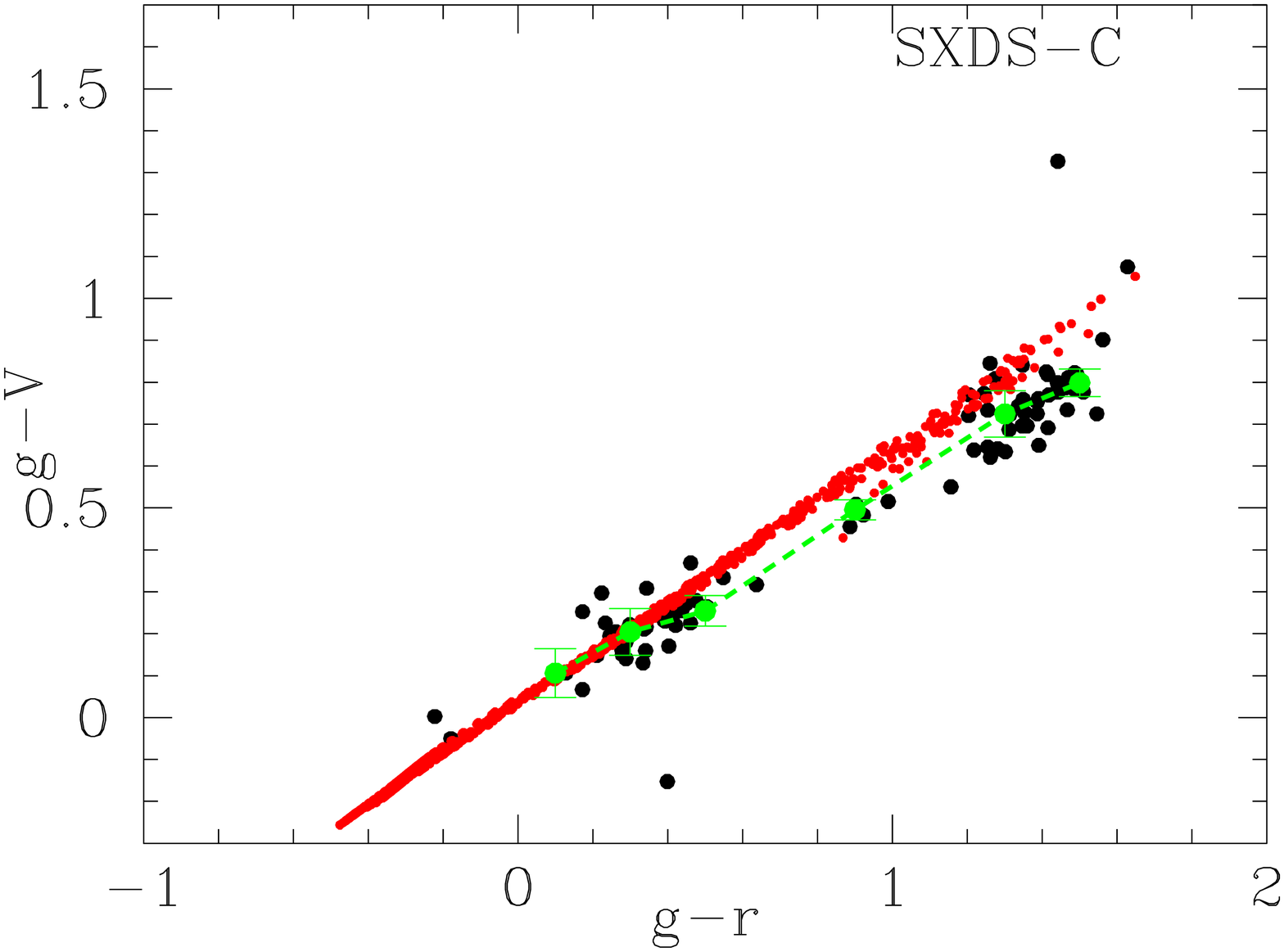}\\
\FigureFile(80mm,60mm){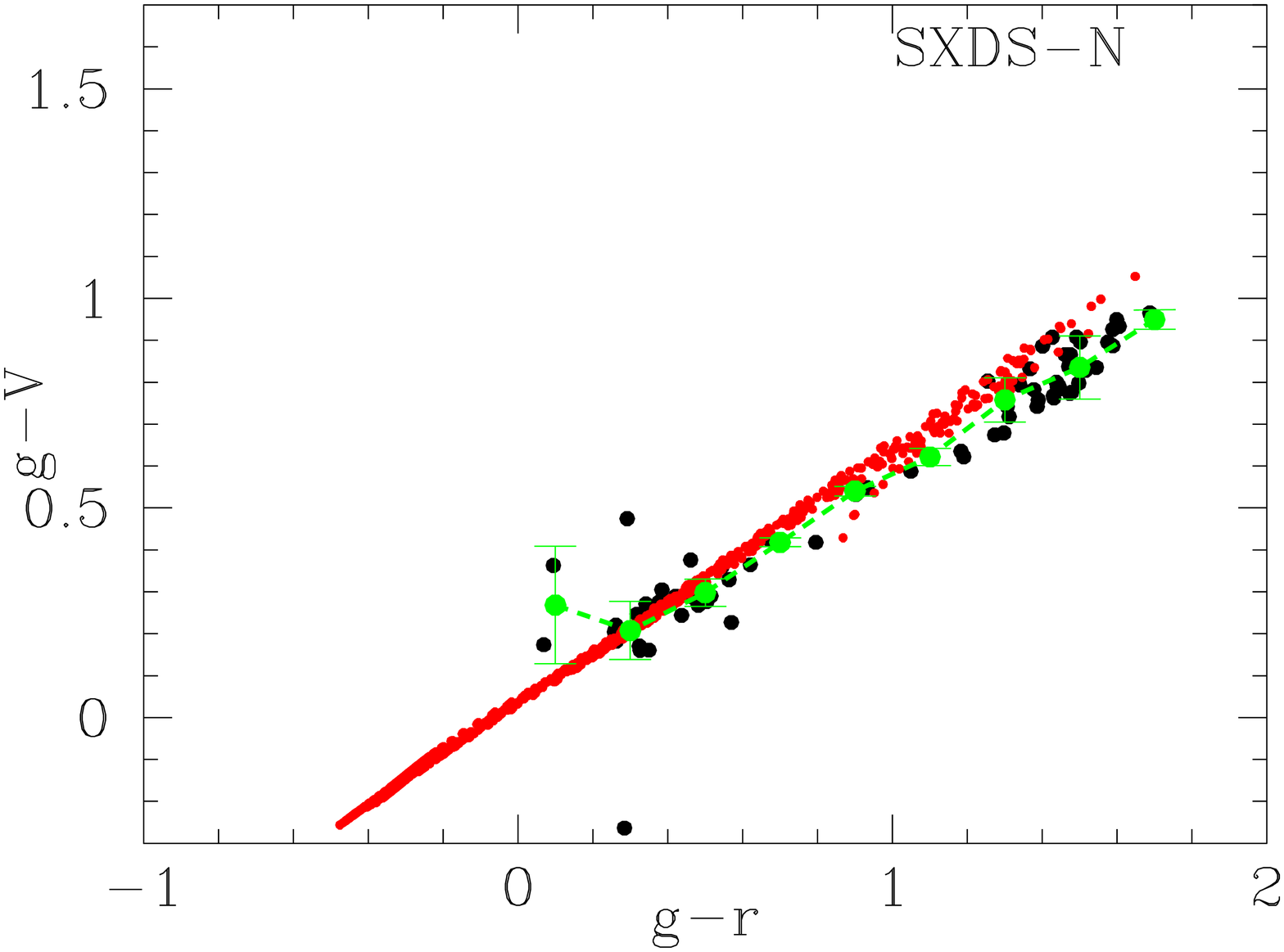}
\FigureFile(80mm,60mm){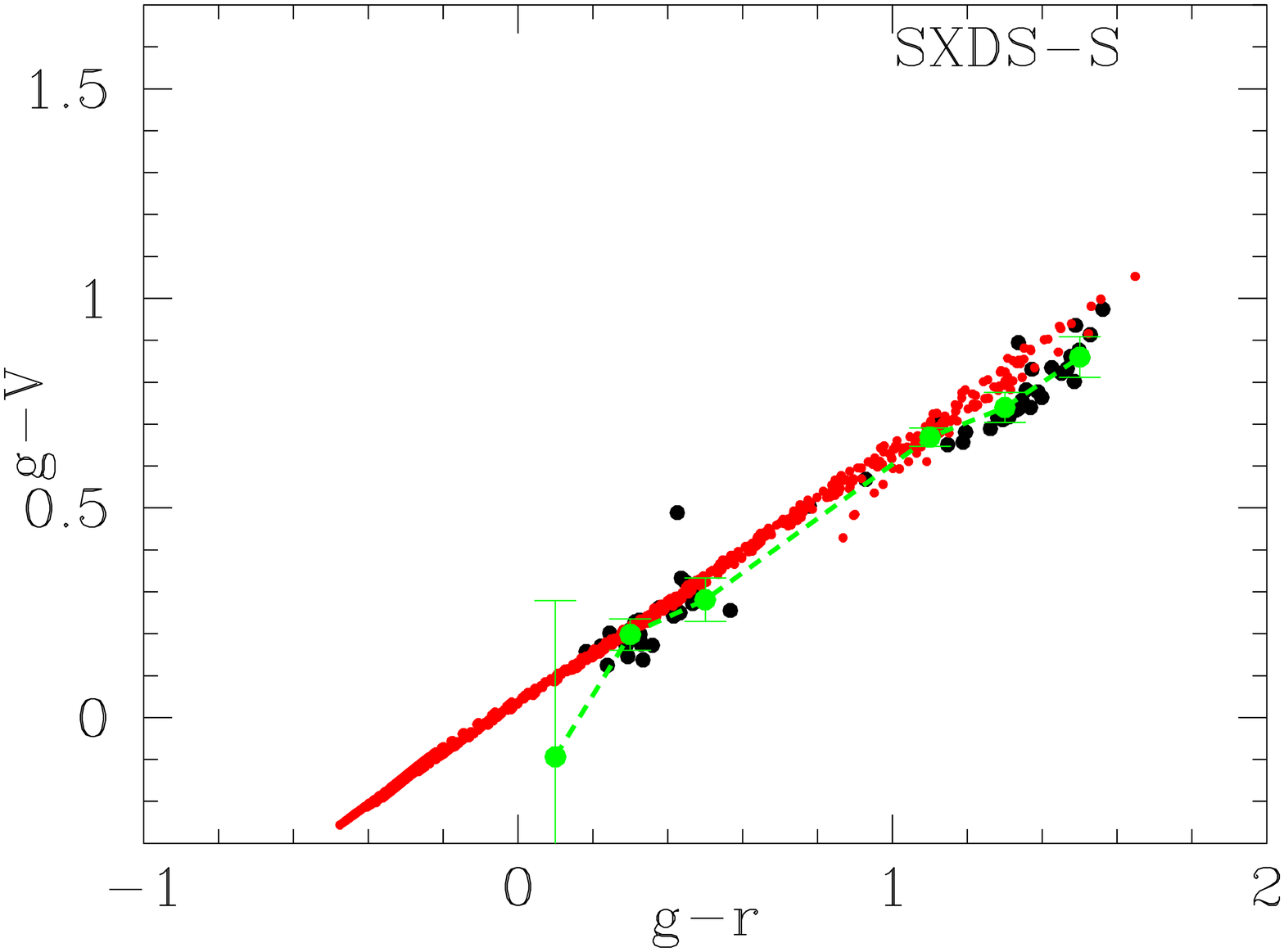}\\
\FigureFile(80mm,60mm){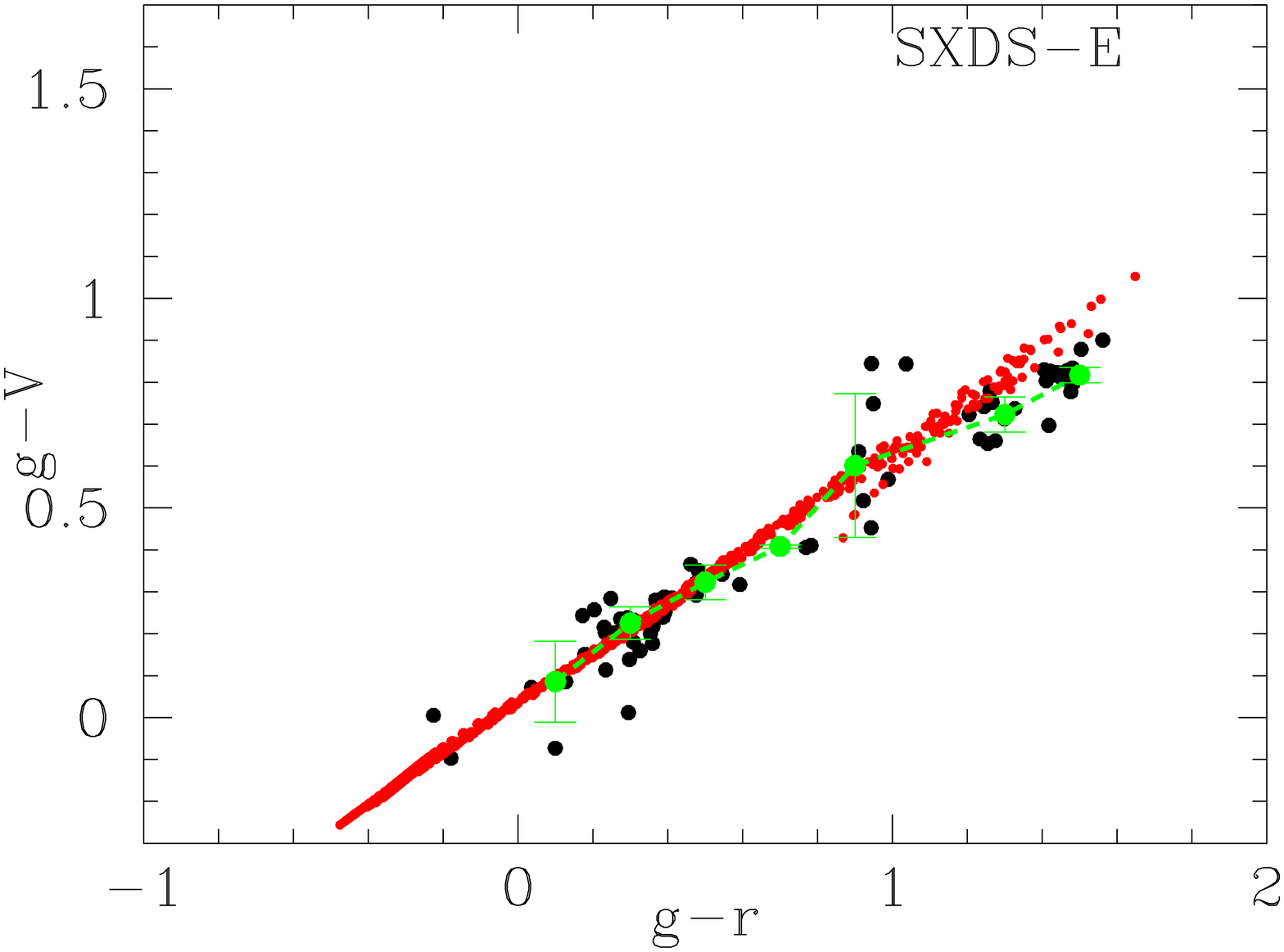}
\FigureFile(80mm,60mm){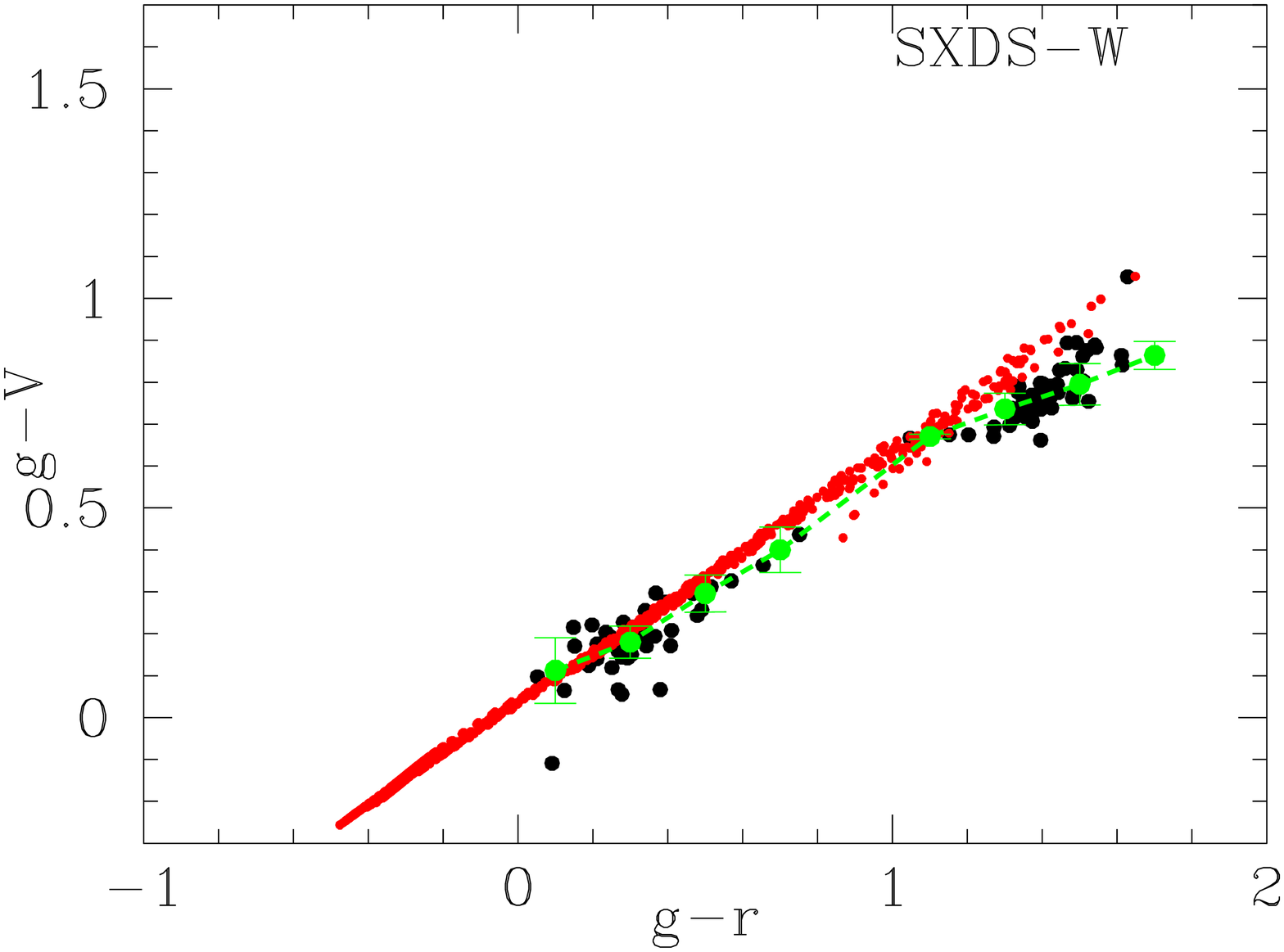}\\
\addtocounter{figure}{-1}
\caption{
Continued...
}
\end{figure}

\clearpage

\begin{figure}
\FigureFile(80mm,60mm){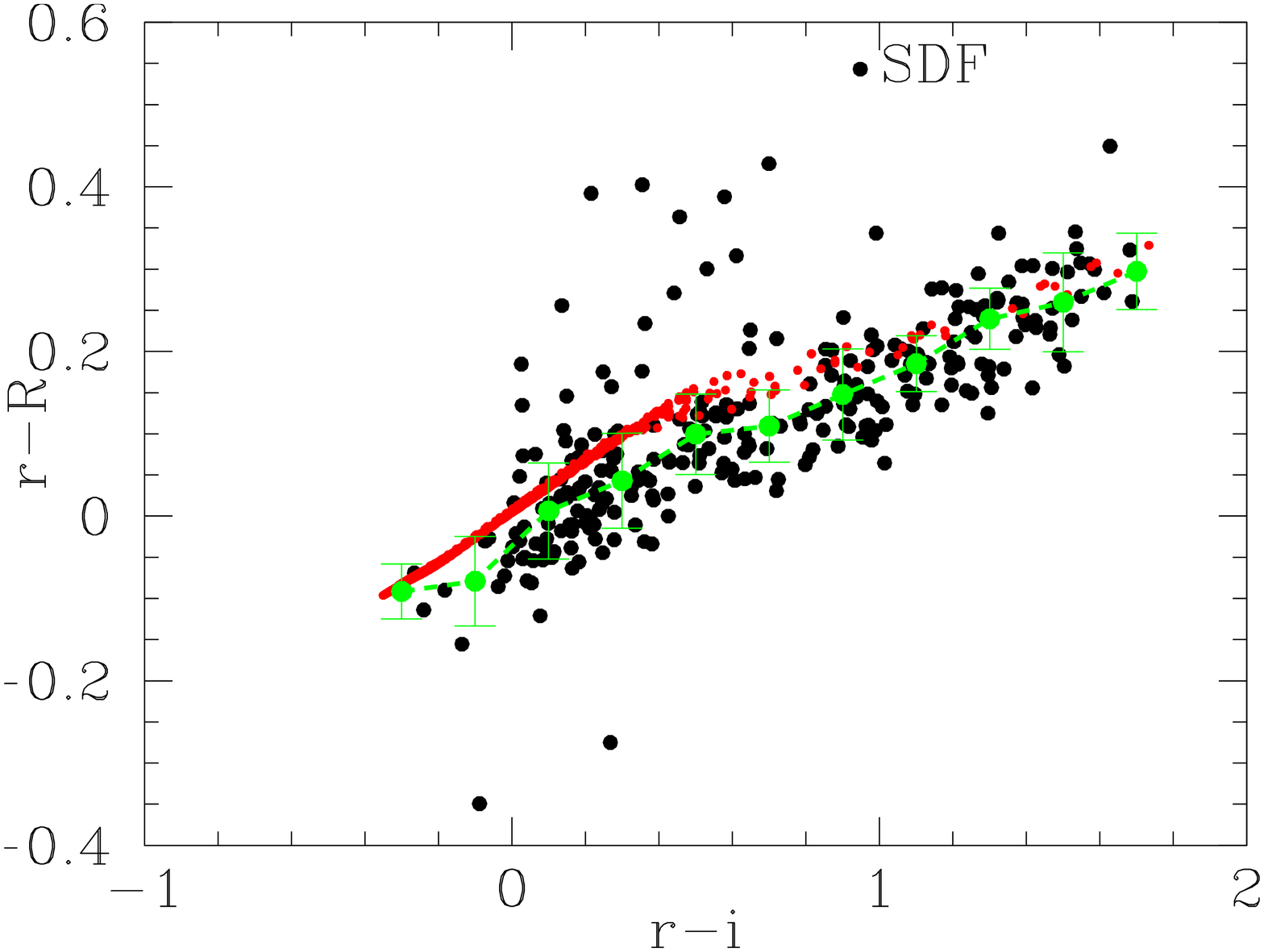}
\FigureFile(80mm,60mm){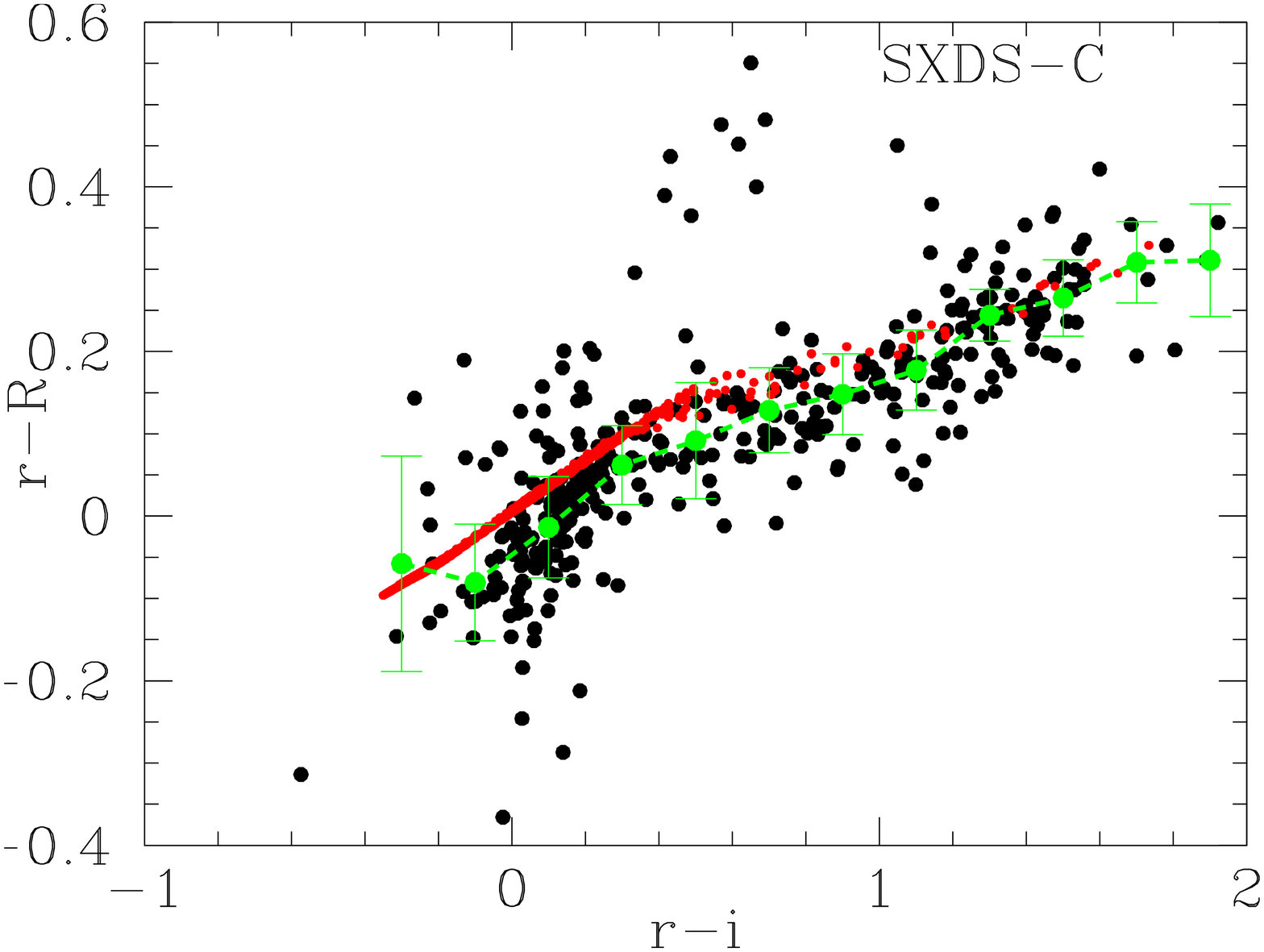}\\
\FigureFile(80mm,60mm){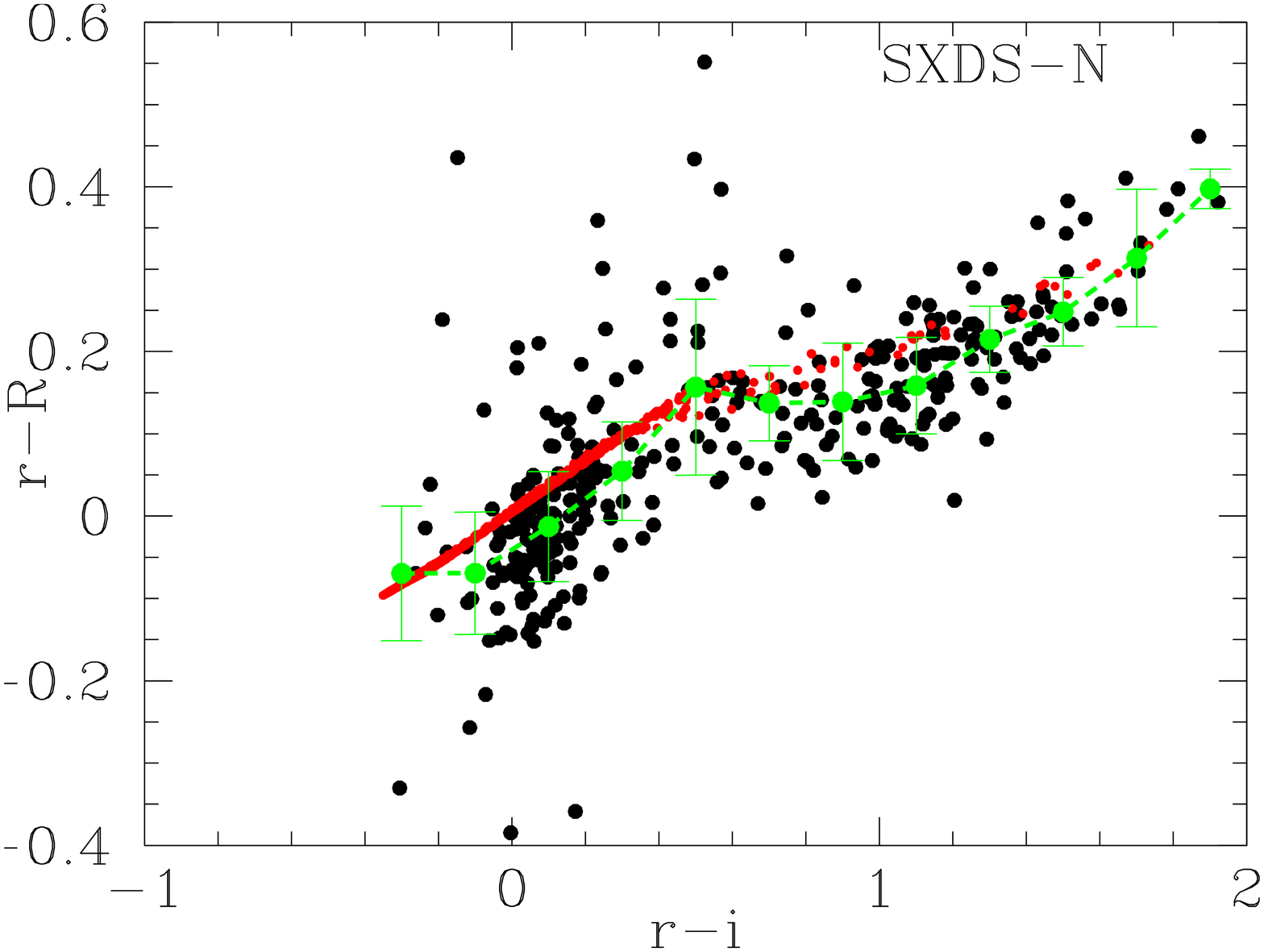}
\FigureFile(80mm,60mm){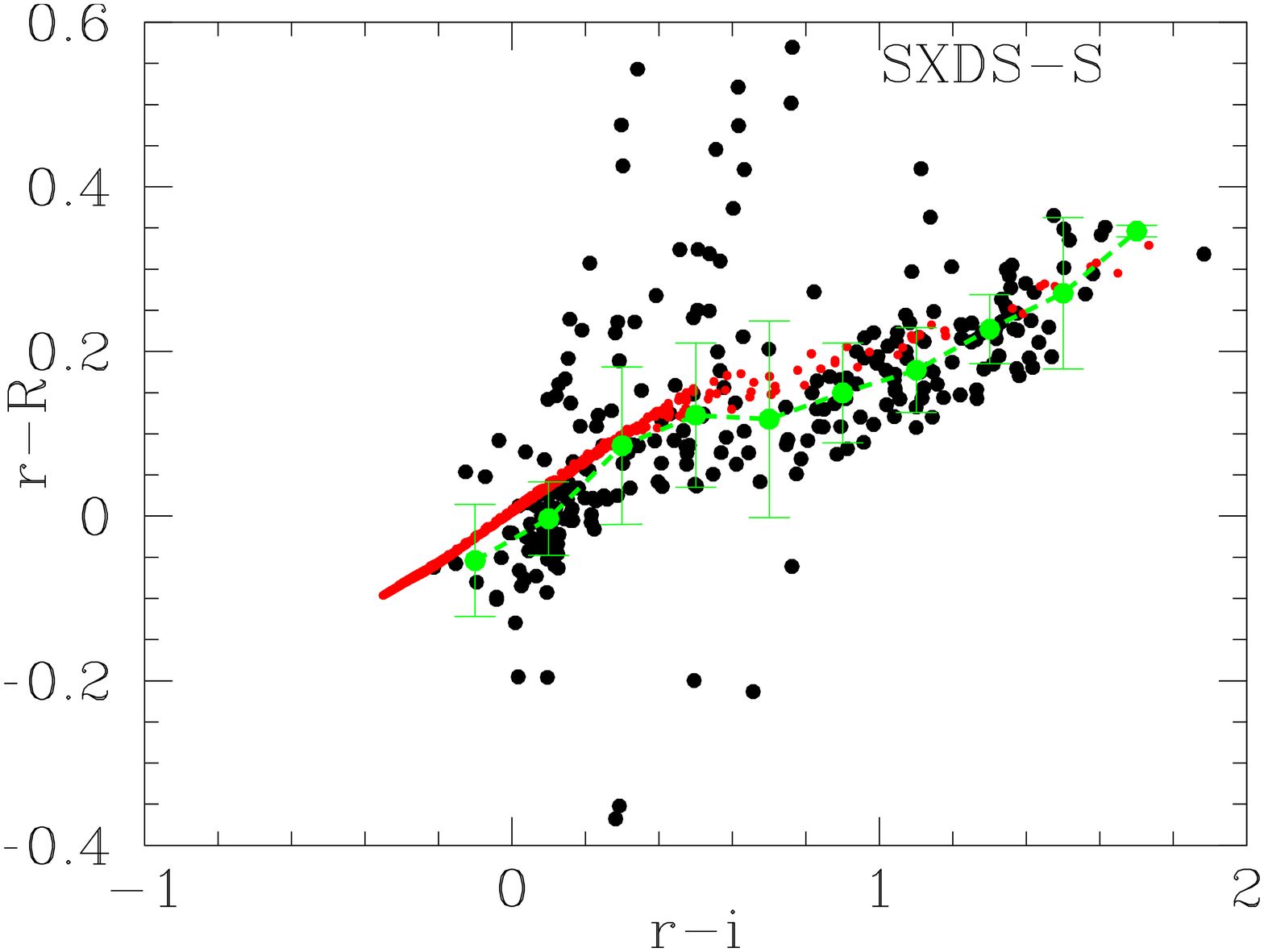}\\
\FigureFile(80mm,60mm){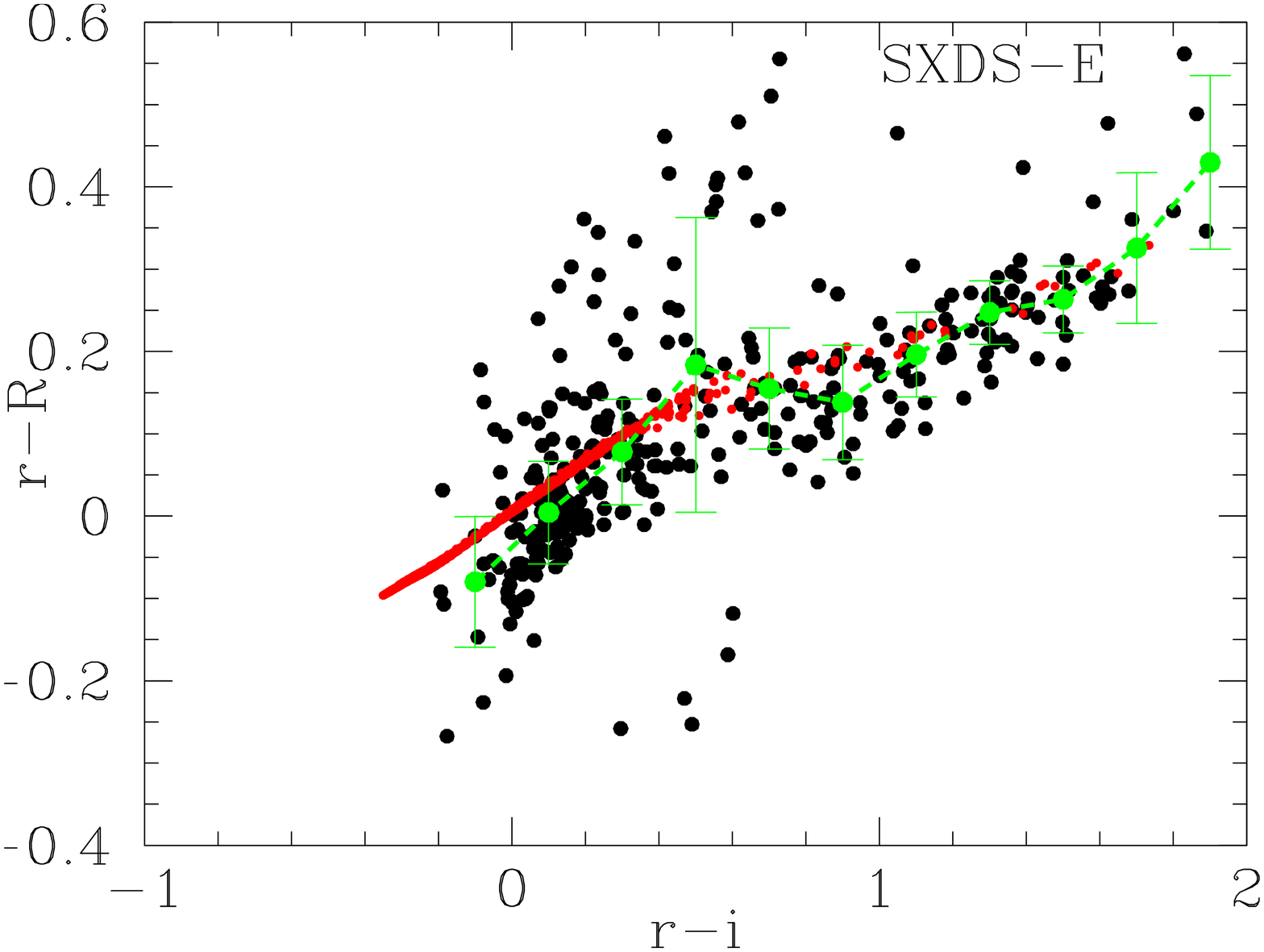}
\FigureFile(80mm,60mm){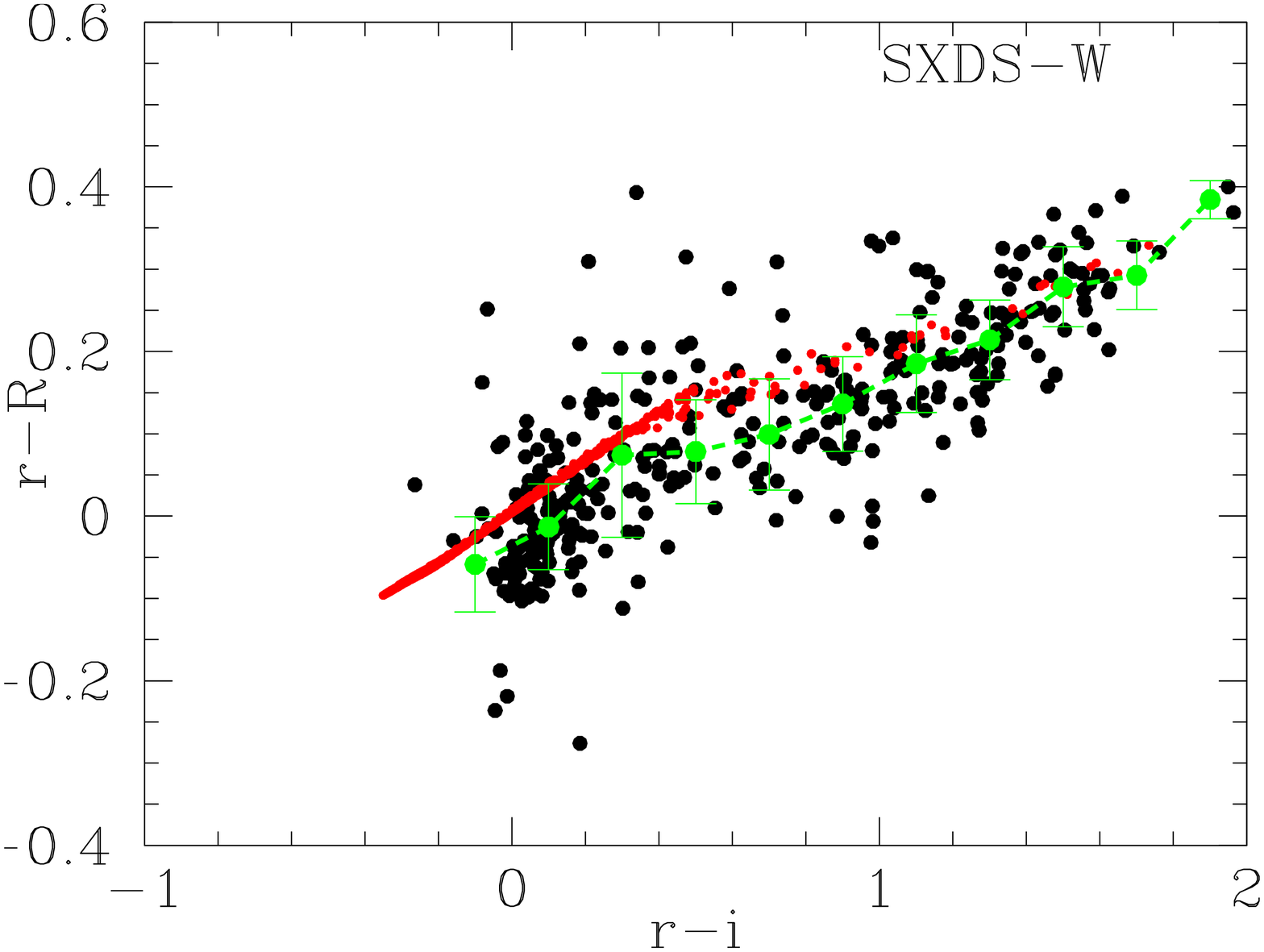}\\
\addtocounter{figure}{-1}
\caption{
Continued...
}
\end{figure}

\clearpage

\begin{figure}
\FigureFile(80mm,60mm){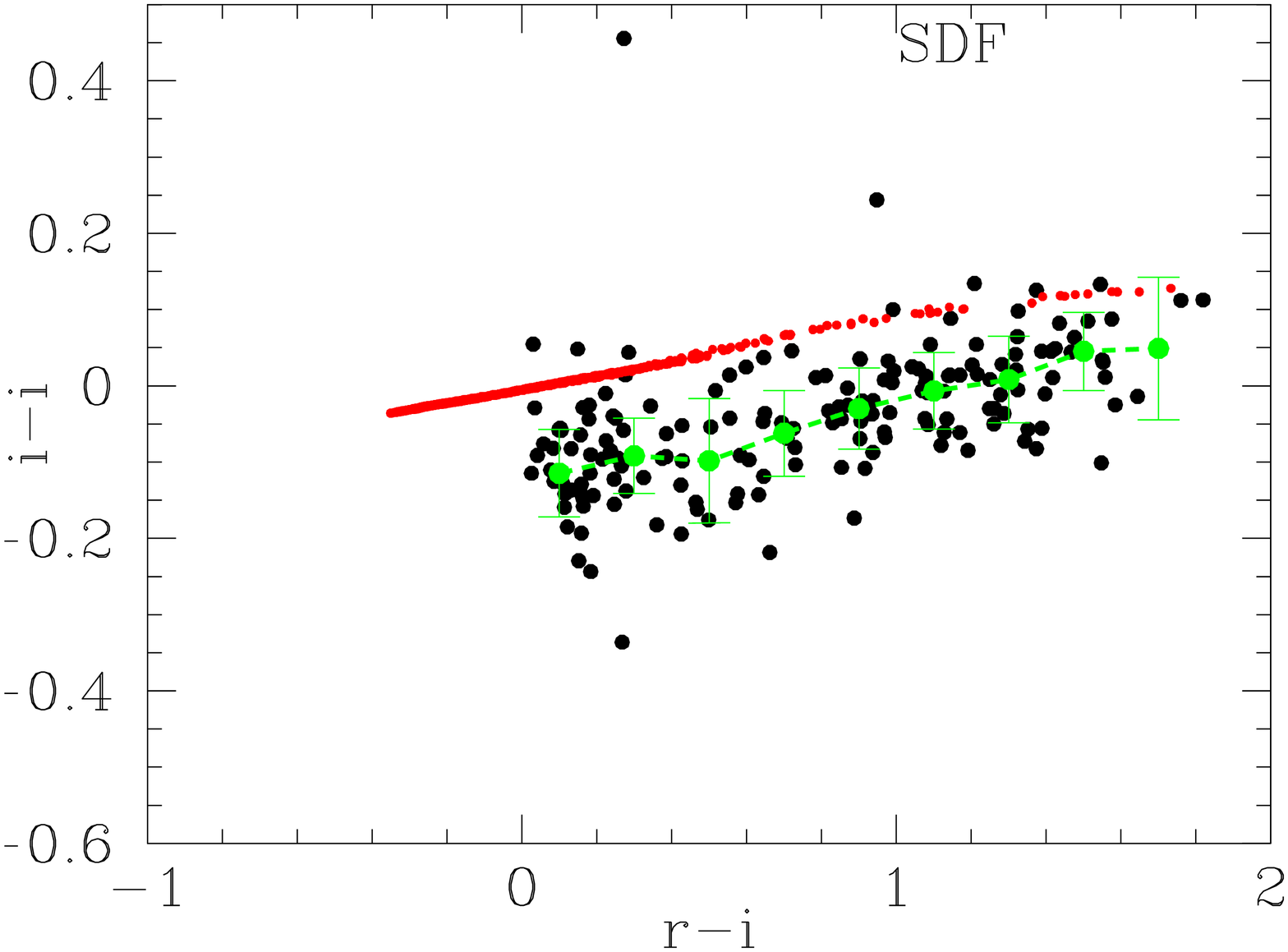}
\FigureFile(80mm,60mm){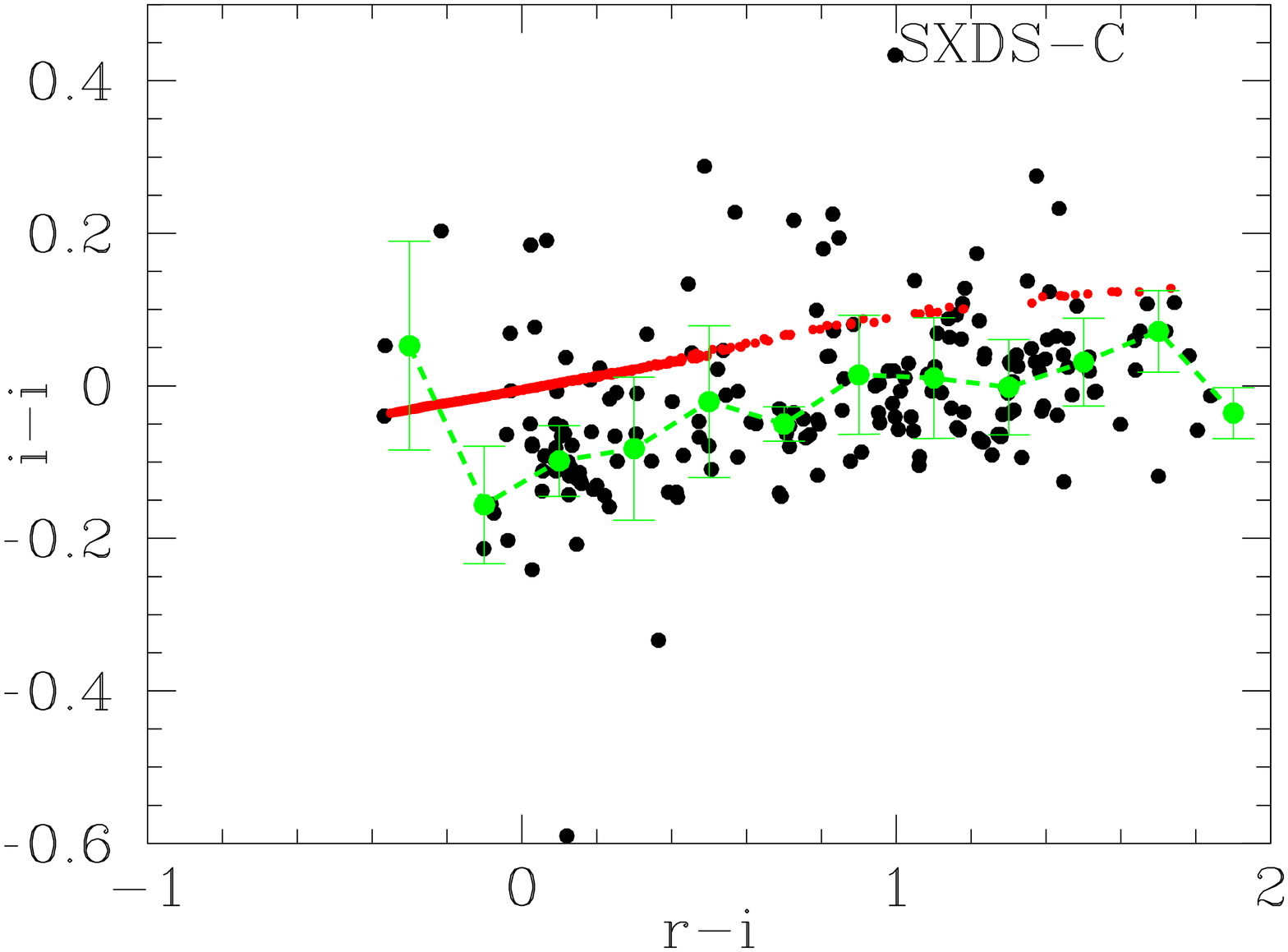}\\
\FigureFile(80mm,60mm){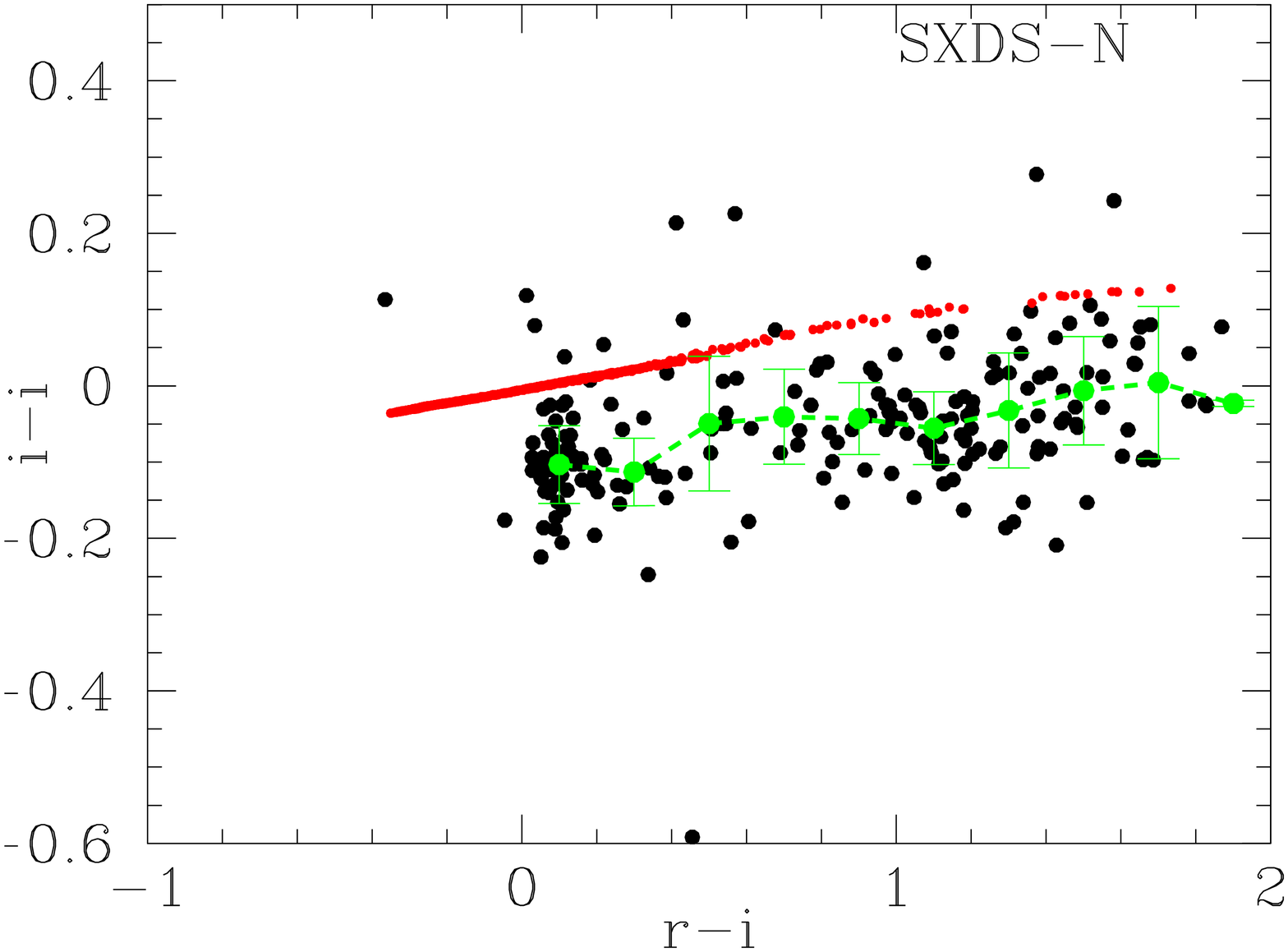}
\FigureFile(80mm,60mm){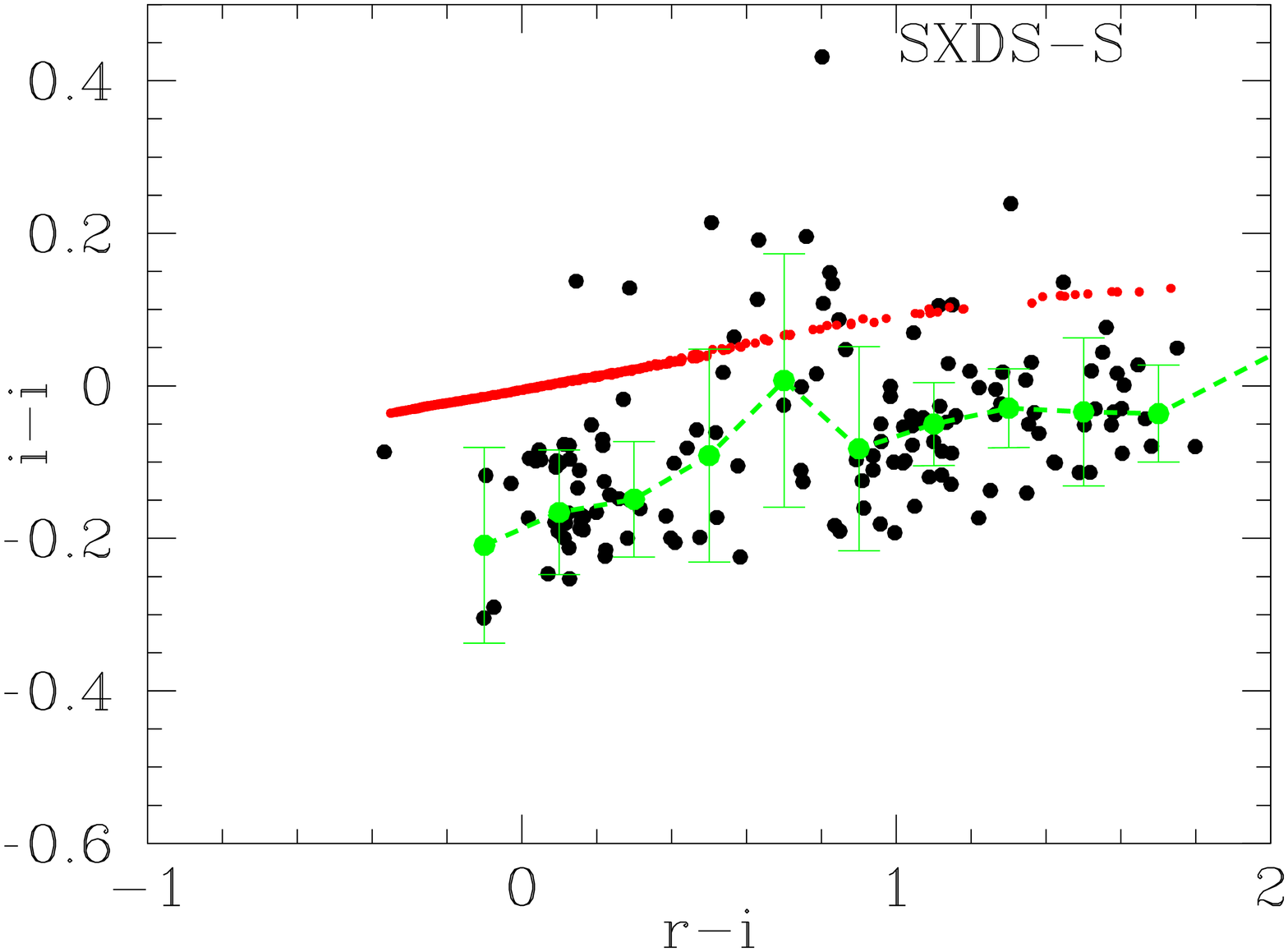}\\
\FigureFile(80mm,60mm){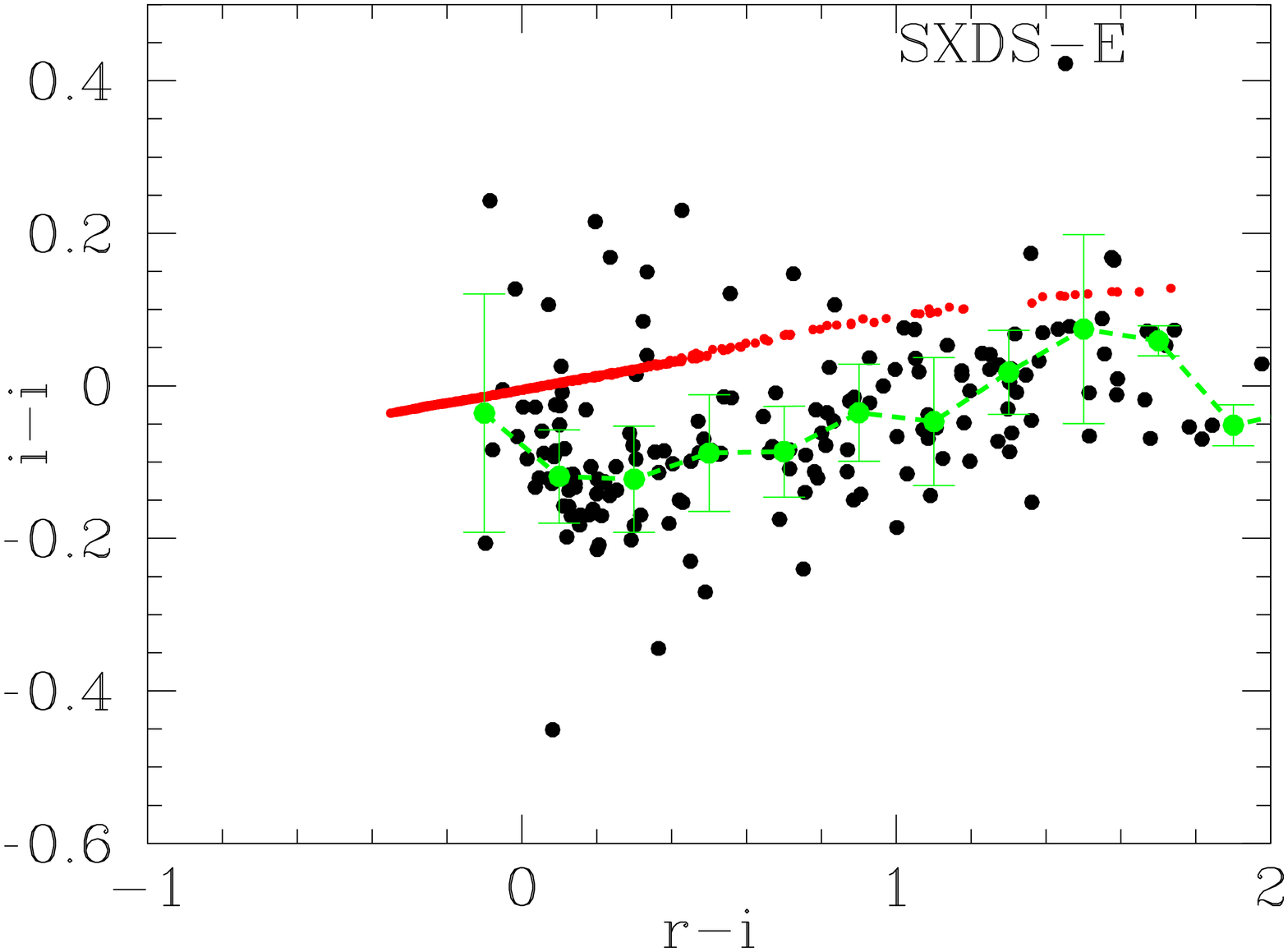}
\FigureFile(80mm,60mm){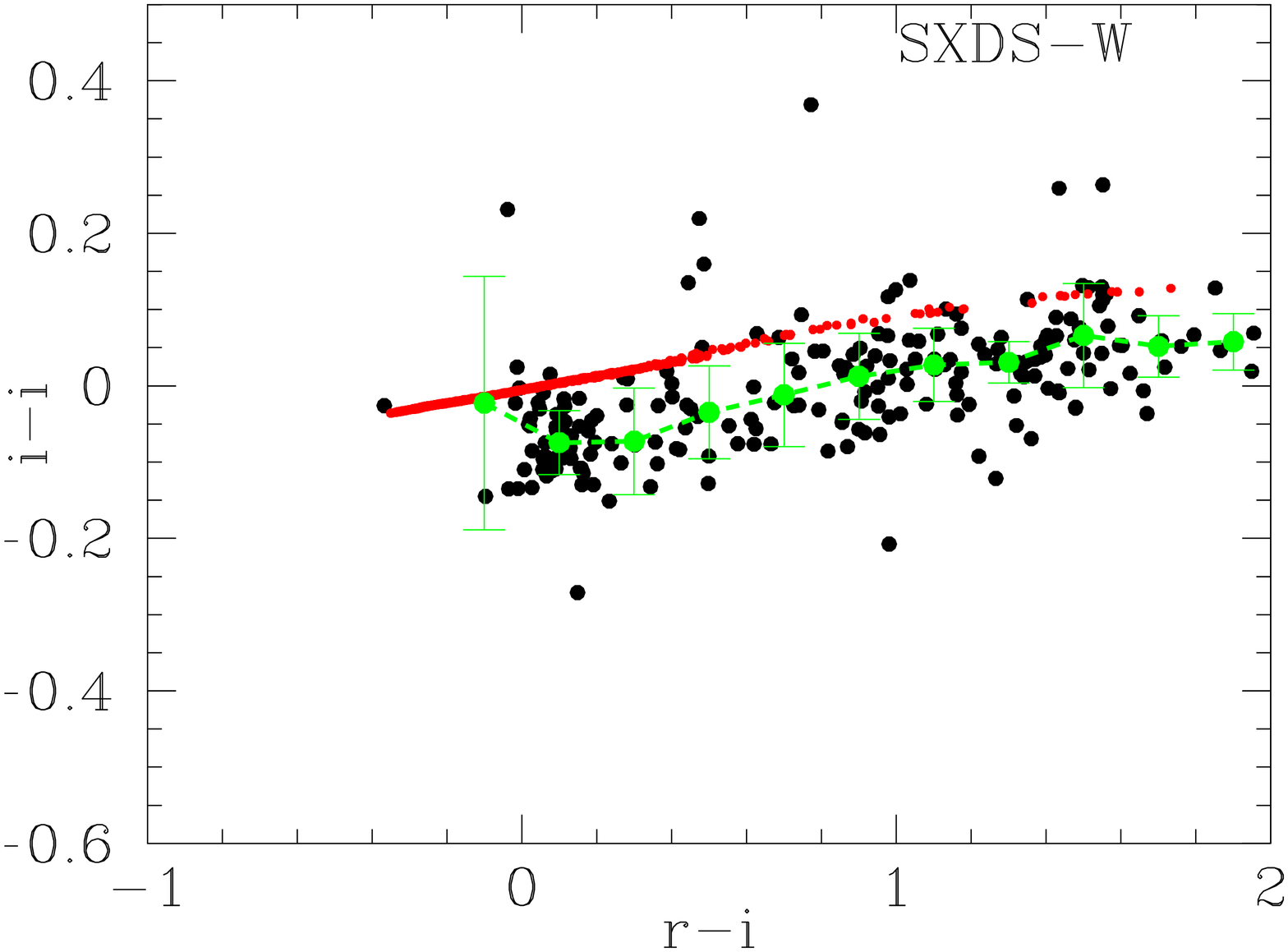}\\
\addtocounter{figure}{-1}
\caption{
Continued...
}
\end{figure}

\clearpage

\begin{figure}
\FigureFile(80mm,60mm){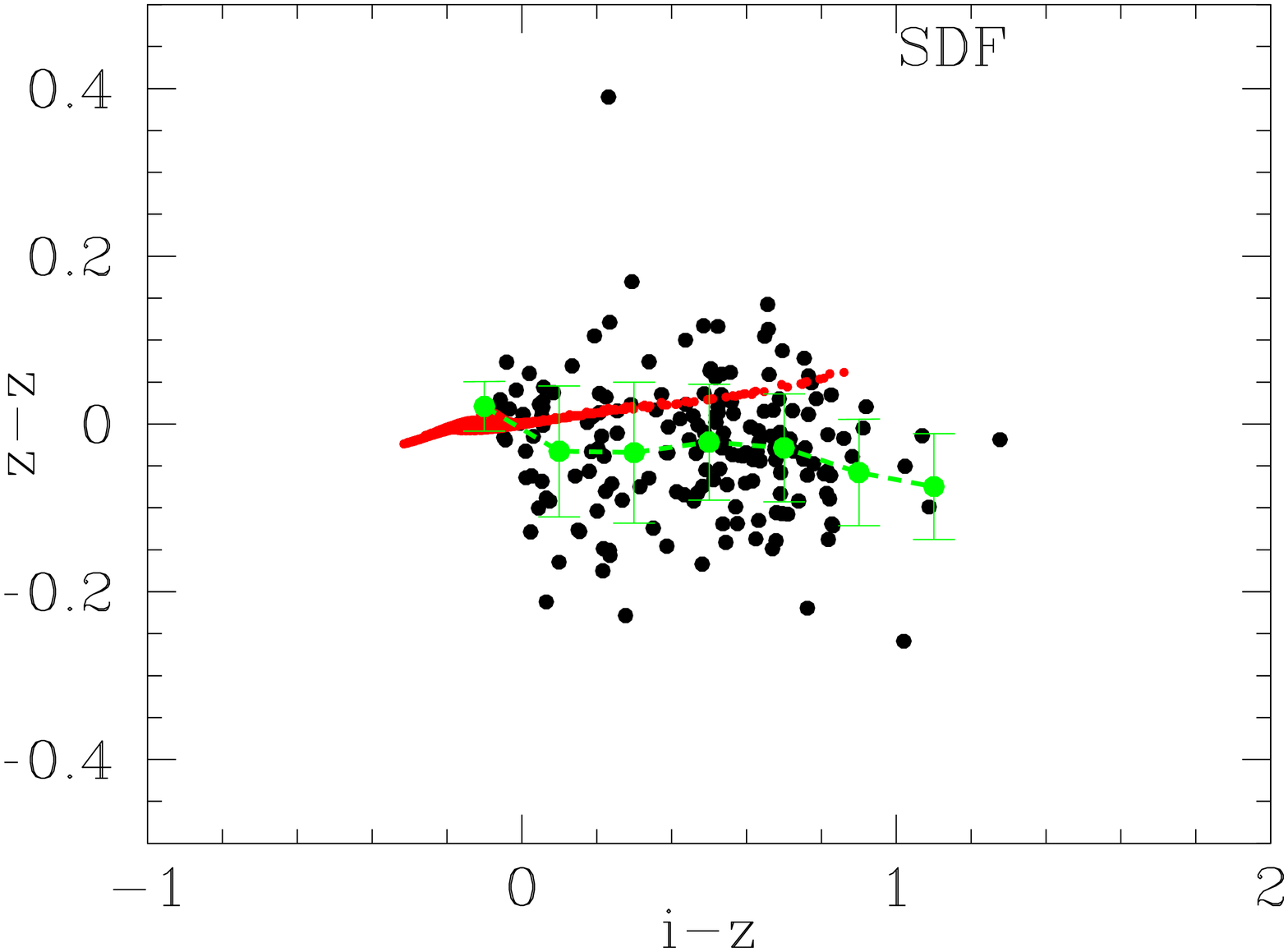}
\FigureFile(80mm,60mm){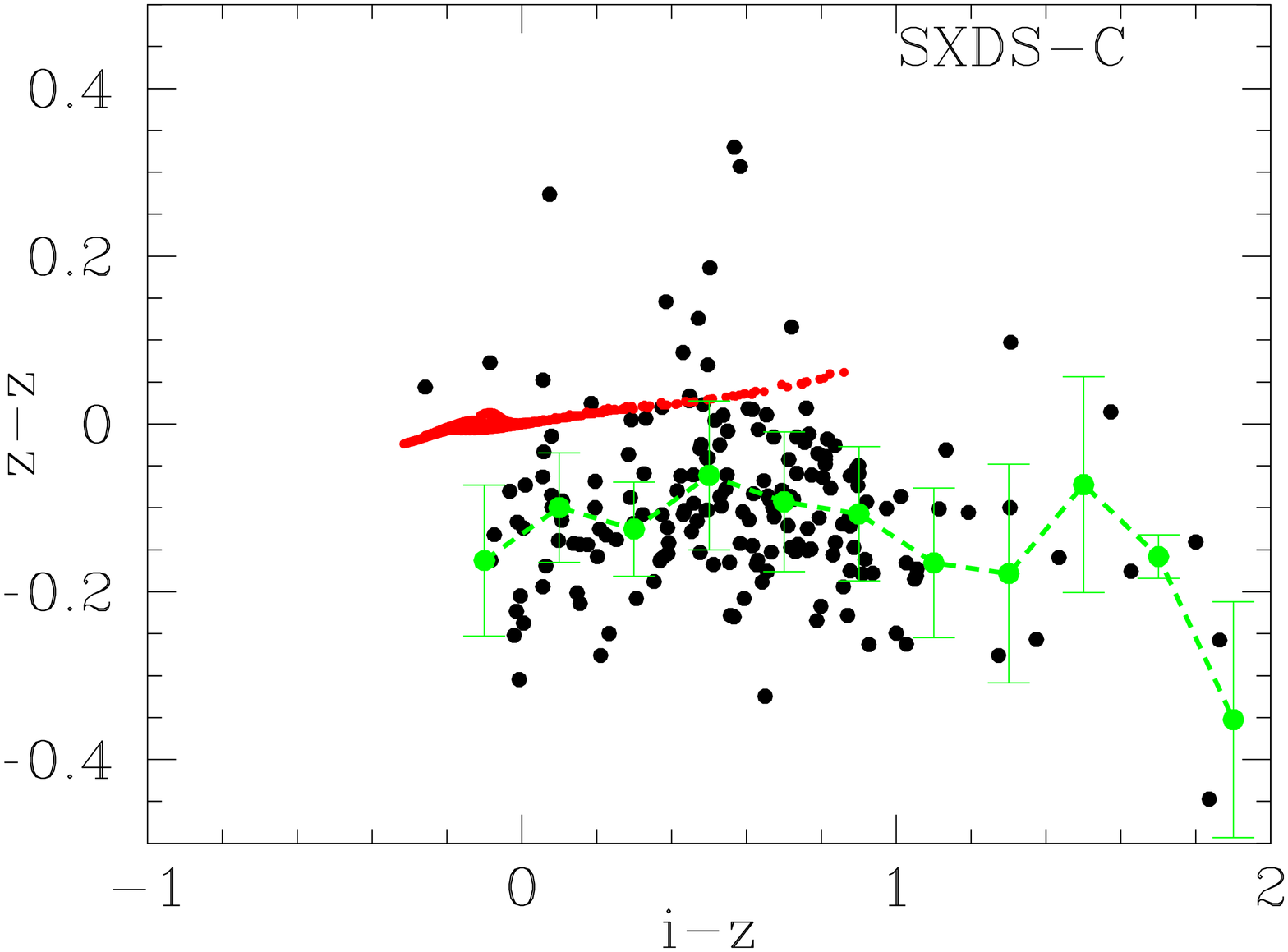}\\
\FigureFile(80mm,60mm){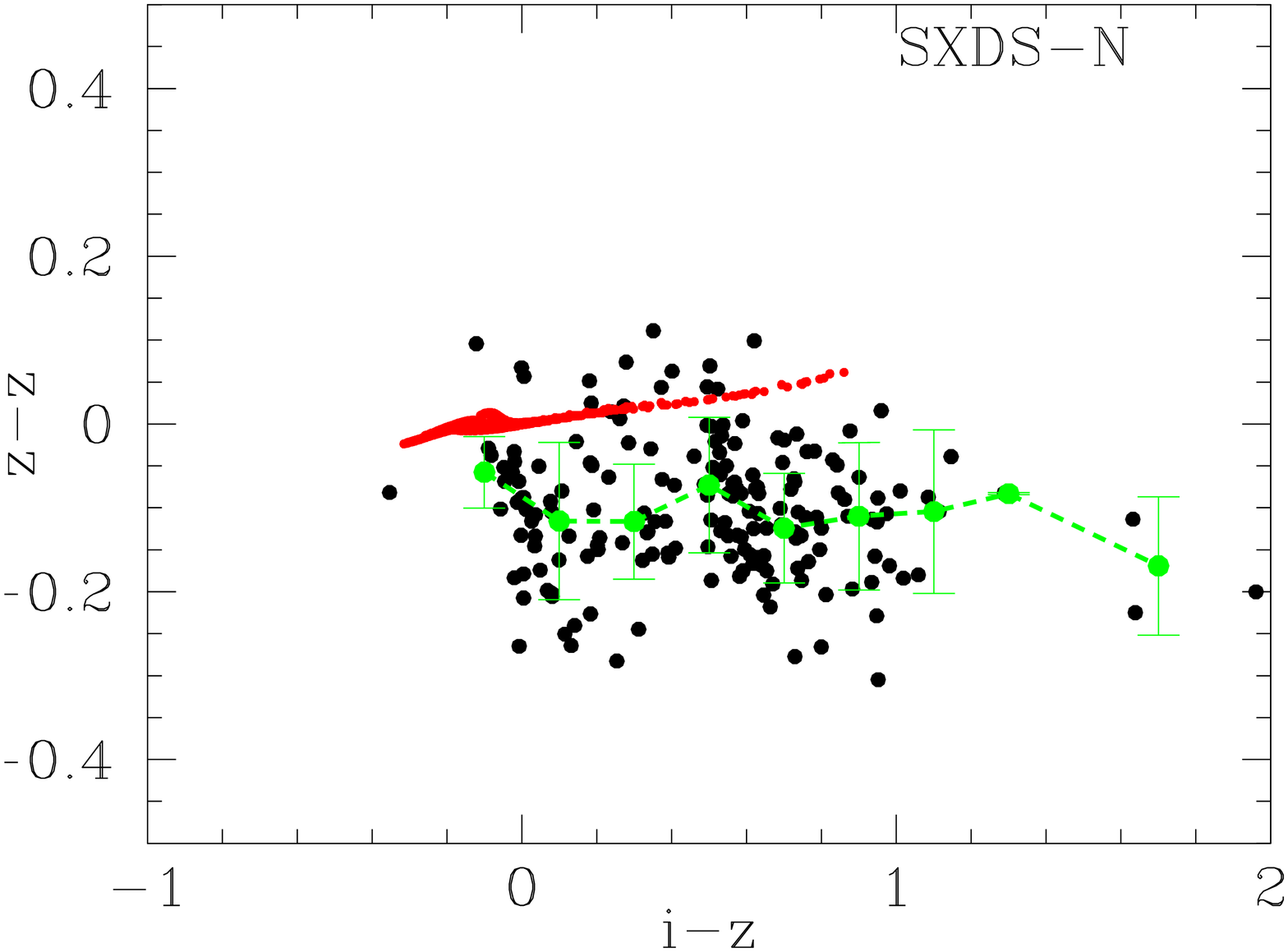}
\FigureFile(80mm,60mm){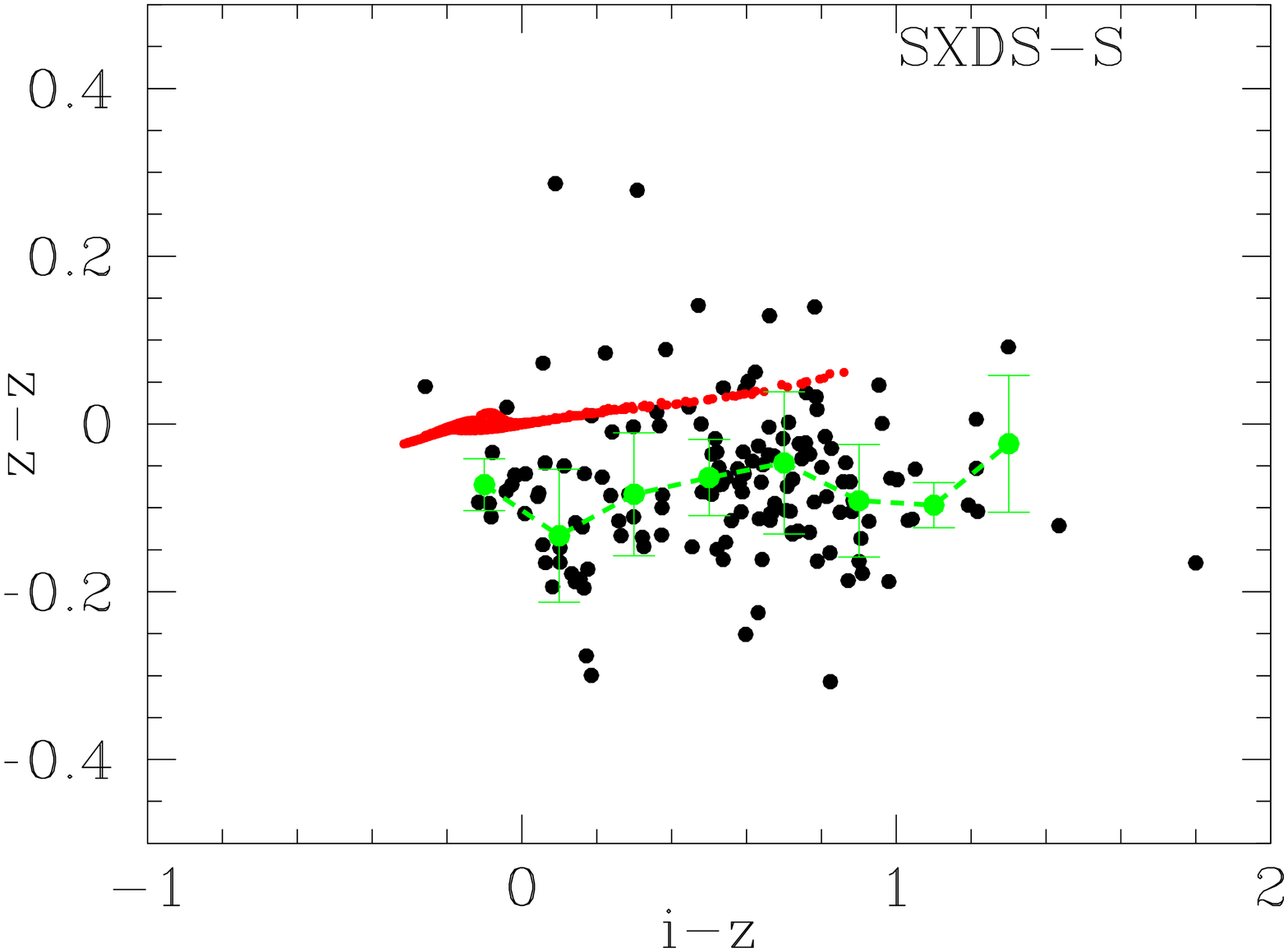}\\
\FigureFile(80mm,60mm){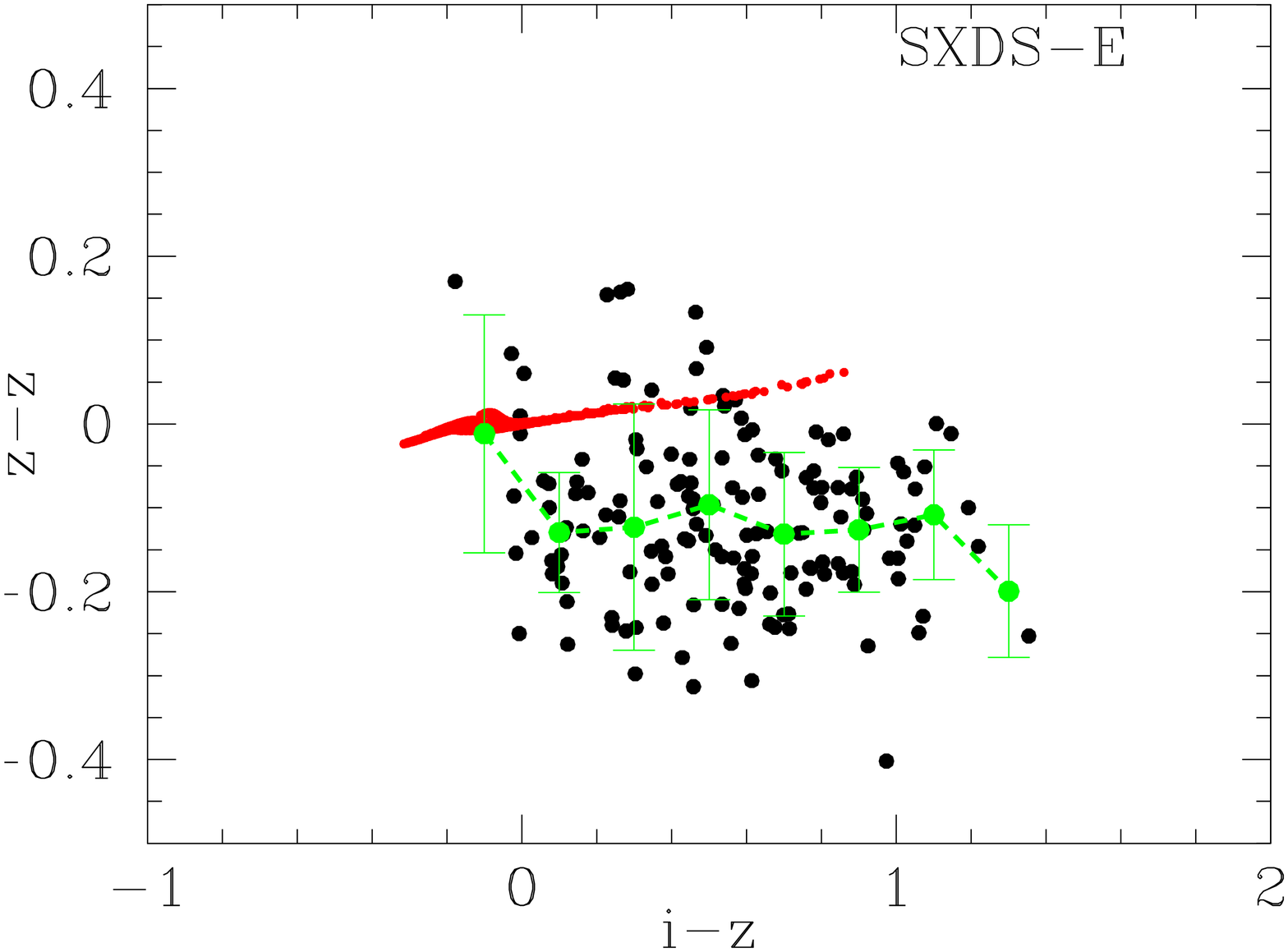}
\FigureFile(80mm,60mm){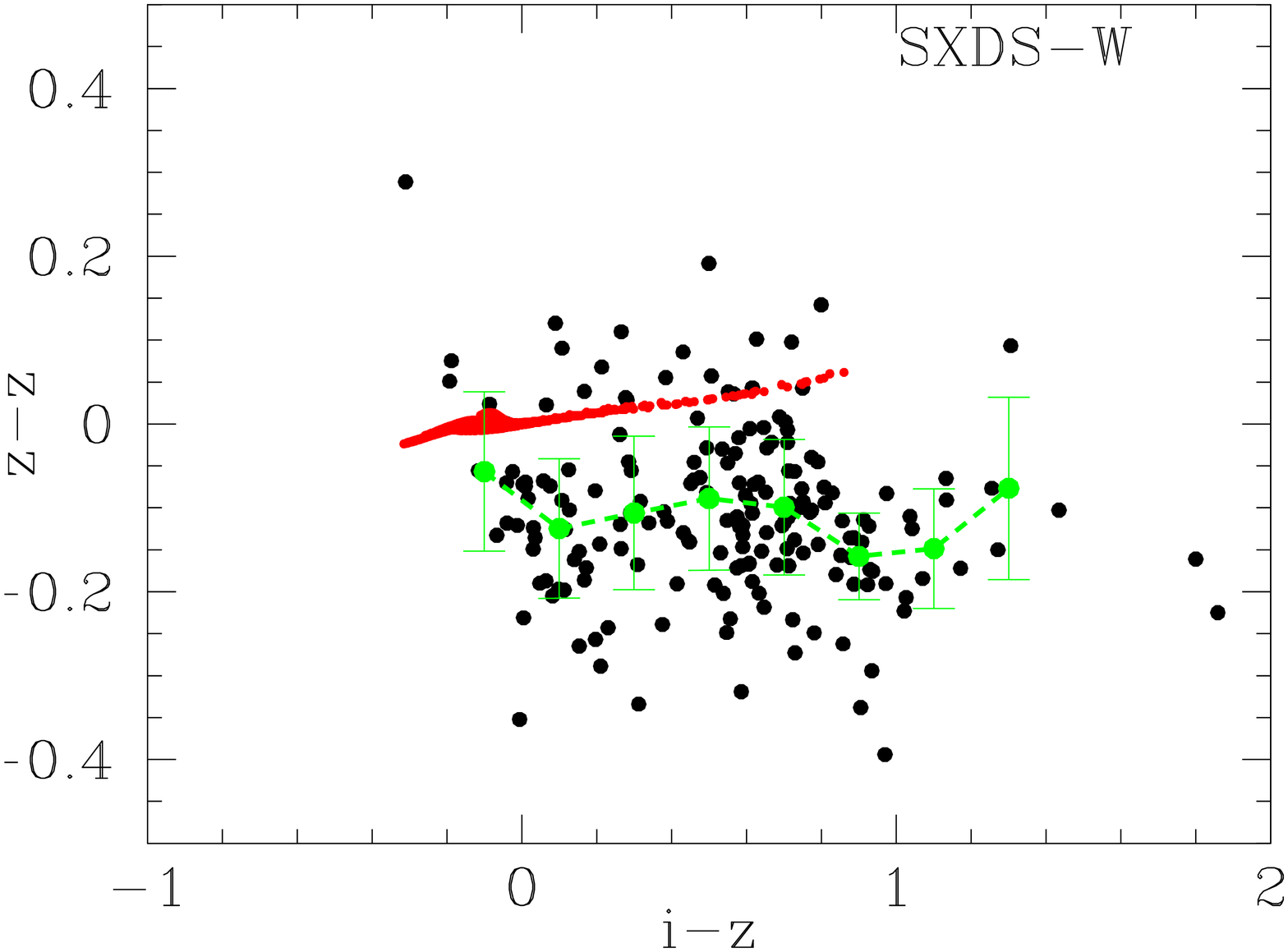}\\
\addtocounter{figure}{-1}
\caption{
Continued...
}
\end{figure}

\clearpage

\begin{figure}
\FigureFile(80mm,60mm){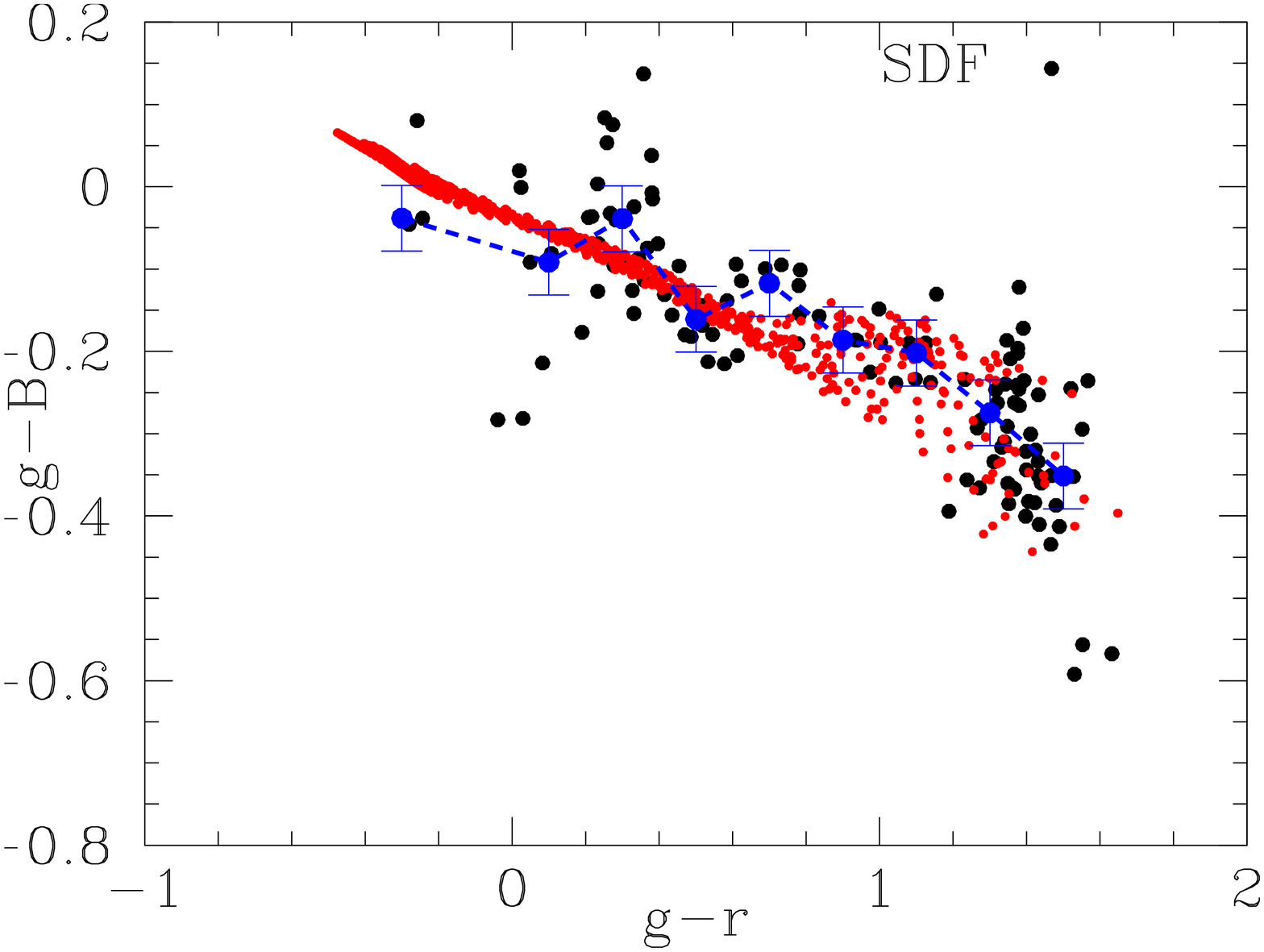}
\FigureFile(80mm,60mm){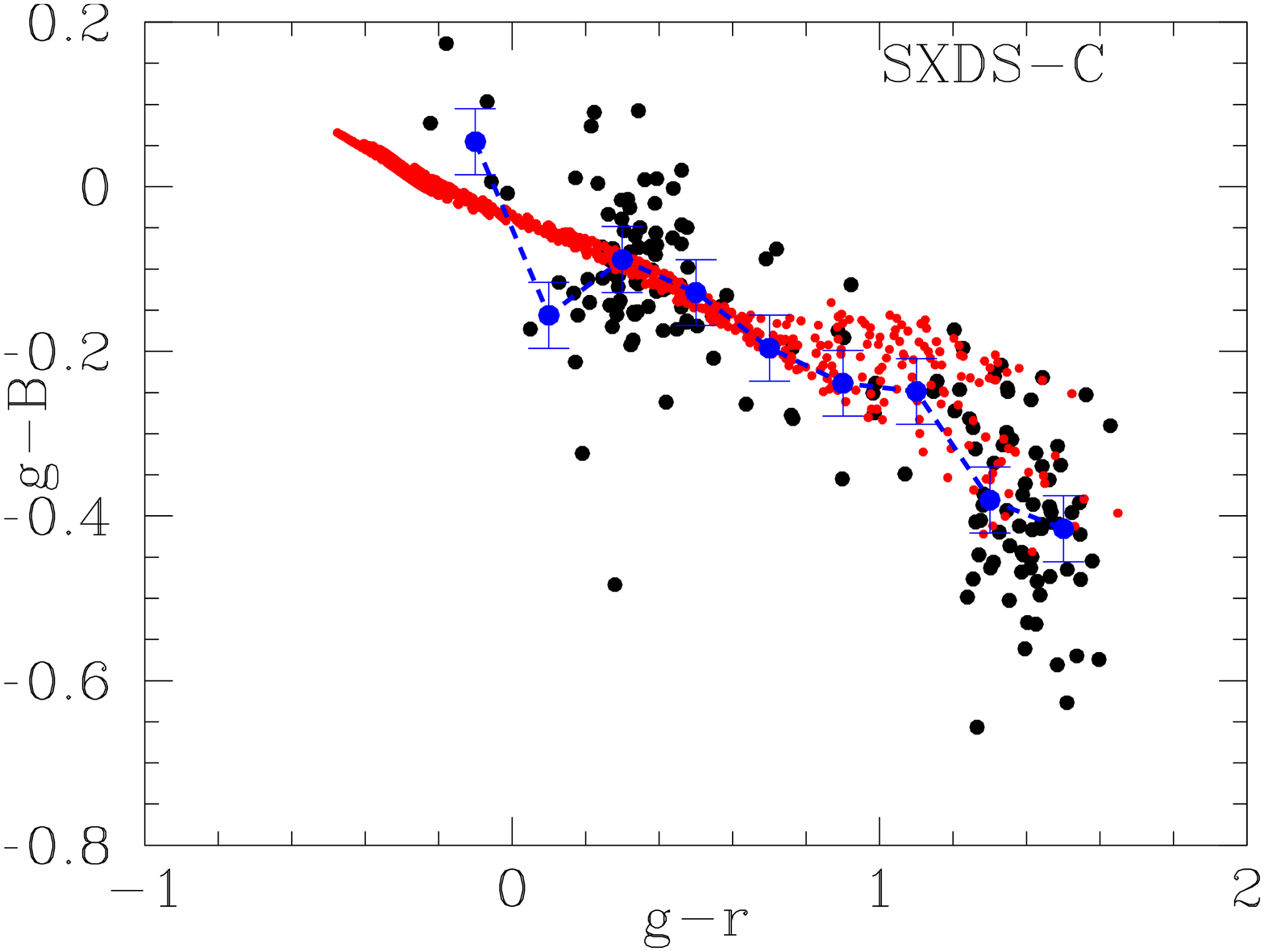}\\
\FigureFile(80mm,60mm){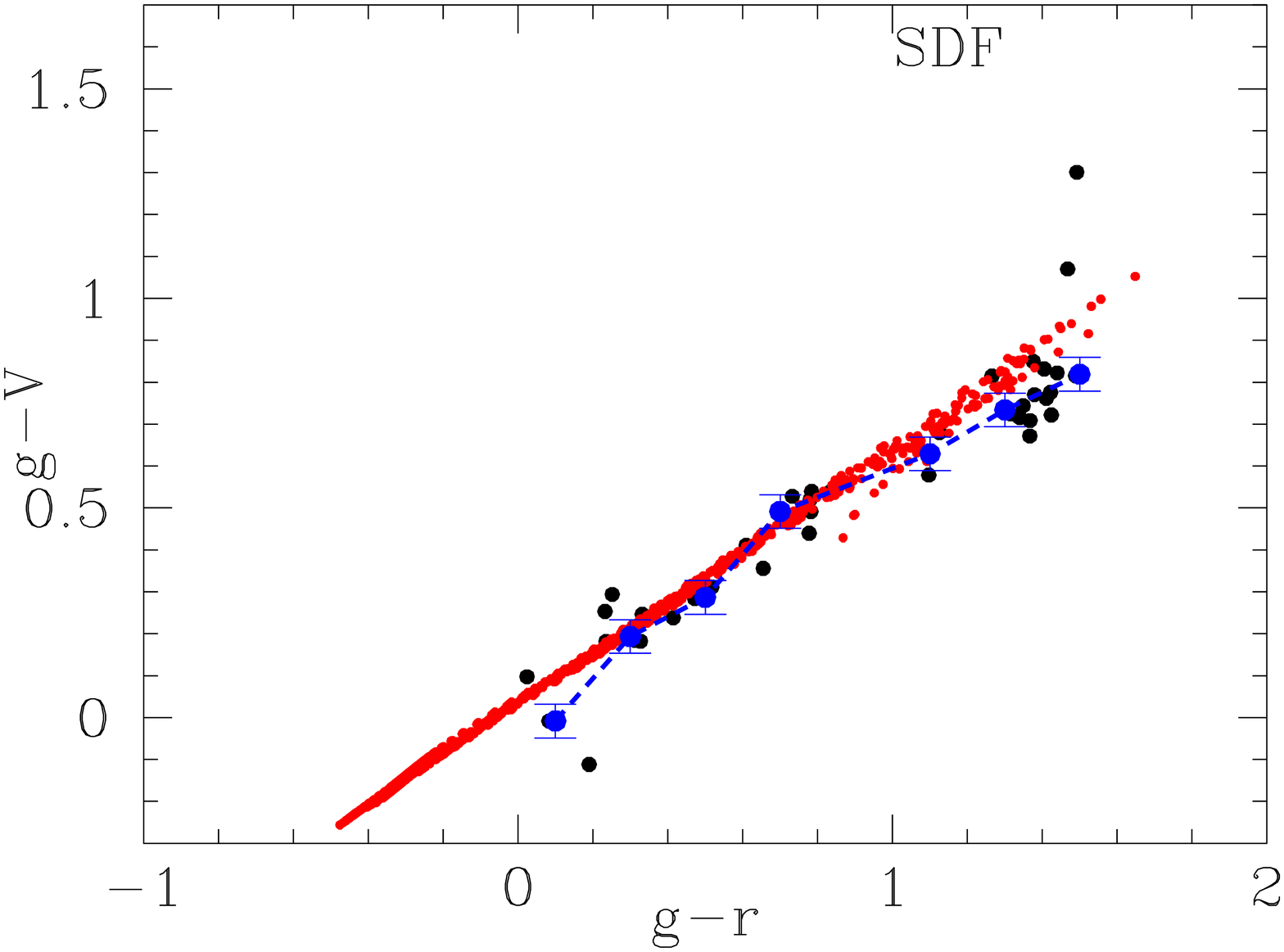}
\FigureFile(80mm,60mm){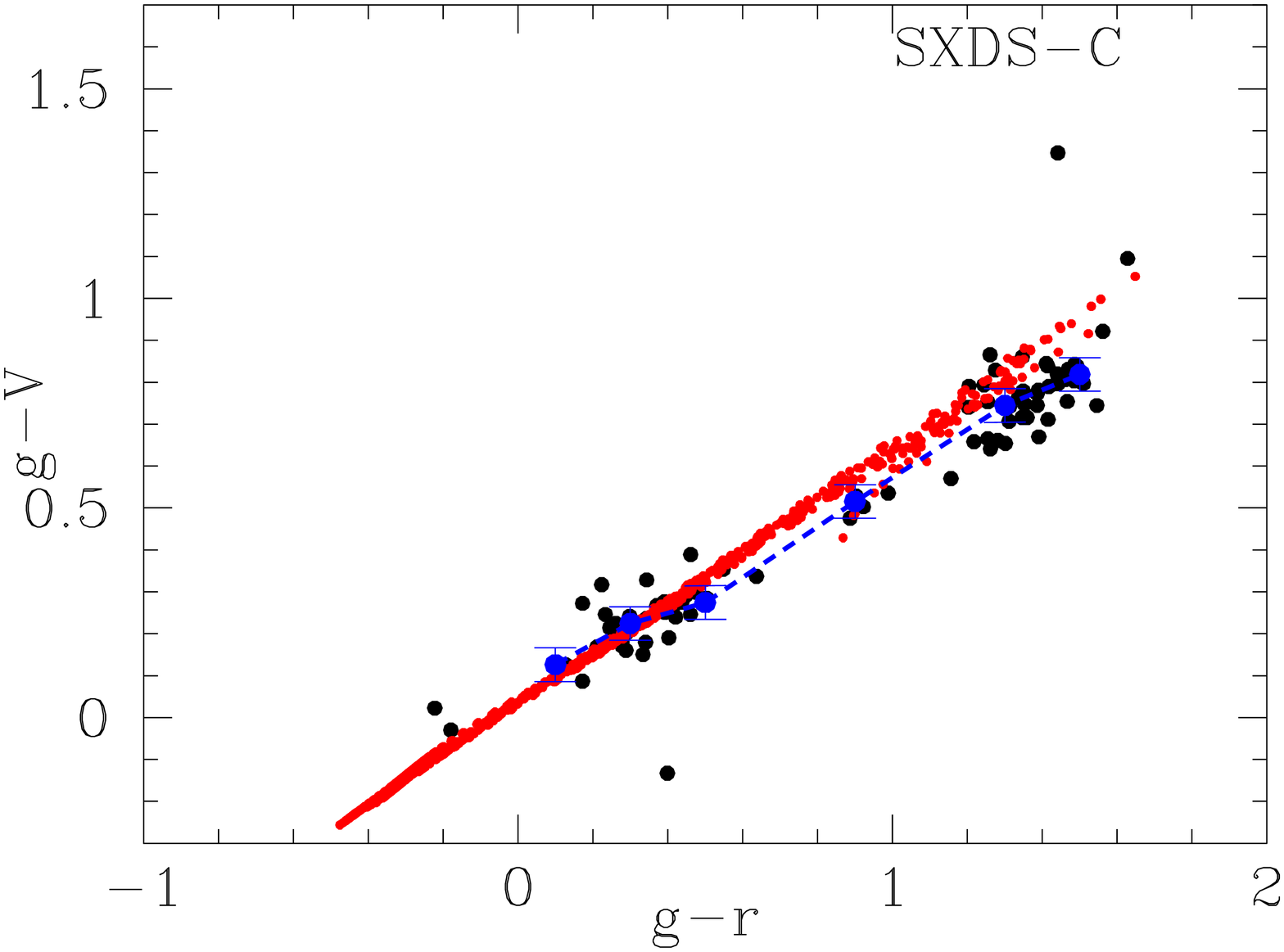}\\
\caption{
Same as figure \ref{fig:cal0},
but after the ZP offset correction, 
and errorbars are fixed as 0.04 mag.
Only SDF and SXDS-C are shown.
}
\label{fig:cal1}
\end{figure}

\clearpage
\begin{figure}
\FigureFile(80mm,60mm){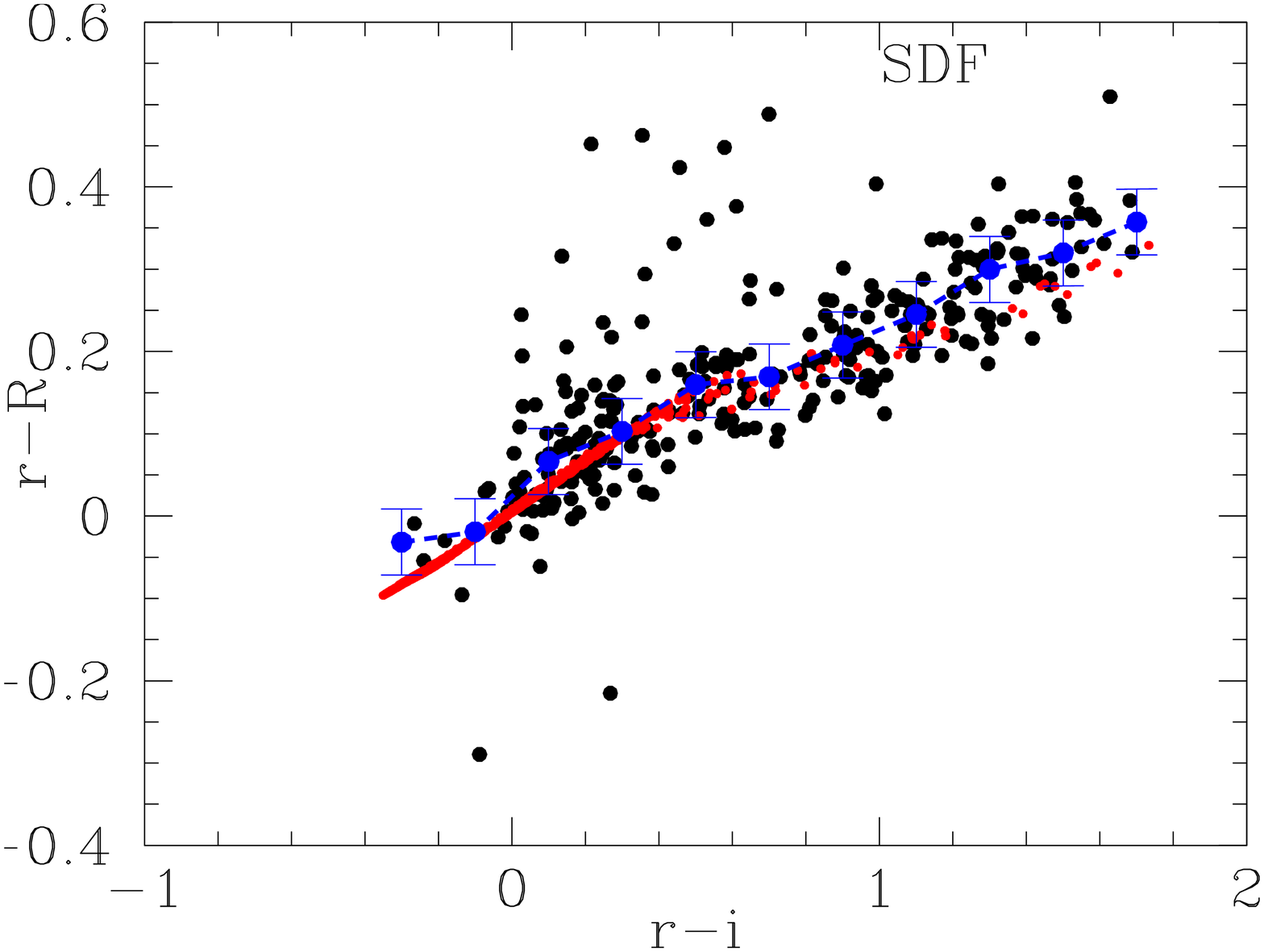}
\FigureFile(80mm,60mm){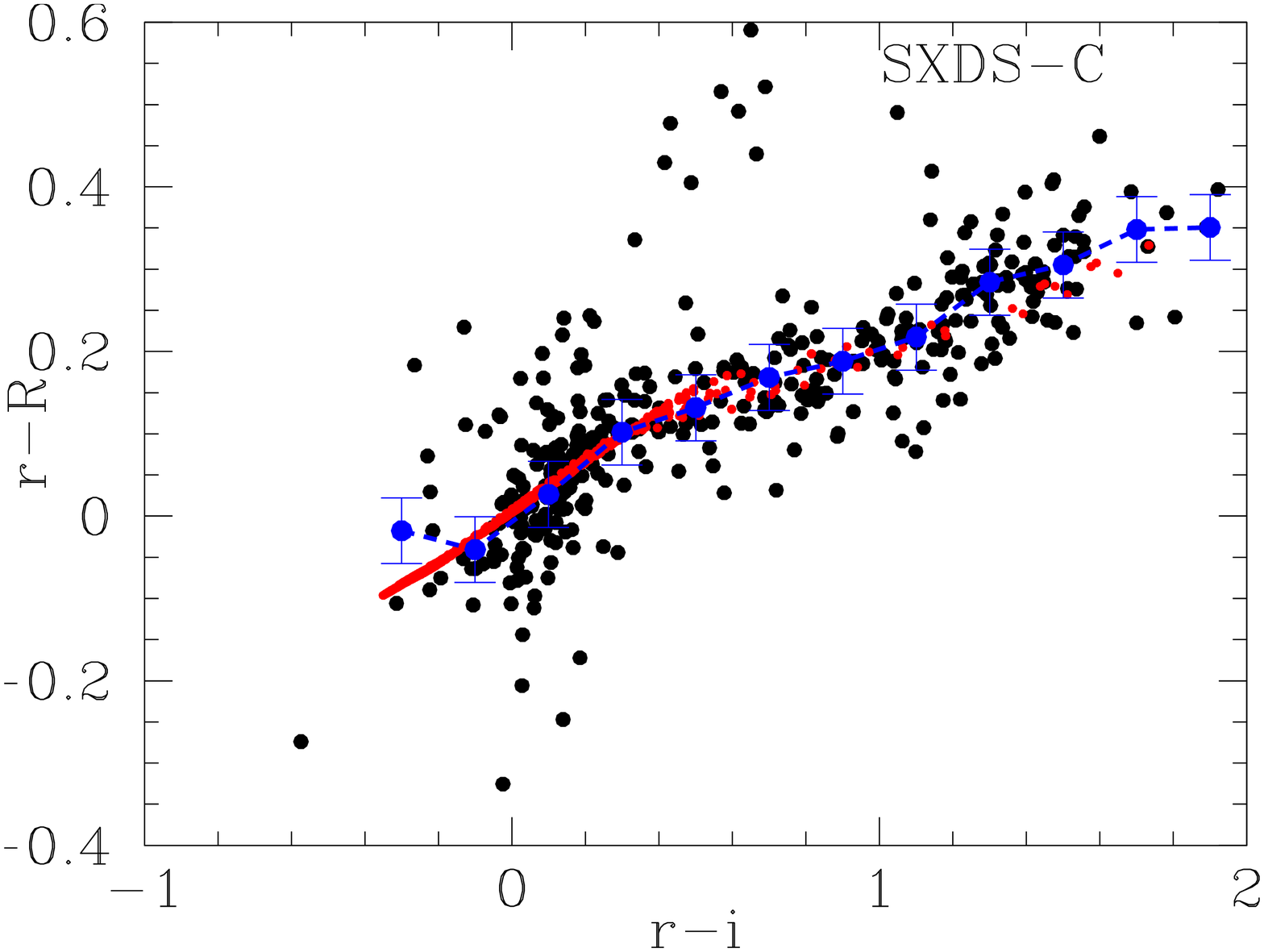}\\
\FigureFile(80mm,60mm){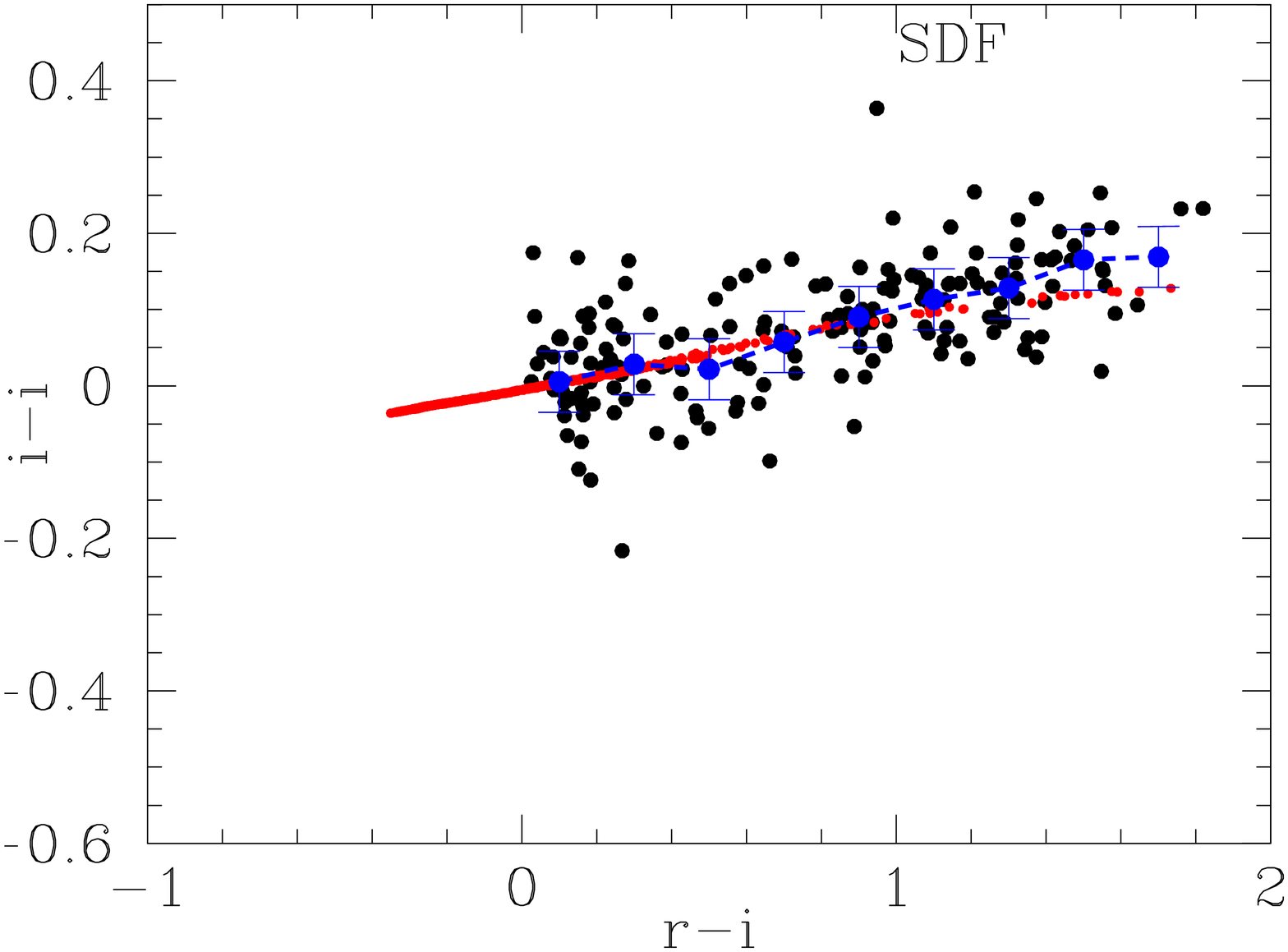}
\FigureFile(80mm,60mm){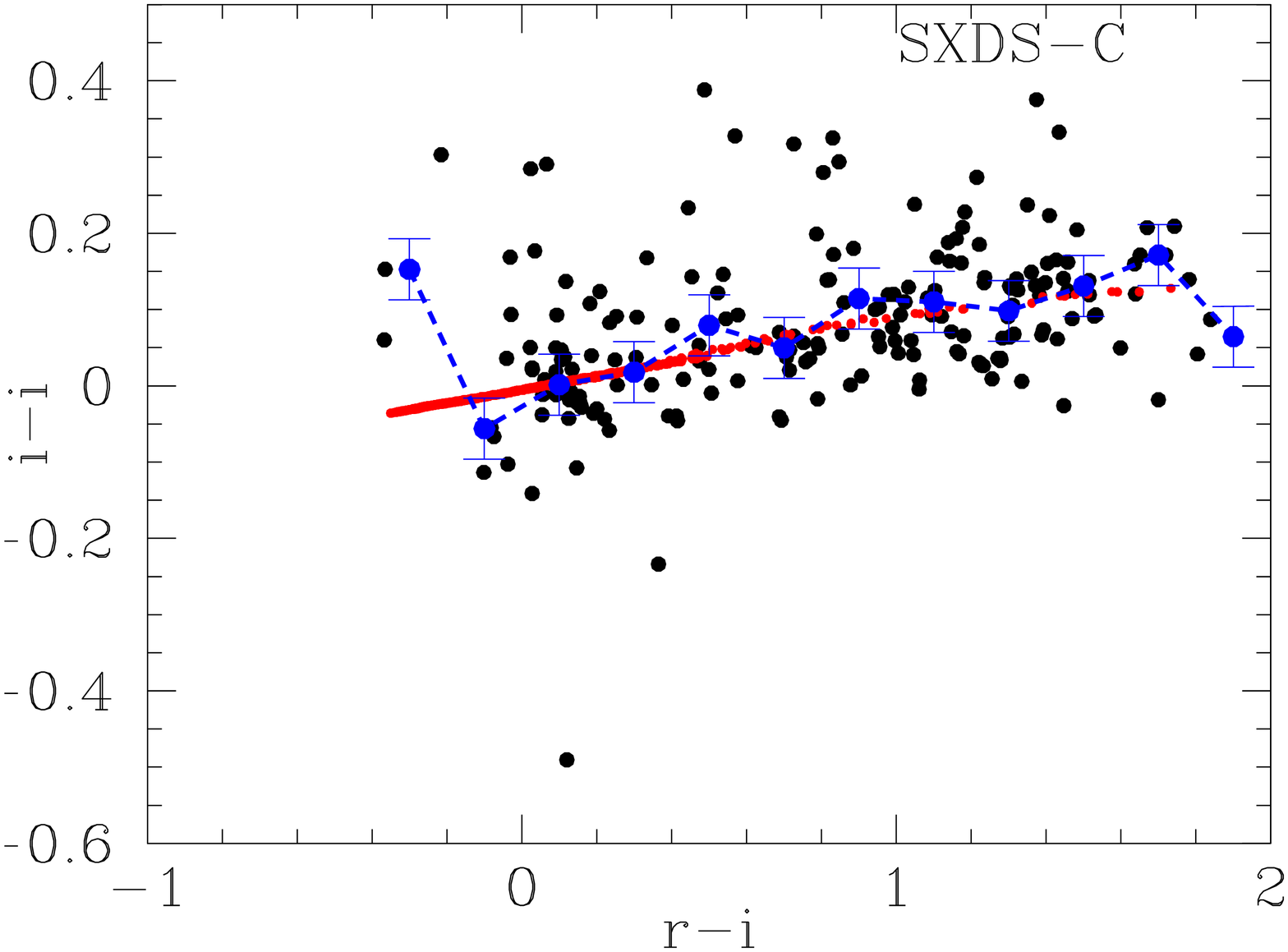}\\
\FigureFile(80mm,60mm){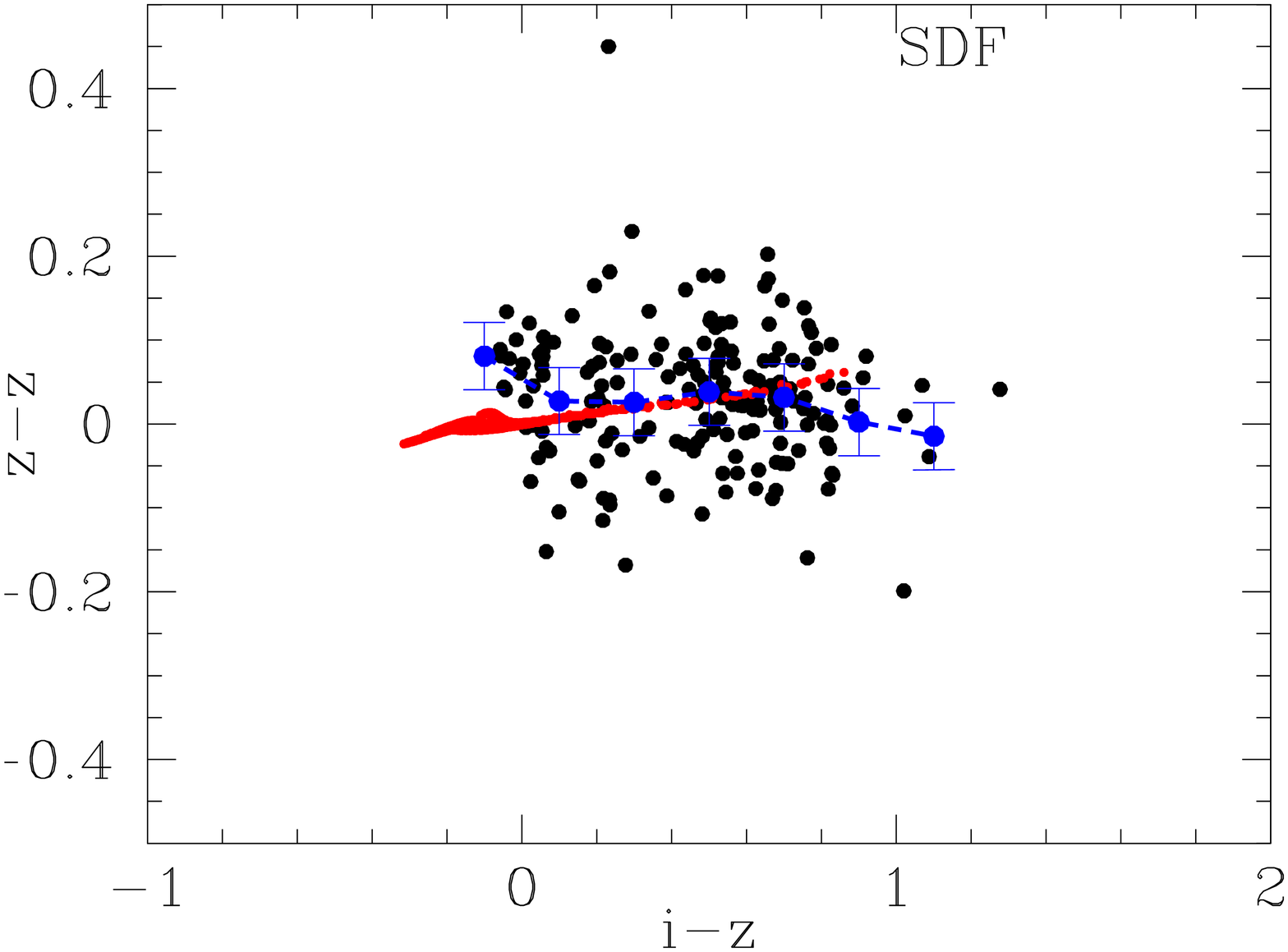}
\FigureFile(80mm,60mm){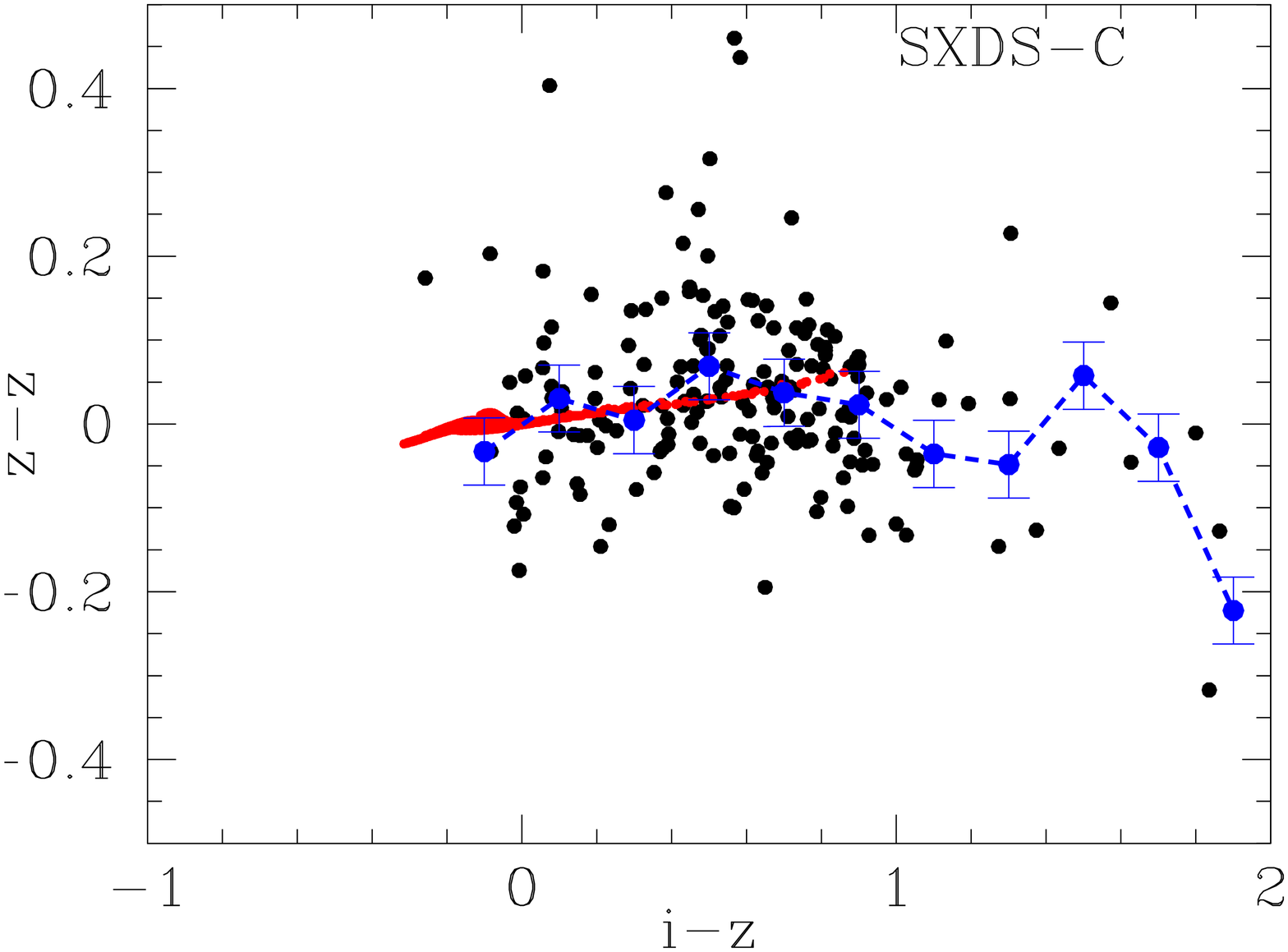}\\

\addtocounter{figure}{-1}
\caption{
Continued...
}
\end{figure}

\clearpage

\clearpage

\clearpage 

\clearpage

\begin{figure}
\FigureFile(80mm,60mm){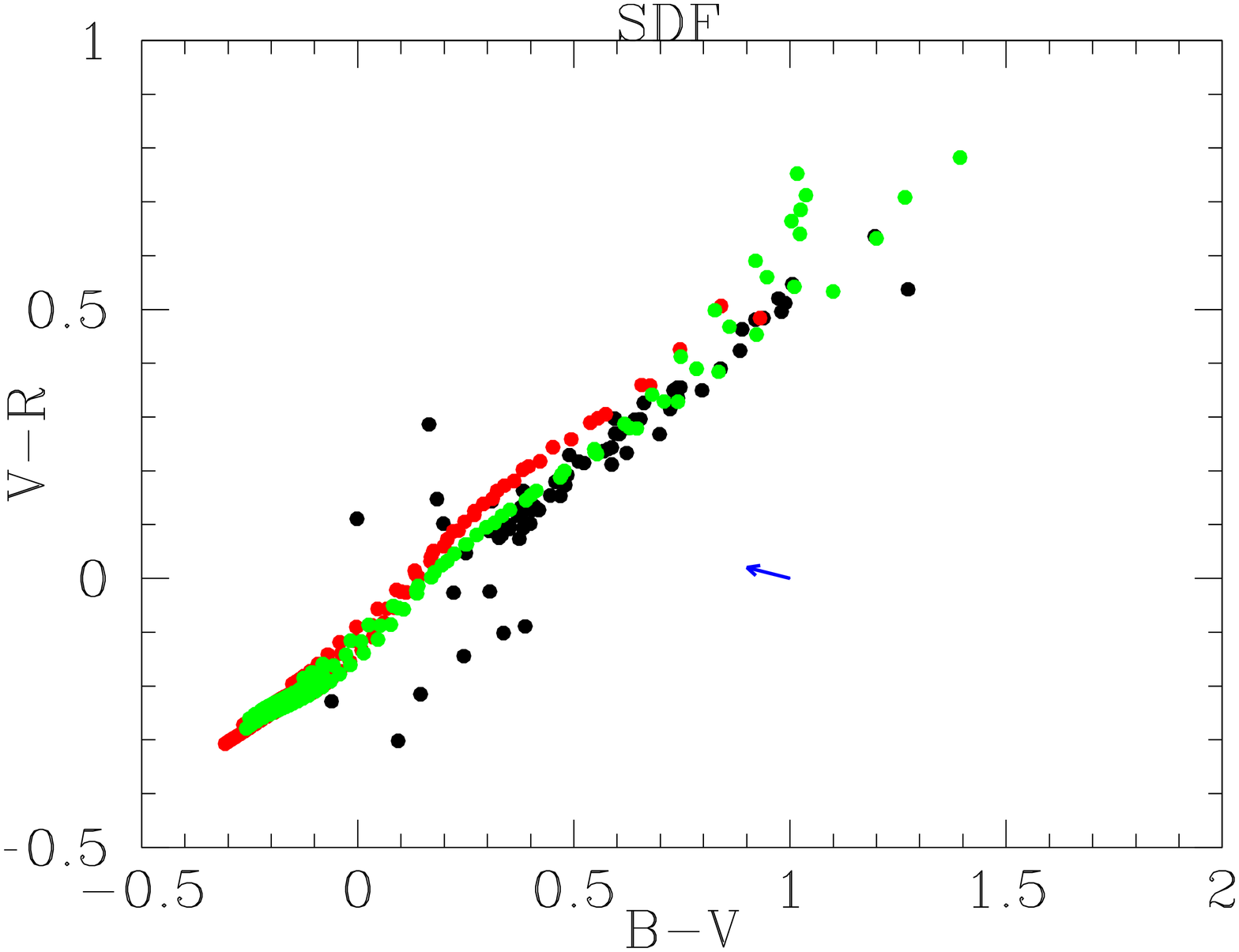}
\FigureFile(80mm,60mm){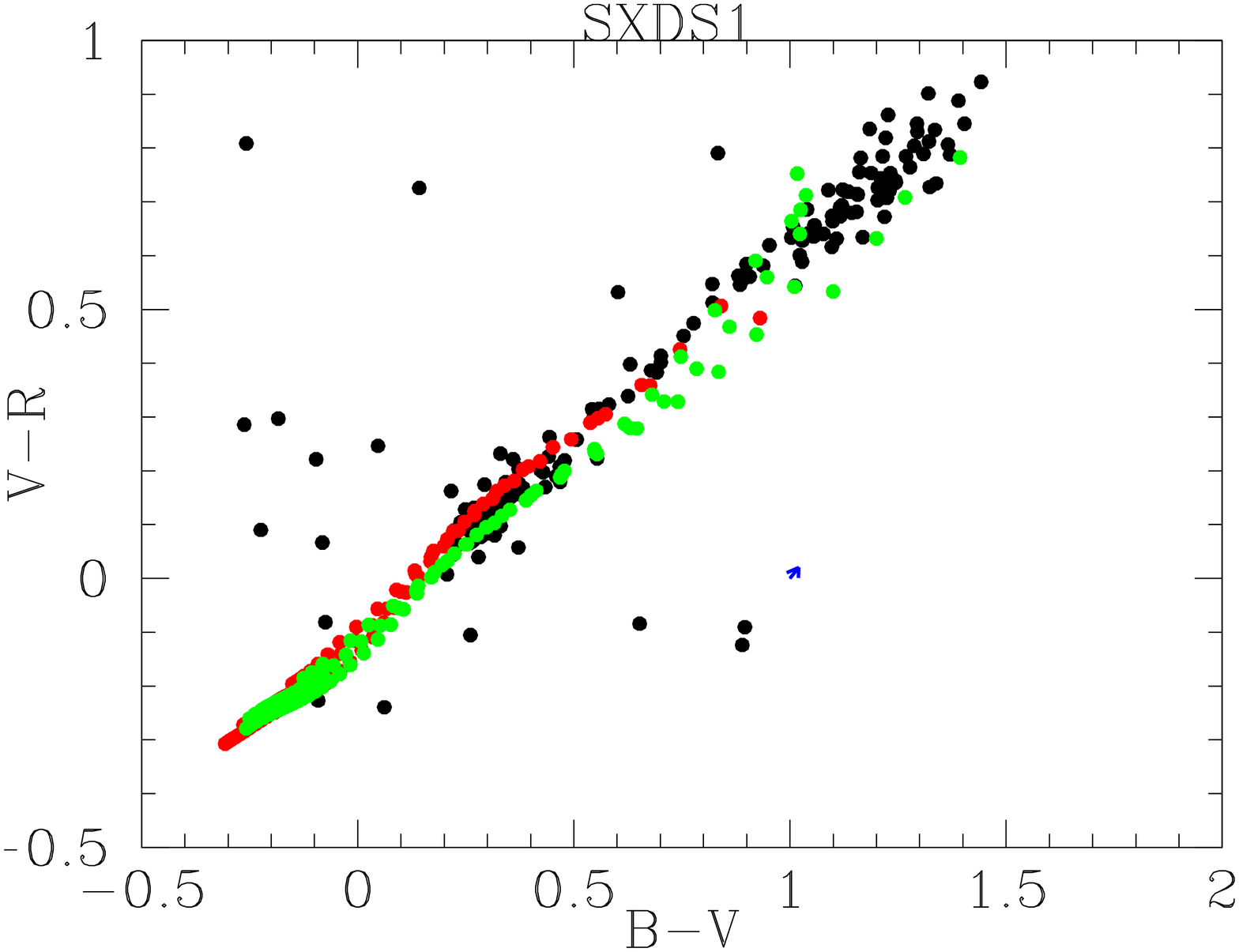}\\
\FigureFile(80mm,60mm){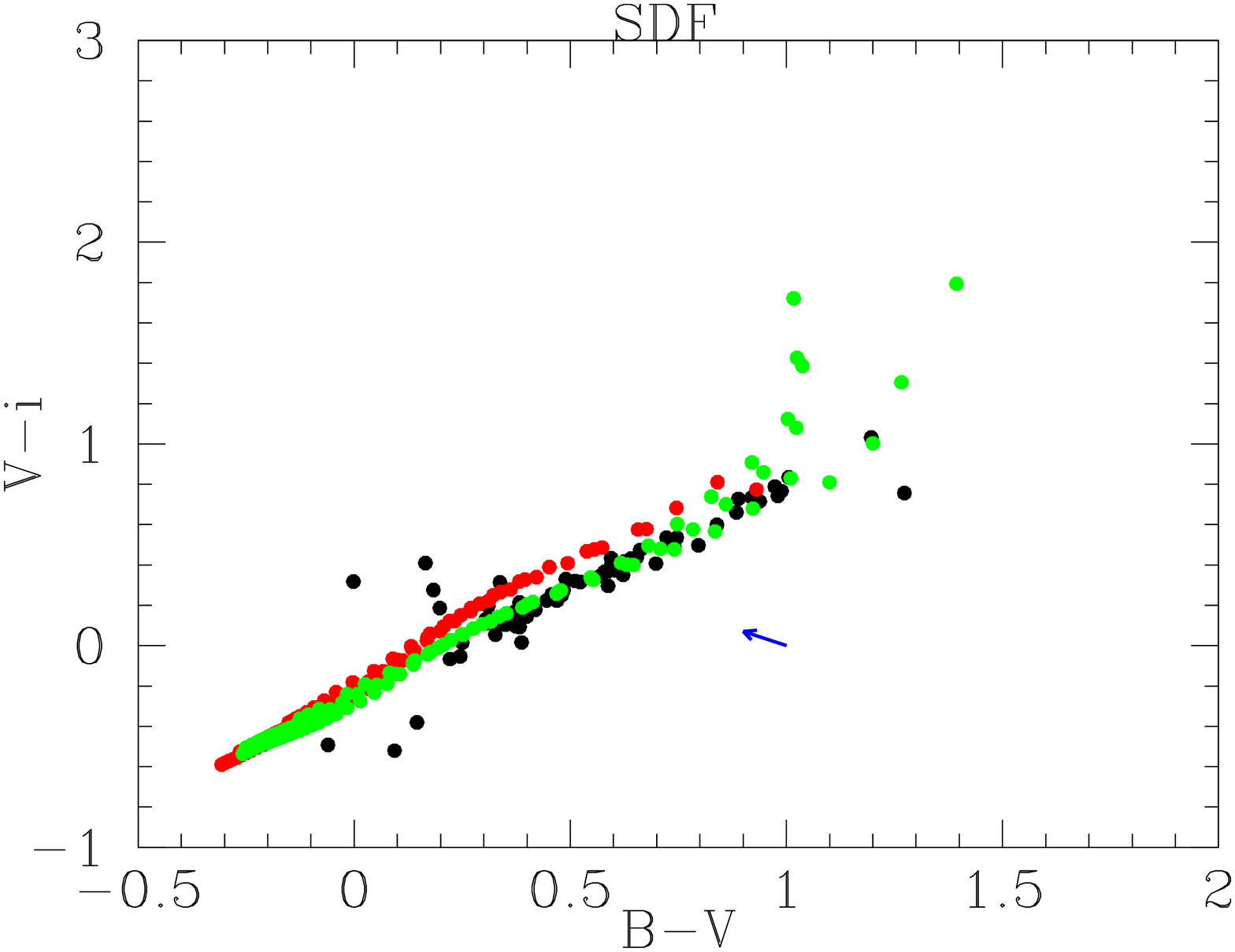}
\FigureFile(80mm,60mm){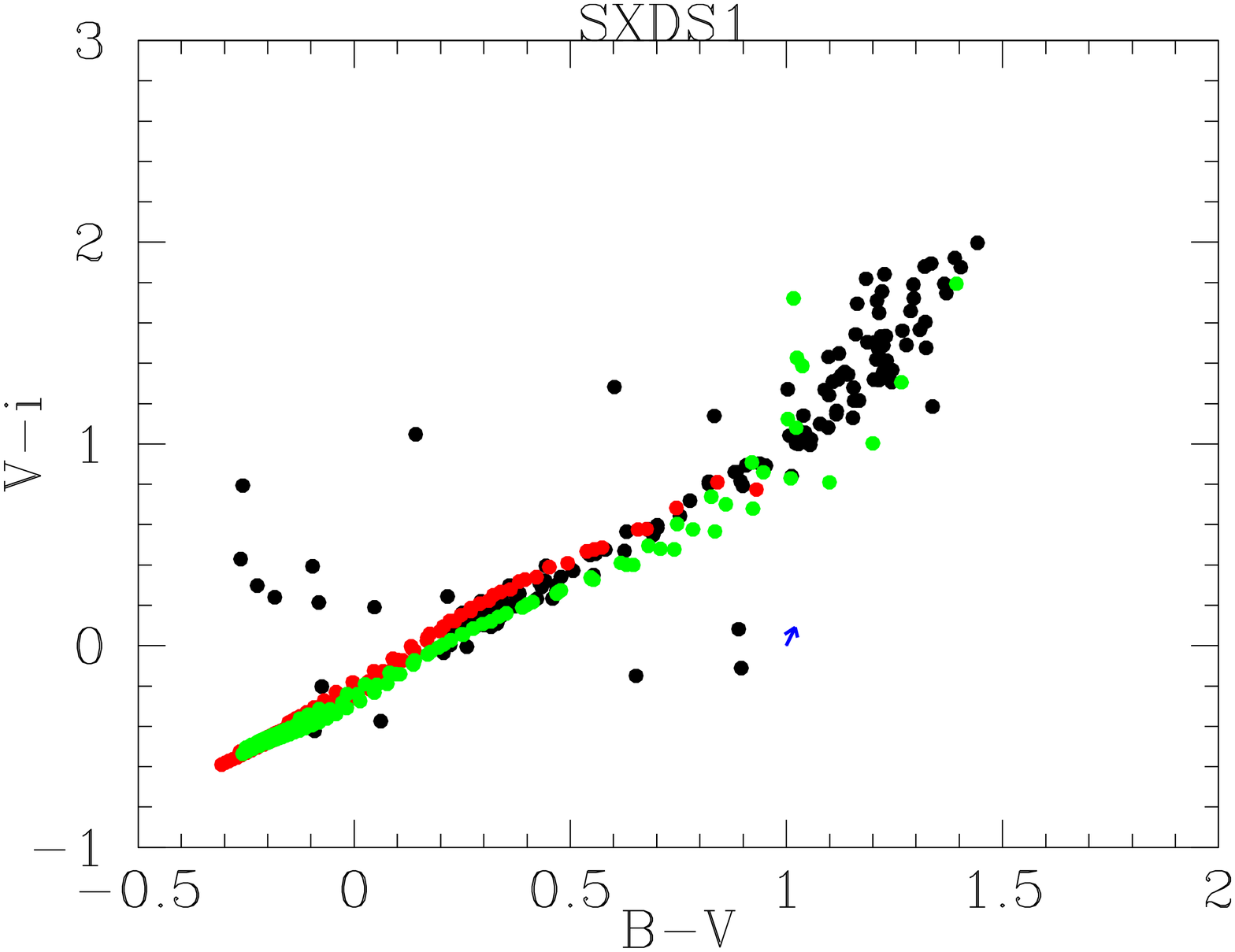}\\
\FigureFile(80mm,60mm){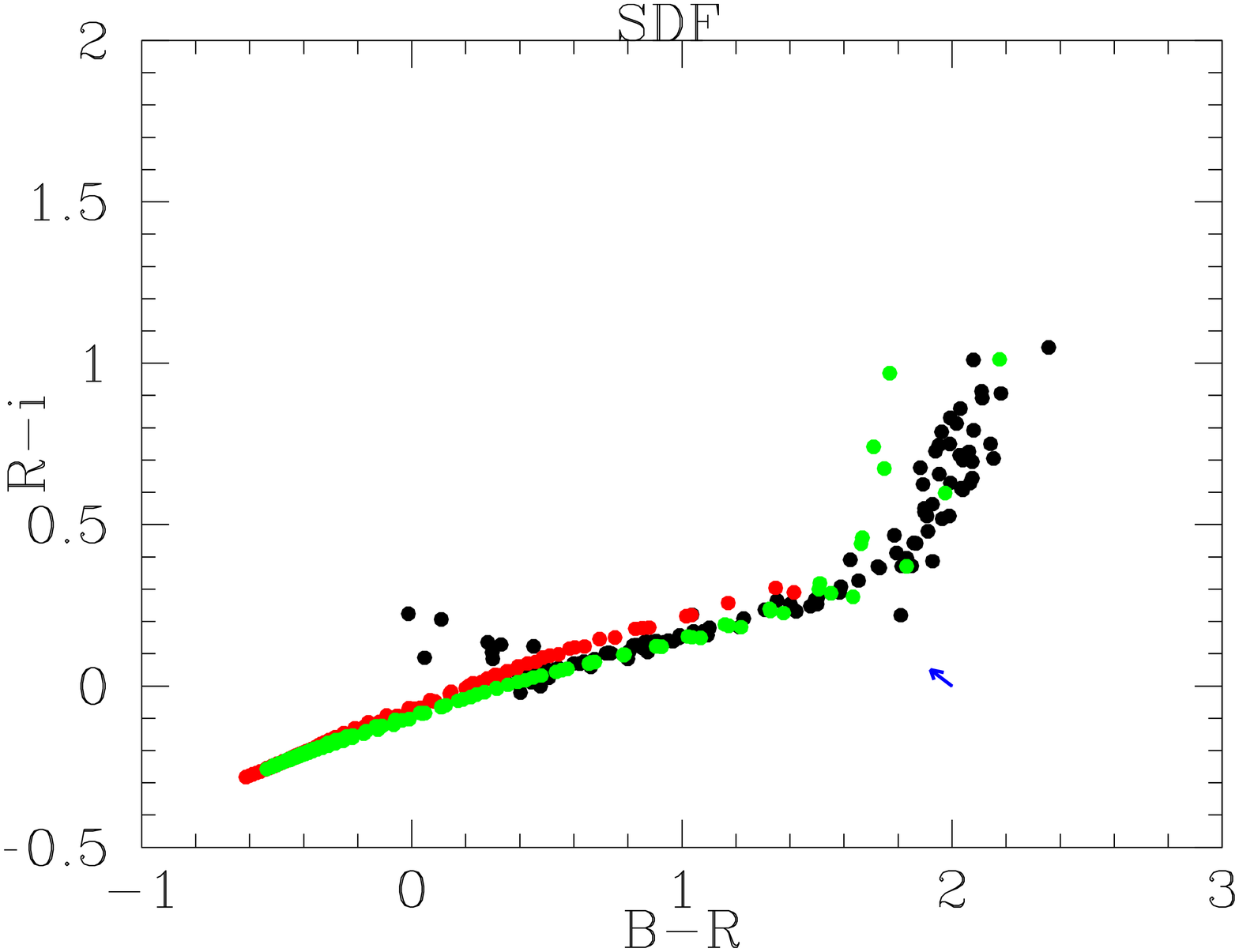}
\FigureFile(80mm,60mm){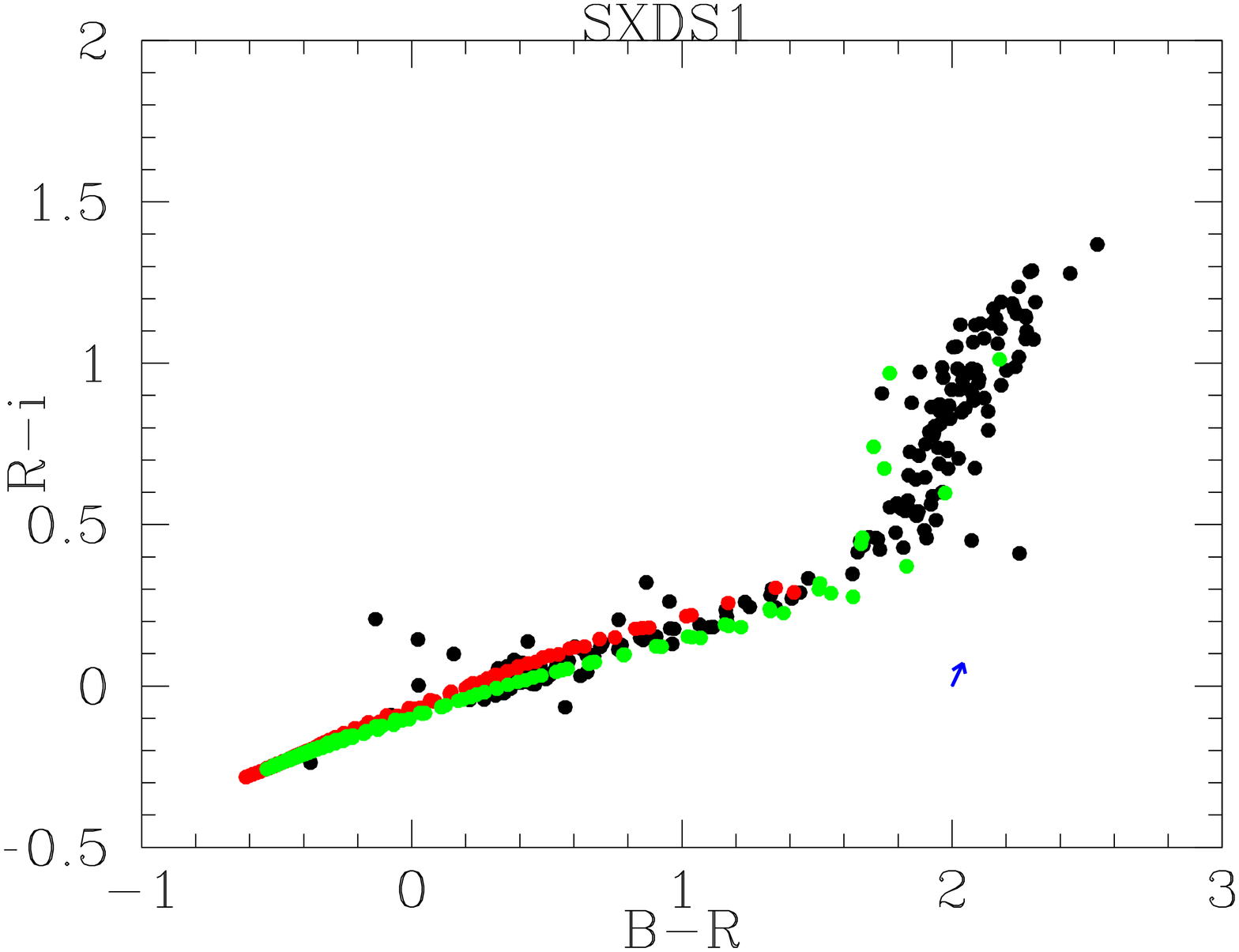}\\
\caption{
Color-color diagram of stars in the SDF/SXDS public catalogs
which match SDSS. 
The magnitude ranges are 
$20.5<g<21.5$ for (X-V) vs (V-Y),
$20<r<21$ for (X-R) vs (R-Y), and 
$19.5<i<20.5$ for (X-i) vs (i-Y), where X=(B,V,R) and Y=(R,i,z).
The filled circles are the values in the original SDF/SXDS catalog, 
and the blue arrow shows our estimation of the offset.
The red and green dots are synthetic colors 
of [Fe/H]=-2.5a and [Fe/H]=0 models, respectively.
}
\label{fig:Supcolcol1}
\end{figure}

\clearpage 

\begin{figure}
\FigureFile(80mm,60mm){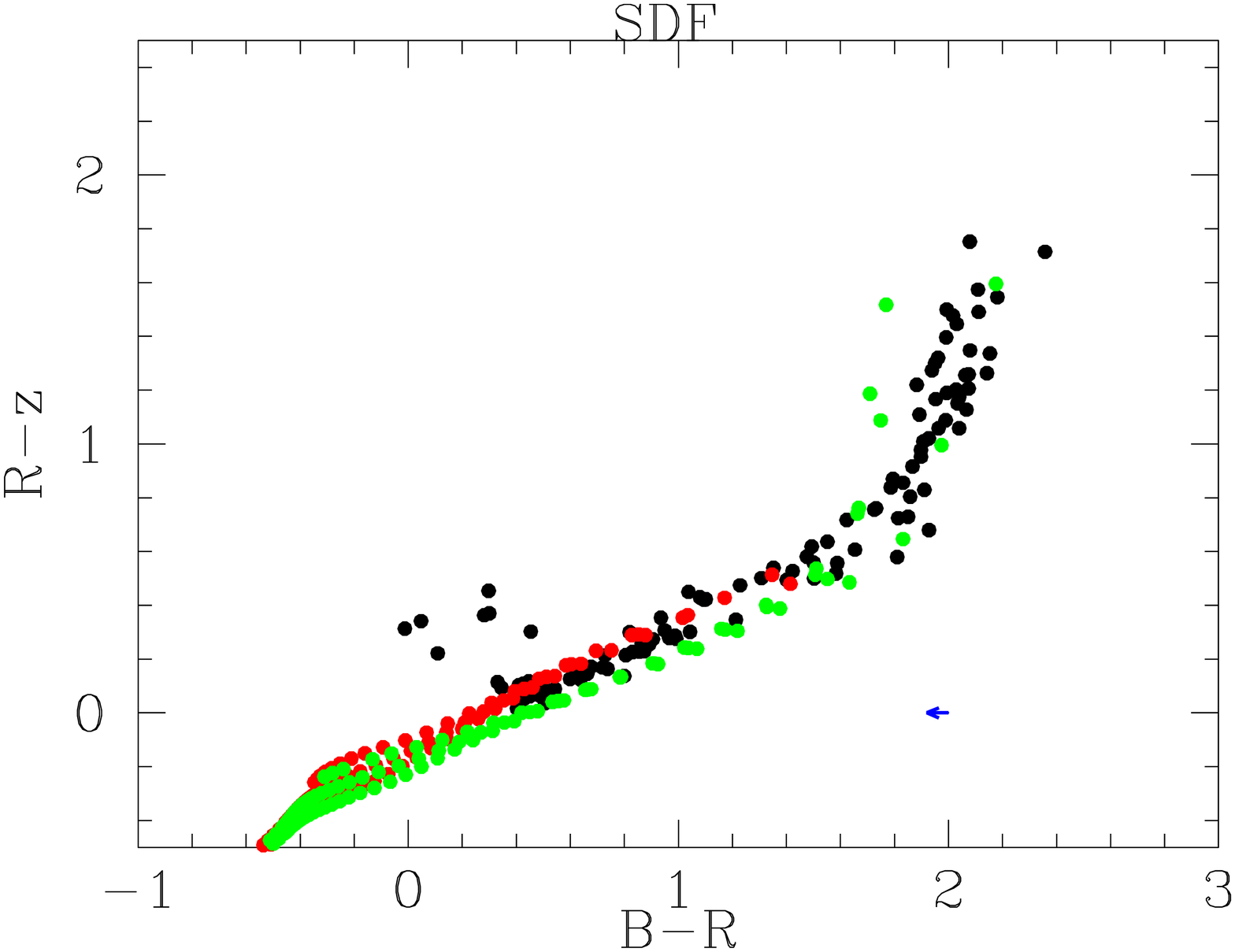}
\FigureFile(80mm,60mm){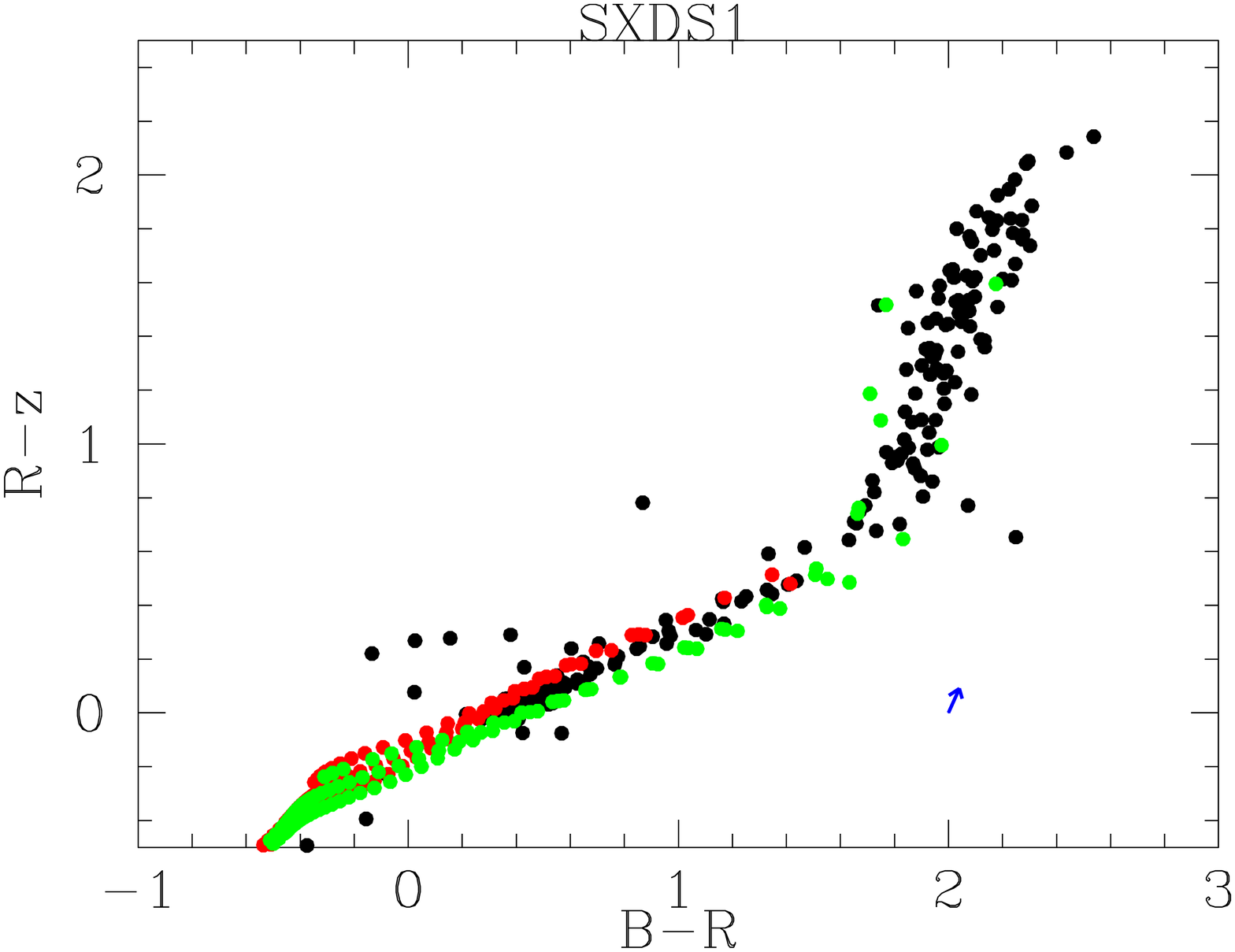}\\
\FigureFile(80mm,60mm){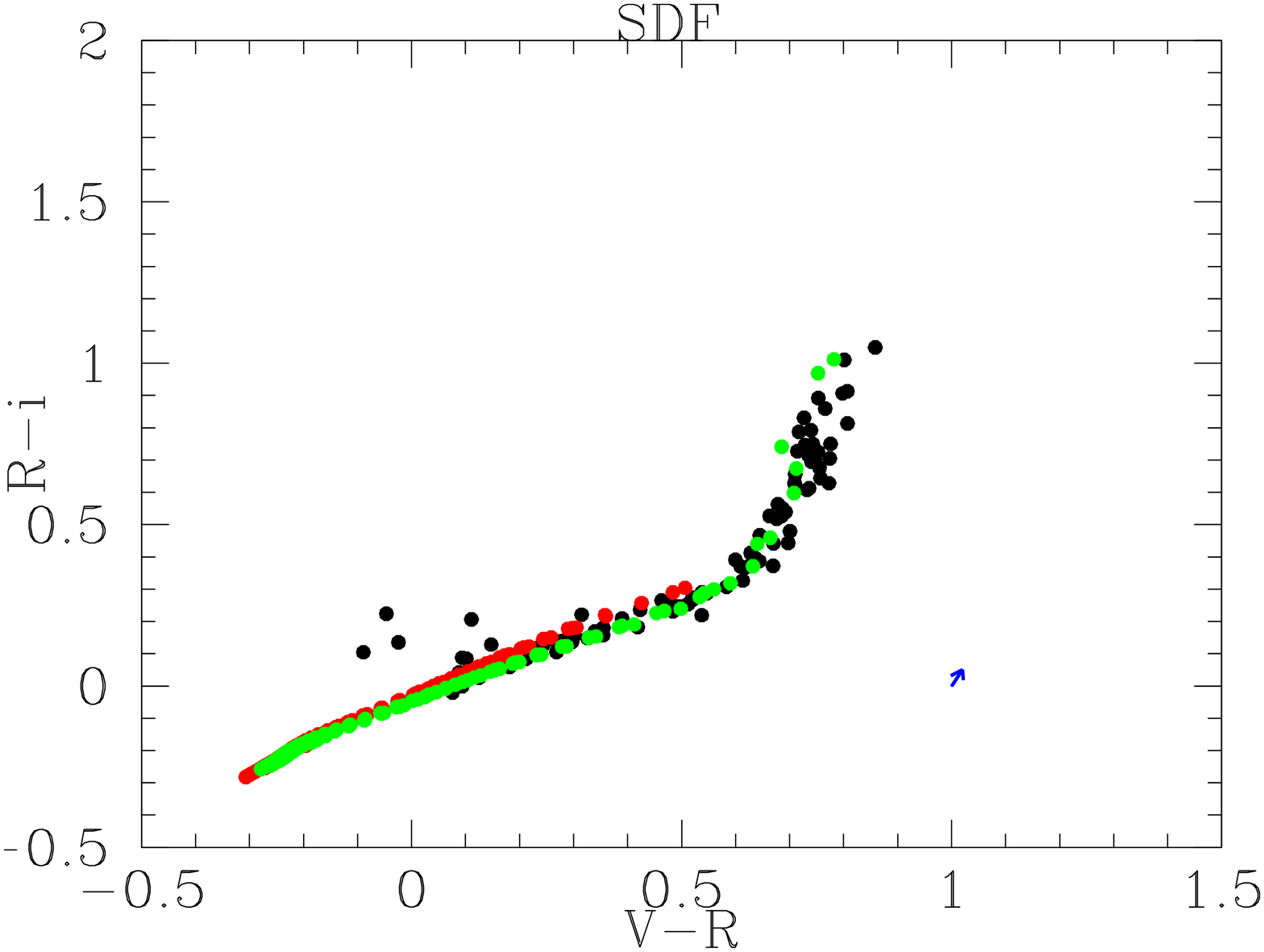}
\FigureFile(80mm,60mm){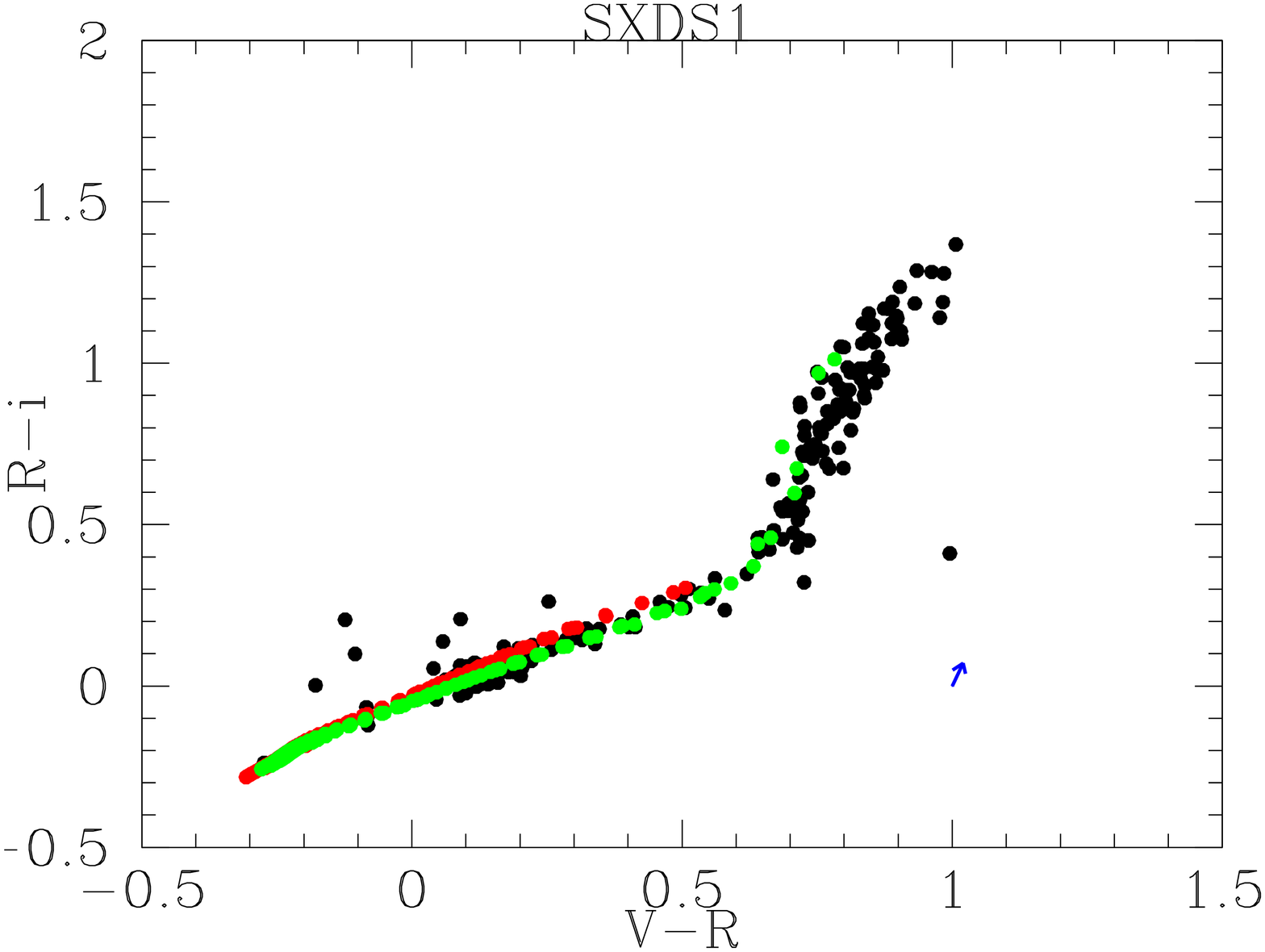}\\
\FigureFile(80mm,60mm){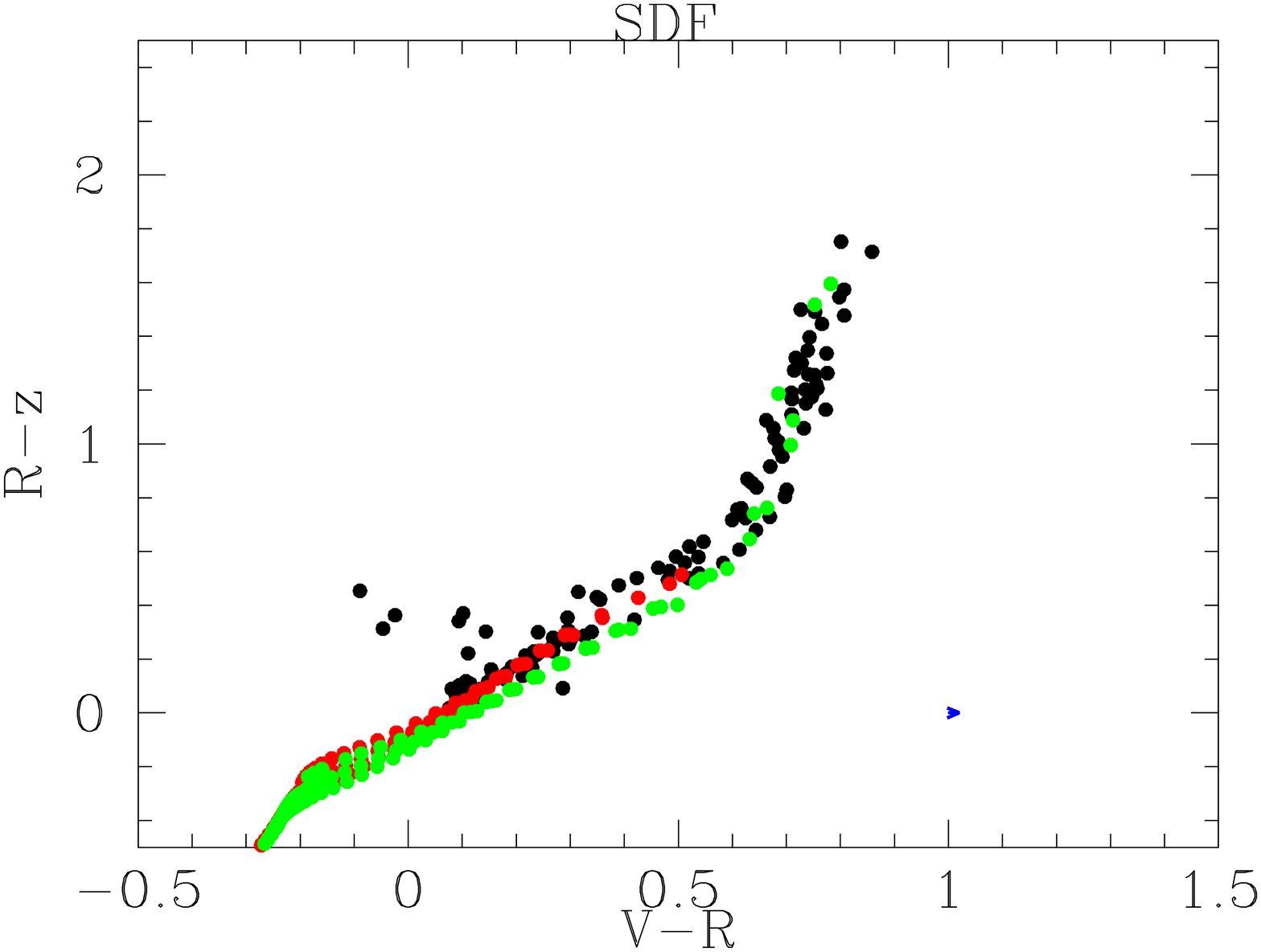}
\FigureFile(80mm,60mm){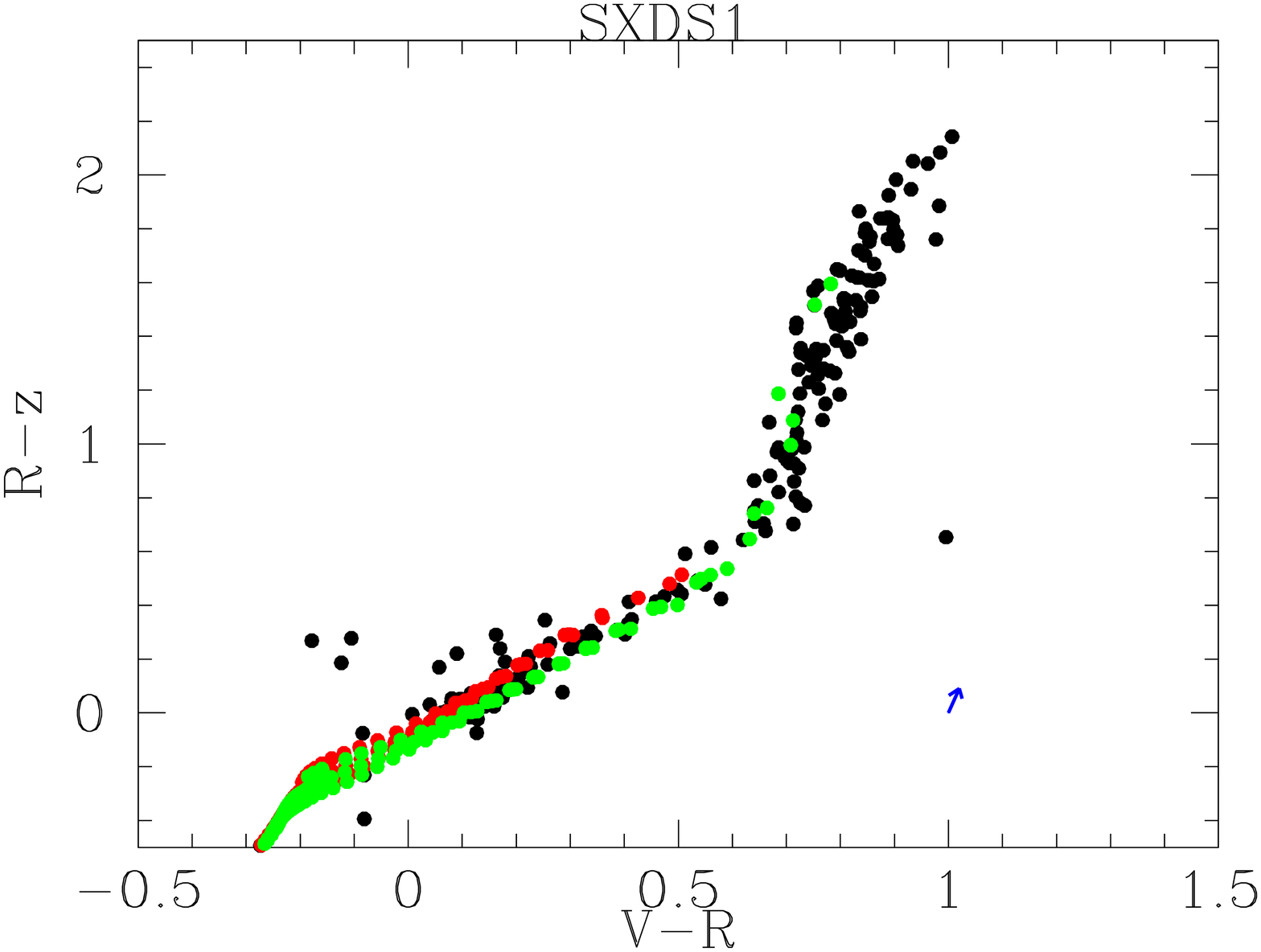}\\
\addtocounter{figure}{-1}
\caption{
Continued...
}
\end{figure}

\clearpage

\begin{figure}
\FigureFile(80mm,60mm){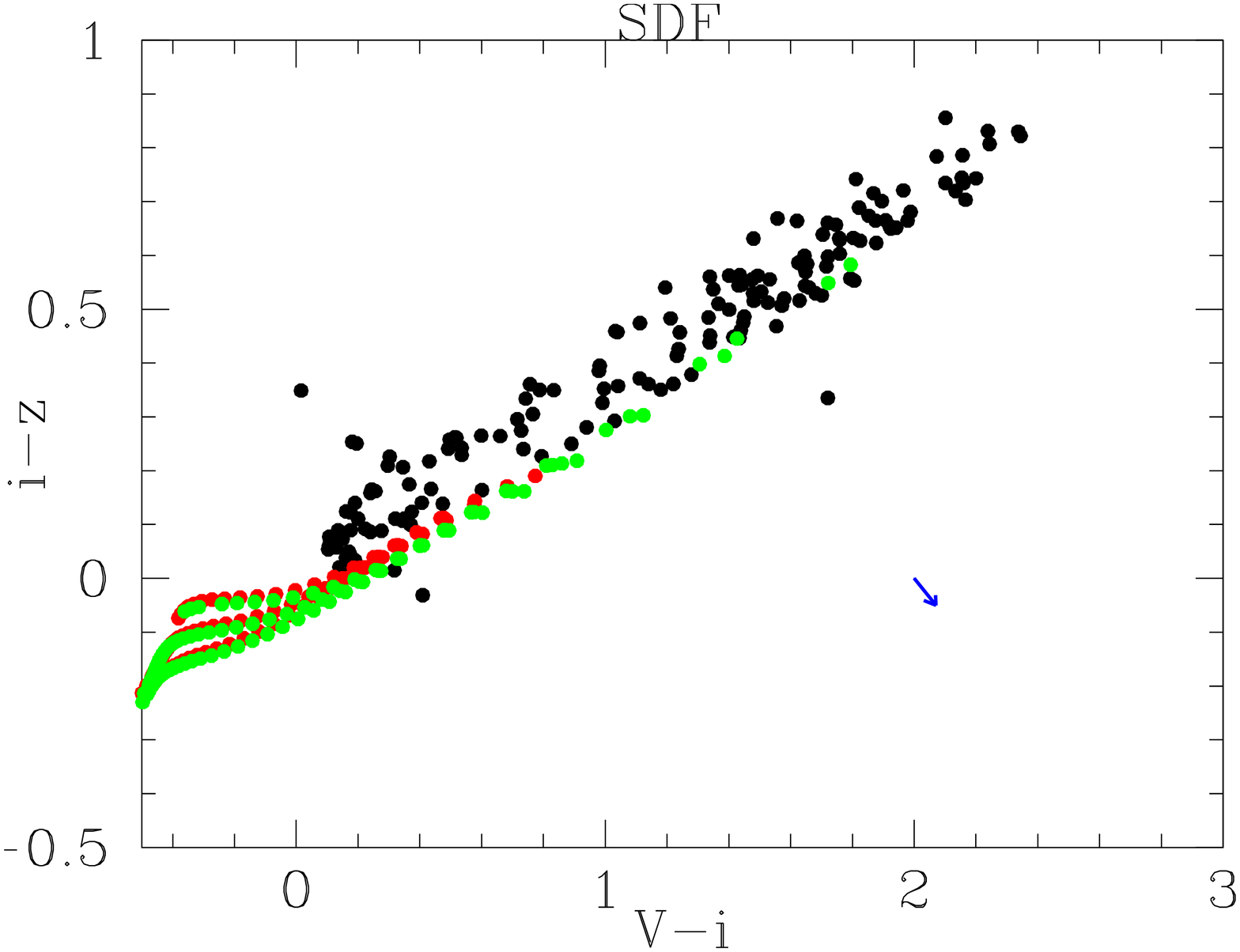}
\FigureFile(80mm,60mm){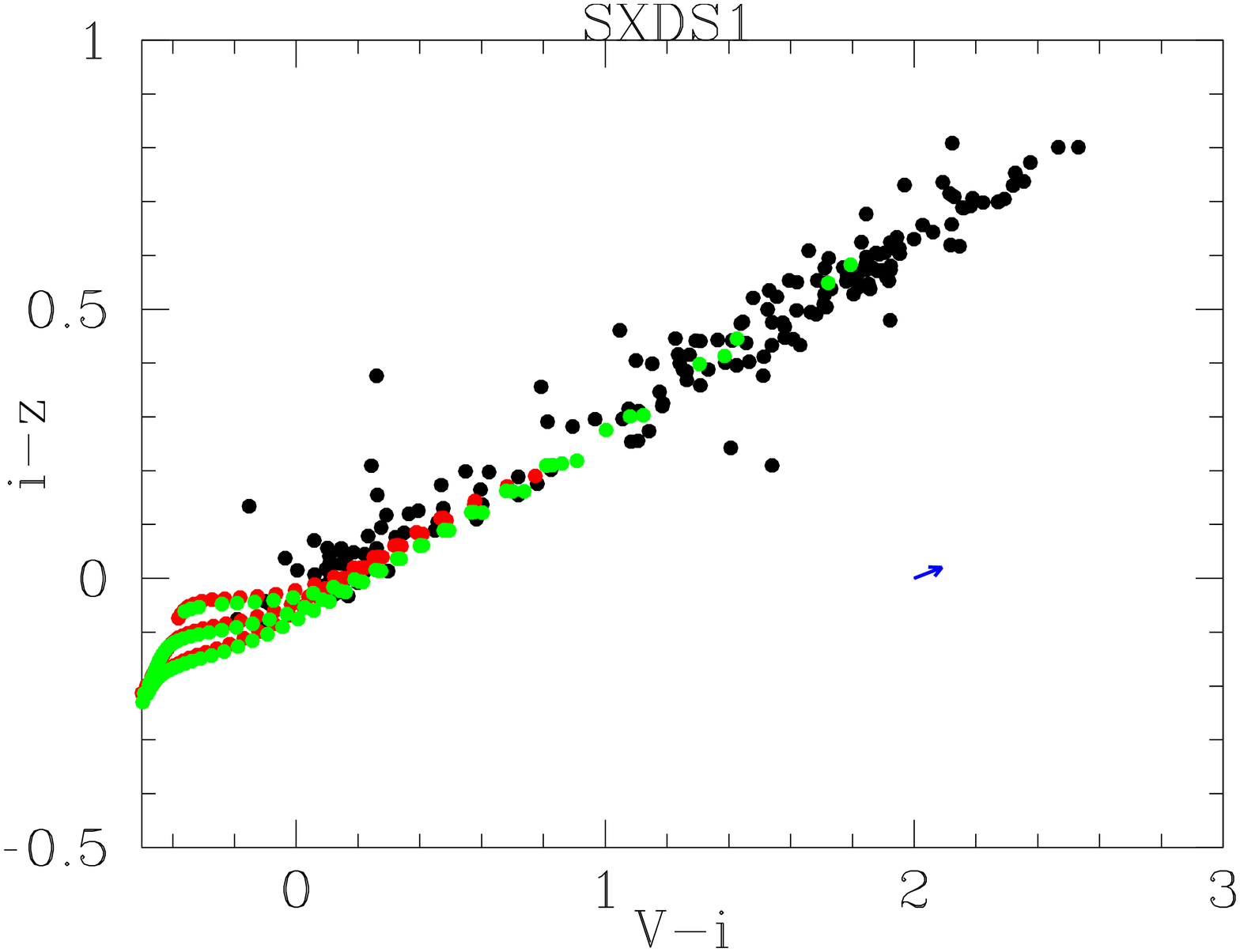}\\
\FigureFile(80mm,60mm){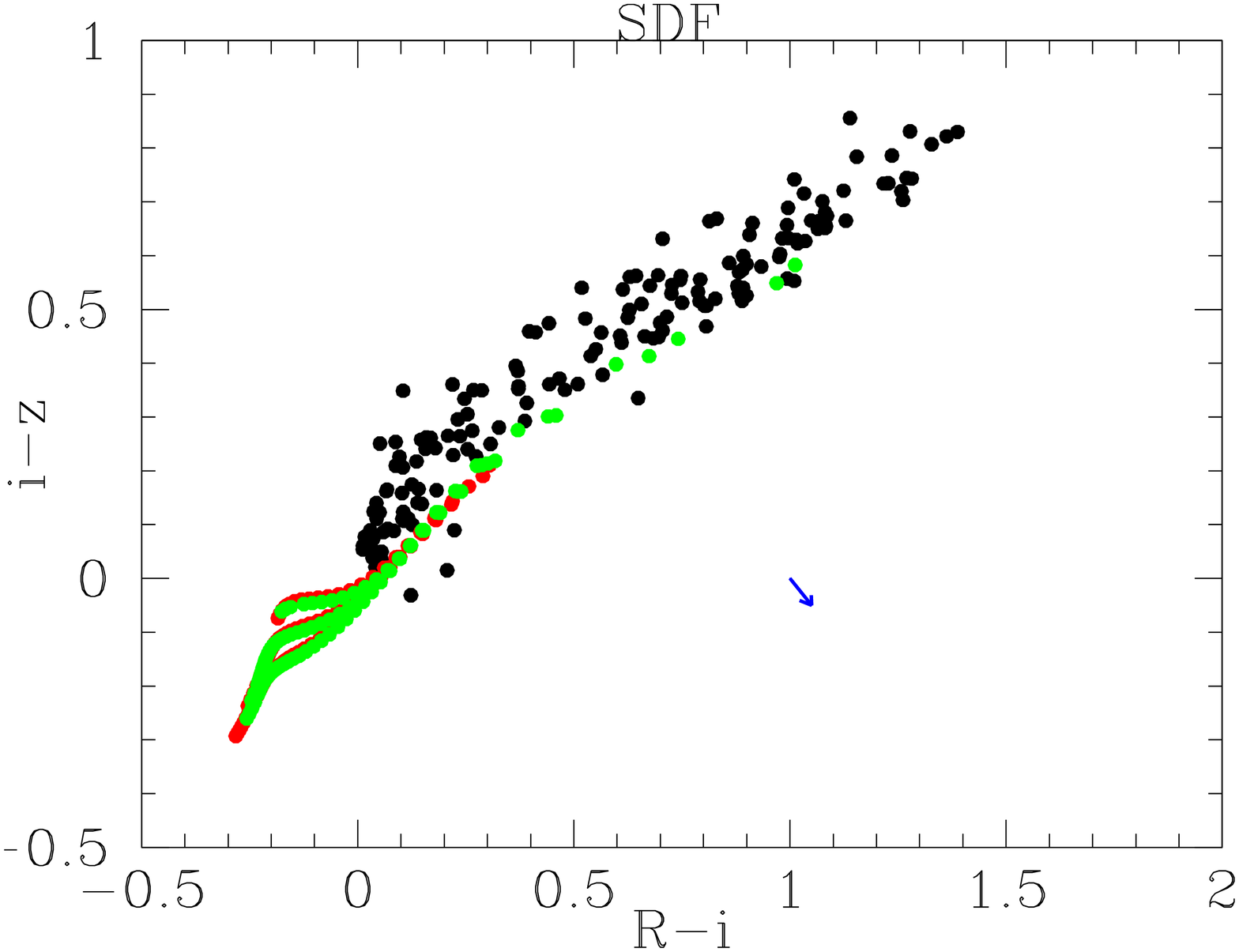}
\FigureFile(80mm,60mm){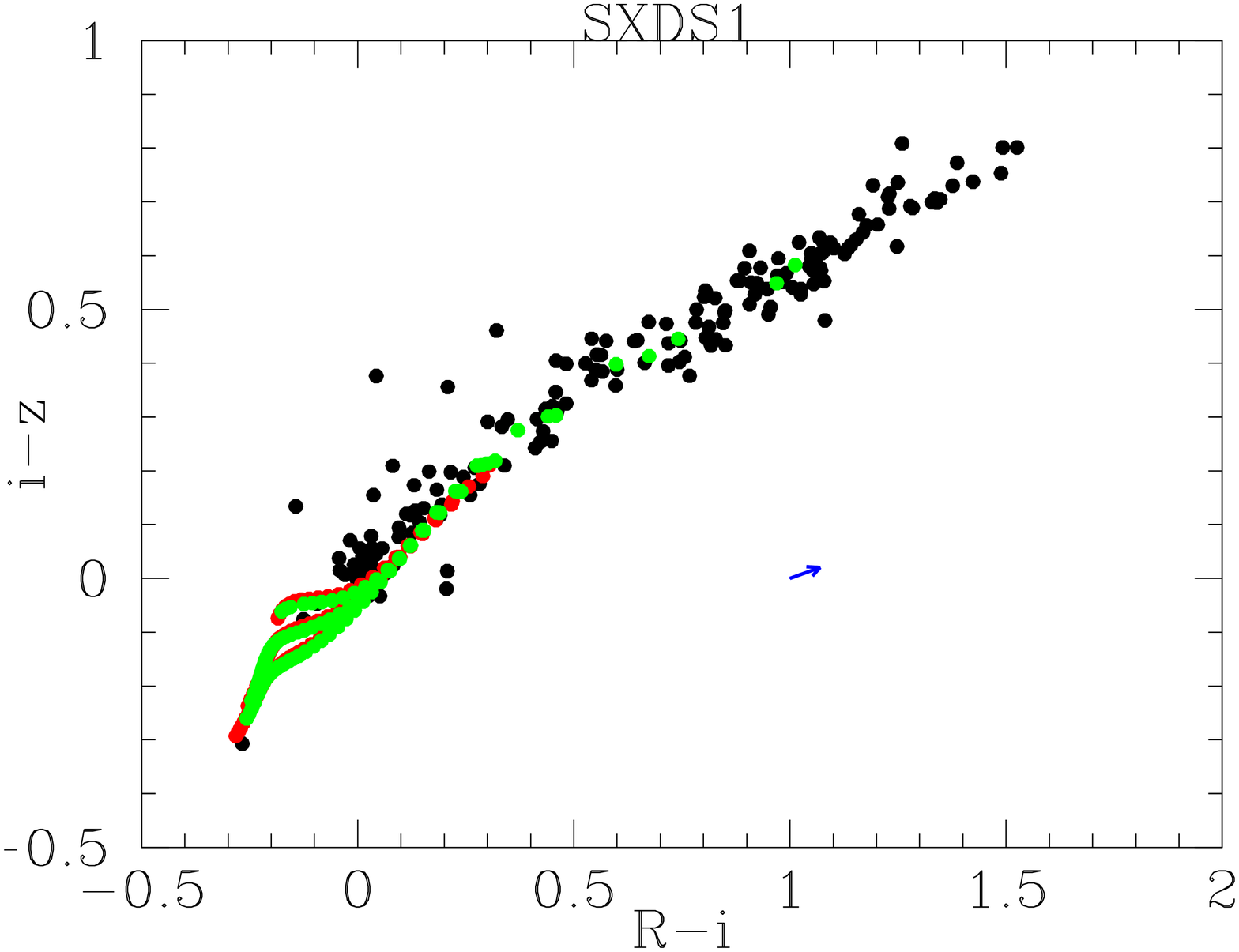}\\
\addtocounter{figure}{-1}
\caption{
Continued.
}
\end{figure}

\clearpage 
\begin{figure}
\FigureFile(80mm,60mm){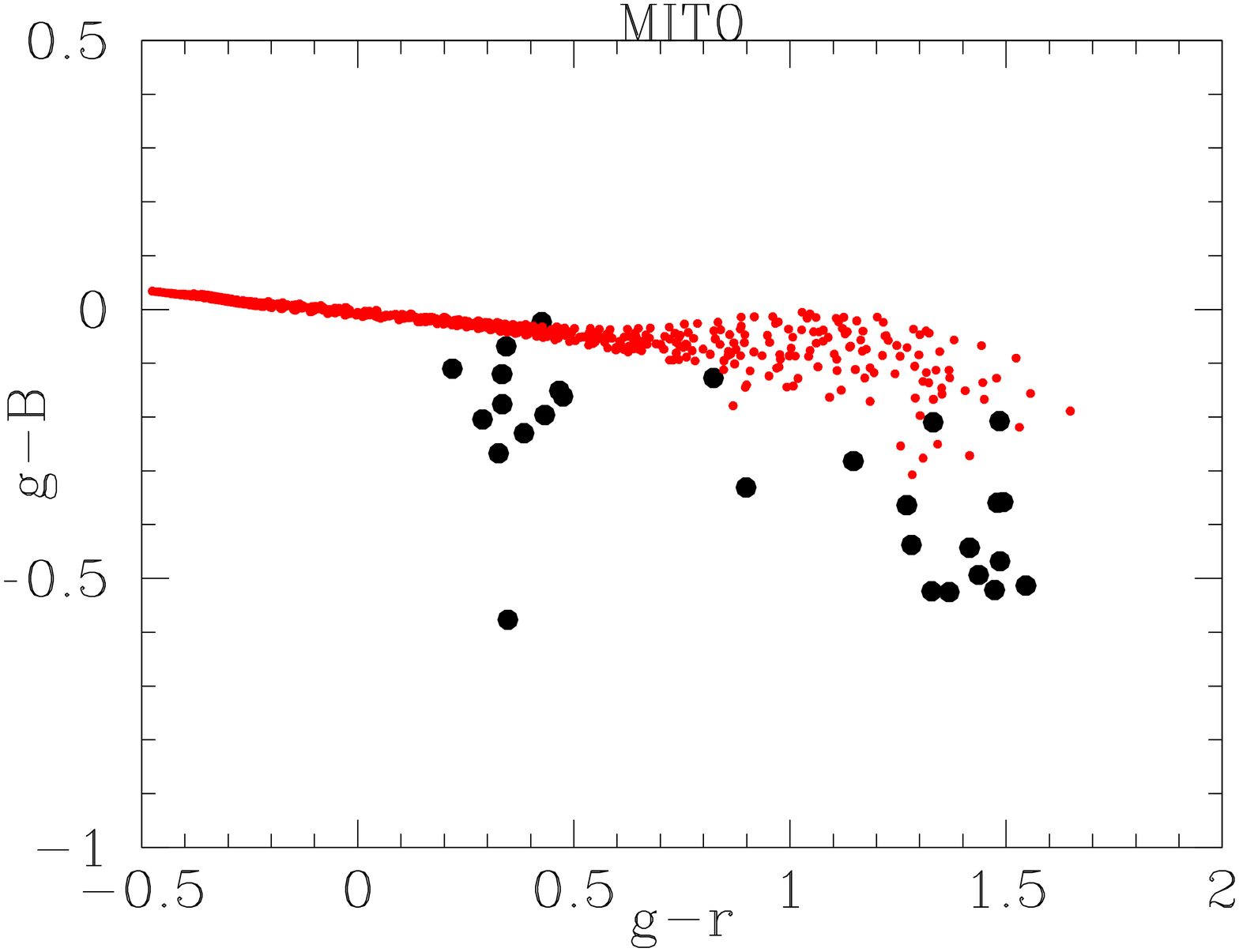}\FigureFile(80mm,60mm){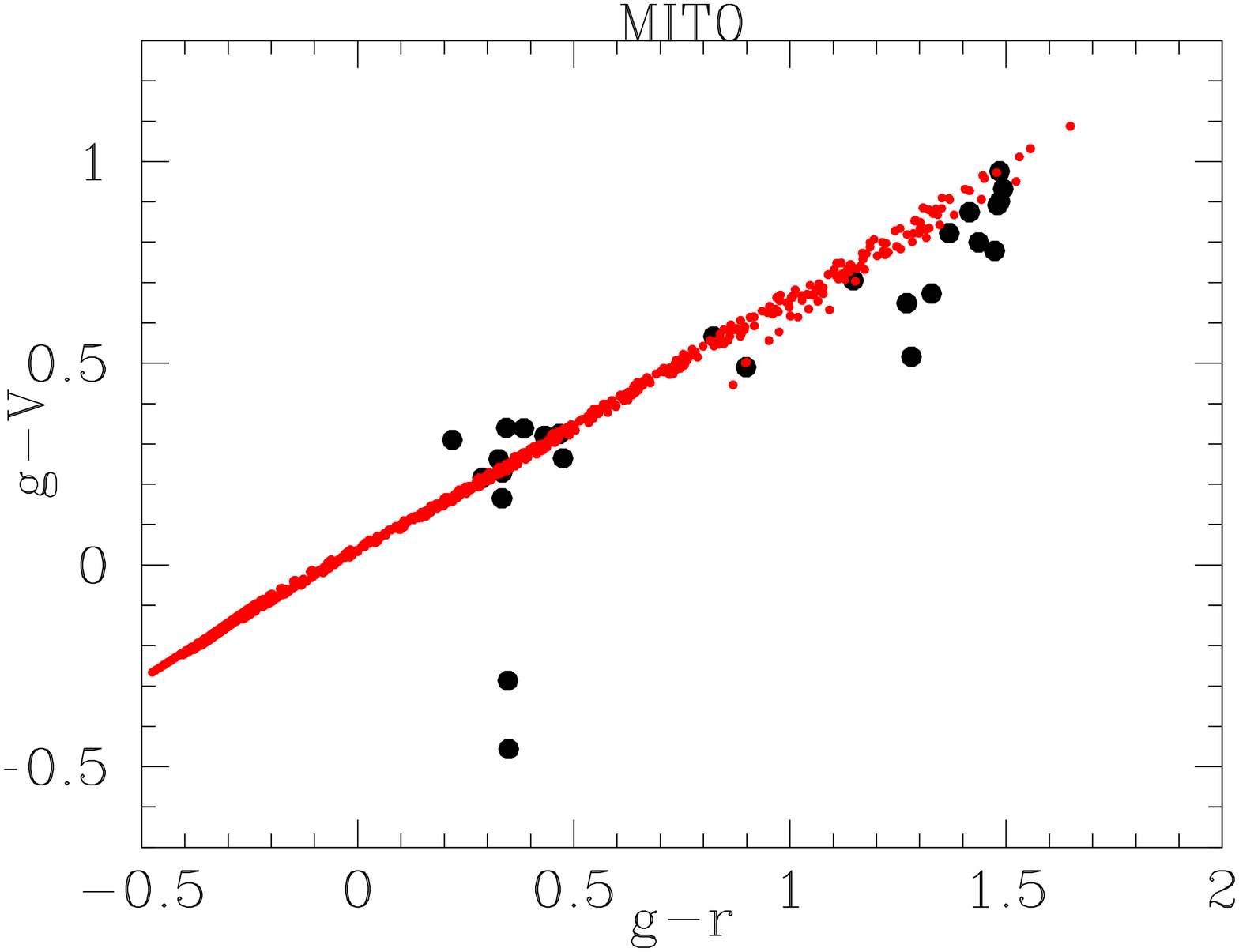}\\
\FigureFile(80mm,60mm){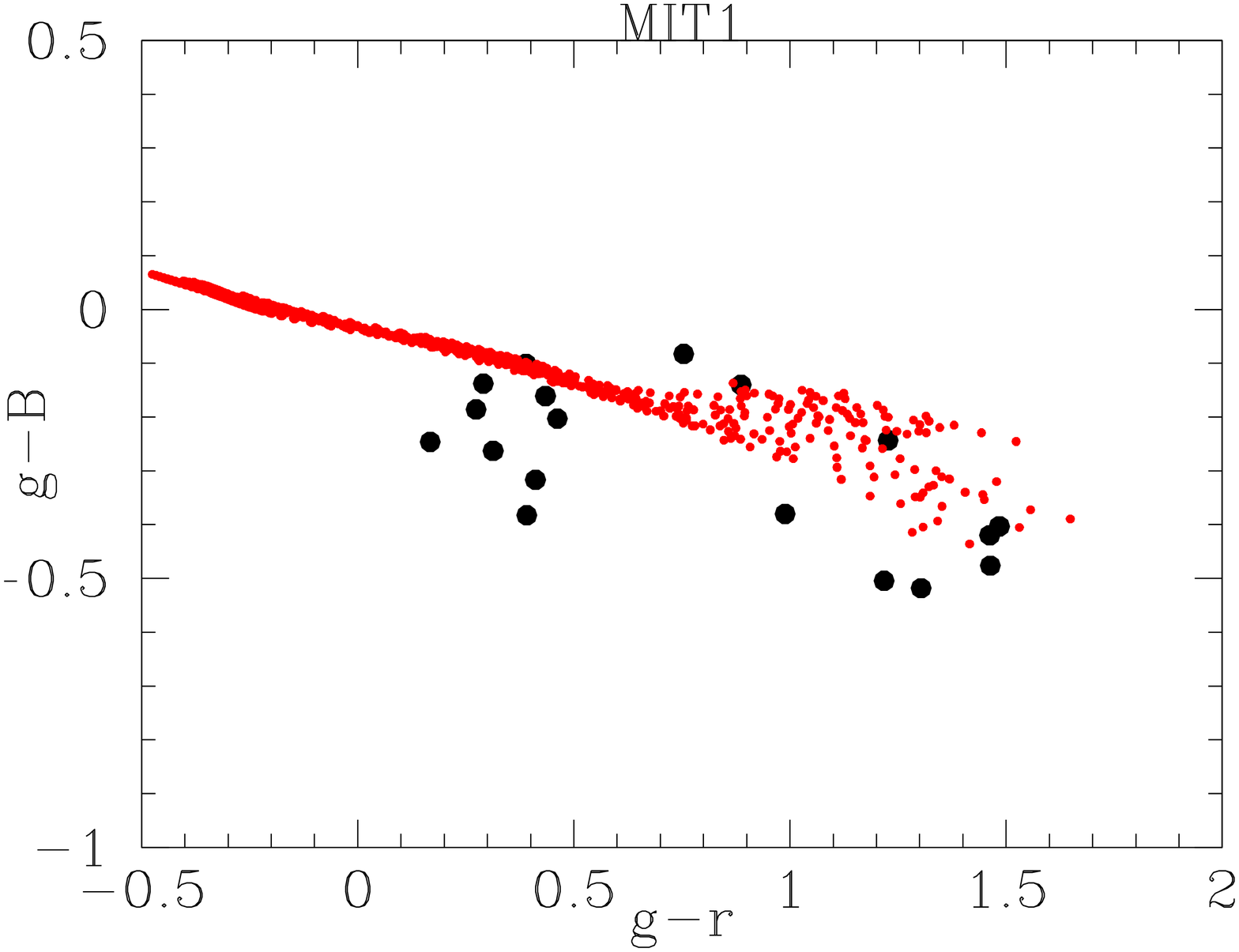}\FigureFile(80mm,60mm){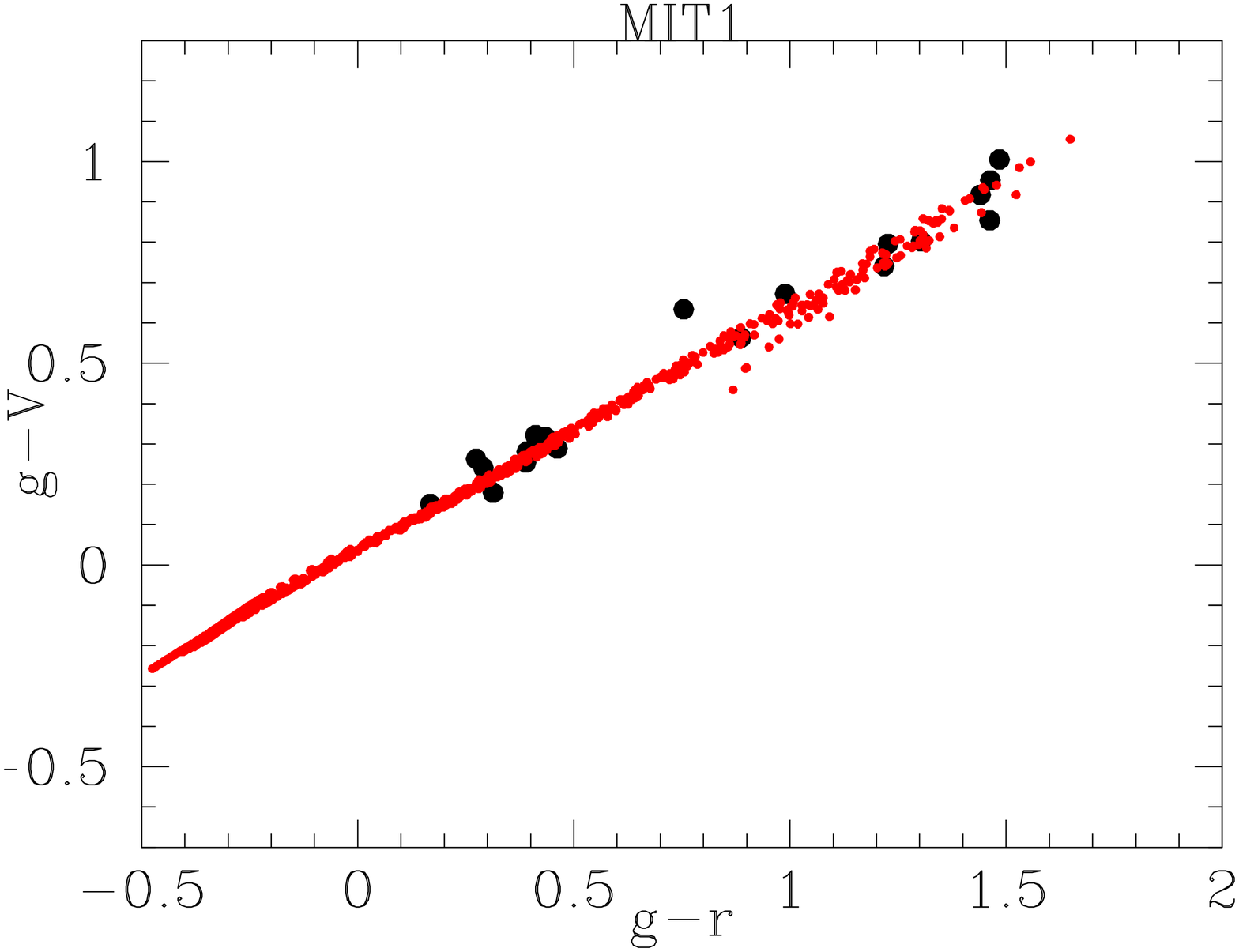}\\
\FigureFile(80mm,60mm){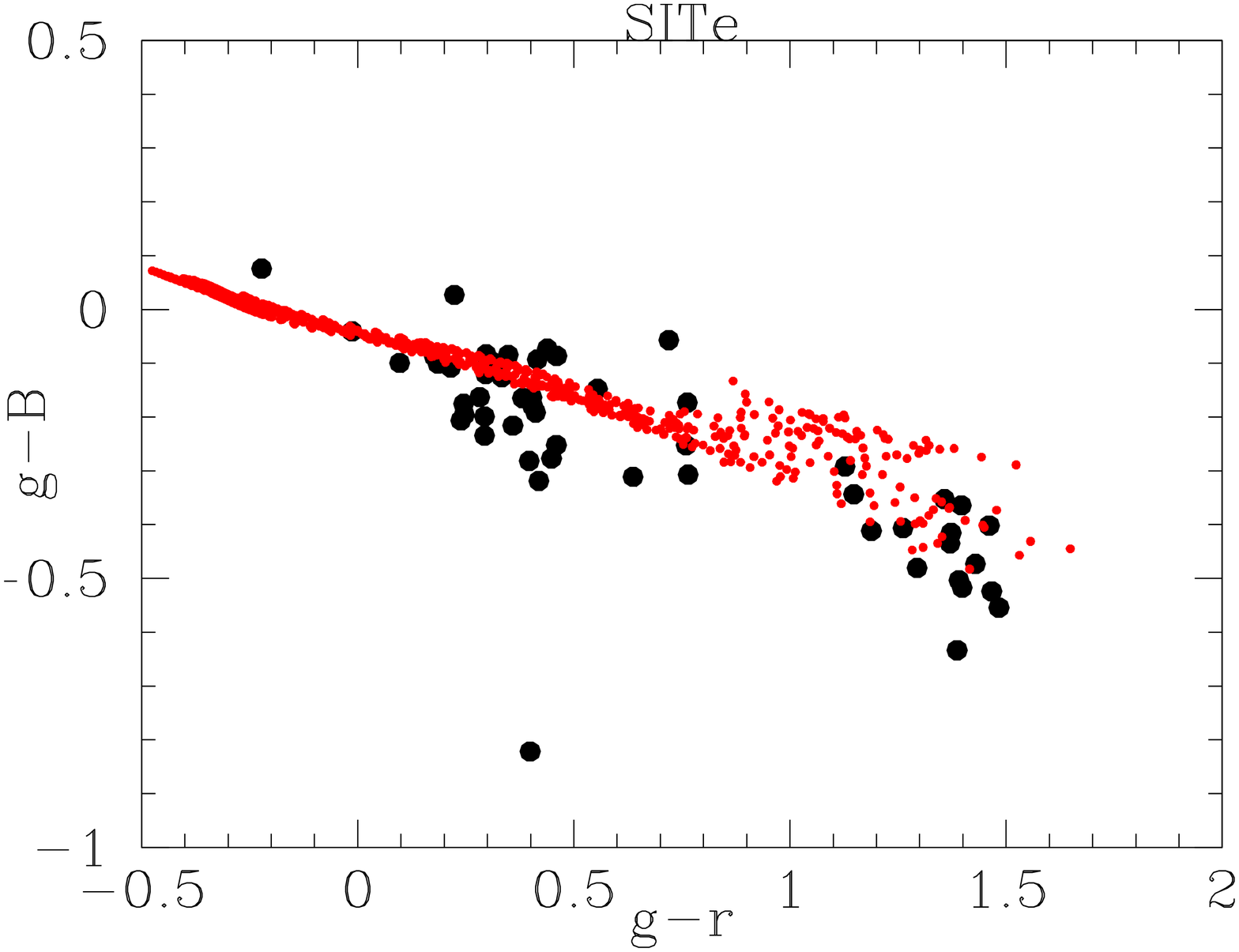}\FigureFile(80mm,60mm){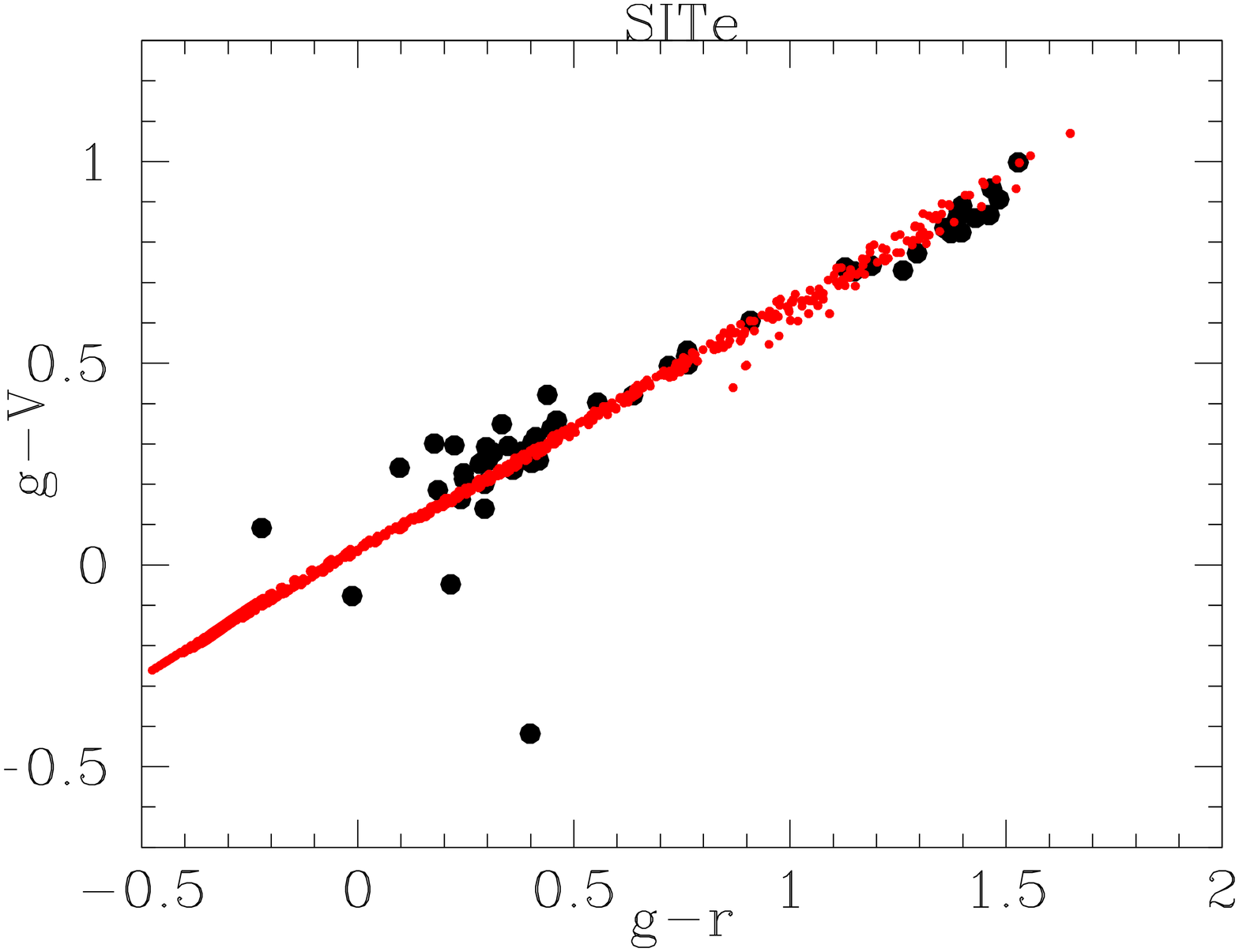}\\
\caption{
SDSS color versus the (SDSS)-(Suprime-Cam) color
of GT-SXDF-catalog data used in SXDS calibration.
The filled red circles represents model colors.
In R-band figures, MIT1 model is overplotted, 
though the difference is indistinguishable.
}
\label{fig:GT0}
\end{figure}

\clearpage

\begin{figure}
\FigureFile(80mm,60mm){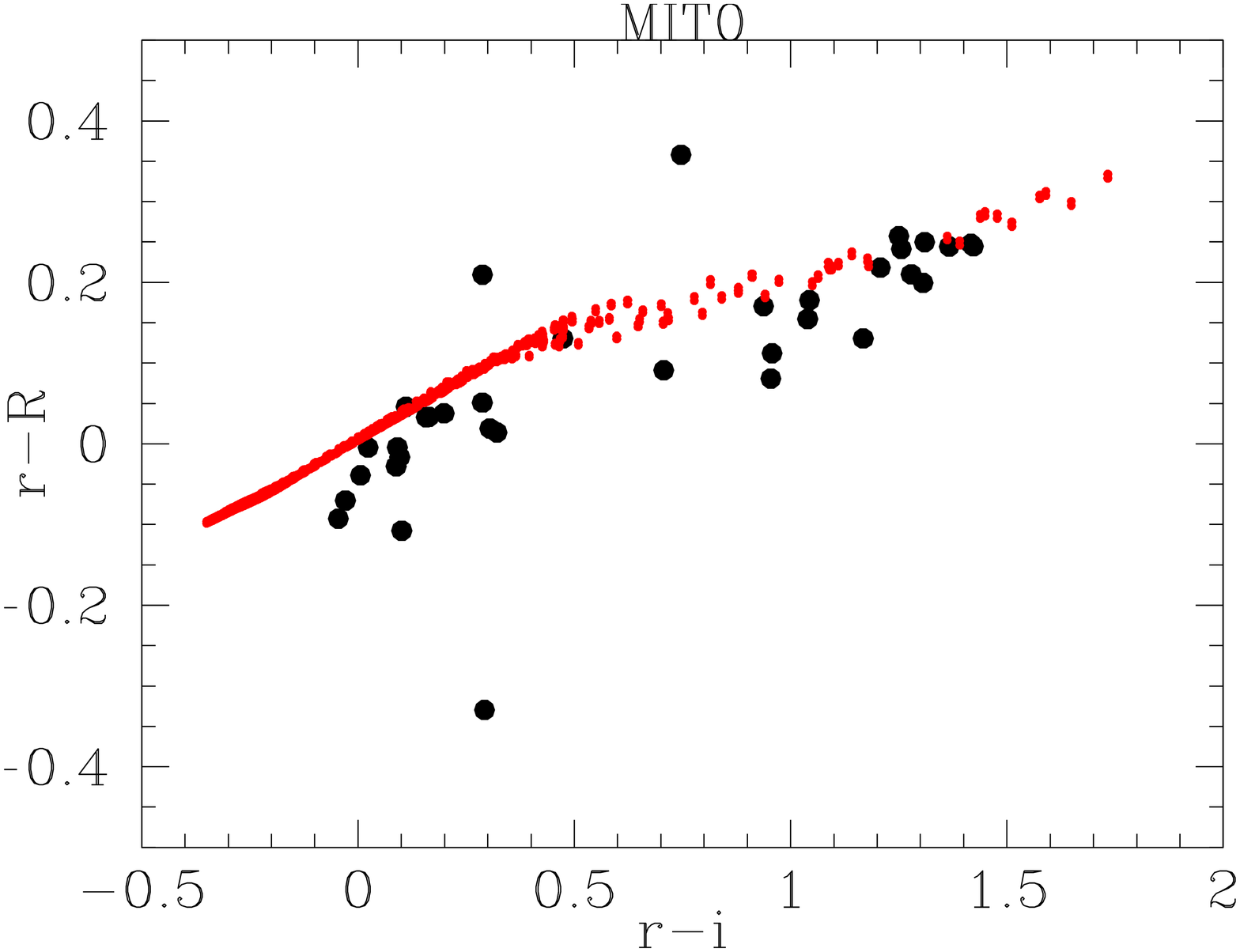}\FigureFile(80mm,60mm){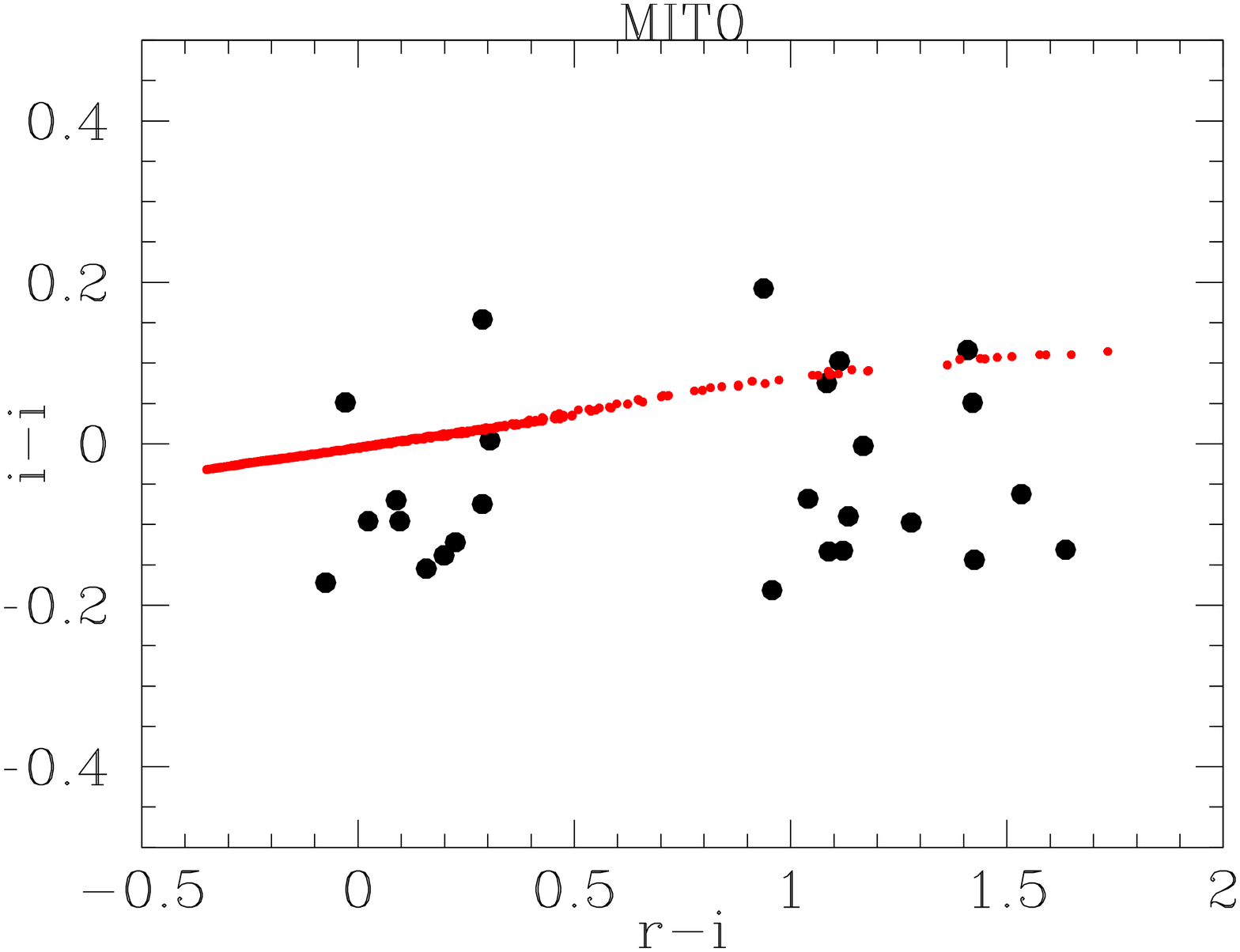}\\
\FigureFile(80mm,60mm){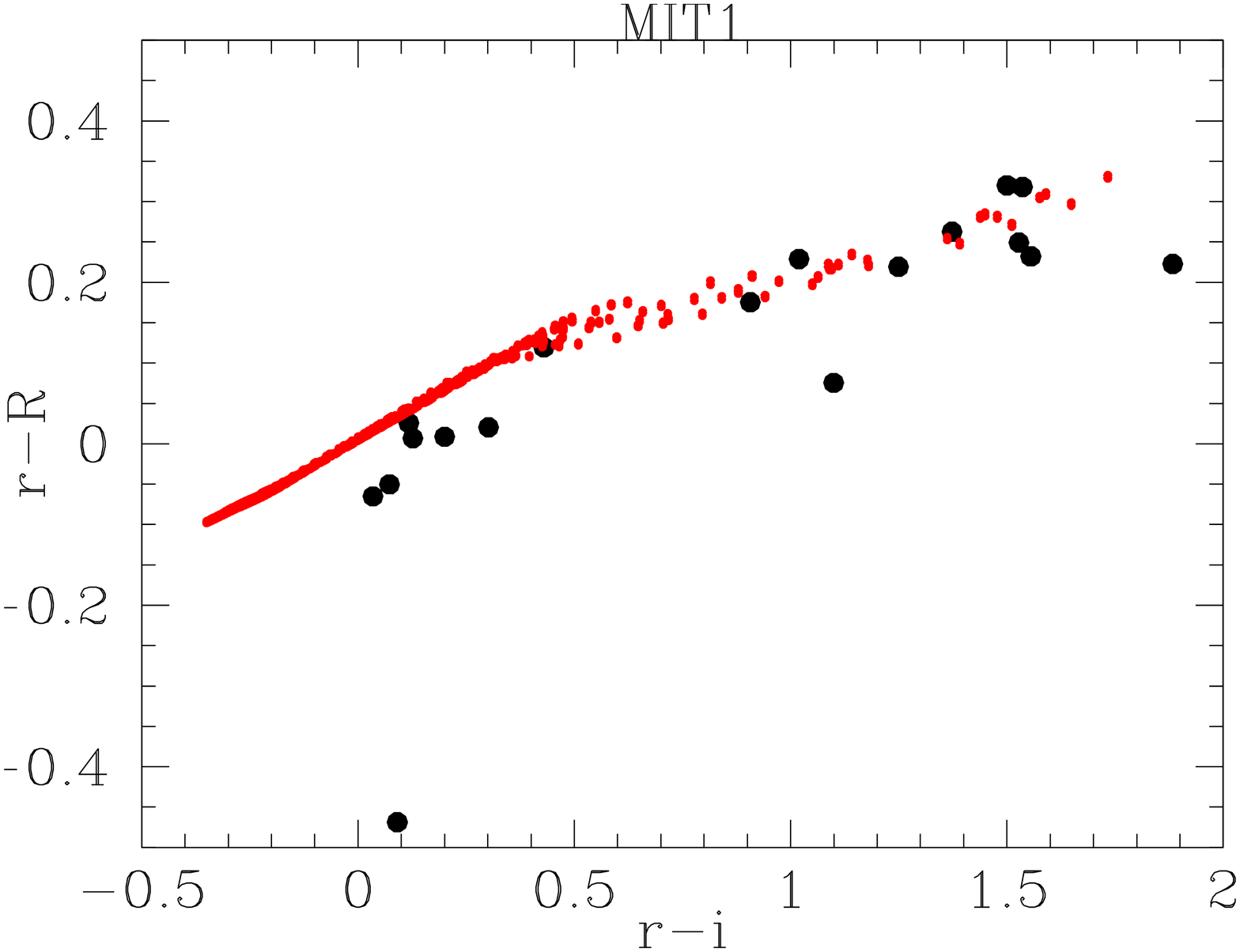}\FigureFile(80mm,60mm){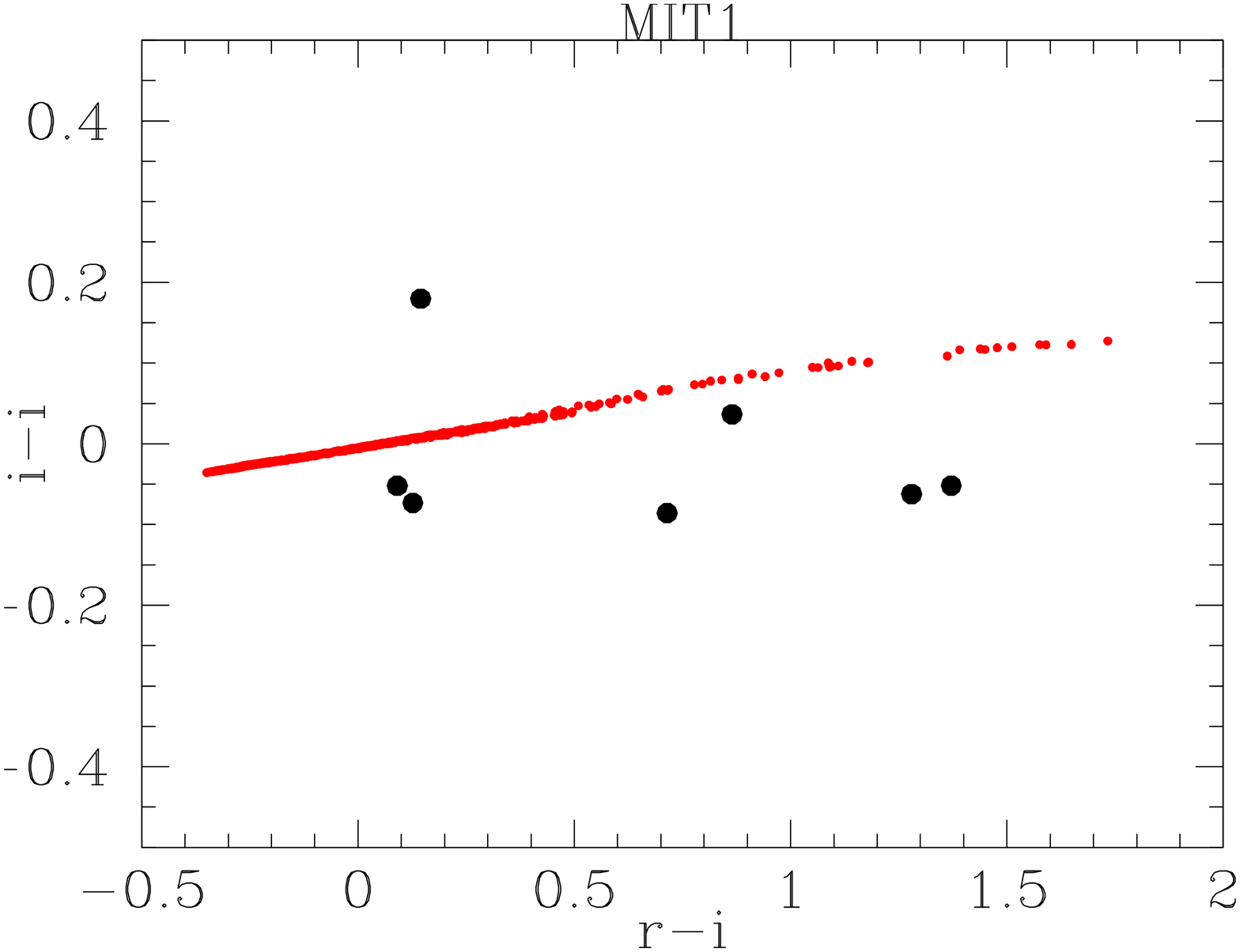}\\
\FigureFile(80mm,60mm){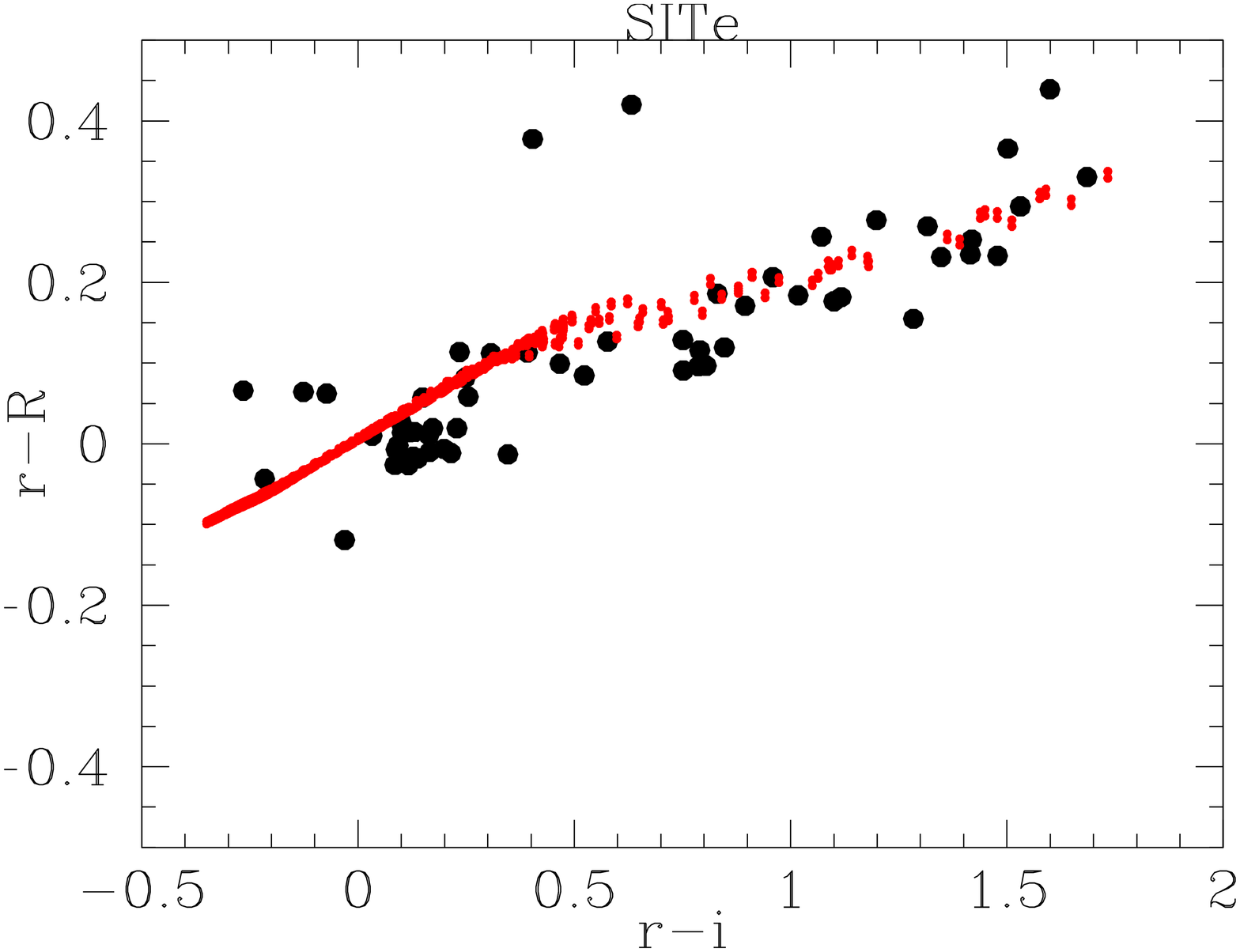}\FigureFile(80mm,60mm){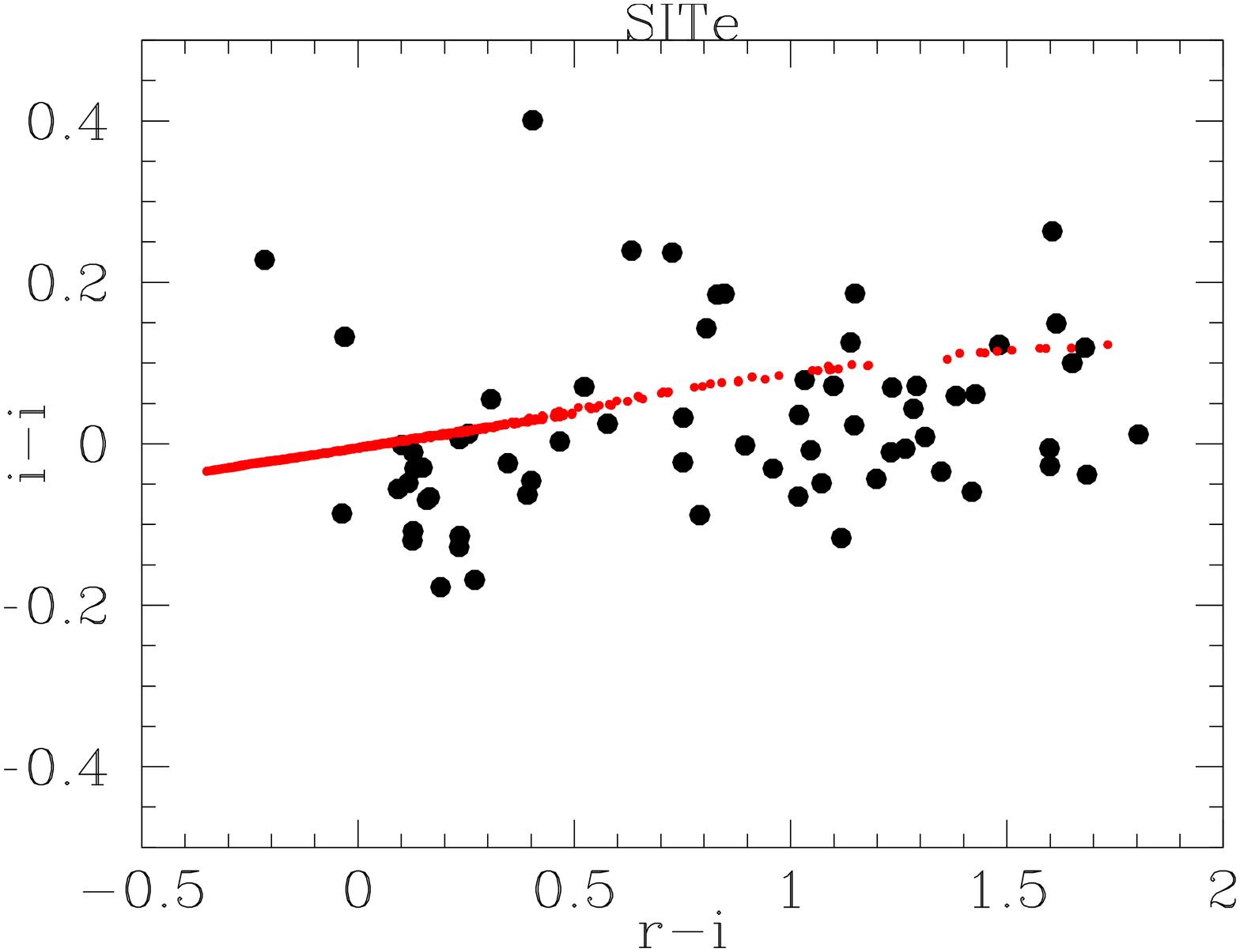}\\
\addtocounter{figure}{-1}
\caption{
Continued...
}
\end{figure}

\clearpage

\begin{figure}
\FigureFile(80mm,60mm){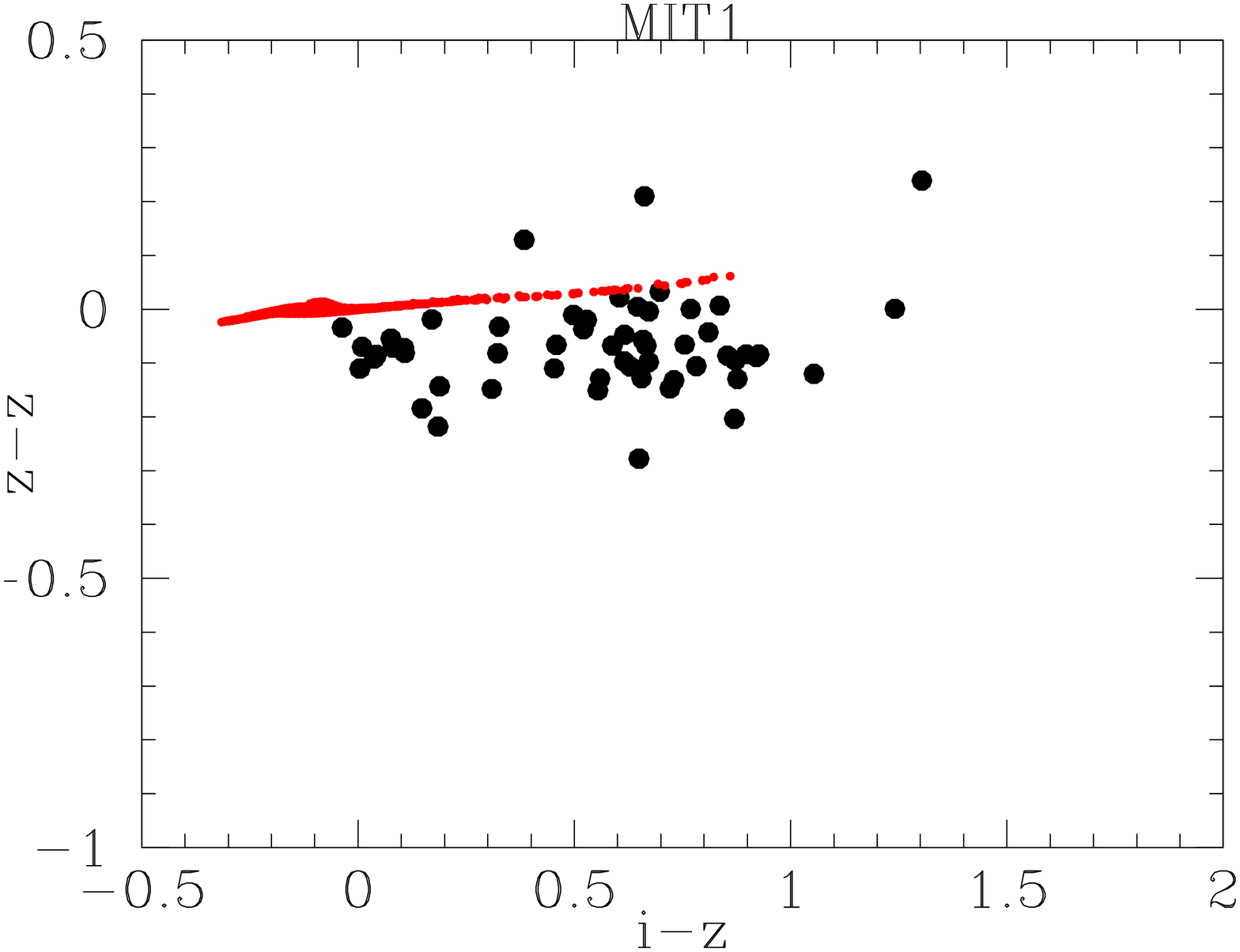}
\addtocounter{figure}{-1}
\caption{
Continued...
}
\end{figure}

\clearpage 

\begin{table}
\begin{tabular}{|c|c|c|}
\hline
field & R.A.(J2000) & Dec(J2000)\\
\hline
SDF    & \timeform{13h24m38.9s}&\timeform{+27D29'25.9''}\\
SXDS-C & \timeform{02h18m00.00s}&\timeform{-05D00'00.0''}\\
SXDS-N & \timeform{02h18m00.00s}&\timeform{-04D35'00.0''}\\
SXDS-S & \timeform{02h18m00.00s}&\timeform{-05D25'00.0''}\\
SXDS-E & \timeform{02h19m47.07s}&\timeform{-05D00'00.0''}\\
SXDS-W & \timeform{02h16m12.93s}&\timeform{-05D00'00.0''}\\
\hline
\end{tabular}
\caption{Center Position of SDF and SXDS (from \cite{SDF,SXDS2})}
\label{tab:field}
\end{table}

\begin{table}
\begin{tabular}{|c|c|c|c|c|c|c|c|c|c|c|}
\hline
SDSS-Suprime& SDSS color& range &$c_0$&$c_1$&$c_2$&$c_3$&$c_4$&$c_5$&$c_6$&$c_7$\\
\hline
$g-B$ & $g-r$ & -0.4$<g-r<$0.7 & -0.037 & -0.160 & 0.009 & -0.307 & 0.246 &--- & --- & --- \\
\hline
$g-V$ & $g-r$ & -0.4$<g-r<$0.8 & 0.038 & 0.565 & -0.024 & 0.260 & -0.218 & --- & --- & --- \\
\hline
$r-R$ & $r-i$ & -0.4$<r-i<$0.6 & 0.006 & 0.317 & -0.065 & -0.157 & 1.667 & -1.179 & -8.202 & 9.857 \\
\hline
$i-i$ & $r-i$ & -0.4$<r-i<$0.8 & -0.005 & 0.087 & 0.006 & 0.011 & 0.020 & -0.015 &  --- & --- \\
\hline
$z-z$ & $i-z$ & 0$<i-z<$0.9 & -0.001 & 0.092 & -0.116 & 0.109 &  --- & --- & --- & ---   \\
\hline
\end{tabular}
\caption{The coefficients of best-fit color conversion polynomials}
\label{tab:coeff0}
\end{table}

\begin{table}
\begin{tabular}{|c|c|c|c|c|c|}
\hline
field & B &V & R& i & z\\ 
\hline
SDF & 1.05 & 1.25 & 1.33 & 1.36 & 1.32 \\
SXDS-C & 1.13 & 1.31 & 1.27 & 1.37 & 1.44 \\
SXDS-N & 1.24 & 1.26 & 1.35 & 1.41 & 1.42 \\
SXDS-S & 1.19 & 1.32 & 1.33 & 1.37 & 1.52 \\
SXDS-E & 1.25 & 1.24 & 1.42 & 1.61 & 1.38 \\
SXDS-W & 1.18 & 1.50 & 1.59 & 1.50 & 1.46 \\
\hline
\end{tabular}
\caption{The exposure time weighted mean of airmass}
\label{tab:airmass}
\end{table}

\begin{table}
\begin{tabular}{|c|c|c|c|c|c|}
\hline
band&$k_1$&$k_2$&color\\
\hline
B & 0.188 & -0.016 & g-r \\
V & 0.110 & -0.001 & g-r \\
R & 0.070 & -0.000 & r-i \\
i & 0.068 &  0.000 & r-i \\
z & 0.102 &  0.008 & i-z \\
\hline
\end{tabular}
\caption{The atmospheric extinction coefficients}
\label{tab:extcoeff}
\end{table}

\begin{table}
\begin{tabular}{|c|c|c|c|c|c|}
\hline
field & band & N & median & $\sigma$ & $\sigma_{\rm SDSS}$\\
\hline
SDF & B & 55 & {\bf -0.14} & 0.07 & 0.03 \\
SDF & V & 58 & -0.04 & 0.05 & 0.03 \\
SDF & R & 80 & {\bf -0.06} & 0.04 & 0.03 \\
SDF & i & 85 & {\bf -0.11} & 0.07 & 0.03 \\
SDF & z & 167 & -0.06 & 0.07 & 0.05 \\
\hline
SXDS-C & B & 102 & 0.00 & 0.08 & 0.05 \\
SXDS-C & V & 106 & -0.02 & 0.04 & 0.05 \\
SXDS-C & R & 117 & -0.04 & 0.05 & 0.04 \\
SXDS-C & i & 95 & {\bf -0.11} & 0.06 & 0.04 \\
SXDS-C & z & 152 & {\bf -0.13} & 0.08 & 0.06 \\
\hline
SXDS-N & B & 99 & 0.02 & 0.07 & 0.05 \\
SXDS-N & V & 101 & -0.02 & 0.06 & 0.05 \\
SXDS-N & R & 113 & -0.05 & 0.05 & 0.04 \\
SXDS-N & i & 89 & {\bf -0.11} & 0.06 & 0.04 \\
SXDS-N & z & 146 & {\bf -0.14} & 0.08 & 0.07 \\
\hline
SXDS-S & B & 65 & -0.02 & 0.08 & 0.04 \\
SXDS-S & V & 67 & -0.02 & 0.04 & 0.04 \\
SXDS-S & R & 81 & {\bf -0.05} & 0.04 & 0.03 \\
SXDS-S & i & 70 & {\bf -0.14} & 0.08 & 0.03 \\
SXDS-S & z & 121 & {\bf -0.11} & 0.07 & 0.05 \\
\hline
SXDS-E & B & 98 & 0.00 & 0.08 & 0.05 \\
SXDS-E & V & 103 & 0.00 & 0.05 & 0.05 \\
SXDS-E & R & 122 & -0.04 & 0.06 & 0.04 \\
SXDS-E & i & 105 & {\bf -0.13} & 0.07 & 0.04 \\
SXDS-E & z & 127 & {\bf -0.15} & 0.11 & 0.07 \\
\hline
SXDS-W & B & 99 & -0.01 & 0.09 & 0.05 \\
SXDS-W & V & 102 & -0.03 & 0.06 & 0.05 \\
SXDS-W & R & 116 & -0.05 & 0.05 & 0.04 \\
SXDS-W & i & 103 & {\bf -0.08} & 0.05 & 0.04 \\
SXDS-W & z & 148 & {\bf -0.14} & 0.09 & 0.07 \\
\hline
\end{tabular}
\caption{
The ZP difference between the catalog and the
estimated value from SDSS. Difference larger than $\sigma$ is shown 
as bold.}
\label{tab:ZPdiff}
\end{table}

\begin{table}
\begin{tabular}{|c|c|c|c|l|}
\hline
star & DR8 objID & i$_{\rm cat}$-i$_{\rm syn}$ & z$_{\rm cat}$-z$_{\rm syn}$ & SED name \\
\hline
HZ21  & 1237665330925535274 &0.034$\pm$0.016 & 0.035$\pm$0.020 & hz21\_001,hz21\_002 \\
      &                     &-0.002$\pm$0.016 & 0.018$\pm$0.020  & hz21\_003 \\
\hline
HZ44  & 1237664672717865012 &-0.012$\pm$0.000 & 0.074$\pm$0.022  & hz44\_001, hz44\_002\\ 
      &                     &-0.014$\pm$0.000 & 0.044$\pm$0.022  & hz44\_003\\ 
\hline
GD153 & 1237667735049470025 &0.141$\pm$0.001 & 0.004$\pm$0.021 
& gd153\_mod\_002 \\
      &                     &0.143$\pm$0.001 & 0.006$\pm$0.021 & gd153\_mod\_003 \\
      &                     &0.136$\pm$0.001 & -0.001$\pm$0.021 & gd153\_mod\_004 \\
      &                     &0.132$\pm$0.001 & -0.005$\pm$0.021 & gd153\_mod\_005 \\
\hline
P177D & 1237655464320237608 &0.005$\pm$0.001 & 0.017$\pm$0.022 & p177d\_001 \\
      &                     &0.007$\pm$0.001 & 0.027$\pm$0.022 & p177d\_stisnic\_001 \\
      &                     &0.002$\pm$0.001 & 0.023$\pm$0.022 & p177d\_stisnic\_002 \\
\hline
P330E & 1237662505371303976 &0.008$\pm$0.000 & -0.002$\pm$0.015 & p330e\_001\\
      &                     &0.019$\pm$0.000 & -0.031$\pm$0.015 &p330e\_stisnic\_002 \\
      &                     &0.015$\pm$0.000 & -0.026$\pm$0.015 &p330e\_stisnic\_002 \\
\hline
\end{tabular}
\caption{Difference between 
SDSS DR8 catalog magnitude and the synthetic magnitude of 
standard stars used for SDF}
\label{SDFstandards}
\end{table}

\end{document}